\documentclass[11pt, a4paper]{article}
\usepackage{amsmath, amssymb, amsthm}
\usepackage{dsfont}   
\DeclareFontFamily{U}{BOONDOX-calo}{\skewchar\font=45 }
\DeclareFontShape{U}{BOONDOX-calo}{m}{n}{<-> s*[1.05] BOONDOX-r-calo}{}
\DeclareMathAlphabet{\mathbcal}{U}{BOONDOX-calo}{m}{n}
\usepackage{graphicx}
\graphicspath{{figures/}}
\usepackage{booktabs}
\usepackage{array}
\usepackage{rotating}
\newcolumntype{R}[1]{>{\raggedright\arraybackslash}p{#1}}
\usepackage{tabularx}
\usepackage{longtable}
\usepackage[margin=1in]{geometry}
\usepackage{hyperref}
\hypersetup{bookmarksdepth=1}
\usepackage{bm}
\usepackage{empheq}
\usepackage{float}
\usepackage[section]{placeins}  
\usepackage[font=small,labelfont=bf]{subcaption}


\makeatletter
\renewcommand\paragraph{\@startsection{paragraph}{4}{\z@}%
  {3.25ex \@plus1ex \@minus.2ex}%
  {-1em}%
  {\normalfont\normalsize\itshape}}
\makeatother

\newcommand{\Prob}{\mathbb{P}}
\newcommand{\dd}{\mathrm{d}}

\title{\textbf{\Large Boom-Busty Marketron as a Model of Market Dynamics: \\
How Investor-Market Reflexivity Amplifies Non-Equilibrium}}

\author{Igor Halperin\thanks{Opinions expressed here are author's only,
and not of his employer. All
calculations, numerical analysis, and manuscript
preparation were performed by Claude Code, successively with
Opus 4.8, Opus 5, Fable 5.1 and Opus 5.5, working as an AI assistant under author's supervision, and an AI review of the manuscript was provided by GPT-6 Astra. I would like to thank Miquel Noguer i Alonso, Andres Bagnasco,
Ernest Baver, Eric Berger, Paul Burchard,
Alejandro Rodriguez Dominguez,
Andrey Itkin, Yinsen Miao, and Alexander Vigodner
for very helpful discussions and comments on the manuscript. Special thanks to
Waron Mann and David Audi for comments about potential connections of OMD
findings with the Soros reflexivity theory. All remaining errors are my own.
}}

\date{\today}

\begin{document}

\emergencystretch=2em

\maketitle

\begin{center}
Email: ighalp@gmail.com
\end{center}

\begin{abstract}
We introduce the Boom-Busty Marketron, a model of a multi-asset market in which
the demand of a single collective investor moves prices and itself depends on
market performance. This creates an investor-market reflexive
loop in the spirit of Soros's theory. The market state is stationary without being
at equilibrium: a non-equilibrium steady state (NESS), with broken detailed balance and a
probability current circulating between mispricing, demand and volatility. While
stochastic volatility with the leverage effect already breaks detailed balance and
makes the market dynamics irreversible, the investor-market loop pushes the market
further away from equilibrium. Calibrated to the S\&P 500 from 1996 to 2024 by simulated
moments, the model reproduces the stylized facts of returns together with
quantitative metrics of the market, including cross-sectional ones, and its return and
cycle statistics stay stationary for a century and a half on the 1996--2010 inputs. Its central
emergent behavior is a boom-bust cycle, whose frequency
meets the S\&P's within one-sigma ranges. The model also produces a default rate of the order of
that of investment-grade issuers, prices inside Fisher Black's factor of two
of value, the dumb-money effect of Frazzini and Lamont, and a measurable arrow of
time. Standard long-short factors built from its cross section rotate through the
cycle, and so do its industry groups along the defensive axis.
\end{abstract}

\setcounter{tocdepth}{1}
\tableofcontents

\section{Introduction}
\label{sec:intro}

As is well known to academics and practitioners alike, real financial markets
are \textbf{reflexive} and \textbf{excitable}. While admittedly a few would
operate in these literal terms, most would immediately agree when presented
qualitative explanations of both terms. A market is \emph{reflexive} when the
perceptions, actions and beliefs of its participants shape the market, and the
market's own changes shape those perceptions and actions in turn. The concept of
\emph{excitability} was introduced in 1946 by Wiener and Rosenblueth, who studied
the conduction of impulses in cardiac muscle \cite{wiener1946}. An excitable
system rests in a stable state where small perturbations decay into small
fluctuations, and responds to any perturbation past a critical size with a sudden
large excursion before returning to rest. Excitable systems
have since found wide application, to gene regulatory networks, chemical
reactions, lasers and semiconductors, and are now one of the working principles of
theoretical neuroscience \cite{lindner2004,fhnreview2024}. At this stage, we offer
market excitability as a qualitative analogy, and the four phases a boom-bust
cycle is read in are the states an excitable system passes through. The calm
market is the rest state, absorbing what reaches it. The boom is the excursion,
begun by a disturbance but then sustained by the reflexive loop rather than by
whatever started it, until the money committed, which is the slow variable of the
system, catches up with the price and stops supporting it. The bust is the return
stroke, the same rule that bought now selling, and nothing then holds the market
back from the next one: its busts cluster rather than wait for a recovery.

While we return to market excitability below,\footnote{Section~\ref{sec:excitability}
shows that the model is self-excitable, crossing its threshold without an
external stimulus.} we begin with market reflexivity,
arguably the more familiar of the two. The same loop has been noticed repeatedly
under other names, as Merton's self-fulfilling prophecy \cite{merton1948},
Popper's Oedipus effect \cite{popper1957}, Buck's reflexive prediction
\cite{buck1963} and Morgenstern's back-coupling \cite{morgenstern1972}.
A particular view of market reflexivity is widely known in the form suggested by
George Soros \cite{soros1987,soros2013}: participants act on a fallible
understanding of a market they are inside, their actions alter the thing they were trying to price,
and the two directions of influence interfere, so that the market tracks a moving
target instead of converging on a fixed one. A later literature on performativity
argues that economic theory does not merely describe markets but shapes them
\cite{mackenzie2006}. For all that recognition,
both orthodox and heterodox financial economics have found market reflexivity
hard to capture. A recent account of how models fail under the complexities of
reflexivity and of multi-directional causal loops is given by Polakow, Gebbie and
Flint \cite{polakow2026}, and we follow their account of the historical
development in our next few paragraphs below.

The orthodox account is built on equilibrium. Prices aggregate the available
information, no forecast beats them systematically, what is left over is noise,
and the market has a resting state to be perturbed around with a price that
responds elastically to the money arriving. Causation runs one way, from the
market to the participants who read it, and the market is priced off a single
representative investor whose marginal utility values every asset, a construction
exact when investors are identical and otherwise needing conditions strong enough
for them to aggregate as though they were. As Cochrane explains
\cite{cochrane2005}, perfectly rational investors with quadratic utility imply the
celebrated CAPM, which performs badly in its original form, in the cross section
and the time series alike.

Historically, the demonstration that reflexivity need not matter turned into the
assumption that it does not. In 1954, the year Arrow and Debreu proved that a
general equilibrium exists \cite{arrowdebreu1954}, two influential papers answered
the reflexivity predictions of Merton \cite{merton1948}. Grunberg and Modigliani
\cite{grunberg1954} and Simon \cite{simon1954} showed that it was \emph{possible}
for reflexivity not to spoil the accuracy of an economic prediction, which is what
\emph{can} happen rather than what \emph{does}. Muth's rational expectations of
1961 then \emph{assumed} it \cite{muth1961}, and the possibility hardened into a
premise with nothing new established in between. The orthodox finance theorists
thus never modeled market reflexivity, but \emph{assumed it away}.

A physicist would recognize the manoeuvre. Richard Feynman, in the lecture for
his 1965 Nobel Prize for quantum electrodynamics (QED), remarked that ``the
renormalization theory is simply a way to sweep the difficulties of the
divergences of electrodynamics under the rug'' \cite{feynman1965}, of a
construction that predicts to twelve figures. Reflexivity went under a much worse
rug, being ignored almost entirely, on considerably weaker grounds, and unlike the
divergences it was never taken back out. Soros put the
objection at its sharpest, calling the postulate of a single optimum view of the
future absurd and kept only so that economics could imitate Newtonian
physics.\footnote{``Economics ended up with the theory of rational expectations,
which maintains that there is a single optimum view of the future, that which
corresponds to it, and eventually all the market participants will converge around
that view. This postulate is absurd, but it is needed in order to allow economic
theory to model itself on Newtonian Physics.'' G.~Soros, \emph{The Soros Lectures
at the Central European University}, PublicAffairs, New York (2010).} The
construction also sits badly beside plain facts. A market that no forecast can
beat coexists with a professional trading industry paid, persistently and by
informed clients, to beat it, and the models describe benign periods moderately
well and crises badly, which is the opposite of the order of importance.

The heterodox response relaxes the assumptions of the orthodox school one at a
time. Agents become
boundedly rational, populations heterogeneous, trend followers are set against
fundamentalists, and expectations feed back on the prices they are formed from
\cite{hommes2013}. These models buy a great deal, producing bubbles and crashes
the orthodox ones cannot. What they usually do not buy is an escape from the
causal architecture, as remarked by Lawson and cited in \cite{polakow2026},
because the loop is generally cut somewhere to keep the model tractable and a
one-way chain is restored at another level. Lawson's complaint is
aimed at the whole family: mainstream economics is on his reading ``awash with
assumptions of perfect foresight or rational expectations, of efficient markets,
and of market equilibrium'', all premised on the predictability of future outcomes
and all treating an open system as though it were closed
\cite{lawson1997,lawson2013}.

In a reflexive market, as Polakow et al. point out, the direction of causation is
ambiguous, so that inference drawn on the assumption of a stable direction is
misapplied rather than merely imprecise, and reflexivity destabilizes calibration
itself because what is being fitted is endogenously tied to its own
representation. Their verdict on the field this produces is blunt: empirical
finance ``is categorically a pathological science where it is able to act as one
at all (i.e.\ where not sufficiently reflexive to matter), and patently not a
science elsewhere, for example at times of crises'' \cite{polakow2026}. They also
conclude that causal loops have to be imposed \emph{top-down}, within a specific
\textbf{structural model}. Causal modeling in finance after Pearl
\cite{pearl2009}, which assumes controllable interventions and one-way causation,
is frail and potentially doomed to fail, because causal loops are
multi-directional and no controllable intervention on a real market is
possible.

The model built in this paper, the Boom-Busty Marketron, or BBM for short, is a
structural model of exactly that kind. It puts the reflexive loop between the
investor and the market at the center of the construction rather than at its
edge, and the investor's demand drives the whole dynamics. Nothing outside the system is
ever needed,\footnote{Except the Gaussian white noises. The dynamics is of the
Langevin kind, so the market is closed in the sense that no external factor such
as news is ever supplied to it, and no money either apart from a small inflow of
new money at about $0.6$ percent of the market a year, and open in the sense
that it is driven by thermal noise. See below for more details.} so there are no
macroeconomic drivers, no exogenous cycles or other enforced schedules, and the
loop is closed in both directions. The construction also avoids the
self-referencing framework that has dominated much of the literature since the
self-fulfilling and self-defeating prophecies of Merton, Simon and others
\cite{merton1948,simon1954}, in which a system's description forms part of what it
describes and the causal loops it generates are recurrent and resist analysis.
The BBM model instead treats the market as a two-sided system: a single
collective investor, keeping the ten largest names and the rest of the index in
two sleeves of its portfolio, and a capital market of many stocks, each with its own state variables and the two coupled through their differential equations.
Causal loops are then read off the structure of those equations rather than off
the description.

Closing the investor-market loop needs no historical data on fundamentals. Value is set by a process internal to the model, whose parameters are calibrated
against market data alone, so no earnings, book values or analyst estimates are
read in at any point. This makes it possible to run a market for
centuries and ask what its cycles look like. The construction builds on the
single-asset Marketron model of Halperin and Itkin \cite{marketron2025}, which already has the essential architecture, a reflexive loop in which an
investor's money moves the market and the market's performance moves the money.
The BBM model adds four components: a multi-asset market, a fundamental value
process, stochastic volatility, and a rational mean-variance investor.

Among the heterogeneous-agent models, the closest relative of this construction is the extended Chiarella model of Majewski, Ciliberti and Bouchaud, a market of trend followers and fundamentalists \cite{majewski2020}. Both models write the log price as a drifting fundamental plus a mispricing, and both move the price by impact rather than by a clearing condition. Both set a trend term, an exponential average of past moves, against a value term that turns nonlinear only far from value. In both the calibrated equity market lies on the stable side of its noise-free dynamics, so its swings around value are driven by noise rather than by a limit cycle, and both keep the price within Black's factor of two of value. The differences lie in what closes the loop. Three groups of traders with fixed weights and no budget become here one investor with wealth, a bond holding, an inflow and an equity cap. Its demand is a state that moves the price on a potential with a barrier, beyond which firms fail. Where theirs has one asset, the BBM simulates an index of $500$ firms, with cross-sectional events a single-asset model cannot show. The two
models also part on the trend effect. Theirs is positive and turns back for large
signals, while on the BBM's index it is a reversal due entirely to the value
effect. Appendix~\ref{sec:comparison} sets the two side by side. An earlier Langevin model, that of Bouchaud and Cont \cite{bouchaud1998}, makes crashes rare activated escapes driven by the feedback of price moves on demand. In the BBM the barrier of the potential governs the failures of firms rather than the timing of busts (Section~\ref{sec:boomsdefaults}).

A bi-directional causal loop becomes a forward-backward loop in our framework,
each component being controlled by its own parameters.
The forward pass is the price impact of the investor's demand, acting through its latent demand memory; its executed trades move its holdings and cash and reach the
price only through later demand.
The backward pass is the decision. The market's performance enters the investor's
view of what it will return next, through a momentum term of weight $\theta$ that
depends on the recent path and a value term of weight $c$ that depends on the gap between
price and value through the marketron potential, and the allocation implied by
that view sets the flow of the following step.

Because the loop is specified rather than inferred, the reflexive market is at
once modellable and measurable. The most important feature of this reflexive market is that it has
{\bf no equilibrium} in the statistical sense.\footnote{There are three different notions of equilibrium. An
economic equilibrium is a set of prices at which the participants' optimal demands
clear, with absence of arbitrage and informational efficiency as related but
distinct conditions. A deterministic equilibrium is a rest point of the noise-free
dynamics. A statistical equilibrium is a stationary law satisfying detailed
balance, under which the stationary process looks the same run backwards. We do not address the first two and focus only on the third.} Its
stationary distribution is a non-equilibrium steady state, stationary in law but with a probability current that never vanishes, and its entropy
production stays positive, giving it a measurable arrow of time. While
irreversibility also arises in stochastic volatility models with a leverage effect,
the investor-market reflexive loop amplifies it at the calibrated parameters by up
to two orders of magnitude: its entropy production, estimated in a linear approximation, rises forty- to a hundredfold against a demand that does not respond to prices, and four- to sixfold
against a market without impact (Section~\ref{sec:reflexverdict}).

The market produced by our model is {\bf noise-driven} and {\bf excitable}, in
different senses at its two levels. A single name has the
geometry: a rest state, a threshold that is a saddle's stable manifold, an
all-or-none excursion past it, a short refractory period of a few months
after its own fall, and far past the threshold a region it does not come back from. The index has the behavior without the geometry, never approaching that
threshold, its booms and busts being large noise-driven excursions about one rest
point while a probability current circulates in the plane of mispricing against demand.
What returns the index to
rest is its own accounting, the membership rule retiring the names that failed
and seating replacements, so the index's excitability is behavioral rather than
the dynamics of a system pushed past a critical point.

\begin{figure}[htbp]
\centering
\includegraphics[width=0.42\textwidth]{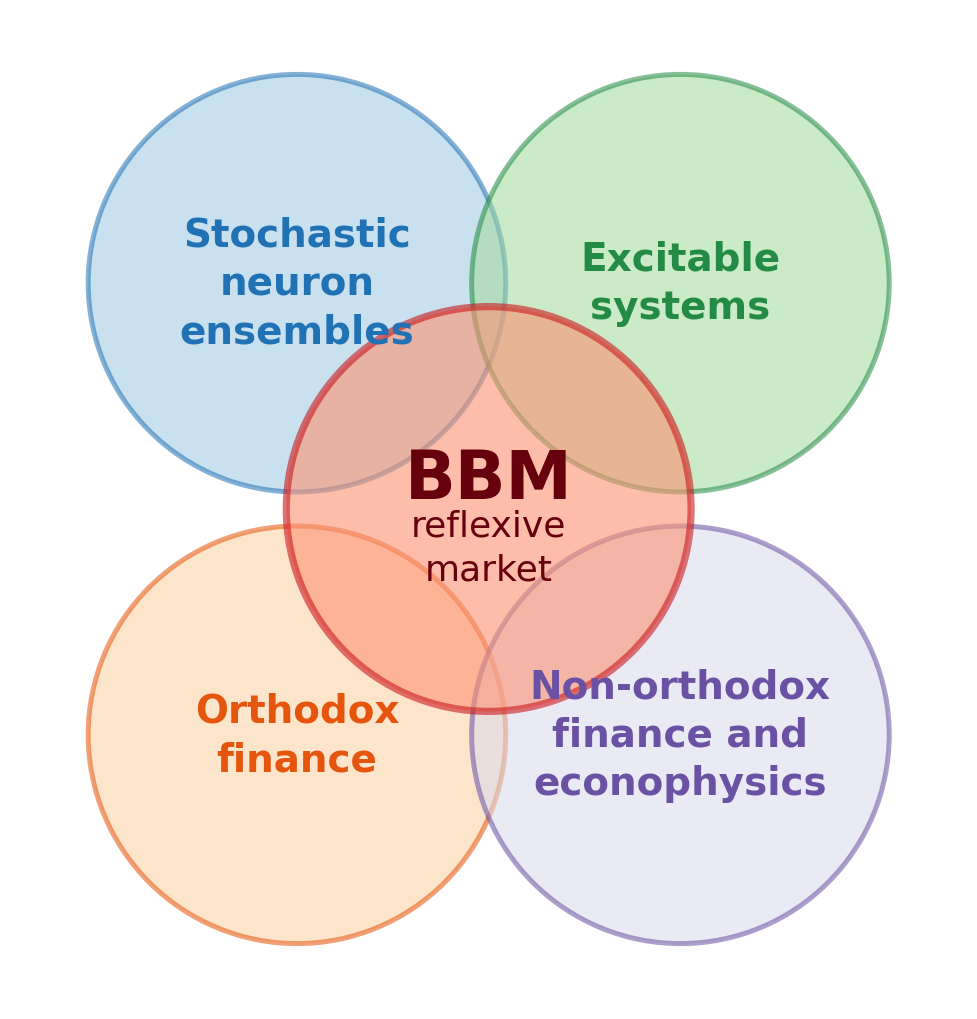}
\caption{Relation of the BBM to the financial, neuroscience and physics
literatures.}
\label{fig:bbmvenn}
\end{figure}

We read the BBM model in two ways, as a financial model and as a physics model
of a multi-particle system of noise-driven marketrons. The Venn diagram of
Figure~\ref{fig:bbmvenn} places it among the classical, non-orthodox and
econophysics finance literatures and the neuroscience and physics ones.

As a financial model, the BBM takes an explicit stand on several questions where
it differs from the standard accounts, set out in Section~\ref{sec:birdseye}. The
first is a chicken-and-egg question the literature has not settled: is the US
equity market stationary, or does the data imply regime changes and structural
breaks? One group of researchers finds the equity premium and the coefficients of
return prediction models unstable over time, by Bayesian estimation of the break
dates and by sequential break tests on predictive regressions respectively
\cite{pastor2001,paye2006}. Another holds that fixed structural parameters can
produce very different local behavior through a persistent state, so that the
same data asks for no break at all \cite{campbell1999}. A third traces part of the
apparent instability to the variables the market is described in rather than to
the process itself, though it still admits shifts in the steady state
\cite{lettau2008}. Every one of these verdicts rests on the model used to reach
it, and the choice of model rests on a prior view of whether breaks occur,
hence the chicken and the egg. We side with the second group, on grounds supplied
by the model itself.

The claim is that in the right variables and with the right dynamics the
process is stationary at the horizon of a window and of a cycle, and that the
calibrated model is consistent
with the absence of a break. One parameter vector describes both calibration
windows at once, and its state does not drift once the market has settled. Counted in observables matched, the vector fine-tuned to the earlier window describes the later one nearly as well as that window's own fine-tune does, while the later window's fine-tune describes the earlier one markedly worse, so the two periods differ in their moments without needing different parameters.

Stationary as it is, the BBM produces booms and busts like the actual market's, without any macroeconomic forcing. Falls of twenty percent in the index arrive $30$ times a century in both windows against $27$ and $14$ observed, the one-sigma ranges of model and market meeting in both, and $26.4 \pm 1.9$ a century over the settled market's long run. The S\&P's own count from January 1975 to August 2026, $51.6$ years and longer than the 1996--2024 data the model is calibrated on, is eight bear markets by the standard dating rule, $15.5 \pm 5.5$ a century with a Poisson error. A history of the settled market of the same length counts $24.8 \pm 6.4$, the spread over sixteen such histories, a range that meets the S\&P's though its center sits above it.

A generator of equity histories that needs no macroeconomic input and does not run out of crises as it ages is useful in its own right, above all for simulating long-horizon equity portfolios, where the length of the history that can be generated is the binding constraint.

As a physics model, the BBM reaches stationarity by a route we have not seen
proposed in the financial literature: excitable dynamics returning to a
non-equilibrium steady state, a boom and the bust that ends it being one large
excursion. A single firm has a threshold in its own stochastic differential
equations, but a firm that crosses it in earnest does not come back, so it is a
threshold unit with an absorbing far side rather than a spiking one. What returns
the index to rest is not in those equations at all but in its accounting: a name
whose capitalization has fallen far enough behind its own recent high is retired,
and an entrant is seated at a size drawn relative to the market's current median.
The analogy with an excitable neural system is therefore behavioral rather than
structural, the index passing the tests by which such systems are identified
without sharing the mechanism behind a spike. The closer structural relative is the growth-with-killing of
Reed and Hughes \cite{reed2002}, where killing proportionally growing units
turns lognormal sizes into a power law.

We note that the BBM is \emph{not} a fundamental model of the market at the level of a limit
order book with tick data, but a long-horizon \emph{phenomenological} model of
market dynamics at the scale of a day and beyond. In physics such constructions
are called effective, or infra-red, theories. They describe mesoscopic scales,
where the details of the microscopic motion have become irrelevant and its net effect is captured by a few parameters. The nonlinear potential of the BBM is of
this kind. Its functional form and its parameters should be derivable from a
high-frequency analysis, which we do not attempt here.

The paper is organized as follows. Section~\ref{sec:birdseye} closes this
introduction with a bird's-eye view of the model.
Section~\ref{sec:reflexivity} takes reflexivity more broadly than Soros's own
statement of it, and sets out the causal ambiguity a closed loop creates for
anyone inferring from data. Section~\ref{sec:marketron} builds the model and
Section~\ref{sec:calibration} fits it to the S\&P 500 in two periods, beside the
standard alternatives on the same data. Section~\ref{sec:rmmresults} begins the
results, giving the model's and the market's together so that agreement can be
told from accommodation, and runs from the cycle and its stages to its geometry,
the market's own crises and the rotation of factors and sectors.
Section~\ref{sec:factorcycle} follows the same cycle through money and firms,
and Section~\ref{sec:excitability} tests the market's excitability
directly, one experiment at a time. Section~\ref{sec:nesssec} measures the
circulating current and the entropy production that show it to be a
non-equilibrium steady state. Section~\ref{sec:discussion} draws the synthesis and records where
the model fails, and Section~\ref{sec:conclusion} concludes. The appendices set the belief function against the
market, compare the model with others, and test for parameter drift.

\subsection{A Bird's-Eye View of the Boom-Busty Marketron}
\label{sec:birdseye}

The BBM model takes a stand on a number of questions where the literature
divides, and they are set out here in turn. Calibrated to the S\&P 500 by
simulated moments, it reproduces the stylized facts of returns and the
quantitative metrics of the market, including cross-sectional ones, and stays
stationary at horizons involving multiple cycles. It holds the default rate to investment-grade
values the market shows, and several quantities no calibration touched land near
their observed values: a price inside Fisher Black's factor of two, the dumb-money effect, a hierarchy of relaxation
rates none of which is a period, the roughness of the log volatility as read through a realized estimate, although no
fractional driver appears anywhere in the construction, and a measurable arrow of
time, although a weaker one than the market's. They are by-products of the fit rather than achievements of it.

\emph{Stationarity.} The data are consistent with a stationary market, though
twenty-nine years cannot prove one. A single parameter vector, fitted to the
1996--2010 and 2011--2024 halves of the S\&P sample at once, describes both. Run on
the first half's inputs, the simulated market keeps the same return and cycle
statistics for a century and a half once it has settled. Run on the second half's
inputs, the mispricing and the share of the ten largest names still drift slowly
(Section~\ref{sec:stationarity}). The case for a structural break is weak. The halves differ in their moments, but a vector tuned to 1996--2010 fits 2011--2024
almost as well as that period's own fit, while a vector tuned to the calmer
2011--2024 does clearly worse on 1996--2010. That is what one expects when a quiet
sample sits inside a more varied one, not when the market changes its rules. A test calibrated on the model's own stationary runs finds only one sign of
drift among seven moments, a rise in the level of returns, which we treat as
exploratory. The claim can be tested. If longer or richer data showed a
genuine break, the model would be refitted or given an external driver, which would
change its parameter values but not its construction.

\emph{How the stationarity is achieved.} Stationarity is achieved not
through an equilibrium but through a non-equilibrium steady state (NESS), with
probability currents that never vanish. The irreversibility has two sources,
stochastic volatility with a leverage effect, which breaks detailed balance on its
own, and the coupling of the investor's demand to the price, which multiplies it by
the factors given above (Section~\ref{sec:reflexverdict}).
Most models of equity returns, the statistical ones in particular, have no investor whose flows act on the price, and so lack the second source. While the agent-based models
this paper compares against, the extended Chiarella model
\cite{majewski2020,chiarella1992}, the structural stochastic volatility model of
Franke and Westerhoff \cite{franke2012} and the adaptive belief systems of Brock
and Hommes \cite{brock1998}, all have demand acting on the price, they lack a
conserved budget and a flow of money with this structure
(Table~\ref{tab:checkpoints}). As in the extended Chiarella model, prices move
by impact rather than by a clearing condition, and at no date is supply required
to meet demand, but the impact here is that of the investor's latent demand on a
nonlinear potential. Earlier work in this line found
that, read through a distance matrix and
two ranking chains, the market does not learn
\cite{halperin2026omdstocks,halperin2026omdportfolio}. Its spectrum stays
un-relaxed, its effective dimension stays high and its geometry keeps moving,
which is what a learning system looks like \emph{long before} it learns anything.
That inverts the usual reading. The orthodox market is stationary because it has
already learned everything, and this one is stationary because it never learns at
all.

\emph{No macroeconomic forcing.} Booms and busts come out of the dynamics of an
autonomous system. No macroeconomic factor, no news process and no schedule
appears anywhere in the specification, and every coefficient is a constant, so the
cycle is not imposed by a time-varying input. All randomness in the model is exogenous, and amounts to the standard Brownian motions and the jump drivers of the model specification. Nor is an
exotic price process needed to reach it. The model uses Brownian factors, the
variance being the exponential of an Ornstein-Uhlenbeck process, and finite-activity
jumps in the fundamental, which the price inherits, with no fractional driver and no
assumed stable-L\'evy law for returns. These three produce the statistical texture of the market, its cycle included.

\emph{Dynamically broken scale invariance.} The model has no constant with the
dimension of a price or any other currency unit: its states are dimensionless and
its constants are rates or pure numbers, so its stochastic differential equations
are unchanged when prices are restated in another unit. The one dollar amount, the
ten-cent default floor, enters the engine as a fraction of each name's entry price.
The same constants therefore describe a firm at any price and the market at any
size, which lets one calibration hold while the market grows for a century. While
the driving equations are scale-invariant, the dynamics they produce is not: the
reversion $\kappa_x$ and the noise $\sigma_x$ of the mispricing combine into a width
$\sigma_x/\sqrt{2\kappa_x}$, a scale the dynamics generates. It is a statistical scale, compatible with the covariance of the equations under a change of currency units. As a heuristic analogy, the thermal noise of the classical Langevin stochastic dynamics replaces here the quantum noise of quantum field theory \cite{coleman1985} in generating a scale unit from an initially scale-invariant dynamics. Power laws are the signature of a scale symmetry that fluctuations restore, as at a critical point where fluctuations of every size make the correlation length diverge. Here the noise does {\bf not} restore the symmetry but fixes the scale that breaks it, and over the ranges tested the
model shows no clear power law, each observable tested on its own
\cite{clauset2009}: the waiting times between crashes are close to exponential but
cluster (Section~\ref{sec:bustlaw}), a lognormal is preferred for the body of firm
sizes, whose tail a likelihood ratio cannot separate from a power law
(Section~\ref{sec:sizedist}), and for the tails of returns
(Appendix~\ref{sec:returnshape}).

\emph{A single rational investor.} The money is moved by one collective investor
holding a fixed mean-variance rule, optimal given the return and the risk it
estimates, so nothing in the decision is biased in the behavioral sense. That is
the classical assumption rather than the econophysics one. Behavioral and
multi-agent accounts recover the cycle by populating the market with types,
chartists against fundamentalists or beliefs revised by imitation, and letting the
shares of the types vary
\cite{brock1998,alfarano2008,franke2012,chiarella1992}, so that the complexity is
produced by the heterogeneity itself. Here the two forces those models assign to
two populations are the path term $+\theta s$ and a value term set by the potential, both belonging to a single rule and both always active. What heterogeneity the model has is in the assets and not in the
agents.

\emph{Black's factor of two.} Fisher Black proposed calling a
market efficient when price stays within a factor of two of value at least ninety
percent of the time, and was explicit that the factor was an intuition rather
than a measurement \cite{black1986}. Three decades later Bouchaud and co-authors
tested it and found it about right, with markets trending over months,
mean-reverting over years, and prices off by roughly a factor of two
\cite{bouchaud2017}. Our market meets that bound without having been asked to,
and meets it inside a dynamics with no resting point, which is Soros's picture
and not Black's. Black bounds a distance without giving a mechanism and Soros
gives a mechanism without bounding anything, and here the two are the same
object, so both can be tracked at once.

\emph{Market inelasticity.} The model relates to the two best-known findings about
money flowing into equities, and for different reasons. Gabaix and Koijen argue that the aggregate market is
inelastic \cite{gabaix2021}, a dollar of demand raising its value by about five. In the model the mispricing responds to a permanent
rise in the investor's latent demand with a susceptibility of $0.64$, the right
direction, and the response disappears when the price impact $c$ is switched off.
It is not directly comparable with their multiplier, which is the response to an
executed dollar (Section~\ref{sec:inelastic}). Frazzini and Lamont find that the names taking in the most money
go on to underperform \cite{frazzini2008}. The model reproduces the sign: the names
its investor buys most go on to trail those it buys least. Over three years the
shortfall, $8$ to $11$ percent, is close to their $9$, but in the model it arrives
within months rather than building over years (Section~\ref{sec:inelastic}). This effect survives with the price impact
switched off, since it comes from what an extrapolating investor chooses to hold
rather than from its demand moving prices. Neither a flow nor an elasticity was a
calibration target, so both are untargeted diagnostics.

\emph{Defaultable equity without a credit model.} A name here can fail at a price floor, a proxy for default rather than a credit event, with no separate model of corporate mortality. A firm fails when
its mispricing escapes over the barrier of the marketron potential and its price
reaches the ten-cent floor, mostly after the index has removed it, so the default
rate is an output of the same equations that price the survivors rather than a
process fitted beside them. It is held to the investment-grade rates the market
shows, and the dependence of that rate on the barrier, on the variance and on the
phase of the cycle is untargeted (Section~\ref{sec:boomsdefaults}).

\emph{An index built from its names.} The model simulates five hundred
individual stocks, the ten largest and the rest, and the index is built from them
as the S\&P is: each name has a capitalization that compounds with its own
return, the index return is the capitalization-weighted average of the names'
returns, and a name whose capitalization falls far enough behind is removed and
replaced, the removed firm being followed until it is sold or defaults.
Simulating the names makes correlations, size distributions and default rates measurable at all. Each group is fitted on its
own capitalization-weighted basket of the names that stay in it, and the index,
which combines the two groups and includes every entrant and exit, is never fitted on its own, so every index observable is an untargeted
diagnostic.

\emph{Why such a market model is wanted.} Long-horizon asset allocation is tested
against histories that do not exist. The usable history of a single equity market
runs to a few decades of daily data and a century or so of coarser series,
holding a handful of crises, and it cannot say how often a market of a given kind
should have them, how deep they should go, or how long a portfolio might have to
wait through one. A model calibrated on twenty-nine years that can be run for
centuries, its slow relaxation to a settled state measured rather than
assumed away, supplies many independent histories with the observed moments, correlation structure and cadence of busts.
What such a simulator has to get right is not the next return but the frequency,
depth and spacing of the events a long-dated allocation must survive, which is
why the cycle statistics and the tail quantities of this paper are the ones we
hold the model to.

\section{Reflexivity and Causal Loops}
\label{sec:reflexivity}

The core of reflexivity is that a market participant runs two functions at
once. The cognitive function, $y=f(x)$, is the effort to understand the market,
mapping the objective situation $x$ into a subjective perception $y$. The
manipulative function, $x=\varphi(y)$, is the effort to act on the market,
mapping the perception $y$ into actions that change the objective situation
$x$. In the natural sciences the cognitive function stands alone, since an
observer's beliefs do not alter physical law. In a market both functions run
together, the independent variable of one is the dependent variable of the
other, and the constant baseline that a deterministic outcome would need is
gone. Perception and reality co-determine a path rather than settling to a
point.

Soros's loop runs in two places at once, and the theory becomes testable once the
two are separated. The first is between prices and
fundamentals. Value determines price in the direction every valuation model
assumes, and price determines value in the direction Soros insists on, since a
company whose shares are expensive can raise capital cheaply, acquire rivals with
paper, and hire on options, so the fundamentals that justified the rise are
improved by the rise itself. The second is between the participants and the
market. Prices form the perception, the perception sets the position, and the
position moves the price. The first loop makes the anchor itself move, and the
second leaves it fixed and lets the market wander around it. The two divide the
work of this line of papers. The first, an empirical question about a market
whose fundamentals are observed, has been examined on real data in OMD-Stocks and
OMD-Portfolio, where transfer entropy is measured between firm fundamentals and
market returns to ask which direction carries information
\cite{halperin2026omdstocks,halperin2026omdportfolio}. The second is the subject
here, and a generative model can address it, since it needs no fundamental data and
closes inside the model.

Each loop has feedback of both signs, and the balance between them decides
whether a market comes to rest. The negative, stabilizing branch is the pull of
value: the further price rises above what a stock is worth, the lower its
expected return and the weaker the reason to hold it. The positive,
destabilizing branch is extrapolation, since a rise that has happened is evidence
to the participants that more rise is coming. A market in which the first
dominates has a resting state and returns to it, and one in which the second
dominates leaves. What Soros describes is neither, a market where the two are of
comparable strength, so the state neither settles nor escapes but circulates.
That is the sense in which reflexivity, in Soros's account, prevents
equilibrium, and it is a
statement about a ratio of gains rather than about the presence of feedback,
which no one disputes.

Two principles anchor the theory. Fallibility, inherited from Popper, holds
that the participants' understanding is permanently incomplete and distorted,
because they are embedded in the market they price and cannot step outside it. The human uncertainty principle holds
that because agents act on distorted perceptions, their actions have unintended consequences, injecting a structural uncertainty
distinct from the probabilistic risk of standard financial engineering.
Together they deny the market a resting state and give it an
asymmetric boom-bust dynamics. An unrecognized trend is recognized and becomes
self-reinforcing as rising prices improve the
fundamentals that justified
the rise. A testing phase either breaks the trend or confirms it. Acceleration pushes valuations above any historical mean. A twilight phase stalls the
price while the fundamentals still look strong. The crash then feeds falling
prices back into solvency through margin calls and forced liquidation, and a
low baseline is set for the next cycle.

Reflexivity has been influential among practitioners and largely resisted by
academic finance, in part because it stayed verbal and never wore
the falsifiable form the Efficient Market and Rational Expectations hypotheses
wear. The work that gave it quantitative content came mostly from
outside economics. The econophysics literature measures
how much of a market's activity it generates itself, using self-exciting Hawkes
processes whose conditional intensity
$\lambda(t)=\mu(t)+\int_{-\infty}^{t}\phi(t-s)\,\dd N(s)$ splits an exogenous
background $\mu$ from an endogenous self-excitation with memory kernel $\phi$. The
branching ratio $n=\int_0^\infty \phi(\tau)\,\dd\tau$ is the fraction of activity
generated internally, so $n=0$ is an exogenous market and $n\to 1$ is a
critical market poised for cascades. Calibrations put the S\&P 500 near that
ceiling, with $n$ rising toward $0.8$ and beyond over recent decades
\cite{filimonov2012,hardiman2013}. A separate literature traces particular
reflexive mechanisms rather than an aggregate index. Funding and market liquidity
reinforce one another, so that losses tighten funding, which withdraws liquidity,
which deepens losses \cite{brunnermeier2009}. Van der Beck, Bouchaud and Villamaina \cite{vanderbeck2024}
measure directly how flows into funds can lift the prices of the assets those
funds already hold, generating the performance that attracts the next flow, which
they call a Ponzi flow. Each is a closed loop of the kind this
paper builds, measured in the market.

A different line keeps the participants and lets them adapt. Lo's adaptive markets
hypothesis treats efficiency not as a property a market has or lacks but as
something that waxes and wanes as populations of strategies are selected on
\cite{lo2004,lo2017}. As Polakow and co-authors observe, it is in essence a
feedback control formulation of the problem, and it is the piece of the
orthodox-heterodox landscape our model most resembles. Here too a controller reads
the market and acts on it, and efficiency is an outcome of the control rather than
an assumption about it. The difference is one of population: the adaptive markets
hypothesis has many strategies under selection, and this paper a single
collective investor whose optimal equity allocation is the control variable,
which lets the loop be written down and solved rather than described.

A third line is methodological rather than quantitative, and it bears on what can
be measured at all. Polakow, Gebbie and Flint set the conventional causal chain,
$\Prob[B \mid \mathrm{do}(A)] > \Prob[B]$, beside its reverse,
$\Prob[A \mid \mathrm{do}(B)] > \Prob[A]$, and argue that the feedback dynamics
of the two are indistinguishable unless an intervention is available
\cite{polakow2026}. A reflexive market is on their reading an open system, and no
stable causal graph can be recovered from its history. That is a limit on the
empirical half of this paper, and it is the argument for the generative half,
since a simulated market supplies the intervention a market withholds.

Our methodological answer to Polakow, Gebbie and Flint is that the causal loop
need \emph{not} be analyzed in the self-referencing paradigm. A self-referencing
system, in the sense that runs from Spencer-Brown's laws of form through
second-order cybernetics to the eigenforms of markets, is one whose description is
part of what it describes, and such systems are notoriously hard to analyze
because there is no outside vantage from which to write the equations down. The
market built here is instead a structural model of a bi-partite system, a capital
market of many stocks and a single collective rational investor, each with its own
state and equations of motion, coupled in both directions. The reflexivity is in
the coupling and not in the description. That resolves the logic of the causal
loop top-down: the loop is closed, but because both halves are written
explicitly, the ambiguity Polakow and co-authors identify becomes a question of
calibration and inference rather than an obstruction in principle. Every causal
relation the model implies can be simulated, intervened upon and measured, which
is what the rest of this paper does.

\section{The Boom-Busty Marketron Model}
\label{sec:marketron}

The BBM model builds on the investor-market reflexive loop of the Marketron model
of Halperin and Itkin \cite{marketron2025}, where a stock is viewed as a particle whose position (log-price) is driven by a memory of the money flow into it, and that flow in turn
responds to the price. To each stock we add a fundamental value process, a
stochastic volatility and a rational mean-variance investor, and we then build a
market of such boom-busty marketrons sharing one collective investor with a
bounded budget, the fourth extension.

The construction is built in five steps. We write the log price of a single stock
as three components separated by how long a shock to each survives. We give the
investor that holds it, a mean-variance allocator whose perception of return is
built from the path of the market. We write the stochastic dynamics of the two
together, for a market of $N$ such stocks, and make that market an index
by drawing the parameters across names and adding a membership rule. We
calibrate it to the ten largest names and to the rest separately, on
capitalization-weighted baskets. Finally we collapse the names and
treat the whole index as a single asset, one marketron standing for the market,
calibrated to the index series itself. The last two steps yield two calibrated
objects rather than one, the BBM and the I-BBM, and
Section~\ref{sec:twomodels} explains what each of them is used for.

\subsection{Value, mispricing and the price}
\label{sec:setup}

The log price $\log S_t$ of a single stock is decomposed into three pieces, separated by how long a shock to
each survives,
\begin{equation}
\log S_t = \log F_t + x_t + \eta_t .
\label{eq:decomp}
\end{equation}
Here $F_t$ is the fundamental value the stock would have if it
were priced correctly, $x_t$ the mispricing, meaning the gap between
the log price and the log value once the transitory error is averaged out, and
$\eta_t$ that transitory error, which decays within days. The mispricing is zero for a correctly
priced stock, positive when it is dear and negative when it is cheap. Only the left
side is observed in real markets. Replacing one measured process by three
unmeasured ones buys two things. The first is a hierarchy of timescales
written in from the start, since a one-dimensional diffusion cannot describe the reversion of a mispricing over months, the decay of a transitory error over days
and the permanence of value at once. Those rates end up nearly three orders of
magnitude apart. The second is that the dynamics is then written in
stationary or nearly-stationary processes. The price itself cannot be made
stationary, since value compounds, so a dynamics in $\log S_t$ has no rest point
and no equilibrium distribution, whereas every question asked here is a question
about a stationary state.

Eq.~\eqref{eq:decomp} belongs to the permanent-plus-transitory family of
decompositions \cite{summers1986,poterba1988}, which splits a price into a
permanent component and stationary deviations from it. Our version differs from
the standard one in having two deviations rather than one, decaying more than
an order of magnitude apart.

The permanent piece and the mispricing play opposite parts in the variance ratio
of Lo and MacKinlay \cite{lomackinlay1988}, the variance of an $h$-day return
divided by $h$ times the variance of a one-day return. It equals one at $h = 1$ by construction and at every horizon for a random walk, falls below one when the price mean-reverts, and for a stationary price falls toward zero as $h$ grows, since a stationary process adds no variance at long horizons.

Over 1996--2024 the S\&P's ratio drops from one to $0.76$ at a month and $0.67$ at a quarter, then rises again to $0.79$ at a year. The rise comes from pooling two periods that behave differently. In 1996--2010 the ratio stays near $0.73$ up to a quarter and climbs back to one by a year, while in 2011--2024 it keeps falling, to $0.44$ at a year (Appendix~\ref{sec:fittests}). In the model the reversion at these horizons comes from the mispricing, which reverts with a half-life of nine weeks for the ten largest names and four and a half months for the rest, so the ratio falls over a month to a year. The permanent part of $F_t$, the part the market shares, keeps it from falling to zero, as it would if the log price were a trend plus a stationary mispricing. The other two reversions, of $F_t$ and $\eta_t$, with half-lives of about seven years and four trading days, are too slow and too fast to matter at these horizons.

By Eq.~\eqref{eq:decomp} every price move is either a revaluation of the
fundamental or price moving against it. In the model the split is known, since
both parts are simulated, while a price path alone does not identify the split without the
timescale assumptions that define it. The specification $S_t = F_t e^{x_t +
\eta_t}$ also lets an unbounded price coexist with either bounded or unbounded
mispricing. The BBM model can produce both, and the calibration determines which
(unbounded, or bounded in the spirit of Black's claim) is realized in the
market.

\subsection{The Reflexive Investor model}
\label{sec:investor}

The system has two halves that cannot be written apart: the capital market, whose
prices respond to the investor's latent demand, and the single collective Reflexive
Investor (RI for short) who shops there for equity, forming that demand by
looking at what the market has done.

Of the three components of the price in Eq.~\eqref{eq:decomp}, the investor reads
the mispricing $x^{(i)}_t$ directly, where $i = 1, \ldots, N = 500$ enumerates the
stocks, and the transitory error $\eta^{(i)}_t$ only through the realized returns
its variance filters average. Reading $x^{(i)}_t$ directly is a simplification. Up
to the fast transitory noise $\eta^{(i)}_t$, it is equivalent to assuming that the
investor observes both the market price and the fundamental value, and turns them
into the two quantities it uses: the mispricing $x^{(i)}_t$ and the average growth
rate $\gamma$ of the fundamental. The current value of $F^{(i)}_t$ itself does not
enter its decisions. It also observes the demand memory $y^{(i)}_t$ defined below,
and adds its own states, the holdings $\varphi^{(i)}_t$, the bonds $b_t$ and the
wealth $\mathcal{W}_t$, and two quantities built from the price history. Its
momentum signal $s^{(i)}_t$, a memory of the recent moves in the name's mispricing,
enters the trend term of the perceived return. Its short and long trailing
variances $\hat v^{s,i}_t$ and $\hat v^{l,i}_t$ average squared realized returns, and
their ratio sets the risk penalty that return is divided by. The one hidden
variable is the stochastic volatility $v^{(i)}_t$. The state per stock is therefore
nine-dimensional:
\begin{equation}
Y^{(i)}_t = \big(\,x^{(i)}_t,\; y^{(i)}_t,\; \eta^{(i)}_t,\;
   \varphi^{(i)}_t,\; b_t,\; \mathcal{W}_t,\;
   s^{(i)}_t,\; \hat v^{s,i}_t,\; \hat v^{l,i}_t\,\big) ,
\label{eq:investorstate}
\end{equation}
of which $b_t$ and $\mathcal{W}_t$ are one number for each of the investor's two
sleeves (sub-portfolios for the large and the small stocks, below) and the other
seven are per name. The first three are market states driven by the stochastic
dynamics of Section~\ref{sec:lmsv}, and the remaining six are the investor's own,
obeying ordinary differential equations that read the market's path. The wealth is
kept in the vector for readability, although Eq.~\eqref{eq:sharewealth} fixes it by
the holdings, the bonds and the prices. The demand memory $y^{(i)}_t$, derived
below, is the one thing the investor returns to the market and the only place its
decisions enter the dynamics.

\enlargethispage{2pt}Holdings are counted in shares.\footnote{We assume that the total number of shares for each stock in the market is one, and holdings are approximated as continuous values on the unit interval.} With one share outstanding, a firm's capitalization is its price, and the fraction $1 - \varphi^{(i)}_t$ not held by the investor is the equity still available for purchase in the capital market. The investor keeps its money in two sleeves, one holding the ten largest names of the index and the other the rest. Each sleeve is a separate book with its own bond holding and wealth. The investor owns $\varphi^{(i)}_t$ shares of name
$i$ and $b^{(k)}_t$ bonds of price $B_t = e^{r_b t}$ in sleeve $k$, so the wealth of each sleeve is
\begin{equation}
\mathcal{W}^{(k)}_t \;=\; \sum_{i \in N_k} \varphi^{(i)}_t\,S^{(i)}_t
   \;+\; b^{(k)}_t\,B_t , \qquad k = 1, 2 ,
\label{eq:sharewealth}
\end{equation}
where $N_k$ is the set of names in sleeve $k$. The operations are the same on both
sleeves, so to lighten the formulas we drop the sleeve index $k$ from here on. The
change of a sleeve's wealth is then
\begin{equation}
d\mathcal{W}_t = \sum_{i=1}^{N} \varphi^{(i)}_t\,dS^{(i)}_t
   + b_t\,dB_t
   + \sum_{i=1}^{N} S^{(i)}_t\,d\varphi^{(i)}_t + B_t\,db_t .
\label{eq:moneywealth}
\end{equation}
The investor receives a flow of new money at the rate $\varsigma\,S^{\rm tot}_t$,
where $S^{\rm tot}_t = \sum_i S^{(i)}_t$ is the total capitalization of the names a
sleeve holds, so that summed over the two sleeves the inflow is $\varsigma$ times
the capitalization of the whole index, and it pays a cost proportional to the traded volume,
\begin{equation}
\mathrm{TC}\big(d\varphi_t\big)
  \;=\; \chi \sum_{i=1}^{N} S^{(i)}_t\,\big|\dot\varphi^{(i)}_t\big|\,dt ,
\qquad \dot\varphi^{(i)}_t = \frac{d\varphi^{(i)}_t}{dt} ,
\label{eq:tccost}
\end{equation}
with $\chi$ the cost of moving a unit of value, five basis points throughout.
Each sleeve receives the new money on its own names, and its self-financing constraint is
\begin{equation}
\sum_{i=1}^{N} S^{(i)}_t\,d\varphi^{(i)}_t + B_t\,db_t
  \;=\; \Big(\varsigma\,S^{\rm tot}_t
     \;-\; \chi \sum_{i=1}^{N} S^{(i)}_t\,\big|\dot\varphi^{(i)}_t\big|\Big)dt .
\label{eq:cashflow}
\end{equation}
Writing the bond leg of
Eq.~\eqref{eq:moneywealth} as $b_t\,dB_t = \big(\mathcal{W}_t -
\sum_i \varphi^{(i)}_t S^{(i)}_t\big) r_b\,dt$ and substituting
Eq.~\eqref{eq:cashflow}, we obtain the SDE for the wealth process:
\begin{equation}
d\mathcal{W}_t
  = \Big(r_b\,\mathcal{W}_t + \varsigma\,S^{\rm tot}_t
     - \chi \sum_{i=1}^{N} S^{(i)}_t\big|\dot\varphi^{(i)}_t\big|\Big)dt
  \;+\; \sum_{i=1}^{N} \varphi^{(i)}_t
     \big(dS^{(i)}_t - r_b\,S^{(i)}_t\,dt\big) .
\label{eq:wealth}
\end{equation}
Re-arranging Eq.~\eqref{eq:cashflow}, we obtain the ODE for $b_t$ in terms of
the control variables $\varphi^{(i)}_t$:
\begin{equation}
db_t = \frac{1}{B_t}\Big[\Big(\varsigma\,S^{\rm tot}_t
     - \chi\sum_{i} S^{(i)}_t\big|\dot\varphi^{(i)}_t\big|\Big)dt
     - \sum_{i} S^{(i)}_t\,d\varphi^{(i)}_t\Big] ,
\label{eq:bondleg}
\end{equation}

The investor neither shorts nor borrows, and it holds no more than a fraction
$\hat\phi$ of its portfolio in equity, with $\hat\phi = 0.80$, the equity weight
of a balanced fund. Written against the wealth of Eq.~\eqref{eq:sharewealth} and
rearranged, the cap becomes a bound on the equity book in units of the bond book,
\begin{equation}
\sum_{i=1}^{N} \varphi^{(i)}_t\,S^{(i)}_t \;\le\; \psi\;b_t\,B_t ,
\;\;\; \psi \;\equiv\; \frac{\hat\phi}{1 - \hat\phi} ,
\label{eq:levcap}
\end{equation}
so the equity book may stand at four times the bond book and no higher.

Each sleeve meets the ceiling of Eq.~\eqref{eq:levcap} on its own, and no money moves between the sleeves. They start at the index's
weights when a window opens and are not rebalanced toward them afterwards, so the
split follows what each sleeve earns. The beliefs set out below are formed name by
name and do not depend on which sleeve holds a name.

Of the investor's six private components, three are the wealth, the holdings
and the bond holding above. The first of the remaining three is the momentum signal $s^{(i)}_t$, a memory of the recent moves in the mispricing $x^{(i)}_t$ kept as an exponentially weighted moving average,
\begin{equation}
ds^{(i)}_t \;=\; \kappa\,\big(dx^{(i)}_t - s^{(i)}_t\,dt\big) ,
\qquad\text{equivalently}\qquad
s^{(i)}_t \;=\; \kappa \int_{-\infty}^{t} e^{-\kappa\,(t-u)}\,dx^{(i)}_u ,
\label{eq:memdef}
\end{equation}

The subjective expected return constructed by the investor, its \emph{belief function},
is linear in the momentum signal $s^{(i)}_t$ and in the demand memory
$y^{(i)}_t$, and has a value term that is non-linear in the mispricing
$x^{(i)}_t$:
\begin{equation}
m^{(i)}_t\big(x, y, s\big)
  \;=\; \gamma \;+\; \theta\,s^{(i)}_t
  \;+\; \mu_i\,\Lambda\,\omega^{(i)}_t\,\big(\bar y_i - y^{(i)}_t\big)
  \;-\; c_i\,\Lambda\,\omega^{(i)}_t\,V\big(x^{(i)}_t\big) ,
\label{eq:belief}
\end{equation}
with $\Lambda$ the price of risk, $\mu_i$ a relaxation rate and
$\omega^{(i)}_t$ the perceived risk of Eq.~\eqref{eq:omega} below.

The third term of Eq.~\eqref{eq:belief} depends on the departure of the demand memory $y^{(i)}_t$ from its
target level $\bar y_i$, a parameter fixed below, and the law of motion of
$y^{(i)}_t$ itself is derived in Eq.~\eqref{eq:yaccum}. It captures the dumb-money, or crowding, effect of Frazzini and Lamont \cite{frazzini2008}: when
the demand accumulated in a name stands above its normal level, the investor
marks the expected return down, and up when the name has been left alone. The
last term, the value term, is non-linear in the mispricing and takes its sign and
magnitude from the current $x^{(i)}_t$ through the potential $V(x)$, to which $c_i$
couples the name. The potential is the marketron potential of Halperin and Itkin
\cite{marketron2025},
\begin{equation}
V(x) = e^{-x}\Big(1 - \frac{1}{\bar\epsilon}\Big)
      + \frac{\log\!\big(1 + \bar\epsilon\,g\,e^{-x}\big)}{\bar\epsilon^{2}\,g},
\label{eq:leanpot}
\end{equation}
normalized so $V(+\infty) = 0$, with $g$ the coupling constant and
$\bar\epsilon$ a regularization parameter controlling the behavior at
strongly negative $x$.  The right panel of Figure~\ref{fig:leanpot} below shows $V(x)$ at the calibrated parameters, and the other panels show the calibrated joint potential $\mathcal{V}(x,y)$ introduced below.

The potential also drives the SDE for the mispricing $x_t$:
\begin{equation}
dx^{(i)}_t = \big[\,a_i - \kappa_x\,x^{(i)}_t
      - c_i\,y^{(i)}_t\,V'(x^{(i)}_t)\,\big]\,dt
      + \sigma_x\,h_i\,d\xi^{(i)}_t ,
\label{eq:xcopy}
\end{equation}
which is Eq.~\eqref{eq:lmsvx} below, and whose remaining symbols are those of
Section~\ref{sec:lmsv}.

The last two terms of Eq.~\eqref{eq:belief} enter as risk-adjusted excess
returns, hence the factor $\Lambda\omega^{(i)}_t$. The risk $\omega^{(i)}_t$ is
the investor's own reading and not the market's, and what it can observe decides
why it takes two variances. It cannot see the variance factor $v^{(i)}_t$, and on
the daily grid of step $\Delta t$ it receives one squared return per step,
\begin{equation}
q^{(i)}_t \;=\; \frac{\big(\Delta \log S^{(i)}_t\big)^{2}}{\Delta t} ,
\label{eq:qest}
\end{equation}
built from the log price of the name itself. As this quantity is very noisy
for a single time step, the investor uses it to construct two exponential moving
averages (EMAs) $\hat v^{s,i}_t$ and $\hat v^{l,i}_t$ over two windows, four
months and four years. The ODEs driving these EMAs read
\begin{equation}
d\hat v^{s,i}_t = \kappa_{\rm sh}\big(q^{(i)}_t - \hat v^{s,i}_t\big)dt , \qquad
d\hat v^{l,i}_t = \kappa_{\rm lg}\big(q^{(i)}_t - \hat v^{l,i}_t\big)dt ,
\label{eq:vshortlong}
\end{equation}
and the decision depends on them only through their ratio, which is the penalty the
perceived return is divided by,
\begin{equation}
\omega^{(i)}_t\big(v^{s}, v^{l}\big)
   \;=\; \big(\hat v^{s,i}_t / \hat v^{l,i}_t\big)^{\mathbcal{w}} .
\label{eq:omega}
\end{equation}
Because both variances are linear filters of the same input, replacing
$q^{(i)}_t$ by a constant multiple of itself leaves $\omega$ unchanged, so the
scale of the variance never enters the decision. What alarms this investor is not
a high variance but \emph{a variance high against the one it has grown used to},
whether the rise came from the diffusion or from a wave of revaluations. The
exponent $\mathbcal{w}$ in Eq.~\eqref{eq:omega} is a calibrated parameter of the
investor.

Over one step the investor maximizes a mean-variance utility using its belief
function and enforcing the cap with a Lagrange multiplier $\vartheta_t$. The cap
is read against the wealth before the trade at the opening prices,
$\mathcal{W}_t = \sum_i \varphi^{(i)}_t S^{(i)}_t + b_t B_t$, which a purchase paid
out of the bond leaves unchanged up to its cost, so that the portfolio after the
trade satisfies Eq.~\eqref{eq:levcap},
\begin{equation}
J\big(\varphi_t\big)
  = \sum_{i} \varphi^{(i)}_t S^{(i)}_t
      \big(m^{(i)}_t(x, y, s) - r_b\big)
  - \frac{\Lambda}{2}\sum_{i} S^{(i)}_t\,\omega^{(i)}_t
      \big(\varphi^{(i)}_t\big)^{2}
  + \vartheta_t\Big(\hat\phi\,\mathcal{W}_t
      - \sum_{i} \varphi^{(i)}_t S^{(i)}_t\Big) .
\label{eq:mvobjective}
\end{equation}
For $\Lambda > 0$, $S^{(i)}_t > 0$ and $\omega^{(i)}_t > 0$, which hold in every state
of the model, the objective is strictly concave in the holdings, so the maximizer
exists and is unique. The penalty is a subjective, diagonal regularizer: it
charges each name its own price-scaled perceived risk and ignores covariances,
so it is not the variance of the investor's wealth under Eq.~\eqref{eq:wealth}.
The KKT (slackness) conditions are
\begin{equation}
\vartheta_t \ge 0 , \qquad
\sum_{i} \varphi^{(i)}_t S^{(i)}_t \le \hat\phi\,\mathcal{W}_t ,
\qquad
\vartheta_t\Big(\sum_{i} \varphi^{(i)}_t S^{(i)}_t
   - \hat\phi\,\mathcal{W}_t\Big) = 0 ,
\label{eq:kkt}
\end{equation}
At fixed $\vartheta_t$ the optimal holding is
\begin{equation}
\varphi^{(i)}_t\big(\vartheta_t\big) \;=\;
   \Big[\frac{m^{(i)}_t(x, y, s) - r_b - \vartheta_t}
        {\Lambda\,\omega^{(i)}_t}\Big]_{0}^{1} ,
\qquad \big[z\big]_{0}^{1} = \min\{\max(z, 0),\,1\} ,
\label{eq:phiofmult}
\end{equation}
the two bounds enforcing the constraints that the investor never shorts and never
holds more of a name than the market has.

By the third condition of Eq.~\eqref{eq:kkt}, either the multiplier is zero,
admissible where the unconstrained holdings satisfy the cap, or the cap binds,
which is one equation in $\vartheta_t$,
\begin{equation}
\sum_{i} \Big[\frac{m^{(i)}_t(x, y, s) - r_b - \vartheta_t}
              {\Lambda\,\omega^{(i)}_t}\Big]_{0}^{1}\;S^{(i)}_t
  \;=\; \hat\phi\,\mathcal{W}_t .
\label{eq:multeq}
\end{equation}
Its left side is non-increasing in $\vartheta_t$, so a root exists, positive exactly
when the unconstrained book breaches the cap, and the holdings it gives are unique.
The two together give the target allocation in two branches,
\begin{equation}
\varphi^{(i)}_{*}(t) =
\begin{cases}
\;\Big[\dfrac{m^{(i)}_t(x, y, s) - r_b}{\Lambda\,\omega^{(i)}_t}\Big]_{0}^{1} ,
   & \text{if } \displaystyle\sum_{i} \varphi^{(i)}_{*} S^{(i)}_t
     \;\le\; \hat\phi\,\mathcal{W}_t , \\[3ex]
\;\Big[\dfrac{m^{(i)}_t(x, y, s) - r_b - \vartheta_t}
        {\Lambda\,\omega^{(i)}_t}\Big]_{0}^{1} ,
   & \text{if } \displaystyle\sum_{i} \varphi^{(i)}_{*} S^{(i)}_t
     \;=\; \hat\phi\,\mathcal{W}_t ,
\end{cases}
\label{eq:alloc}
\end{equation}
with $\vartheta_t$ the root of Eq.~\eqref{eq:multeq} in the second branch. Every
$\varphi^{(i)}_t$ therefore lies in $[0, 1]$, from none of the firm's shares to
all of them. The first branch has the form of a Merton share, an excess return
divided by a price of risk, with $\Lambda$ the risk aversion and $\omega^{(i)}_t$
its state dependence, in the spirit of the mean-variance rules of Markowitz
\cite{markowitz1952} and Merton \cite{merton1969}. It differs from them in that the drift it
optimizes against is not a constant but the belief of Eq.~\eqref{eq:belief}, which
the market it is trading has just supplied. The second branch is the same rule with the equity cap active.
Subtracting the multiplier from every name's excess return drops the names whose
belief does not clear it, and the harder the cap binds the fewer names survive,
until only those with the highest excess belief are held.

The set $\{\varphi^{(i)}_{*}(t)\}$ is the allocation the investor is aiming at
and not the one it holds. Money reaches a market at a finite rate, so the
holdings relax toward the target rather than jumping to it,
\begin{equation}
d\varphi^{(i)}_t \;=\; \big(\varphi^{(i)}_{*}(t) - \varphi^{(i)}_t\big)\,dt ,
\label{eq:partialadj}
\end{equation}
In simulation we use a daily time step, $\Delta t = 1/252$, and approximate the
continuous-time SDEs and ODEs by their discretized versions
(Section~\ref{sec:numerics}). Over a day Eq.~\eqref{eq:partialadj} thus proposes to change the holding of name $i$ by
$(\varphi^{(i)}_{*} - \varphi^{(i)}_t)\,\Delta t$ shares, traded at the opening prices. Sales are made in full, and purchases are scaled by the
factor $f$ constructed to keep the constraint~\eqref{eq:levcap} valid post-trade:
\begin{equation}
f = \min\Big\{1,\, \max\Big(0,\, \frac{H}{(1 + \hat\phi\chi)\,\Delta^{+}}\Big)\Big\},
\qquad H = \hat\phi\,\big(\mathcal{W}_t - \chi\,\Delta^{-}\big) - E^{-} ,
\label{eq:fscale}
\end{equation}
where $\Delta^{+}$ and $\Delta^{-}$ are the values of the proposed purchases and
sales, $E^{-}$ is the equity after the sales, and $f = 1$ when nothing is bought.
If the sales alone leave the book above the cap, $H < 0$, nothing is bought and
nothing more is sold, so the breach persists until later sales and prices remove
it. In the audited runs no sleeve opens above the cap: the opening share reaches at
most $0.793$ against $\hat\phi = 0.80$ over $240$ sleeve-years (Supplement,
Table~S9).\footnote{The Supplement is available from the author upon reasonable request.} The inflow and the interest are credited to
the bond account after the trade.
The one-step utility of Eq.~\eqref{eq:mvobjective} with a policy that amounts
to trading toward its target according to Eq.~\eqref{eq:partialadj} is a
myopic, one-step version of a method proposed in \cite{sciphy2026} for a more
general multi-period portfolio optimization.

The executed change of the holding, $d\varphi^{(i)}_t$, is the proposed one
when it is a sale and $f$ times it when it is a purchase. The money traded in name
$i$, positive for a purchase, and the bond account after the trade are then
\begin{equation}
u^{(i)}_t\,dt = S^{(i)}_t\,d\varphi^{(i)}_t ,
\qquad
b^{+}_t B_t = b_t B_t - \sum_i \big(u^{(i)}_t + \chi\,\big|u^{(i)}_t\big|\big)\,dt .
\label{eq:uderived}
\end{equation}
These executed trades move the holdings and the bond account. The price does
not respond to them directly. It responds to a latent demand state, the demand
memory $y^{(i)}_t$, whose drift is the same rule without its constraints: the first
branch of Eq.~\eqref{eq:alloc} without the bounds and the multiplier, with
Eq.~\eqref{eq:belief} substituted,
\begin{equation}
\frac{\gamma + \theta\,s^{(i)}_t - r_b}{\Lambda\,\omega^{(i)}_t}
    + \mu_i\big(\bar y_i - y^{(i)}_t\big) - c_i\,V\big(x^{(i)}_t\big)
    - \varphi^{(i)}_t
  = \mu_i\big(\hat y_i - y^{(i)}_t\big) - c_i\,V\big(x^{(i)}_t\big) .
\label{eq:flowpot}
\end{equation}
The demand memory is therefore the investor's desired, unconstrained demand, and
it departs from the executed trades whenever the cap binds or the target is
clipped: in the calibrated market by a quarter to a third of its drift among the
ten largest names and by about two thirds in the rest of the index. The last
equality collects the two terms that do not already have the Marketron form into a
state-dependent target,
\begin{equation}
\hat y_i \;\equiv\; \bar y_i \;+\; \frac{1}{\mu_i}
   \bigg(\frac{\gamma + \theta\,s^{(i)}_t - r_b}
              {\Lambda\,\omega^{(i)}_t\big(v^{s}, v^{l}\big)}
   \;-\; \varphi^{(i)}_t\bigg) .
\label{eq:yhat}
\end{equation}
The SDE of the memory variable adds noise to this drift:
\begin{equation}
dy^{(i)}_t \;=\; \Big[\,\mu_i\big(\hat y_i - y^{(i)}_t\big)
      - c_i\,V\big(x^{(i)}_t\big)\Big]dt
   \;+\; \sigma_y\,dW^{y,(i)}_t ,
\label{eq:yaccum}
\end{equation}
which is Eq.~\eqref{eq:lmsvy} below, and $y^{(i)}_t$ is the {\bf demand memory}
of name $i$. The added noise represents the volatility of demand for the
stock. Demand in the market has sources other than the mean-variance
calculation of our investor, and $\sigma_y$ is the dispersion of the realized
flow around the rule that the drift describes. It is a shock to the demand memory
and not to the perception.

Comparing the two SDEs, for $y_t$ in Eq.~\eqref{eq:yaccum} and for $x_t$ in
Eq.~\eqref{eq:xcopy}, we see that they together form a two-dimensional Langevin
dynamics with the joint potential, see Fig.~\ref{fig:leanpot},
\begin{equation}
\mathcal{V}(x, y) \;=\; \frac{\kappa_x}{2}\Big(x - \frac{a}{\kappa_x}\Big)^{2}
  \;+\; \frac{\mu}{2}\,\big(y - \hat y\big)^{2}
  \;+\; c\,y\,V(x) .
\label{eq:jointpot}
\end{equation}

\begin{figure}[!ht]
\centering
\includegraphics[width=\textwidth]{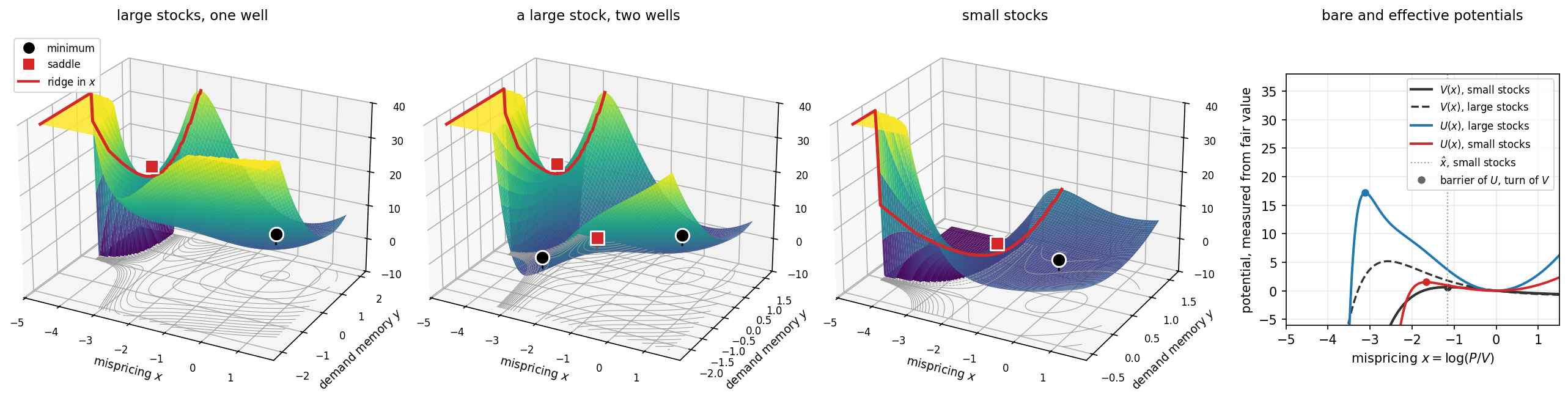}
\caption{The joint potential $\mathcal{V}(x,y)$ of Eq.~\eqref{eq:jointpot} with
its critical points, on a common vertical scale: the large stocks at their
group's fitted constants, one simulated large stock whose own constants give a
second minimum and a second saddle, and the small stocks. Both types are
encountered among the ten largest stocks of the simulated market. Right, the
bare $V(x)$ of Eq.~\eqref{eq:leanpot} and the effective $U(x)$ of
Eq.~\eqref{eq:effpot} for each group.}
\label{fig:leanpot}
\end{figure}

\label{sec:potential}%
The investor's action enters the market dynamics only through the target level
$\hat y$ of Eq.~\eqref{eq:yhat}. Figure~\ref{fig:leanpot} draws each group's
landscape for a name of unit size at its baseline rest level $\bar y$, with the
investor's contribution to $\hat y$ set to zero.

At the calibrated parameters both groups' potentials have one minimum and one
saddle. A larger coupling $c$ can add a second minimum and a second saddle, and
since each name draws its own $c$ around its group's value, one of the ten
large stocks of the simulated market has them (Figure~\ref{fig:leanpot},
second panel). Along the valley floor the demand memory is
$y^{*}(x) = \hat y - cV(x)/\mu$, and a large enough $c$ turns it negative near the
top of $V$, reversing the sign of the force $-c\,y^{*}V'(x)$ on the mispricing.

The demand noise is small beside the mispricing noise, $\sigma_y = 0.054$ against
$\sigma_x = 0.159$ for the large names and $0.266$ for the rest. Dropping it, and in
the limit where the demand memory relaxes fast against the mispricing, it follows
$y^{*}(x)$, and integrating the remaining drift gives the
effective potential
\begin{equation}
U(x) = -\int^{x}\!\Big[\,a - \kappa_x u - c\,y^{*}(u)\,V'(u)\,\Big]\,du ,
\label{eq:effpot}
\end{equation}
drawn beside the bare $V$ in the right panel of Figure~\ref{fig:leanpot}.

\subsection{Stochastic dynamics of the investor-market system}
\label{sec:lmsv}
\label{sec:leancross}

The other half of the system is the capital market. It supplies the shares of
$N$ companies, forms their capitalization-weighted index as the S\&P 500 is
formed, provides the liquidity the investor trades against, and sets the prices,
each driven by shocks to its fundamental, by market noise and by the investor's
demand. 

The fundamental process $F_t$ jumps via three compound jump processes.  A jump is a revaluation of what a firm is worth, and
the construction is deliberately silent about what caused it. Some causes are
outside the market, as when a firm fails and the failure is reported, and some
are not, as when the re-evaluation of credit portfolios in 2007 and 2008 wrote
down the book values of many firms at once with nothing arriving from outside to
do it. One object serves both, since the model describes a revaluation by its size
and its clustering rather than by its origin. Each process has the form
\begin{equation}
J_t \;=\; \sum_{k=1}^{N_t} Y_k ,
\label{eq:compound}
\end{equation}
with $N_t$ counting arrivals and the sizes $Y_k$ drawn independently. Two of the
processes are market-wide and downward, and one is firm-specific and two-sided,
\begin{equation}
\begin{aligned}
dJ^{M}_t &= -\,Y^{M}\,dN^{M}_t , \qquad
d\lambda^{M}_t = \beta_H\big(\lambda_M - \lambda^{M}_t\big)\,dt + \alpha\,\beta_H\,dN^{M}_t , \\
dJ^{S}_t &= -\,Y^{S}\,dN^{S}_t , \qquad
dJ^{(i)}_t = Y^{(i)}\,dN^{(i)}_t ,
\end{aligned}
\label{eq:jumpprocs}
\end{equation}
where $N^{M}_t$ arrives at the stochastic intensity $\lambda^{M}_t$, $N^{S}_t$ and
$N^{(i)}_t$ are Poisson at the constant intensities $\lambda_S$ and $\lambda_I$,
the sizes $Y^{M}$ and $Y^{S}$ are exponential with means $\jmath_M$ and $\jmath_S$,
and $Y^{(i)}$ is normal with mean zero and standard deviation $\sigma_{JI}$, so that
one firm can be revalued upward. The firm processes are independent across names.
$J^{M}$ is rare, large and clustered. It is a Hawkes process: each event raises the
intensity by $\alpha\beta_H$ and the excess decays at rate $\beta_H$, so $\alpha$ is
the fraction of events descended from earlier ones and $\lambda_M/(1-\alpha)$ the
stationary intensity, the reading of a Hawkes intensity used to measure reflexivity
in order flow \cite{filimonov2012,hardiman2013}. $J^{S}$ is its routine
counterpart, frequent, small and without clustering. The drift of the fundamental
subtracts the jumps' mean contribution through a
constant compensator $\bar\jmath_i$ defined below. Writing
$l^{(i)}_t$ for the log of the fundamental, so that $F^{(i)}_t = e^{l^{(i)}_t}$
and $\overline{l}_t$ is its mean over the index's members, name $i$ obeys four stochastic differential equations, the first two of which,
for $x^{(i)}_t$ and $y^{(i)}_t$, were introduced above, driven by the noises
defined after them,
\begin{align}
dx^{(i)}_t &= \big[\,a_i - \kappa_x\,x^{(i)}_t
      - c_i\,y^{(i)}_t\,V'(x^{(i)}_t)\,\big]\,dt
      + \sigma_x\,h_i\,d\xi^{(i)}_t , \label{eq:lmsvx} \\
dy^{(i)}_t &= \big[\,\mu_i\,\big(\hat y_i - y^{(i)}_t\big)
      - c_i\,V(x^{(i)}_t)\,\big]\,dt
      + \sigma_y\,dW^{y,(i)}_t , \label{eq:lmsvy} \\
dl^{(i)}_t &= \big(\gamma_i - \bar\jmath_i
      - \kappa_F\,(l^{(i)}_t - \overline{l}_t)\big)\,dt
      + \sigma_F\,h_i\,d\xi^{F,(i)}_t
      + h_i\Big(\beta^{J}_i \sum_{k \in \{M, S\}} dJ^{k}_t
      + \psi^{J}_i\,dJ^{(i)}_t\Big) , \label{eq:fundamental} \\
d\eta^{(i)}_t &= -\kappa_\eta\,\eta^{(i)}_t\,dt
      + \sigma_\eta\,h_i\,dW^{\eta,(i)}_t . \label{eq:lmsveta}
\end{align}
Here $W^{y,(i)}_t$ and $W^{\eta,(i)}_t$ are standard Brownian motions,
$\xi^{(i)}_t$ and $\xi^{F,(i)}_t$ are stochastic drivers defined below, and
$\beta^{J}_i$ and $\psi^{J}_i$ are jump loadings defined in
Section~\ref{sec:jumpspot}. The mean $\overline{l}_t$ in Eq.~\eqref{eq:fundamental}
couples the names through a mean field of McKean-Vlasov type
(Section~\ref{sec:interactions}). The noise scales $\sigma_x$, $\sigma_F$ and
$\sigma_\eta$ and the rates $\kappa_x$ and $\kappa_\eta$ are one value for each size
group, and the dimensionless scale $h_i$ models how the volatilities of individual
names spread around that single tuple $(\sigma_x, \sigma_F, \sigma_\eta)$. It is
drawn once and held for the name's life: log-normal with log standard deviation
$d_\sigma$ around a level fixed for the group, times the volatility multiplier of
its industry group (Section~\ref{sec:interactions}). Its center being fixed, only
its dispersion $d_\sigma$ is fitted, so it does not compete with the group's noise
scales. It multiplies every shock the name receives, so a volatile name is volatile
in its value as well as in its price.

The compensator $\bar\jmath_i$ in Eq.~\eqref{eq:fundamental} is the stationary
mean of the jump drift,
\begin{equation}
\bar\jmath_i = -\,h_i\,\beta^{J}_i\Big(\frac{\lambda_M\,\jmath_M}{1-\alpha}
   + \lambda_S\,\jmath_S\Big) ,
\label{eq:jbar}
\end{equation}
the mean of the two downward market arrivals, with $\lambda_M/(1-\alpha)$ the
stationary intensity of the self-exciting one, the firm jumps having mean zero. It
holds the drift of the log fundamental at $\gamma_i$ on average rather than path by
path, since the market intensity moves while the compensator does not, and in the
calibrated market the residual mean jump drift is $-0.05 \pm 0.16$ percent a year
on the index.

The drivers $\xi^{(i)}_t$ and $\xi^{F,(i)}_t$ are made of common Brownian motions,
for the market and for a sector, and an idiosyncratic one, additionally scaled by
stochastic volatility factors,
\begin{align}
d\xi^{(i)}_t &= \sqrt{v^{M}_t}\,\Big(\beta_i\,dW^{M}_t
      + \sum_{s=1}^{3}\beta^{\rm sec}_{i,s}\,dW^{{\rm sec},s}_t\Big)
      + \sqrt{v^{(i)}_t}\;\psi_i\,dW^{(i)}_t ,
      \label{eq:beta} \\
d\xi^{F,(i)}_t &= \sqrt{v^{M}_t}\,\Big(\beta_i\,dW^{M,F}_t
      + \sum_{s=1}^{3}\beta^{\rm sec}_{i,s}\,dW^{{\rm sec},F,s}_t\Big)
      + \sqrt{v^{(i)}_t}\;\psi_i\,dW^{F,(i)}_t ,
      \label{eq:betaF} \\
\beta^{\rm sec}_{i,s} &= \beta^{s}_{\rm sec}\,\mathds{1}\{s = s(i)\} , \qquad
\psi_i = \Big(1 - \beta_i^{2} - \sum_{s=1}^{3}\big(\beta^{\rm sec}_{i,s}\big)^{2}\Big)^{1/2} . \notag
\end{align}
Here $\beta_i$ is the loading of name $i$ on the market factor and
$\beta^{\rm sec}_{i,s}$ its loading on the factor of industry group $s$, one of
three (Section~\ref{sec:interactions}), whose Brownian motions are
$W^{{\rm sec},s}_t$ and $W^{{\rm sec},F,s}_t$. Each name belongs to one industry
group $s(i)$ and loads only on its factor, with the loading $\beta^{s}_{\rm sec}$
of that group. In the implementation we take the three loadings the same,
$\beta^{s}_{\rm sec} = \beta_{\rm sec}$ for every $s$, with one value of
$\beta_{\rm sec}$ for each size group, so the industry groups differ in their
Brownian motions and in the names that load on them, not in the loading. The weight $\psi_i$ of the idiosyncratic shock gives each driver unit
variance when the volatility factors are at their mean of one. The Brownian motions $W^{M}_t$,
$W^{M,F}_t$, $W^{{\rm sec},s}_t$, $W^{{\rm sec},F,s}_t$, $W^{(i)}_t$ and
$W^{F,(i)}_t$ are independent, and $v^{M}_t$ and $v^{(i)}_t$ are the market and
idiosyncratic stochastic volatility (SV) factors.
The latter are constructed as exp-OU SV models
\cite{scott1987,wiggins1987,fouque2000}. The market factor and a name's own
factor are each a sum of a fast mode, reverting at the rate $\kappa_v$, and a slow
mode, reverting at $\kappa_s$ or $\kappa_{ns}$. They are therefore two-scale, or
equivalently two-factor, SV models in the sense of \cite{fouque2003multiscale},
while the common factor has the fast mode only. The factors are defined by the
following SDEs:
\begin{equation}
\begin{aligned}
dz^{M}_t &= -\kappa_v z^{M}_t dt + \nu_M dW^{v,M}_t , \qquad
dz^{S}_t = -\kappa_s z^{S}_t dt + \nu_S dW^{v,S}_t , \\
dz^{c}_t &= -\kappa_v z^{c}_t dt + \nu_c dW^{v,c}_t , \\
dz^{n,(i)}_t &= -\kappa_v z^{n,(i)}_t dt + \nu_n dW^{v,(i)}_t , \qquad
dz^{ns,(i)}_t = -\kappa_{ns} z^{ns,(i)}_t dt + \nu_{ns} dW^{vs,(i)}_t ,
\end{aligned}
\label{eq:volfactors}
\end{equation}
\begin{equation}
v^{M}_t = e^{\,z^{M}_t + z^{S}_t - s_M^{2}/2 - C_M} , \qquad
v^{(i)}_t = \exp\Big(\lambda_i\,z^{c}_t + z^{n,(i)}_t + z^{ns,(i)}_t
      - \tfrac12\big(\lambda_i^{2} s_c^{2} + s_n^{2}\big) - C_n\Big) .
\label{eq:volmult}
\end{equation}
The tunable parameters of the volatility factors are not the noise scales $\nu$
but the stationary standard deviations of the exponents, $s_M$ for the market's
$z^{M}_t + z^{S}_t$, $s_c$ for the common $z^{c}_t$ and $s_n$ for a name's own
$z^{n,(i)}_t + z^{ns,(i)}_t$, together with the shares $w$ and $w_n$ of the
market's and a name's variance due to the slow mode. Since an OU process with
rate $\kappa$ and noise scale $\nu$ has stationary variance $\nu^{2}/2\kappa$,
\[
\nu_M^{2} = 2\kappa_v (1 - w)\,s_M^{2} , \quad
\nu_S^{2} = 2\kappa_s\, w\, s_M^{2} , \quad
\nu_c^{2} = 2\kappa_v\, s_c^{2} , \quad
\nu_n^{2} = 2\kappa_v (1 - w_n)\,s_n^{2} , \quad
\nu_{ns}^{2} = 2\kappa_{ns}\, w_n\, s_n^{2} .
\]
A name's loading $\lambda_i$ on the common factor is drawn around one with
standard deviation $s_\lambda$. These six are fitted (Section~\ref{sec:leancal}),
while the fast rate $\kappa_v$ is held from the first-stage calibration and the
slow rates $\kappa_s$ and $\kappa_{ns}$ are set by hand
(Section~\ref{sec:measured}). The subtracted terms in Eq.~\eqref{eq:volmult} remove
the stationary variance of each mode taken alone and, through $C_M = {\rm Cov}(z^{M}, z^{S})$ and $C_n = {\rm Cov}(z^{n,(i)}, z^{ns,(i)})$, the stationary covariance of a factor's two modes, which are
driven by correlated noises (below). Every volatility multiplier therefore averages exactly one at the stationary state, and $s_M$ is the standard deviation the two
market modes would have if they were independent.

The volatility noises are built from the price shocks,
\begin{equation}
\begin{aligned}
dW^{v,M}_t &= \rho_M\,dW^{M}_t + \sqrt{1-\rho_M^{2}}\,dB^{v,M}_t , \quad
dW^{v,S}_t = \rho_M\,dW^{M}_t + \sqrt{1-\rho_M^{2}}\,dB^{v,S}_t , \\
dW^{v,c}_t &= \rho_c\,dW^{M}_t + \sqrt{1-\rho_c^{2}}\,dB^{v,c}_t , \\
dW^{v,(i)}_t &= \rho_n\,dW^{(i)}_t + \sqrt{1-\rho_n^{2}}\,dB^{v,(i)}_t , \quad
dW^{vs,(i)}_t = \rho_{ns}\,dW^{(i)}_t + \sqrt{1-\rho_{ns}^{2}}\,dB^{vs,(i)}_t , \\
dW^{y,(i)}_t &= \rho_y\,\frac{d\xi^{(i)}_t}{\sqrt{q^{(i)}_t}}
  + \sqrt{1-\rho_y^{2}}\,dB^{y,(i)}_t ,
\end{aligned}
\label{eq:corr}
\end{equation}
with every $B$ an independent Brownian motion and $q^{(i)}_t$ the variance rate of
$\xi^{(i)}_t$, so that $\rho_y$ is a correlation of standardized innovations,
$d\langle \xi^{(i)}, W^{y,(i)}\rangle_t = \rho_y\sqrt{q^{(i)}_t}\,dt$. The
construction is a valid covariance by design and induces correlations between the
volatility noises themselves, $\rho_M^{2}$ between the two market modes,
$\rho_M\rho_c$ between each of them and the common factor and $\rho_n\rho_{ns}$
between a name's two modes. Across names the only shared price drivers are the
market and sector ones,
\begin{equation}
\begin{aligned}
d\big\langle \xi^{(i)}, \xi^{(j)}\big\rangle_t
  &= c_{ij}\,dt, \qquad
  c_{ij} = \Big(\beta_i\beta_j + \sum_{s=1}^{3}\beta^{\rm sec}_{i,s}\,\beta^{\rm sec}_{j,s}\Big)
     v^{M}_t , \\
d\big\langle W^{y,(i)}, W^{y,(j)}\big\rangle_t
  &= \rho_y^{2}\,\frac{c_{ij}}{\sqrt{q^{(i)}_t q^{(j)}_t}}\,dt , \qquad
d\big\langle \xi^{(i)}, W^{y,(j)}\big\rangle_t
  = \rho_y\,\frac{c_{ij}}{\sqrt{q^{(j)}_t}}\,dt , \\
d\big\langle W^{\eta,(i)}, W^{\eta,(j)}\big\rangle_t &= 0
\end{aligned}
\label{eq:crossname}
\end{equation}
for $i \neq j$. The demand noises inherit the price's cross-name correlation
through Eq.~\eqref{eq:corr}, as the engine builds them, two names' demand noises
correlating at $\rho_y^2$ times their prices', about $0.03$ to $0.15$ at the market's
pairwise correlations. The fundamental's driver has the form of $\xi$, with the same
market and sector arrivals in the jumps. Apart from these and the correlations of Eq.~\eqref{eq:corr}, the
drivers are independent, in particular the price's three drivers $\xi^{F,(i)}$, $\xi^{(i)}$ and
$W^{\eta,(i)}$. Names are also tied through the shared volatility factors $z^{M}$
and $z^{c}$ of Eq.~\eqref{eq:volfactors} and through each sleeve's wealth and bond
holding. The investor's decisions enter the market dynamics only through the
dependence of $\hat y_i$ on its private state variables.

Eqs.~\eqref{eq:lmsvx} to \eqref{eq:volmult}, complemented by Eq.~\eqref{eq:yhat}, form a closed system of autonomous SDEs without any time-dependent parameters, Eq.~\eqref{eq:yhat} itself depending on the full investor sub-model. Writing
\begin{equation}
X^{(i)}_t = \big(x^{(i)}_t,\; y^{(i)}_t,\; \eta^{(i)}_t,\; l^{(i)}_t,\;
   z^{n,(i)}_t,\; z^{ns,(i)}_t\big)
\label{eq:state}
\end{equation}
for the market's part of the state of one stock, the fundamental is part of it
because its current value sets the price, and through the price the investor's
wealth, its inflow and the squared return $q^{(i)}_t$. With the investor state
$Y^{(i)}_t$ of Eq.~\eqref{eq:investorstate}, $(X^{(i)}_t, Y^{(i)}_t)$ is the
complete state of the model for the $i$-th stock. Netting the three components the
two vectors share and the wealth, which is a function of the others, a name has ten states, six of the market and four of the investor
($\varphi^{(i)}_t$, $s^{(i)}_t$, $\hat v^{s,i}_t$ and $\hat v^{l,i}_t$). Each
sleeve adds its bond holding $b_t$, and the market as a whole adds four states
shared by every name, the volatility exponents $z^{M}_t$, $z^{S}_t$ and $z^{c}_t$
and the excess intensity of the self-exciting market jumps of
Section~\ref{sec:jumpspot}. The quantities $m$, $\omega$ and $\varphi^{*}$ are
functions of the state, and that is the whole specification. The market's states are diffusions, and the fundamental has jumps as well, while the
momentum signal and the two trailing variances are driven by the realized
increment and its square, so conditional on the market's path they are ordinary
differential equations.

The subsections that follow take the dependence across names first and then the
jumps and the potential.

\subsection{Dependence across names}
\label{sec:interactions}

Names depend on one another in three ways. The first is the mean-field coupling of
the log fundamentals in Eq.~\eqref{eq:fundamental}. The drift
$-\kappa_F\,(l^{(i)}_t - \overline{l}_t)$ draws each name toward the mean over the
index's members, so the names form an interacting particle system whose large-$N$
limit is of McKean-Vlasov type. The same drift appears in the systemic-risk model of
Carmona, Fouque and Sun \cite{carmona2015}, where banks' log reserves are drawn
toward their own average, and the stationary spread it produces is what Bouchaud
and M\'ezard \cite{bouchaud2000} use to obtain a heavy-tailed wealth distribution
from multiplicative growth. Both of those are diffusions, whereas the coupling here acts on a jump-diffusion, which lets a name be revalued sharply and then drawn back
rather than displaced for good. At $\kappa_F = 0.1$ a year, a half-life of about
seven years, the coupling binds the levels of the names rather than their daily
returns, and what it decides is the distribution of relative size, which it keeps
stationary (Section~\ref{sec:sizedist}).

The second is the jumps. A market event of Eq.~\eqref{eq:jumpprocs} moves every
name on the same day, by an amount graded through the jump loadings, while the
firm events are independent across names (Section~\ref{sec:jumpspot}). The third
is the correlated Brownian drivers of Eq.~\eqref{eq:beta}, which correlate both the
returns and the volatilities of the names. The rest of this subsection treats the
third, after the industry groups that shape it.

In the model a sector is defined by its industry Brownian motion in
Eq.~\eqref{eq:beta}, together with a few numbers taken from its S\&P counterpart:
its share of names, the size of its members and the spread of their sizes, its mean
market loading, its volatility multiplier in $h_i$ and its growth rate. The
industry factors share one loading, so this structure is not rich enough to tell
the eleven GICS sectors apart, and we aggregate them into the three groups that
index providers use. Materials, consumer
discretionary and financials are cyclical, energy, industrials, technology and
telecommunications are sensitive, and staples, health care and utilities are
defensive. The spread of sizes matters because under
capitalization weights it decides how much of a group its largest members speak
for. The measured market loadings of the groups are rescaled to the fitted mean
loading $\bar\beta$, their relative structure left alone. Cyclicality, a group's
beta on the worst decile of days against its ordinary beta, enters through the
jump loadings of Section~\ref{sec:jumpspot}, which scale a name's market loading
by its group's measured ratio.

The Brownian drivers correlate the names in two ways: their returns, through the
shared Brownian motions of Eq.~\eqref{eq:beta}, and their volatilities, through
the shared volatility factors. The market factor $v^{M}$ scales
both the market and the industry terms of Eq.~\eqref{eq:beta}, and the name's own
factor $v^{(i)}$ scales the idiosyncratic term, so the instantaneous variance of
$d\xi^{(i)}$ is
$D_i = \big(\beta_i^{2} + \sum_{s=1}^{3}\big(\beta^{\rm sec}_{i,s}\big)^{2}\big) v^{M}_t + \psi_i^{2}\,v^{(i)}_t$.
From Eq.~\eqref{eq:crossname}, the returns of two names in the same industry group
and of two names in different groups correlate at
\begin{equation}
\rho_{\rm within} = \frac{\big(\beta_i\beta_j + \sum_{s=1}^{3}\beta^{\rm sec}_{i,s}\,\beta^{\rm sec}_{j,s}\big)
   v^{M}_t}{\sqrt{D_i D_j}} , \qquad
\rho_{\rm between} = \frac{\beta_i\beta_j\,v^{M}_t}{\sqrt{D_i D_j}} ,
\label{eq:sectorcorr}
\end{equation}
and the industry structure is the difference between them,
$\sum_{s}\beta^{\rm sec}_{i,s}\beta^{\rm sec}_{j,s}\, v^{M}_t / \sqrt{D_i D_j}$.
For a single asset standing for the whole market the driver collapses to
$\sqrt{v_t}\,dW^{M}_t$, the form the I-BBM uses.

The volatility factors make these correlations move. The share of a name's
variance common to all names,
\begin{equation}
\frac{\big(\beta_i^{2}+\sum_{s=1}^{3}\big(\beta^{\rm sec}_{i,s}\big)^{2}\big)\,v^{M}_t}{D_i} ,
\label{eq:commonshare}
\end{equation}
rises when the market factor rises against the names' own, and both correlations
of Eq.~\eqref{eq:sectorcorr} rise with it. Two facts about the S\&P support this
construction. The first is why two factors are needed. In the rest of the index,
the average pairwise correlation of returns varies across years with a standard
deviation of $0.11$ and $0.12$, and rises with the volatility of their basket, the
two correlating at $0.80$ and $0.86$. A rise in $v^{M}$ raises the common share of
every name's variance and so the correlations, which one factor scaling whole names
cannot do. The second concerns the correlation between the names' volatilities: the
average cross-name correlation of monthly changes in log realized volatility is
$0.27$ and $0.33$, a first principal component explaining $0.29$ and $0.35$ of the
variation, which separates the common scale $s_c$ of $v^{(i)}$ from each name's own
$s_n$.

Table~\ref{tab:sectors} sets the built groups against the measured ones, the
model's spread over its markets against the market's bootstrap error. The groups
separate in the market as their names promise, the defensive group with the lowest beta and name volatility and, over 2011--2024, the highest downside beta relative to its ordinary one, and the model reproduces that ordering in both windows. Its downside
ratios meet the market's in five of the six cases and miss the sensitive group of
1996--2010 by a band width, which checks that the measured ratios
built into the jump loadings carry through the simulation rather than predicting
them. What it overdoes is the spread of the
betas, too high in the sensitive and cyclical groups and too low in the defensive
one. Its name volatilities run high in the second window, where the market's names
were quieter, and its shares of capitalization miss in three of the six cases, by
at most $1.4$ band widths.

\begin{figure}[htbp]
\centering
\includegraphics[width=\textwidth]{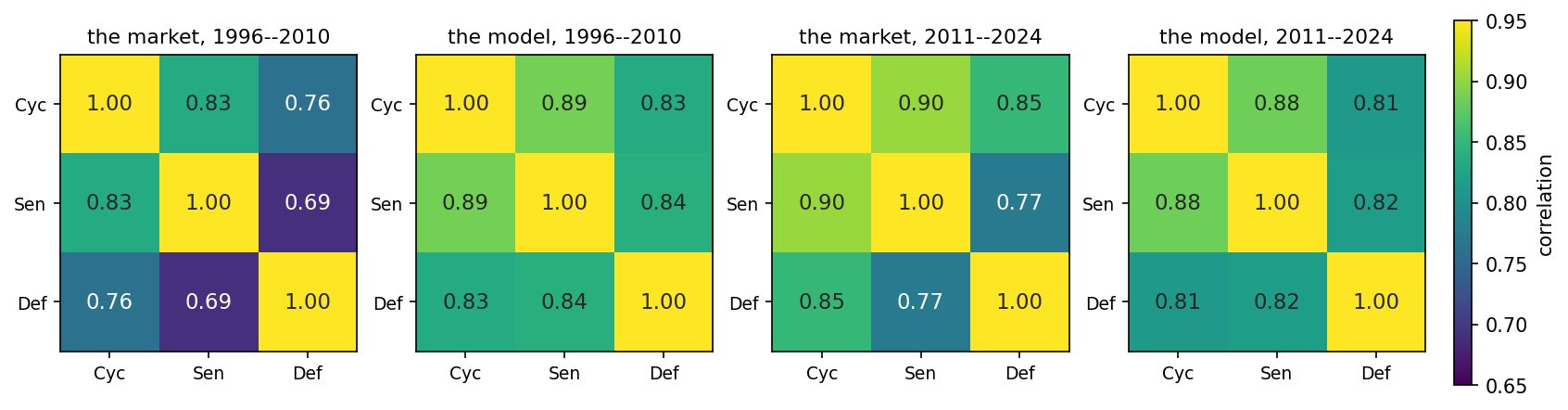}
\caption{Correlations between the three group indices, the market and the model
in each window, taken from the covariance matrix of the indices. The color scale
is set by the off-diagonal entries. Model matrices are averaged over the sixteen
markets of each window that no calibration read.}
\label{fig:seccorr}
\end{figure}

The whole matrix of correlations between the three group indices
(Figure~\ref{fig:seccorr}) is a further check, since neither its entries nor the
volatilities of the group indices are targets. The model keeps the cyclical and
sensitive groups the closest pair in both windows, as the market does. Its groups
move together too much in the first window, where no correlation meets the market's
range, and all three meet in the second. The group volatilities meet in
every group of both windows.

\begin{table}[htbp]
\centering
\scriptsize
\begin{tabular}{l rr rr}
\toprule
& \multicolumn{2}{c}{share of cap} & \multicolumn{2}{c}{market beta} \\
\cmidrule(lr){2-3}\cmidrule(lr){4-5}
Sector & market & model & market & model \\
\midrule
\multicolumn{5}{l}{\textit{1996--2010}} \\
Sensitive & $0.381$ & $0.372$\,\tiny[0.346,0.397] & $1.06$\,\tiny[1.04,1.08] & $1.16$\,\tiny[1.14,1.18]$^{\bullet}$ \\
Cyclical & $0.338$ & $0.378$\,\tiny[0.348,0.409]$^{\bullet}$ & $1.07$\,\tiny[1.05,1.09] & $1.14$\,\tiny[1.13,1.16]$^{\bullet}$ \\
Defensive & $0.281$ & $0.250$\,\tiny[0.226,0.274]$^{\bullet}$ & $0.66$\,\tiny[0.63,0.68] & $0.57$\,\tiny[0.56,0.58]$^{\bullet}$ \\
\textit{largest gap, band widths} & \multicolumn{2}{c}{1.3} & \multicolumn{2}{c}{2.4} \\
\addlinespace
\multicolumn{5}{l}{\textit{2011--2024}} \\
Sensitive & $0.426$ & $0.387$\,\tiny[0.359,0.416]$^{\bullet}$ & $1.09$\,\tiny[1.08,1.10] & $1.20$\,\tiny[1.18,1.22]$^{\bullet}$ \\
Cyclical & $0.380$ & $0.405$\,\tiny[0.375,0.434] & $1.04$\,\tiny[1.03,1.05] & $1.06$\,\tiny[1.05,1.08]$^{\bullet}$ \\
Defensive & $0.194$ & $0.208$\,\tiny[0.175,0.242] & $0.71$\,\tiny[0.70,0.73] & $0.63$\,\tiny[0.61,0.66]$^{\bullet}$ \\
\textit{largest gap, band widths} & \multicolumn{2}{c}{1.4} & \multicolumn{2}{c}{3.1} \\
\addlinespace
\midrule
& \multicolumn{2}{c}{downside / beta} & \multicolumn{2}{c}{name volatility} \\
\cmidrule(lr){2-3}\cmidrule(lr){4-5}
Sector & market & model & market & model \\
\midrule
\multicolumn{5}{l}{\textit{1996--2010}} \\
Sensitive & $0.88$\,\tiny[0.85,0.90] & $0.94$\,\tiny[0.90,0.98]$^{\bullet}$ & $0.44$\,\tiny[0.42,0.46] & $0.42$\,\tiny[0.38,0.45] \\
Cyclical & $1.06$\,\tiny[1.03,1.09] & $1.05$\,\tiny[1.01,1.08] & $0.41$\,\tiny[0.38,0.43] & $0.40$\,\tiny[0.37,0.43] \\
Defensive & $0.99$\,\tiny[0.94,1.04] & $1.02$\,\tiny[0.97,1.06] & $0.33$\,\tiny[0.32,0.35] & $0.31$\,\tiny[0.29,0.32] \\
\textit{largest gap, band widths} & \multicolumn{2}{c}{1.0} & \multicolumn{2}{c}{0.8} \\
\addlinespace
\multicolumn{5}{l}{\textit{2011--2024}} \\
Sensitive & $0.97$\,\tiny[0.96,0.99] & $0.96$\,\tiny[0.93,0.99] & $0.33$\,\tiny[0.31,0.35] & $0.42$\,\tiny[0.39,0.46]$^{\bullet}$ \\
Cyclical & $1.02$\,\tiny[1.01,1.04] & $1.01$\,\tiny[0.98,1.03] & $0.30$\,\tiny[0.28,0.32] & $0.38$\,\tiny[0.35,0.41]$^{\bullet}$ \\
Defensive & $1.10$\,\tiny[1.06,1.13] & $1.09$\,\tiny[1.05,1.13] & $0.26$\,\tiny[0.24,0.27] & $0.32$\,\tiny[0.31,0.34]$^{\bullet}$ \\
\textit{largest gap, band widths} & \multicolumn{2}{c}{0.5} & \multicolumn{2}{c}{2.0} \\
\addlinespace
\bottomrule
\end{tabular}
\caption{The index's three industry groups as measured, against the model's as built. Brackets give one standard error either side, the market's from a block bootstrap and the model's the spread over the sixteen markets of each window that no calibration read; a bullet marks ranges that do not meet, and the last row of each period is the largest gap in band widths. The market's share of capitalization is read on the window's first day.}
\label{tab:sectors}
\end{table}

\subsection{Jumps, leverage and two-scale volatility}
\label{sec:jumpspot}

The two processes of Eq.~\eqref{eq:jumpprocs} account for different aspects of dynamics. The rare clustered jumps
are the crises, which a symmetric diffusion cannot produce, and the routine jumps produce the everyday asymmetry, so the skewness and the tails can be set separately
where one process would tie them. The market shows the same split: outside its
stress episodes falls of the index beyond two percent arrive about six times a
year and average $2.6$ percent, while inside the twelve public episodes since 1996
they arrive at $42$ a year, average $3.8$ and reach $12.7$. Nor are the two one
process with two sizes, since only $J^{M}$ breeds further events.

A market event is common to every name, and its effect is graded at three levels.
Every name moves on the same day. Within a size group, the event reaches name $i$
through its jump loading $\beta^{J}_i = \tau_{s(i)}\beta_i$, the ordinary loading
scaled by a factor of its industry group set from the group's beta on the market's
worst days (Section~\ref{sec:interactions}). Each name's share is further scaled by
$(1 + \zeta\epsilon_i)$ with $\epsilon_i$ standard normal, floored at zero, so that
no two names move by quite the same amount. A firm event reaches only its own name,
through
$\psi^{J}_i = \big(1 - (\beta^{J}_i)^2 - \sum_{s=1}^{3}\big(\beta^{\rm sec}_{i,s}\big)^{2}\big)^{1/2}$.
Between the size groups, both sizes and all three rates are each group's own,
fitted with the rest of its vector, while the market events stay common, the group
with the lower rate receiving a share of them (Section~\ref{sec:leancal}). The ten
largest take rare and firm events more often, and the rest of the index larger
market jumps.

Three further elements of the specification set the dependence between returns
and volatility, and between price and demand. The correlations $\rho_n$, $\rho_M$ and $\rho_c$ of Eq.~\eqref{eq:corr} produce the leverage effect, a stock's most volatile months being its falls. They are fitted
against the correlation of a week's and a month's return with the next change in
log realized volatility, for single names and for each group's basket.

Volatility also needs a second speed. With every factor reverting at
$\kappa_v = 16.85$ a year, a half-life of ten trading days, the index would forget
a crisis within a month, where the average log volatility of the market's names
carries over from one month to the next with a correlation of $0.84$ over
1996--2010 and $0.66$ over 2011--2024. A slow mode reverting at $\kappa_s = 1$ a
year, a half-life of eight months, accounts for over half of the market factor's
variance, and a name's own slow mode a fifth to a third of its log variance, with
a leverage of its own. The shares and the slow leverage are fitted and the rates
set (Section~\ref{sec:measured}).

The correlation $\rho_y$ between a name's price shock and its demand shock in
Eq.~\eqref{eq:corr} is held at $-0.51$ from the first-stage calibration, so that a
rise in price comes with selling and a fall with buying. No calibration target
measures the investor's flows, and over $\rho_y$ from $-0.25$ to $-0.75$ no
observable moves by more than half a band. What it acts on is the relation between
flows and prices of Section~\ref{sec:factorcycle}.

\subsection{The BBM as an information engine}
\label{sec:twoviews}

The construction admits a reading from outside finance, which names a literature
that has already studied the object.

The state vectors of Eqs.~\eqref{eq:state} and \eqref{eq:investorstate} overlap in
the mispricing, the demand memory and the transitory error. The investor's own
states are the holding $\varphi^{(i)}_t$, the bond position $b_t$, the wealth
$\mathcal{W}_t$, the momentum signal $s^{(i)}_t$ and the two realized variances
$\hat v^{s,i}_t$ and $\hat v^{l,i}_t$. The market's own are the fundamental, which
the investor reads only through the price, and the volatility exponents and the
jump intensity, which it cannot observe.

The division suggests, loosely, the picture of an information engine
\cite{Parrondo_2015,Ehrich_2023}. There a controller reads a tape, encodes what
it read, and from that state defines a protocol driving the tape's next move, and
the thermodynamics of the pair rests on a bipartite structure in which the two
sub-systems have independent noise. Here the market plays the tape and the
investor the controller, but the analogy is no more than that: the investor's
momentum signal is driven by the same price noise as the mispricing, the demand
noise is correlated with the price noise, and the alternation of reading and
acting is a property of the daily scheme rather than of the continuous-time
dynamics, so none of the identities of that theory is imported here. The investor reads the market
into $s^{(i)}_t$, $v^{s}_t$ and $v^{l}_t$, converts that into a perceived return
and a perceived risk, and returns the result through the target $\hat y_i$ of
Eq.~\eqref{eq:yhat}, which is the only combination of its private states that
reaches the market. Soros's two functions map onto the two halves of the picture,
the cognitive being the read and the participating the control.

\subsection{The index and its membership}
\label{sec:index}

The split of the investor's money across names needs no rule beyond
Eq.~\eqref{eq:alloc}. Each name is held at its own expected excess return over its
perceived risk, clipped to the shares outstanding, with one common shift
$\vartheta_t$ when the budget binds. There is no separate chasing between names. A
name feeds back on itself through the $-c_i V(x_i)$ of Eq.~\eqref{eq:flowpot},
which draws money toward it the further its price sits from value, and the holding
reaches the dynamics through the target $\hat y_i$ of Eq.~\eqref{eq:yhat}, moving
the level the memory relaxes toward and leaving the loop itself alone.

The observables this paper tests are those of an index built on that cross
section, constructed as the S\&P is: each name has a capitalization that
compounds with its own return, and the index return on a day is the average of
the day's returns weighted by the previous day's capitalizations, so no weight is
set by a return it is about to be applied to. The parameters are calibrated
against the capitalization-weighted baskets of the two size groups, each built
from the names that stay in it through the window, on the model's side as on the
market's, so the index itself is never fitted and every index observable is an
untargeted diagnostic.

A name leaves the index for one of two reasons, and we call the two together
\emph{attrition}. \emph{Removal} is a question of membership rather than
solvency, the name having fallen behind while going on trading afterwards.
\emph{Default} is failure: the name's price closes under ten cents a share,
which is how the market's own failures are counted. Each name draws a price on
entry from the market's own distribution of prices at entry to the index, so the
floor sits as far below a member's price as it does in the S\&P. A removed name
is not dropped, because of what the calibration has to match. The index's
turnover is counted over its members, but a firm's failure is counted over firms,
and the market's own count follows a firm past its removal for as long as CRSP lists it, since most of the firms that fail have left the index first. A model
that dropped a name at removal would see almost none of the failures. So the
removed firm is followed, its default counted whether it falls in the index or
after it, and turnover and default are measured on the same populations on both
sides. The investor's book follows a rule of its own. A removed firm may be held
or sold but not bought, as an index investor treats a deleted name, and a position
still held five years after the removal is sold at that day's price, paying the
cost every sale pays. A firm that defaults before then is written off at the floor,
the investor recovering ten cents a share with no cost. The five-year horizon is
the model's and not the market's count. Without it the followed firms accumulate
over a long run into thousands, and a market-wide fall then fails them together.
These rules conserve the investor's money by design. Every exit moves value only
between its shares and its cash: a removal, a sale at the end of the five years and
a default each turn the shares into cash at the stated price, and the book changes
by nothing but the cost of the sale or, at a default, the difference between the
floor recovered and the price last marked.
In the market six of the eight floor defaults of 1996--2010 fall while the firm is
a member or within five years of its removal, and none of the three of
2011--2024. A member that defaults books its fall to the floor in the index. Both exits are fitted, the removal threshold so
that names leave at the rate the index shows and the default rate against the
rate at which the market's own firms reach the floor, the first over the
membership and the second over every firm in the index or followed past its
removal.

The removal test is applied at every close. It computes a name's capitalization
relative to that of the median member, and removes the name when this falls
below a fraction $\delta_{\rm exit}$ of the highest relative size it held over the trailing
fifteen years. Reading size relative to the median rather than in levels is what
makes the rule about falling behind the market rather than falling with it: read
in levels it would empty the index in a crash, where the S\&P removes a few percent of its names a year and not appreciably more in 2008 than in 2005. A
trailing high rather than a lifetime one matters for the same kind of reason,
since a lifetime maximum only ratchets and the removal rate then climbs without
limit as the market ages. The entrant takes the departing name's sector and
constants, with its market loading, its volatility scale and its loading on the
common volatility factor drawn afresh as an initial member's are, and its
mispricing at fair value. It is seated at the median member's capitalization
times a ratio drawn from the index's own entrants, whose median is $0.82$ and
$0.68$ of the median member in the two windows, and its demand memory rests at
its own relative size through Eq.~\eqref{eq:anchor}.
Section~\ref{sec:sizedist} shows what these choices do to the size
distribution.

The index's turnover sets $\delta_{\rm exit}$. Counted from the membership
itself, the S\&P lost $5.77$ percent of its names a year to removals and member
defaults in 1996--2010 and $4.02$ in 2011--2024, with the membership near $490$
names throughout. That rate is one of the calibrated quantities, so $\delta_{\rm exit}$ is
fitted alongside the rest, at $0.41$, and the rate it delivers, $6.3 \pm 0.9$
and $5.3 \pm 0.5$ percent a name-year over the model's markets, meets the
market's $5.8 \pm 0.7$ in 1996--2010 and runs above its $4.0 \pm 0.3$ in 2011--2024. Removals arrive in merger waves and after crises
rather than independently, so a block bootstrap over calendar years in blocks of
three gives a standard error of $0.66$ points where a Poisson count would give
$0.28$. The ten largest are exempt, with one removal between them in the first
window and none in the second.

The rule differs from the growth-with-killing pair of Reed and Hughes
\cite{reed2002}. Theirs kills at a rate independent of size, which is what
delivers a heavy-tailed size distribution with no critical point behind it, where
ours kills what has fallen behind and spares what has not. Together with the
reverting fundamental of Eq.~\eqref{eq:fundamental} it gives a stationary
distribution of relative size, which Section~\ref{sec:sizedist} measures against
the market's own.

\subsection{Simulating the system}
\label{sec:numerics}

The system is advanced on a daily step, $\Delta t = 1/252$, where the naive
Euler-Maruyama scheme fails. The drifts of the mispricing and of the demand memory contain $-c_i\,y_t V'(x_t)$ and $-c_i V(x_t)$, which below the turnover of the
potential grow exponentially in $-x$. Neither is globally Lipschitz, the case in
which an explicit Euler step is known to diverge \cite{hutzenthaler2011}, and here
the impact flow $\dot x = -K V'(x)$, $K = c_i y_t$, reaches $x = -\infty$ in finite
time: for the rest of the index at its calibrated constants and rest memory, within
about forty-eight trading days from $x = -1.6$, twenty-one from $-2$ and five from
$-3$. An explicit step through this region overshoots by an amount proportional to
$V(x)\,\Delta t$, which does not shrink with the step but moves to other paths.
Taming the drift \cite{hutzenthaler2012} would keep the moments finite but change
the escape over the barrier, the event the default rate counts.

The scheme used instead splits each step \cite{mclachlan2002}. The impact part of
the mispricing's drift is integrated exactly, the demand memory held at its value at
the start. With $u = e^{x}$ and $w = u - g(1-\bar\epsilon)$ the flow separates into
$w + g\log|w| = Kt + C$, so a step is one scalar root found by Newton's method, and
a name whose flow reaches $x = -\infty$ inside the step is recorded as absorbed
within it rather than carried past it. The linear reversion and the noise of the
mispricing are added by an Euler step, while the demand memory, an
Ornstein-Uhlenbeck process given the mispricing, and the log-volatility factors
advance by their exact transitions \cite{gillespie1996}. Over step $n$, from $t_n$
to $t_{n+1}$, a log-volatility mode $z$ with rate $\kappa$ and stationary standard
deviation $s$, the mispricing and the demand memory thus advance as
\begin{equation}
\begin{aligned}
z_{n+1} &= e^{-\kappa\Delta t}\,z_n + s\sqrt{1-e^{-2\kappa\Delta t}}\,
   \big(\rho_z\,\epsilon_n + \sqrt{1-\rho_z^{2}}\,\epsilon^{z}_n\big) , \\
x_{n+1} &= \Phi_{\Delta t}\big(x_n;\,c_i y_n\big) + (a_i - \kappa_x x_n)\,\Delta t
   + \sigma_x\,h_i(z_n)\sqrt{\Delta t}\,\epsilon_n , \\
y_{n+1} &= e^{-\mu_i\Delta t}\,y_n + \big(1-e^{-\mu_i\Delta t}\big)\,\tilde y_{i,n}
   + \sigma_y\sqrt{\tfrac{1-e^{-2\mu_i\Delta t}}{2\mu_i}}\,\big(\rho_y\,\epsilon_n + \sqrt{1-\rho_y^{2}}\,\epsilon^{y}_n\big) ,
\end{aligned}
\label{eq:recursion}
\end{equation}
with $\epsilon_n$ the standardized price shock of step $n$ (the market's for the
market modes, the name's own for a name's), $\epsilon^{z}_n$ and $\epsilon^{y}_n$
independent standard normals, $\Phi_{\Delta t}$ the exact impact flow above and
$\tilde y_{i,n} = \hat y_{i,n} - c_i V(x_n)/\mu_i$ the demand memory's target. The
price over step $n$ uses the variance known at its start, $h_i(z_n)$, while the same
shock moves the variance from step $n+1$: the adapted form of the correlations of
Eq.~\eqref{eq:corr}, an index shift and not a lagged noise, which makes a fall raise
the next day's volatility. Two correlated modes are given the fixed correlation
times their two scales rather than the exact joint transition, whose factor
$\frac{2\sqrt{\kappa_a\kappa_b}}{\kappa_a+\kappa_b}(1-e^{-(\kappa_a+\kappa_b)\Delta t})/\sqrt{(1-e^{-2\kappa_a\Delta t})(1-e^{-2\kappa_b\Delta t})}$
differs from one by less than a tenth of a percent at the calibrated rates. The
calibrated object is this daily model. The equations of Section~\ref{sec:marketron}
are its motivation, and the entropy production of Section~\ref{sec:ness}
is computed from its daily paths, taken as an approximation to them. Jumps arrive as events within the
step, and the mispricing is unbounded: a name that falls far reaches the default
floor, which retires it.

Halving and quartering the step, on ninety-six markets of forty years for the BBM
(two hundred and twenty of thirty years for its default rate) and forty-eight of
two hundred years for the I-BBM, leaves the index volatility, the excess kurtosis
and the bear markets within their errors. The 20\% falls run $27.0 \pm 1.1$ a
century at $\Delta t$ and $26.3 \pm 0.9$ at $\Delta t/4$. Three quantities move.
Defaults may rise by up to a seventh, from $5.2$ to $6.0$ per ten thousand
name-years ($+0.8 \pm 0.6$ paired). The skew of daily index returns is not converged, going from $-0.31$
to $-0.61$ ($\pm 0.04$) in the BBM and from $-0.11$ to $-0.23$ ($\pm 0.02$) in the
I-BBM. The entropy production of the simulated BBM market falls from $86$ to $76$ nats a
year at $\Delta t/2$ and no further. These three are therefore properties of the
daily model, at which every table is made.

\subsection{I-BBM: Index-only BBM specification}
\label{sec:twomodels}

The construction yields two calibrated objects, best put side by side. The \emph{BBM} is the market of Section~\ref{sec:leancross},
$N$ marketrons with a common factor structure and a membership rule, calibrated
group by group on the capitalization-weighted survivor baskets of the ten largest
names and of the rest. The \emph{I-BBM} of Section~\ref{sec:indexone} is the index
treated as one marketron, fitted to the index series itself. Neither is a special
case of the other: the first is never fitted to the index, and the second has no
names to weight.

They also differ in what they cost. The BBM fits thirty-one constants for the rest
of the index and twenty-seven for the ten largest (Section~\ref{sec:leancal}),
which generally come out at different values for the two groups. The I-BBM has
thirteen.
Everything else either passes in from a three-stage pre-calibration, which
begins with a single-name version of the model and is described in
Section~\ref{sec:leancal}, or is held, and the sector characteristics are matched
to the S\&P's own rather than fitted. That is the price of a cross
section, and the fine-tunes of Section~\ref{sec:groupsummary} show that the
split into two groups survives a refit on either window.

Table~\ref{tab:modelcompare} compares them where a comparison means anything.
Most parameters cannot be compared at all, because the group fits take them as
given from that pre-calibration rather than estimating them, so agreement would
be arithmetic rather than evidence. One is estimated independently by all three fits, $\kappa_x$, and on the joint
fit the single asset lands at $2.11$ against $4.15$ for the large names and
$1.87$ for the rest. The aggregate relaxes between its parts and nearer the rest,
which hold three quarters of its capitalization.

The I-BBM is fitted with two parameters held at the values of the full BBM: the
size of a market revaluation, a two-point draw over the two groups' fitted sizes
weighted by the capitalization the ten largest hold, and the decay of its
self-excitation. The investor's own constants are taken from the BBM
as well, so the same investor appears in both models and they differ only in
whether the market is one asset or five hundred.

\begin{table}[htbp]
\centering
\small
\setlength{\tabcolsep}{4pt}
\begin{tabular}{l rr r l}
\toprule
& \multicolumn{2}{c}{multi-asset, by group} & single-asset & \\
\cmidrule(lr){2-3}
Parameter & large & bulk & index & \\
\midrule
\multicolumn{5}{l}{\emph{fitted independently in all three}} \\
\quad mispricing mean reversion $\kappa_x$ & $4.148$ & $1.875$ & $2.105$ & between the groups \\
\addlinespace
\multicolumn{5}{l}{\emph{carried over to the single asset, not refitted}} \\
\quad market jump size $\jmath_M$ & $0.0168$ & $0.0239$ & $(0.0168,0.0239)$ & two-point, weight $0.247$ \\
\quad excitation decay rate, a year & $26.97$ & $43.26$ & $39.24$ & held \\
\addlinespace
\multicolumn{5}{l}{\emph{refitted under those held values}} \\
\quad revaluations a year $\lambda_M$ & $3.084$ & $1.831$ & $0.752$ & \\
\quad share bred by earlier events $\alpha$ & $0.294$ & $0.168$ & $0.026$ & \\
\bottomrule
\end{tabular}
\caption{The two models compared where comparison means something. The multi-asset columns are the joint calibration, each group fitted to both windows at once. Only the mispricing mean reversion is estimated independently by all three fits, and the index's rate lies between those of its two parts. The jump sizes are carried over from the groups rather than refitted, since one index series cannot separate how often the market is revalued from how far, and the decay is the capitalization-weighted mean of the two groups' values.}
\label{tab:modelcompare}
\end{table}

How the two groups are put back together completes the construction. The ten
largest names and the rest are simulated as two markets driven by the same
market-wide draws, each at its own calibrated vector, and the index is their
capitalization-weighted combination: the two groups' simulated capitalizations are
weighted so that the window opens at the S\&P's share of the ten largest, and the
weights move thereafter as the groups' capitalizations move. Nothing in the
combination is fitted. Within a group, what belongs to a name varies between
names, while the jump arrivals and the market and common volatility factors
describe one process the whole market shares.

They should agree only approximately, and
the asymmetry between them explains why. The I-BBM spends the parametric
freedom of the BBM specification on the index itself, holding the index to
constraints the BBM places only on its baskets, the skew and the variance ratios
across horizons among them. The BBM spends the same freedom on the
names, on single-name volatility, pairwise correlation and the link
between correlation and index volatility, and reaches the index only by
aggregating names none of which was fitted to it. Its index observables are
therefore untargeted diagnostics in both windows, and Section~\ref{sec:leancal}
reports them as such. Where the two models differ is a measure of what aggregation costs, not of
one of them being wrong.

\section{Model Calibration}
\label{sec:calibration}

\subsection{Parameters fixed before the calibration}
\label{sec:measured}

The parameters fall into three groups: those read from the market data, those
set by hand, both held through the fit (Table~\ref{tab:historical}), and those
fitted to return moments (Section~\ref{sec:leancal}). Each window supplies its
length, its firms' sizes and its industry groups, the ten largest names and the
rest of the index being defined on its first day. The fifty-year warm-up settles
the mispricing, the demand memory, the volatility factors and the investor's book
but not the sizes, so when the window opens each group's members are re-seated by
rank to the market's relative sizes, with total capitalization and the investor's
wealth unchanged. An entrant comes in at the median member's size times a ratio
drawn from the index's own entrants. Each size group contains the three industry
groups of Section~\ref{sec:interactions} with the shares, loadings, volatilities,
growth and sizes of its S\&P counterpart (Table~\ref{tab:sectors}). Read from the
market and common to both windows are the inflow rate $\varsigma$, set between the
fund-flow and Financial Accounts \cite{fedz1} measures of new money reaching US
equities, the bond return $r_b$ and the default floor of ten cents.

Fifteen constants are set by hand, one value each for both windows and both
groups. Ten of them fix the investor's constraints and the way shocks spread
across names: the equity cap, the trading cost, the two windows of the variance
filters, the sale of a removed firm's holding after five years
(Section~\ref{sec:index}), the jitter and span of a market jump across names
(Section~\ref{sec:jumpspot}), and the dispersions of the demand reversion, the
flow impact and the coupling of the potential across names.

The other five are rates the return moments do not identify. The growth reversion
$\kappa_F$ of Eq.~\eqref{eq:fundamental} draws a name's value toward the average
of the index's members and keeps the size distribution stationary
(Section~\ref{sec:interactions}). Over a century any rate from $0.03$ to $0.4$ a year
does so, and it is set at $0.1$, a half-life of about seven years. The slow modes
of the market's and a name's volatility revert at one a year, faster rates fitting
the index worse. The growth rate of value is shared by the two groups, since a
difference compounds without limit: fitted separately in an earlier calibration,
it took the ten largest's share of the index from $0.22$ to $0.94$ over a hundred
and fifty years, a drift no fifteen-year window can see, while the fit was flat
across the shared rate. The relaxation rate $\kappa_\eta$ of the transitory
component, a half-life of about four trading days, is invisible to the fit:
scaled from a quarter to four times its value it moves no observable by more than
half a band.

\begin{table}[htbp]
\centering
\caption{Constants fixed before the moment search, read from each window, read from the market, or set by hand.}
\label{tab:historical}
\vspace{0.4em}
\footnotesize
\begin{tabularx}{\textwidth}{>{\raggedright\arraybackslash}p{0.34\textwidth} >{\raggedright\arraybackslash}X c c}
\toprule
Constant & Source & 1996--2010 & 2011--2024 \\
\midrule
\multicolumn{4}{l}{\textit{read from each window}} \\[2pt]
spread of log size, the ten largest & capitalizations at the window's start & $0.30$ & $0.25$ \\[3pt]
spread of log size, the rest & capitalizations at the window's start & $1.06$ & $0.93$ \\[3pt]
entrant size over the median member, median (mean) & the market's 429 and 282 entrants & $0.82$ ($1.21$) & $0.68$ ($1.11$) \\[3pt]
industry groups & Table~\ref{tab:sectors} & per window & per window \\[3pt]
\addlinespace
\multicolumn{4}{l}{\textit{read from the market, common to both windows}} \\[2pt]
inflow rate $\varsigma$, a year & fund flows and the Financial Accounts & \multicolumn{2}{c}{$0.00617$} \\[3pt]
bond return $r_b$, a year & the nominal short rate & \multicolumn{2}{c}{$0.030$} \\[3pt]
default floor, \$ & the price at which a firm has failed & \multicolumn{2}{c}{$0.10$} \\[3pt]
entry price, median (5--95\%), \$ & CRSP price on the first day of membership & \multicolumn{2}{c}{$43.25$ ($13.38$--$178.78$)} \\[3pt]
\addlinespace
\multicolumn{4}{l}{\textit{set by hand, common to both windows and both groups}} \\[2pt]
growth of value $\gamma$, a year & one value, so neither group's share of the index drifts & \multicolumn{2}{c}{$0.1262$} \\[3pt]
growth reversion $\kappa_F$, a year & keeps the size distribution stationary & \multicolumn{2}{c}{$0.1$} \\[3pt]
transitory decay $\kappa_\eta$, a year & not seen by the fit & \multicolumn{2}{c}{$42.37$} \\[3pt]
slow volatility rates $\kappa_s$, $\kappa_{ns}$, a year & faster rates fit the index worse & \multicolumn{2}{c}{$1.00$, $1.00$} \\[3pt]
jump jitter across names $\zeta$ & spread of a market jump across names & \multicolumn{2}{c}{$0.35$} \\[3pt]
jump delivery span, days & a market jump reaches every name within it & \multicolumn{2}{c}{$1$} \\[3pt]
spreads across names $d_\mu$, $d_c$, $d_g$ & standard deviations of $\log\mu$, $\log c$, $\log g$ & \multicolumn{2}{c}{$0.18$, $0.20$, $0.10$} \\[3pt]
equity cap $\hat\phi$ & equity weight of a balanced fund & \multicolumn{2}{c}{$0.80$} \\[3pt]
transaction cost $\chi$ & per unit of value moved & \multicolumn{2}{c}{$0.0005$} \\[3pt]
variance rates $\kappa_{\rm sh}$, $\kappa_{\rm lg}$, a year & windows of four months and four years & \multicolumn{2}{c}{$3.00$, $0.25$} \\[3pt]
years a removed firm is followed & then sold & \multicolumn{2}{c}{$5$} \\[3pt]
\bottomrule
\end{tabularx}
\end{table}

\subsection{Calibration to market observables}
\label{sec:leancal}

The calibration has two stages. The first fits a homogeneous market, every name
drawn around one vector, by a Tree-structured Parzen Estimator
\cite{optuna2019} over sixteen observables. It supplies the second stage's
starting point and the eight constants the second stage holds from it, those of
the investor's belief and risk, of the demand memory's noise and reversion, the
reversion of the volatility and the price-flow correlation
(Table~\ref{tab:bbmparams}).

The second stage fits thirty-one constants for the rest of the index and
twenty-seven for the ten largest.\footnote{The four constants the ten largest do
not fit are the removal threshold, since they are exempt from removal
(Section~\ref{sec:index}), and the slow share and leverage of the market-wide
volatility and the common leverage, which they take from the rest of the index
because all three act on the market shock that the two groups share.} Each group is a market of its own, held by one sleeve of
the investor of Section~\ref{sec:investor} and scored on its own basket. The two
draw the market-wide shocks, the volatility factors and the market jumps from one
stream, so their capitalizations add up to the index of Table~\ref{tab:index}.
Every constant is one value for both windows, and the BBM is the pair of vectors
this returns. Two constants are derived rather than fitted. The rest level of the
demand memory depends on the name's own relative size $w_i$ through a constant
$m_{g(i)}$ of its size group $g(i)$, fitted with the rest of the group's vector
(Table~\ref{tab:bbmparams}). The constant of the price drift
then cancels the flow term at fair value,
\begin{equation}
\bar y_i \;=\; m_{g(i)}\,w_i \;+\; \frac{c_i}{\mu_i}\,V(0) , \qquad
a_i \;=\; c_i\Big(\bar y_i - \frac{c_i}{\mu_i}\,V(0)\Big)\,V'(0) .
\label{eq:anchor}
\end{equation}
Only the scale $\Lambda$ of Eq.~\eqref{eq:alloc} and one combination of the rest
levels reach the dynamics, so of the remaining investor constants only $b$ is
searched.

\begin{table}[htbp]
\centering
\scriptsize
\begin{tabular}{l rr}
\toprule
Parameter & the ten largest & the rest of the index \\
\midrule
\multicolumn{3}{l}{\textit{fitted, one vector per group, on both windows at once}} \\
g & $0.0934$ & $0.4240$ \\
$\bar\epsilon$ & $0.1857$ & $0.2598$ \\
$\hat x$, derived & $-2.576$ & $-1.159$ \\
$c$ flow impact & $2.439$ & $3.067$ \\
$m$ & $1.1400$ & $0.4092$ \\
$\sigma_x$ & $0.1588$ & $0.2659$ \\
$\kappa_x$ & $4.148$ & $1.875$ \\
$\sigma_\eta$ & $0.0638$ & $0.0694$ \\
$\sigma_F$ & $0.0606$ & $0.0341$ \\
$\bar\beta$ mean market loading & $0.6544$ & $0.4751$ \\
$\lambda_M$ & $3.084$ & $1.831$ \\
$\jmath_M$ & $0.0168$ & $0.0239$ \\
$\alpha$ & $0.2935$ & $0.1678$ \\
$\beta_H$, excitation decay (yr$^{-1}$) & $26.97$ & $43.26$ \\
$\lambda_S$ & $11.788$ & $9.991$ \\
$\jmath_S$ routine jump size & $0.0070$ & $0.0181$ \\
$\lambda_I$ & $4.279$ & $1.298$ \\
$\sigma_{JI}$ firm jump size & $0.0516$ & $0.0362$ \\
$s_M$ & $1.3866$ & $1.2359$ \\
$s_c$ & $0.3163$ & $0.7045$ \\
$s_n$ & $1.5480$ & $1.2799$ \\
$s_\lambda$ & $0.6534$ & $0.4690$ \\
$\mathrm{sd}(\beta)$ & $0.0470$ & $0.0806$ \\
$d_\sigma$ & $0.1061$ & $0.1608$ \\
$\beta_{\rm sec}$ & $0.1158$ & $0.1887$ \\
$\delta_{\rm exit}$ & off & $0.4133$ \\
$\rho_n$ name leverage & $-0.303$ & $-0.176$ \\
$w_n$ slow share of a name's volatility & $0.189$ & $0.324$ \\
$\rho_{ns}$ leverage of that slow mode & $-0.430$ & $-0.286$ \\
\midrule
\multicolumn{3}{l}{\textit{fitted on the rest of the index, held by the ten largest}} \\
$\rho_M$ market leverage & \multicolumn{2}{c}{$-0.586$} \\
$\rho_c$ common leverage & \multicolumn{2}{c}{$-0.444$} \\
$w$ slow share of market volatility & \multicolumn{2}{c}{$0.549$} \\
\midrule
\multicolumn{3}{l}{\textit{held from the first-stage (homogeneous-market) calibration}} \\
$\sigma_y$ demand volatility & \multicolumn{2}{c}{$0.0538$} \\
$\mu$ demand reversion rate & \multicolumn{2}{c}{$8.4112$} \\
$\kappa_v$ volatility reversion, a year & \multicolumn{2}{c}{$16.85$} \\
$\rho_y$ price-flow correlation, rounded & \multicolumn{2}{c}{$-0.51$} \\
$\theta$ momentum weight of the belief & \multicolumn{2}{c}{$1.782$} \\
$\kappa$ rate of the belief's memory, a year & \multicolumn{2}{c}{$0.0607$} \\
$\Lambda$ price of risk & \multicolumn{2}{c}{$0.158$} \\
$\mathbcal{w}$ exponent of the perceived risk & \multicolumn{2}{c}{$0.403$} \\
\bottomrule
\end{tabular}
\caption{The calibrated BBM: one parameter vector for the ten largest names and one for the rest of the index, each fitted to both windows at once. The barrier $\hat x = \ln g(1-\bar\epsilon)$ is derived, and the constants below the rule are held from the first stage; those set by hand are in Table~\ref{tab:historical}.}
\label{tab:bbmparams}
\end{table}

Each group is fitted to two windows at once, the data split at the end of 2010,
when a typical name's volatility fell by a quarter, and each window keeps its own
targets and errors. Every market first runs through a warm-up that removes its
initial conditions, fifty years for the BBM and five for the I-BBM, after which
nothing the paper measures drifts under the first window's inputs
(Section~\ref{sec:stationarity}). Each group is scored on its
capitalization-weighted basket of survivors, computed by the same function in the
market and the model, and the index of Table~\ref{tab:index}, read on
point-in-time membership, is the check on that choice. Beside the basket's return
moments the objective holds the per-name volatility and its spread, the spread of
loadings, the correlations between the three industry sub-baskets, the size
distribution, the dynamics and leverage of volatility, the turnover of the
membership, to which the removal threshold is fitted, and the default rate over
every firm (Section~\ref{sec:index}).

\begin{figure}[htbp]
\centering
\includegraphics[width=\textwidth]{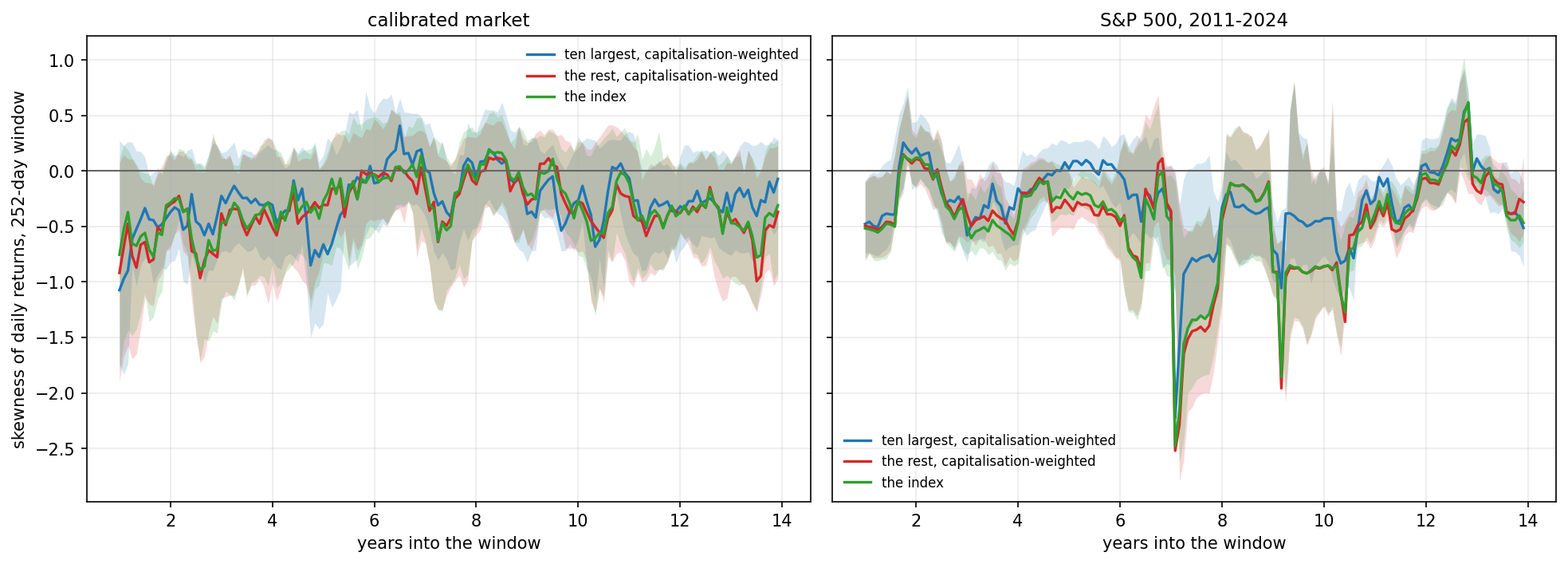}
\caption{Skewness of daily returns on a rolling year for the baskets of the ten
largest and of the rest, and for the index: left the calibrated market on the
second window's inputs, right the S\&P 500 over 2011--2024, with block-bootstrap bands.}
\label{fig:skewgroup}
\end{figure}

A comparison within a window is between two intervals rather than two points,
since a point estimate moves with the window: the market's skewness on a rolling
year swings between $+0.6$ and $-2.5$ over the second window
(Figure~\ref{fig:skewgroup}), so a point comparison would keep or reject a vector
by the year the window closed in.

Write $T_k$ and $S_k$ for the market's estimate of observable $k$ and its
bootstrap error, and $m_k$, $s_k$ for the model's center and spread across
simulated windows, the median and the interquartile range over $1.349$. The
market's error is a stationary block bootstrap \cite{politis1994} of its single
history. The return
and cross-name moments resample trading days in blocks of a year on average,
with the names resampled alongside. The moments of volatility and leverage
resample months in blocks of six months on average, jointly across names, and the
removal rate and the size distribution resample calendar years in blocks of three.
A model that is right has its center within one market error of the target and a spread
equal to the market's, and the loss charges each on its own,
\begin{equation}
z_k = \frac{1}{S_k}\Big[
      \max\!\big(\,|m_k - T_k| - \varepsilon_k,\ 0\,\big)
    + \big|\,s_k - S_k\,\big|\Big],
\qquad \varepsilon_k = S_k .
\label{eq:zhinge}
\end{equation}
A window's loss is the sum,
\begin{equation}
L_w = \sum_{k} z_{w,k} .
\label{eq:totalloss}
\end{equation}
An observable is counted as matching when the two one-sigma ranges meet,
$|m_k - T_k| \le S_k + s_k$, and a miss is sized in band widths, the gap over
$S_k + s_k$. A correct model would meet the criterion about ninety-four percent of the time,
and correlated observables are not independent counts, so the count describes the
fit rather than testing it. The bust count's error is not bootstrapped, since
a window holds two to four busts. It is the Poisson error of the eleven
twenty percent falls in the S\&P from January 1975 to August 2026, $6.4$ a
century, counted as episodes and not by the bear-market dating rule.

Each set of markets holds sixteen markets per window, and four disjoint sets serve
for training, validation, test and report. A candidate proposed near the incumbent
whose training loss is within the scatter of one vector's loss between sets is
scored on the validation and test sets too, and is adopted when it matches more
observables there. The report set enters nothing in the search and gives
Table~\ref{tab:groupjointfifty}. It guards against overfitting the simulation's
seeds and is not a historical holdout. A market whose basket collapses is replaced
by a spare and charged in the loss, and under the adopted vectors no report market
needed one.

The calibrated pair meets the market on twenty-eight and twenty-nine of the ten
largest's thirty-three observables in the two windows, and on twenty-eight and
thirty-three of the rest of the index's forty (Table~\ref{tab:groupjointfifty}).
The misses of the rest of the index share a direction: the market concentrates
within a window as its winners pull away and the model does not, and the model's
volatility is too short-lived in the first window and a name's own volatility too
persistent in the second. The ten largest miss mainly their size distribution.
The two vectors differ where the groups do. The ten largest revert more than twice
as fast with three fifths of the mispricing noise, have more news in their
fundamental, and couple to the potential a fifth as strongly, which puts their
barrier far below fair value. One relaxation rate and noise for every name could
not hold both groups.

The index built from the two baskets is not fitted, so its rows are untargeted
diagnostics. It meets the market on eighteen and nineteen of the twenty-one
observables of Table~\ref{tab:index}, among them the drift, the volatility and
the frequency and depth of twenty percent falls, and no miss exceeds $1.7$ band
widths. Table~\ref{tab:volpredicted} compares the volatility of single names,
eight of its twelve rows being untargeted. The ten largest meet on nine and
twelve rows, with wide ranges, and the rest of the index on six and eleven. In the
first window the market's bursts of volatility were larger and more persistent
than the model's, the gap coming from the part common to the names.

\begin{table}[htbp]
\centering
\footnotesize
\setlength{\tabcolsep}{2.4pt}
\begin{tabular}{l rr rr cc}
\toprule
& \multicolumn{2}{c}{1996--2010} & \multicolumn{2}{c}{2011--2024} & \multicolumn{2}{c}{inside} \\
\cmidrule(lr){2-3}\cmidrule(lr){4-5}\cmidrule(lr){6-7}
Observable & market & model & market & model & $1\sigma$ & $2\sigma$ \\
\midrule
annual drift & $0.066$\,\tiny[0.02,0.12] & $0.111$\,\tiny[0.09,0.14] & $0.130$\,\tiny[0.10,0.16] & $0.116$\,\tiny[0.09,0.14] & Yes & Yes \\
annual volatility & $0.207$\,\tiny[0.18,0.23] & $0.204$\,\tiny[0.16,0.25] & $0.172$\,\tiny[0.15,0.19] & $0.208$\,\tiny[0.16,0.25] & Yes & Yes \\
excess kurtosis, daily & $7.332$\,\tiny[4.79,9.87] & $12.370$\,\tiny[1.37,23.37] & $13.941$\,\tiny[7.41,20.47] & $13.336$\,\tiny[-0.96,27.63] & Yes & Yes \\
falls beyond 2\% a year & $12.610$\,\tiny[9.36,15.86] & $11.167$\,\tiny[7.66,14.67] & $8.586$\,\tiny[6.80,10.37] & $11.400$\,\tiny[7.62,15.18] & Yes & Yes \\
deepest drawdown, log & $-0.793$\,\tiny[-1.08,-0.50] & $-0.512$\,\tiny[-0.84,-0.19] & $-0.409$\,\tiny[-0.50,-0.32] & $-0.522$\,\tiny[-0.85,-0.19] & Yes & Yes \\
20\% falls per century & $26.688$\,\tiny[13.89,39.48] & $30.024$\,\tiny[13.14,46.90] & $14.310$\,\tiny[4.95,23.67] & $29.953$\,\tiny[11.44,48.46] & Yes & Yes \\
variance ratio, 1 month & $0.750$\,\tiny[0.70,0.80] & $0.891$\,\tiny[0.74,1.04] & $0.761$\,\tiny[0.71,0.81] & $0.877$\,\tiny[0.72,1.03] & Yes & Yes \\
variance ratio, 1 quarter & $0.737$\,\tiny[0.66,0.81] & $0.693$\,\tiny[0.51,0.88] & $0.557$\,\tiny[0.49,0.62] & $0.672$\,\tiny[0.49,0.85] & Yes & Yes \\
variance ratio, 6 months & $0.855$\,\tiny[0.71,0.99] & $0.592$\,\tiny[0.37,0.82] & $0.456$\,\tiny[0.36,0.55] & $0.553$\,\tiny[0.34,0.76] & Yes & Yes \\
variance ratio, 1 year & $0.991$\,\tiny[0.79,1.19] & $0.434$\,\tiny[0.25,0.62]$^{\bullet}$ & $0.439$\,\tiny[0.31,0.57] & $0.389$\,\tiny[0.22,0.56] & No & Yes \\
skewness, daily & $-0.180$\,\tiny[-0.28,-0.07] & $-0.380$\,\tiny[-0.99,0.23] & $-0.763$\,\tiny[-1.01,-0.51] & $-0.383$\,\tiny[-1.03,0.26] & Yes & Yes \\
skewness, weekly & $-0.418$\,\tiny[-0.76,-0.08] & $-0.952$\,\tiny[-1.93,0.02] & $-0.420$\,\tiny[-0.61,-0.23] & $-0.964$\,\tiny[-1.91,-0.01] & Yes & Yes \\
skewness, monthly & $-0.665$\,\tiny[-0.88,-0.45] & $-0.809$\,\tiny[-1.65,0.03] & $-1.061$\,\tiny[-1.47,-0.65] & $-0.911$\,\tiny[-1.89,0.06] & Yes & Yes \\
skewness, quarterly & $-0.360$\,\tiny[-0.55,-0.17] & $-0.641$\,\tiny[-1.23,-0.05] & $-1.154$\,\tiny[-1.46,-0.85] & $-0.706$\,\tiny[-1.31,-0.11] & Yes & Yes \\
mean pairwise correlation & $0.272$\,\tiny[0.24,0.31] & $0.292$\,\tiny[0.24,0.34] & $0.335$\,\tiny[0.31,0.36] & $0.292$\,\tiny[0.24,0.35] & Yes & Yes \\
correlation, spread & $0.122$\,\tiny[0.12,0.13] & $0.129$\,\tiny[0.12,0.14] & $0.145$\,\tiny[0.14,0.15] & $0.124$\,\tiny[0.12,0.13]$^{\bullet}$ & No & Yes \\
market beta, spread & $0.440$\,\tiny[0.40,0.48] & $0.374$\,\tiny[0.36,0.39]$^{\bullet}$ & $0.314$\,\tiny[0.29,0.34] & $0.333$\,\tiny[0.32,0.35] & No & Yes \\
name volatility, mean & $0.424$\,\tiny[0.39,0.46] & $0.403$\,\tiny[0.37,0.43] & $0.322$\,\tiny[0.30,0.34] & $0.397$\,\tiny[0.37,0.42]$^{\bullet}$ & No & Yes \\
name volatility, spread & $0.163$\,\tiny[0.14,0.19] & $0.108$\,\tiny[0.09,0.12]$^{\bullet}$ & $0.097$\,\tiny[0.09,0.11] & $0.099$\,\tiny[0.09,0.11] & No & Yes \\
first PC of vol chg. & $0.298$\,\tiny[0.26,0.34] & $0.351$\,\tiny[0.30,0.40] & $0.358$\,\tiny[0.30,0.42] & $0.352$\,\tiny[0.30,0.40] & Yes & Yes \\
corr. of vol chg. & $0.281$\,\tiny[0.24,0.32] & $0.333$\,\tiny[0.28,0.38] & $0.341$\,\tiny[0.28,0.40] & $0.335$\,\tiny[0.28,0.39] & Yes & Yes \\
\midrule
ranges that meet & \multicolumn{2}{c}{18 of 21} & \multicolumn{2}{c}{19 of 21} & \multicolumn{2}{c}{at $1\sigma$} \\
\bottomrule
\end{tabular}
\caption{The capitalization-weighted index, untargeted rather than fitted: the model's two calibrated baskets combined, against the S\&P members of the CRSP panel. Ranges and marks as in Table~\ref{tab:groupjointfifty}.}
\label{tab:index}
\end{table}

\begin{table}[htbp]
\centering
\tiny
\setlength{\tabcolsep}{1.0pt}
\begin{tabular}{l rrc rrc cc}
\toprule
& \multicolumn{3}{c}{1996--2010} & \multicolumn{3}{c}{2011--2024} & \multicolumn{2}{c}{inside} \\
\cmidrule(lr){2-4}\cmidrule(lr){5-7}\cmidrule(lr){8-9}
Observable & market & model & & market & model & & $1\sigma$ & $2\sigma$ \\
\midrule
\multicolumn{7}{l}{\textbf{the ten largest}} \\
annual drift & $0.090$\,\tiny[0.045,0.135] & $0.116$\,\tiny[0.087,0.146] &  & $0.150$\,\tiny[0.123,0.177] & $0.135$\,\tiny[0.096,0.173] &  & Yes & Yes \\
annual volatility & $0.196$\,\tiny[0.177,0.214] & $0.187$\,\tiny[0.142,0.231] &  & $0.178$\,\tiny[0.154,0.201] & $0.215$\,\tiny[0.167,0.263] &  & Yes & Yes \\
excess kurtosis & $6.5$\,\tiny[3.9,9.1] & $9.7$\,\tiny[6.2,13.2] &  & $12.1$\,\tiny[6.5,17.6] & $9.9$\,\tiny[7.2,12.7] &  & Yes & Yes \\
falls beyond 2\% a year & $10.1$\,\tiny[7.9,12.4] & $9.2$\,\tiny[5.9,12.4] &  & $8.7$\,\tiny[6.3,11.0] & $11.6$\,\tiny[7.2,16.0] &  & Yes & Yes \\
mean depth, 20\% falls & $0.301$\,\tiny[0.249,0.354] & $0.269$\,\tiny[0.219,0.319] &  & $0.257$\,\tiny[0.232,0.281] & $0.257$\,\tiny[0.213,0.300] &  & Yes & Yes \\
20\% falls per century & $26.7$\,\tiny[20.3,33.1] & $16.7$\,\tiny[8.0,25.3] &  & $14.3$\,\tiny[7.9,20.7] & $28.6$\,\tiny[12.7,44.5] &  & Yes & Yes \\
variance ratio, month & $0.702$\,\tiny[0.633,0.772] & $0.831$\,\tiny[0.685,0.977] &  & $0.725$\,\tiny[0.688,0.762] & $0.847$\,\tiny[0.740,0.954] &  & Yes & Yes \\
variance ratio, quarter & $0.590$\,\tiny[0.529,0.651] & $0.558$\,\tiny[0.435,0.681] &  & $0.544$\,\tiny[0.471,0.617] & $0.582$\,\tiny[0.437,0.726] &  & Yes & Yes \\
variance ratio, 6 months & $0.642$\,\tiny[0.539,0.746] & $0.479$\,\tiny[0.346,0.613] &  & $0.407$\,\tiny[0.322,0.491] & $0.453$\,\tiny[0.291,0.615] &  & Yes & Yes \\
variance ratio, year & $0.794$\,\tiny[0.613,0.975] & $0.357$\,\tiny[0.239,0.476] & $\bullet$ & $0.382$\,\tiny[0.282,0.483] & $0.334$\,\tiny[0.174,0.494] &  & No & Yes \\
skewness, daily & $0.026$\,\tiny[-0.099,0.151] & $-0.280$\,\tiny[-0.680,0.120] &  & $-0.394$\,\tiny[-0.524,-0.265] & $-0.361$\,\tiny[-0.712,-0.010] &  & Yes & Yes \\
skewness, weekly & $-0.314$\,\tiny[-0.573,-0.056] & $-0.532$\,\tiny[-1.121,0.056] &  & $-0.094$\,\tiny[-0.501,0.314] & $-0.456$\,\tiny[-0.743,-0.169] &  & Yes & Yes \\
skewness, monthly & $-0.155$\,\tiny[-0.443,0.133] & $-0.369$\,\tiny[-0.930,0.191] &  & $-0.751$\,\tiny[-1.288,-0.214] & $-0.373$\,\tiny[-0.764,0.017] &  & Yes & Yes \\
skewness, quarterly & $0.170$\,\tiny[-0.194,0.535] & $-0.496$\,\tiny[-0.904,-0.088] &  & $-0.268$\,\tiny[-0.641,0.106] & $-0.480$\,\tiny[-0.922,-0.038] &  & Yes & Yes \\
mean pairwise correlation & $0.334$\,\tiny[0.273,0.394] & $0.389$\,\tiny[0.276,0.501] &  & $0.379$\,\tiny[0.318,0.440] & $0.391$\,\tiny[0.306,0.476] &  & Yes & Yes \\
correlation, across pairs & $0.073$\,\tiny[0.056,0.090] & $0.076$\,\tiny[0.065,0.087] &  & $0.107$\,\tiny[0.077,0.136] & $0.129$\,\tiny[0.107,0.151] &  & Yes & Yes \\
cyclical against sensitive & $0.548$\,\tiny[0.488,0.608] & $0.606$\,\tiny[0.529,0.684] &  & $0.498$\,\tiny[0.442,0.553] & $0.469$\,\tiny[0.399,0.539] &  & Yes & Yes \\
cyclical against defensive & $0.553$\,\tiny[0.495,0.610] & $0.574$\,\tiny[0.486,0.661] &  & $0.520$\,\tiny[0.468,0.572] & $0.213$\,\tiny[0.139,0.287] & $\bullet$ & No & No \\
sensitive against defensive & $0.572$\,\tiny[0.497,0.647] & $0.716$\,\tiny[0.637,0.795] &  & $0.473$\,\tiny[0.403,0.543] & $0.378$\,\tiny[0.295,0.461] &  & Yes & Yes \\
market beta, spread & $0.232$\,\tiny[0.178,0.286] & $0.206$\,\tiny[0.149,0.262] &  & $0.281$\,\tiny[0.218,0.345] & $0.338$\,\tiny[0.309,0.367] &  & Yes & Yes \\
name volatility, mean & $0.289$\,\tiny[0.264,0.314] & $0.271$\,\tiny[0.244,0.297] &  & $0.235$\,\tiny[0.210,0.261] & $0.312$\,\tiny[0.278,0.347] & $\bullet$ & No & Yes \\
name volatility, spread & $0.039$\,\tiny[0.028,0.049] & $0.036$\,\tiny[0.023,0.049] &  & $0.049$\,\tiny[0.038,0.060] & $0.060$\,\tiny[0.048,0.072] &  & Yes & Yes \\
vol variation, median name & $0.915$\,\tiny[0.853,0.977] & $0.811$\,\tiny[0.758,0.865] &  & $0.758$\,\tiny[0.699,0.818] & $0.814$\,\tiny[0.767,0.861] &  & Yes & Yes \\
vol variation, 90th pct & $0.958$\,\tiny[0.897,1.019] & $0.913$\,\tiny[0.839,0.987] &  & $0.889$\,\tiny[0.807,0.970] & $0.915$\,\tiny[0.862,0.967] &  & Yes & Yes \\
spread of log size & $0.573$\,\tiny[0.516,0.630] & $0.317$\,\tiny[0.266,0.369] & $\bullet$ & $0.645$\,\tiny[0.531,0.759] & $0.276$\,\tiny[0.230,0.323] & $\bullet$ & No & No \\
effective names, share & $0.801$\,\tiny[0.782,0.819] & $0.909$\,\tiny[0.889,0.930] & $\bullet$ & $0.662$\,\tiny[0.586,0.739] & $0.917$\,\tiny[0.892,0.942] & $\bullet$ & No & No \\
name leverage, week & $-0.061$\,\tiny[-0.081,-0.041] & $-0.044$\,\tiny[-0.072,-0.016] &  & $-0.054$\,\tiny[-0.073,-0.034] & $-0.036$\,\tiny[-0.051,-0.020] &  & Yes & Yes \\
name leverage, month & $0.008$\,\tiny[-0.037,0.053] & $-0.025$\,\tiny[-0.100,0.050] &  & $-0.020$\,\tiny[-0.066,0.027] & $-0.020$\,\tiny[-0.096,0.056] &  & Yes & Yes \\
basket leverage, week & $-0.096$\,\tiny[-0.133,-0.060] & $-0.030$\,\tiny[-0.085,0.026] &  & $-0.109$\,\tiny[-0.149,-0.068] & $-0.011$\,\tiny[-0.078,0.055] &  & Yes & Yes \\
basket vol persistence, month & $0.717$\,\tiny[0.652,0.781] & $0.488$\,\tiny[0.374,0.602] & $\bullet$ & $0.595$\,\tiny[0.516,0.674] & $0.476$\,\tiny[0.361,0.592] &  & No & Yes \\
name $|r|$ autocorrelation, day & $0.243$\,\tiny[0.216,0.269] & $0.208$\,\tiny[0.161,0.256] &  & $0.250$\,\tiny[0.180,0.321] & $0.208$\,\tiny[0.174,0.242] &  & Yes & Yes \\
name $|r|$ autocorrelation, month & $0.157$\,\tiny[0.132,0.181] & $0.070$\,\tiny[0.019,0.121] & $\bullet$ & $0.076$\,\tiny[0.052,0.100] & $0.071$\,\tiny[0.033,0.109] &  & No & Yes \\
name own-vol persistence, month & $0.332$\,\tiny[0.291,0.373] & $0.312$\,\tiny[0.261,0.364] &  & $0.262$\,\tiny[0.220,0.304] & $0.318$\,\tiny[0.277,0.359] &  & Yes & Yes \\
\addlinespace
\multicolumn{7}{l}{\textbf{the rest of the index}} \\
annual drift & $0.076$\,\tiny[0.021,0.131] & $0.127$\,\tiny[0.105,0.149] &  & $0.130$\,\tiny[0.104,0.156] & $0.125$\,\tiny[0.105,0.145] &  & Yes & Yes \\
annual volatility & $0.210$\,\tiny[0.183,0.236] & $0.190$\,\tiny[0.136,0.244] &  & $0.170$\,\tiny[0.151,0.189] & $0.192$\,\tiny[0.150,0.234] &  & Yes & Yes \\
excess kurtosis & $6.8$\,\tiny[4.5,9.2] & $8.9$\,\tiny[6.3,11.5] &  & $14.5$\,\tiny[8.2,20.9] & $9.3$\,\tiny[6.5,12.1] &  & Yes & Yes \\
falls beyond 2\% a year & $12.9$\,\tiny[9.6,16.3] & $10.1$\,\tiny[6.1,14.1] &  & $8.2$\,\tiny[6.4,9.9] & $9.3$\,\tiny[5.0,13.7] &  & Yes & Yes \\
mean depth, 20\% falls & $0.360$\,\tiny[0.314,0.405] & $0.273$\,\tiny[0.210,0.335] &  & $0.257$\,\tiny[0.215,0.300] & $0.271$\,\tiny[0.224,0.317] &  & Yes & Yes \\
20\% falls per century & $26.7$\,\tiny[20.3,33.1] & $26.7$\,\tiny[14.3,39.1] &  & $21.5$\,\tiny[15.1,27.9] & $28.6$\,\tiny[12.7,44.5] &  & Yes & Yes \\
variance ratio, month & $0.753$\,\tiny[0.709,0.797] & $0.901$\,\tiny[0.804,0.999] & $\bullet$ & $0.768$\,\tiny[0.713,0.823] & $0.883$\,\tiny[0.782,0.984] &  & No & Yes \\
variance ratio, quarter & $0.738$\,\tiny[0.664,0.812] & $0.678$\,\tiny[0.510,0.846] &  & $0.570$\,\tiny[0.510,0.630] & $0.704$\,\tiny[0.553,0.855] &  & Yes & Yes \\
variance ratio, 6 months & $0.852$\,\tiny[0.709,0.995] & $0.592$\,\tiny[0.435,0.748] &  & $0.481$\,\tiny[0.384,0.578] & $0.530$\,\tiny[0.361,0.698] &  & Yes & Yes \\
variance ratio, year & $0.947$\,\tiny[0.744,1.149] & $0.419$\,\tiny[0.235,0.604] & $\bullet$ & $0.457$\,\tiny[0.317,0.597] & $0.407$\,\tiny[0.267,0.548] &  & No & Yes \\
skewness, daily & $-0.155$\,\tiny[-0.251,-0.058] & $-0.363$\,\tiny[-0.714,-0.012] &  & $-0.801$\,\tiny[-1.043,-0.559] & $-0.374$\,\tiny[-0.729,-0.019] &  & Yes & Yes \\
skewness, weekly & $-0.587$\,\tiny[-0.937,-0.237] & $-0.528$\,\tiny[-1.020,-0.036] &  & $-0.431$\,\tiny[-0.813,-0.048] & $-0.606$\,\tiny[-1.027,-0.185] &  & Yes & Yes \\
skewness, monthly & $-0.274$\,\tiny[-0.809,0.261] & $-0.594$\,\tiny[-1.259,0.072] &  & $-1.147$\,\tiny[-2.029,-0.264] & $-0.592$\,\tiny[-1.247,0.064] &  & Yes & Yes \\
skewness, quarterly & $-0.381$\,\tiny[-0.941,0.180] & $-0.656$\,\tiny[-1.165,-0.147] &  & $-1.396$\,\tiny[-1.863,-0.928] & $-0.764$\,\tiny[-1.188,-0.340] &  & Yes & Yes \\
mean pairwise correlation & $0.300$\,\tiny[0.250,0.350] & $0.289$\,\tiny[0.187,0.392] &  & $0.384$\,\tiny[0.346,0.423] & $0.303$\,\tiny[0.219,0.388] &  & Yes & Yes \\
correlation, across pairs & $0.097$\,\tiny[0.090,0.103] & $0.102$\,\tiny[0.091,0.112] &  & $0.119$\,\tiny[0.112,0.126] & $0.093$\,\tiny[0.076,0.110] & $\bullet$ & No & Yes \\
correlation, across years & $0.113$\,\tiny[0.090,0.137] & $0.108$\,\tiny[0.058,0.159] &  & $0.124$\,\tiny[0.105,0.143] & $0.114$\,\tiny[0.081,0.147] &  & Yes & Yes \\
correlation against index volatility & $0.797$\,\tiny[0.645,0.949] & $0.967$\,\tiny[0.952,0.983] & $\bullet$ & $0.856$\,\tiny[0.801,0.912] & $0.969$\,\tiny[0.958,0.980] & $\bullet$ & No & Yes \\
within less between industry groups & $0.057$\,\tiny[0.049,0.066] & $0.061$\,\tiny[0.046,0.076] &  & $0.060$\,\tiny[0.053,0.067] & $0.060$\,\tiny[0.042,0.077] &  & Yes & Yes \\
cyclical against sensitive & $0.777$\,\tiny[0.735,0.820] & $0.858$\,\tiny[0.835,0.881] & $\bullet$ & $0.902$\,\tiny[0.878,0.927] & $0.855$\,\tiny[0.835,0.876] & $\bullet$ & No & Yes \\
cyclical against defensive & $0.708$\,\tiny[0.665,0.751] & $0.752$\,\tiny[0.719,0.785] &  & $0.816$\,\tiny[0.790,0.843] & $0.797$\,\tiny[0.756,0.839] &  & Yes & Yes \\
sensitive against defensive & $0.629$\,\tiny[0.529,0.728] & $0.734$\,\tiny[0.703,0.764] &  & $0.764$\,\tiny[0.719,0.809] & $0.782$\,\tiny[0.745,0.819] &  & Yes & Yes \\
market beta, spread & $0.307$\,\tiny[0.277,0.337] & $0.333$\,\tiny[0.311,0.355] &  & $0.281$\,\tiny[0.263,0.299] & $0.295$\,\tiny[0.283,0.308] &  & Yes & Yes \\
name volatility, mean & $0.371$\,\tiny[0.335,0.407] & $0.337$\,\tiny[0.307,0.366] &  & $0.285$\,\tiny[0.260,0.310] & $0.341$\,\tiny[0.316,0.365] & $\bullet$ & No & Yes \\
name volatility, spread & $0.099$\,\tiny[0.084,0.115] & $0.065$\,\tiny[0.060,0.070] & $\bullet$ & $0.067$\,\tiny[0.059,0.074] & $0.059$\,\tiny[0.052,0.066] &  & No & Yes \\
first component, vol changes & $0.290$\,\tiny[0.250,0.330] & $0.359$\,\tiny[0.312,0.405] &  & $0.351$\,\tiny[0.296,0.405] & $0.361$\,\tiny[0.320,0.403] &  & Yes & Yes \\
correlation, vol changes & $0.272$\,\tiny[0.232,0.313] & $0.339$\,\tiny[0.288,0.390] &  & $0.333$\,\tiny[0.276,0.390] & $0.346$\,\tiny[0.304,0.387] &  & Yes & Yes \\
vol variation, median name & $0.813$\,\tiny[0.765,0.861] & $0.824$\,\tiny[0.774,0.873] &  & $0.746$\,\tiny[0.683,0.809] & $0.822$\,\tiny[0.778,0.867] &  & Yes & Yes \\
vol variation, 90th pct & $1.019$\,\tiny[0.938,1.101] & $0.931$\,\tiny[0.881,0.981] &  & $0.876$\,\tiny[0.812,0.939] & $0.936$\,\tiny[0.894,0.978] &  & Yes & Yes \\
spread of log size & $1.072$\,\tiny[1.041,1.102] & $0.930$\,\tiny[0.903,0.957] & $\bullet$ & $0.960$\,\tiny[0.945,0.975] & $0.895$\,\tiny[0.879,0.911] & $\bullet$ & No & No \\
effective names, share & $0.312$\,\tiny[0.295,0.328] & $0.412$\,\tiny[0.397,0.426] & $\bullet$ & $0.308$\,\tiny[0.280,0.336] & $0.341$\,\tiny[0.318,0.365] &  & No & No \\
removals per name-year & $0.058$\,\tiny[0.051,0.064] & $0.064$\,\tiny[0.054,0.074] &  & $0.040$\,\tiny[0.037,0.043] & $0.054$\,\tiny[0.044,0.063] & $\bullet$ & No & Yes \\
defaults, bp a name-year & $9.9$\,\tiny[6.3,13.5] & $6.0$\,\tiny[2.6,9.4] &  & $3.892$\,\tiny[1.675,6.109] & $4.744$\,\tiny[1.582,7.906] &  & Yes & Yes \\
name leverage, week & $-0.064$\,\tiny[-0.074,-0.055] & $-0.058$\,\tiny[-0.069,-0.046] &  & $-0.026$\,\tiny[-0.041,-0.011] & $-0.054$\,\tiny[-0.070,-0.038] &  & Yes & Yes \\
name leverage, month & $-0.058$\,\tiny[-0.091,-0.026] & $-0.023$\,\tiny[-0.059,0.013] &  & $-0.052$\,\tiny[-0.093,-0.010] & $-0.029$\,\tiny[-0.067,0.010] &  & Yes & Yes \\
basket leverage, week & $-0.142$\,\tiny[-0.180,-0.104] & $-0.027$\,\tiny[-0.065,0.012] & $\bullet$ & $-0.099$\,\tiny[-0.134,-0.064] & $-0.016$\,\tiny[-0.047,0.014] & $\bullet$ & No & Yes \\
basket vol persistence, month & $0.746$\,\tiny[0.672,0.820] & $0.486$\,\tiny[0.377,0.596] & $\bullet$ & $0.543$\,\tiny[0.455,0.631] & $0.485$\,\tiny[0.384,0.587] &  & No & Yes \\
name $|r|$ autocorrelation, day & $0.228$\,\tiny[0.203,0.252] & $0.248$\,\tiny[0.228,0.269] &  & $0.230$\,\tiny[0.167,0.293] & $0.248$\,\tiny[0.231,0.266] &  & Yes & Yes \\
name $|r|$ autocorrelation, month & $0.154$\,\tiny[0.123,0.184] & $0.078$\,\tiny[0.056,0.099] & $\bullet$ & $0.105$\,\tiny[0.079,0.131] & $0.082$\,\tiny[0.061,0.102] &  & No & Yes \\
name own-vol persistence, month & $0.383$\,\tiny[0.356,0.409] & $0.394$\,\tiny[0.381,0.407] &  & $0.293$\,\tiny[0.271,0.316] & $0.389$\,\tiny[0.374,0.403] & $\bullet$ & No & No \\
\addlinespace
\midrule
the ten largest, ranges that meet & \multicolumn{3}{c}{85\%} & \multicolumn{3}{c}{88\%} & 79\% & 91\% \\
the rest of the index, ranges that meet & \multicolumn{3}{c}{75\%} & \multicolumn{3}{c}{80\%} & 65\% & 92\% \\
\bottomrule
\end{tabular}
\caption{The calibration against each window, each group scored on its capitalization-weighted basket of survivors. Brackets give one standard deviation either side, the market's block-bootstrap error and the model's spread across sixteen markets the search never read. A bullet marks ranges that do not meet, and the last columns read Yes when they meet in both windows, at one and at two standard deviations.}
\label{tab:groupjointfifty}
\end{table}

\begin{table}[htbp]
\centering
\caption{Dynamics of single-name volatility within each size group, monthly realized volatility relative to each name's own median. A dagger marks a row whose counterpart is a calibration target; ranges and marks as in Table~\ref{tab:groupjointfifty}.}
\label{tab:volpredicted}
\vspace{0.4em}
\scriptsize
\setlength{\tabcolsep}{2.4pt}
\begin{tabular}{>{\raggedright\arraybackslash}p{4.3cm} rr rr}
\toprule
& \multicolumn{2}{c}{1996--2010} & \multicolumn{2}{c}{2011--2024} \\
\cmidrule(lr){2-3}\cmidrule(lr){4-5}
Observable & market & model & market & model \\
\midrule
\multicolumn{5}{l}{\textbf{ten largest}} \\
\multicolumn{5}{l}{\emph{size of bursts}} \\
99th pct.\ of monthly vol / own median & $3.36$\,\tiny[2.96,3.75] & $3.36$\,\tiny[2.83,3.89] & $3.09$\,\tiny[2.10,4.08] & $3.49$\,\tiny[2.80,4.18] \\
99.9th pct. & $4.93$\,\tiny[3.97,5.89] & $5.44$\,\tiny[3.48,7.40] & $6.06$\,\tiny[4.33,7.79] & $5.44$\,\tiny[3.11,7.77] \\
months above 5 times, share & $0.0012$\,\tiny[0.0003,0.0020] & $0.0018$\,\tiny[-0.0006,0.0042] & $0.0054$\,\tiny[0.0013,0.0095] & $0.0020$\,\tiny[-0.0011,0.0050] \\
\multicolumn{5}{l}{\emph{persistence of bursts}} \\
names with two months running above 3 times & $0.40$\,\tiny[0.21,0.59] & $0.39$\,\tiny[0.12,0.67] & $0.20$\,\tiny[0.08,0.32] & $0.39$\,\tiny[0.08,0.71] \\
mean run above 3 times, months & $1.30$\,\tiny[1.16,1.44] & $1.25$\,\tiny[1.06,1.45] & $1.20$\,\tiny[1.05,1.35] & $1.28$\,\tiny[1.02,1.53] \\
\multicolumn{5}{l}{\emph{month-to-month autocorrelation of log vol}} \\
a name's & $0.69$\,\tiny[0.64,0.75] & $0.45$\,\tiny[0.35,0.54]$^{\bullet}$ & $0.43$\,\tiny[0.37,0.50] & $0.46$\,\tiny[0.35,0.56] \\
\quad its market-wide part & $0.82$\,\tiny[0.76,0.87] & $0.56$\,\tiny[0.43,0.69]$^{\bullet}$ & $0.61$\,\tiny[0.53,0.69] & $0.56$\,\tiny[0.44,0.69] \\
\quad its name-specific part$^{\dagger}$ & $0.32$\,\tiny[0.29,0.36] & $0.32$\,\tiny[0.28,0.36] & $0.26$\,\tiny[0.22,0.30] & $0.33$\,\tiny[0.29,0.36] \\
\multicolumn{5}{l}{\emph{autocorrelation of a name's $|r|$}} \\
at a day$^{\dagger}$ & $0.23$\,\tiny[0.20,0.26] & $0.23$\,\tiny[0.17,0.28] & $0.25$\,\tiny[0.19,0.31] & $0.24$\,\tiny[0.18,0.30] \\
at a month$^{\dagger}$ & $0.16$\,\tiny[0.13,0.18] & $0.09$\,\tiny[0.04,0.14] & $0.08$\,\tiny[0.06,0.10] & $0.09$\,\tiny[0.04,0.15] \\
at half a year & $0.09$\,\tiny[0.07,0.12] & $0.03$\,\tiny[-0.01,0.06]$^{\bullet}$ & $0.04$\,\tiny[0.02,0.05] & $0.02$\,\tiny[-0.01,0.06] \\
\multicolumn{5}{l}{\emph{price-volatility correlation}} \\
return against next month's change in log vol$^{\dagger}$ & $-0.015$\,\tiny[-0.056,0.027] & $-0.019$\,\tiny[-0.090,0.051] & $-0.020$\,\tiny[-0.067,0.027] & $-0.029$\,\tiny[-0.098,0.039] \\
ranges that meet & \multicolumn{2}{c}{9 of 12} & \multicolumn{2}{c}{12 of 12} \\
\midrule
\multicolumn{5}{l}{\textbf{rest of the index}} \\
\multicolumn{5}{l}{\emph{size of bursts}} \\
99th pct.\ of monthly vol / own median & $3.97$\,\tiny[3.47,4.47] & $3.24$\,\tiny[2.98,3.49] & $3.42$\,\tiny[2.73,4.11] & $3.25$\,\tiny[2.96,3.55] \\
99.9th pct. & $7.87$\,\tiny[6.75,8.99] & $5.12$\,\tiny[4.39,5.85]$^{\bullet}$ & $6.92$\,\tiny[5.70,8.15] & $5.20$\,\tiny[4.24,6.16] \\
months above 5 times, share & $0.0049$\,\tiny[0.0029,0.0068] & $0.0012$\,\tiny[0.0002,0.0022]$^{\bullet}$ & $0.0041$\,\tiny[0.0012,0.0070] & $0.0013$\,\tiny[0.0001,0.0025] \\
\multicolumn{5}{l}{\emph{persistence of bursts}} \\
names with two months running above 3 times & $0.36$\,\tiny[0.24,0.47] & $0.15$\,\tiny[0.06,0.24] & $0.19$\,\tiny[0.11,0.27] & $0.17$\,\tiny[0.06,0.28] \\
mean run above 3 times, months & $1.70$\,\tiny[1.51,1.89] & $1.18$\,\tiny[1.08,1.27]$^{\bullet}$ & $1.26$\,\tiny[1.16,1.36] & $1.19$\,\tiny[1.07,1.31] \\
\multicolumn{5}{l}{\emph{month-to-month autocorrelation of log vol}} \\
a name's & $0.62$\,\tiny[0.57,0.68] & $0.43$\,\tiny[0.38,0.47]$^{\bullet}$ & $0.48$\,\tiny[0.42,0.54] & $0.44$\,\tiny[0.39,0.48] \\
\quad its market-wide part & $0.84$\,\tiny[0.78,0.91] & $0.52$\,\tiny[0.42,0.63]$^{\bullet}$ & $0.66$\,\tiny[0.57,0.74] & $0.52$\,\tiny[0.42,0.62] \\
\quad its name-specific part$^{\dagger}$ & $0.36$\,\tiny[0.34,0.38] & $0.37$\,\tiny[0.36,0.38] & $0.30$\,\tiny[0.27,0.32] & $0.38$\,\tiny[0.36,0.39]$^{\bullet}$ \\
\multicolumn{5}{l}{\emph{autocorrelation of a name's $|r|$}} \\
at a day$^{\dagger}$ & $0.21$\,\tiny[0.19,0.23] & $0.25$\,\tiny[0.23,0.28] & $0.20$\,\tiny[0.16,0.25] & $0.25$\,\tiny[0.23,0.28] \\
at a month$^{\dagger}$ & $0.13$\,\tiny[0.10,0.16] & $0.08$\,\tiny[0.06,0.10] & $0.09$\,\tiny[0.07,0.12] & $0.08$\,\tiny[0.06,0.11] \\
at half a year & $0.07$\,\tiny[0.05,0.08] & $0.01$\,\tiny[-0.00,0.03]$^{\bullet}$ & $0.03$\,\tiny[0.02,0.04] & $0.02$\,\tiny[0.00,0.03] \\
\multicolumn{5}{l}{\emph{price-volatility correlation}} \\
return against next month's change in log vol$^{\dagger}$ & $-0.080$\,\tiny[-0.109,-0.052] & $-0.023$\,\tiny[-0.054,0.009] & $-0.059$\,\tiny[-0.097,-0.022] & $-0.028$\,\tiny[-0.060,0.004] \\
ranges that meet & \multicolumn{2}{c}{6 of 12} & \multicolumn{2}{c}{11 of 12} \\
\bottomrule
\end{tabular}
\end{table}

To explore possible evidence for non-stationarity, we then fine-tuned each group's
vector on one window alone, starting from the joint vector (Tables~\ref{tab:groupparams} and~\ref{tab:grouptuned}). The
fine-tunes change the count of matches by at most three and serve the window
comparisons of Section~\ref{sec:stationarity}, while the model of the rest of the
paper is the joint pair. They also test the split into two
groups.\label{sec:groupsummary} Across the joint vector and the two fine-tunes,
six parameters that separate the groups, among them the reversion $\kappa_x$ and
the mispricing noise $\sigma_x$, keep the ten largest on the same side of the
rest.

\begin{table}[htbp]
\centering
\scriptsize
\begin{tabular}{l rr rr}
\toprule
& \multicolumn{2}{c}{the ten largest} & \multicolumn{2}{c}{the rest of the index} \\
\cmidrule(lr){2-3}\cmidrule(lr){4-5}
Parameter & 1996--2010 & 2011--2024 & 1996--2010 & 2011--2024 \\
\midrule
g & $0.1180$ & $0.0882$ & $0.5345$ & $0.4758$ \\
$\bar\epsilon$ & $0.2727$ & $0.0678$ & $0.2824$ & $0.2710$ \\
$\hat x$, derived & $-2.456$ & $-2.498$ & $-0.958$ & $-1.059$ \\
$c$ flow impact & $2.589$ & $2.287$ & $3.165$ & $2.480$ \\
$m$ & $1.1207$ & $1.4672$ & $0.3177$ & $0.4817$ \\
$\sigma_x$ & $0.1791$ & $0.1476$ & $0.2791$ & $0.2496$ \\
$\kappa_x$ & $3.968$ & $4.266$ & $1.656$ & $1.499$ \\
$\sigma_\eta$ & $0.0532$ & $0.0446$ & $0.0257$ & $0.1081$ \\
$\sigma_F$ & $0.0585$ & $0.0487$ & $0.0402$ & $0.0415$ \\
$\bar\beta$ mean market loading & $0.6433$ & $0.6562$ & $0.4368$ & $0.5213$ \\
$\lambda_M$ & $2.861$ & $3.454$ & $1.799$ & $1.462$ \\
$\jmath_M$ & $0.0184$ & $0.0151$ & $0.0238$ & $0.0293$ \\
$\alpha$ & $0.4514$ & $0.2043$ & $0.2075$ & $0.0478$ \\
$\beta_H$, excitation decay (yr$^{-1}$) & $29.77$ & $32.70$ & $42.45$ & $39.98$ \\
$\lambda_S$ & $11.782$ & $10.460$ & $7.789$ & $10.027$ \\
$\jmath_S$ routine jump size & $0.0063$ & $0.0053$ & $0.0136$ & $0.0162$ \\
$\lambda_I$ & $5.015$ & $4.826$ & $1.239$ & $1.191$ \\
$\sigma_{JI}$ firm jump size & $0.0565$ & $0.0439$ & $0.0468$ & $0.0308$ \\
$s_M$ & $1.2714$ & $1.4192$ & $1.1007$ & $1.7488$ \\
$s_c$ & $0.2931$ & $0.3285$ & $0.6434$ & $0.7500$ \\
$s_n$ & $1.6105$ & $1.0611$ & $0.9804$ & $1.4641$ \\
$s_\lambda$ & $0.5671$ & $0.6376$ & $0.7141$ & $0.7362$ \\
$\mathrm{sd}(\beta)$ & $0.0308$ & $0.0337$ & $0.0783$ & $0.1506$ \\
$d_\sigma$ & $0.1067$ & $0.2807$ & $0.1128$ & $0.2311$ \\
$\beta_{\rm sec}$ & $0.1343$ & $0.0009$ & $0.2134$ & $0.1900$ \\
$\delta_{\rm exit}$ & off & off & $0.3459$ & $0.4214$ \\
$\rho_n$ name leverage & $-0.231$ & $-0.518$ & $-0.112$ & $-0.246$ \\
$w_n$ slow share of a name's volatility & $0.068$ & $0.106$ & $0.327$ & $0.150$ \\
$\rho_{ns}$ leverage of that slow mode & $-0.307$ & $-0.486$ & $-0.281$ & $-0.280$ \\
\bottomrule
\end{tabular}
\caption{The fine-tuned parameters: each group's vector of Table~\ref{tab:bbmparams} refitted on one window alone by the same search as the joint fit, starting from the joint vector. Table~\ref{tab:grouptuned} shows what they produce. These vectors serve the window-specific experiments only; the paper's model is the joint pair.}
\label{tab:groupparams}
\end{table}

\begin{table}[htbp]
\centering
\tiny
\setlength{\tabcolsep}{1.0pt}
\begin{tabular}{l rrc rrc cc}
\toprule
& \multicolumn{3}{c}{1996--2010} & \multicolumn{3}{c}{2011--2024} & \multicolumn{2}{c}{inside} \\
\cmidrule(lr){2-4}\cmidrule(lr){5-7}\cmidrule(lr){8-9}
Observable & market & model & & market & model & & $1\sigma$ & $2\sigma$ \\
\midrule
\multicolumn{7}{l}{\textbf{the ten largest}} \\
annual drift & $0.090$\,\tiny[0.045,0.135] & $0.119$\,\tiny[0.098,0.140] &  & $0.150$\,\tiny[0.123,0.177] & $0.137$\,\tiny[0.107,0.167] &  & Yes & Yes \\
annual volatility & $0.196$\,\tiny[0.177,0.214] & $0.206$\,\tiny[0.160,0.252] &  & $0.178$\,\tiny[0.154,0.201] & $0.194$\,\tiny[0.158,0.231] &  & Yes & Yes \\
excess kurtosis & $6.5$\,\tiny[3.9,9.1] & $9.2$\,\tiny[5.4,13.0] &  & $12.1$\,\tiny[6.5,17.6] & $11.7$\,\tiny[6.2,17.2] &  & Yes & Yes \\
falls beyond 2\% a year & $10.1$\,\tiny[7.9,12.4] & $11.2$\,\tiny[7.2,15.3] &  & $8.7$\,\tiny[6.3,11.0] & $10.1$\,\tiny[6.6,13.6] &  & Yes & Yes \\
mean depth, 20\% falls & $0.301$\,\tiny[0.249,0.354] & $0.289$\,\tiny[0.252,0.326] &  & $0.257$\,\tiny[0.232,0.281] & $0.275$\,\tiny[0.227,0.323] &  & Yes & Yes \\
20\% falls per century & $26.7$\,\tiny[20.3,33.1] & $26.7$\,\tiny[10.6,42.8] &  & $14.3$\,\tiny[7.9,20.7] & $21.5$\,\tiny[13.5,29.4] &  & Yes & Yes \\
variance ratio, month & $0.702$\,\tiny[0.633,0.772] & $0.847$\,\tiny[0.773,0.921] & $\bullet$ & $0.725$\,\tiny[0.688,0.762] & $0.824$\,\tiny[0.697,0.950] &  & No & Yes \\
variance ratio, quarter & $0.590$\,\tiny[0.529,0.651] & $0.654$\,\tiny[0.456,0.853] &  & $0.544$\,\tiny[0.471,0.617] & $0.574$\,\tiny[0.467,0.682] &  & Yes & Yes \\
variance ratio, 6 months & $0.642$\,\tiny[0.539,0.746] & $0.517$\,\tiny[0.319,0.716] &  & $0.407$\,\tiny[0.322,0.491] & $0.426$\,\tiny[0.343,0.508] &  & Yes & Yes \\
variance ratio, year & $0.794$\,\tiny[0.613,0.975] & $0.413$\,\tiny[0.243,0.583] & $\bullet$ & $0.382$\,\tiny[0.282,0.483] & $0.331$\,\tiny[0.281,0.382] &  & No & Yes \\
skewness, daily & $0.026$\,\tiny[-0.099,0.151] & $-0.432$\,\tiny[-0.779,-0.084] &  & $-0.394$\,\tiny[-0.524,-0.265] & $-0.064$\,\tiny[-0.455,0.327] &  & Yes & Yes \\
skewness, weekly & $-0.314$\,\tiny[-0.573,-0.056] & $-0.629$\,\tiny[-1.257,-0.002] &  & $-0.094$\,\tiny[-0.501,0.314] & $-0.475$\,\tiny[-0.921,-0.029] &  & Yes & Yes \\
skewness, monthly & $-0.155$\,\tiny[-0.443,0.133] & $-0.650$\,\tiny[-1.239,-0.060] &  & $-0.751$\,\tiny[-1.288,-0.214] & $-0.370$\,\tiny[-0.614,-0.126] &  & Yes & Yes \\
skewness, quarterly & $0.170$\,\tiny[-0.194,0.535] & $-0.697$\,\tiny[-1.352,-0.041] &  & $-0.268$\,\tiny[-0.641,0.106] & $-0.132$\,\tiny[-0.596,0.333] &  & Yes & Yes \\
mean pairwise correlation & $0.334$\,\tiny[0.273,0.394] & $0.405$\,\tiny[0.296,0.514] &  & $0.379$\,\tiny[0.318,0.440] & $0.399$\,\tiny[0.354,0.444] &  & Yes & Yes \\
correlation, across pairs & $0.073$\,\tiny[0.056,0.090] & $0.069$\,\tiny[0.060,0.079] &  & $0.107$\,\tiny[0.077,0.136] & $0.135$\,\tiny[0.120,0.150] &  & Yes & Yes \\
cyclical against sensitive & $0.548$\,\tiny[0.488,0.608] & $0.606$\,\tiny[0.511,0.701] &  & $0.498$\,\tiny[0.442,0.553] & $0.467$\,\tiny[0.393,0.540] &  & Yes & Yes \\
cyclical against defensive & $0.553$\,\tiny[0.495,0.610] & $0.565$\,\tiny[0.452,0.678] &  & $0.520$\,\tiny[0.468,0.572] & $0.219$\,\tiny[0.134,0.304] & $\bullet$ & No & No \\
sensitive against defensive & $0.572$\,\tiny[0.497,0.647] & $0.731$\,\tiny[0.630,0.833] &  & $0.473$\,\tiny[0.403,0.543] & $0.409$\,\tiny[0.357,0.460] &  & Yes & Yes \\
market beta, spread & $0.232$\,\tiny[0.178,0.286] & $0.214$\,\tiny[0.163,0.266] &  & $0.281$\,\tiny[0.218,0.345] & $0.423$\,\tiny[0.362,0.485] & $\bullet$ & No & Yes \\
name volatility, mean & $0.289$\,\tiny[0.264,0.314] & $0.295$\,\tiny[0.261,0.329] &  & $0.235$\,\tiny[0.210,0.261] & $0.272$\,\tiny[0.236,0.308] &  & Yes & Yes \\
name volatility, spread & $0.039$\,\tiny[0.028,0.049] & $0.036$\,\tiny[0.029,0.044] &  & $0.049$\,\tiny[0.038,0.060] & $0.083$\,\tiny[0.060,0.106] & $\bullet$ & No & Yes \\
vol variation, median name & $0.915$\,\tiny[0.853,0.977] & $0.879$\,\tiny[0.829,0.929] &  & $0.758$\,\tiny[0.699,0.818] & $0.759$\,\tiny[0.724,0.793] &  & Yes & Yes \\
vol variation, 90th pct & $0.958$\,\tiny[0.897,1.019] & $0.947$\,\tiny[0.869,1.025] &  & $0.889$\,\tiny[0.807,0.970] & $0.835$\,\tiny[0.789,0.880] &  & Yes & Yes \\
spread of log size & $0.573$\,\tiny[0.516,0.630] & $0.348$\,\tiny[0.311,0.384] & $\bullet$ & $0.645$\,\tiny[0.531,0.759] & $0.297$\,\tiny[0.262,0.331] & $\bullet$ & No & No \\
effective names, share & $0.801$\,\tiny[0.782,0.819] & $0.894$\,\tiny[0.874,0.914] & $\bullet$ & $0.662$\,\tiny[0.586,0.739] & $0.911$\,\tiny[0.879,0.943] & $\bullet$ & No & No \\
name leverage, week & $-0.061$\,\tiny[-0.081,-0.041] & $-0.058$\,\tiny[-0.098,-0.019] &  & $-0.054$\,\tiny[-0.073,-0.034] & $-0.076$\,\tiny[-0.119,-0.033] &  & Yes & Yes \\
name leverage, month & $0.008$\,\tiny[-0.037,0.053] & $-0.038$\,\tiny[-0.102,0.027] &  & $-0.020$\,\tiny[-0.066,0.027] & $-0.061$\,\tiny[-0.110,-0.012] &  & Yes & Yes \\
basket leverage, week & $-0.096$\,\tiny[-0.133,-0.060] & $-0.071$\,\tiny[-0.149,0.007] &  & $-0.109$\,\tiny[-0.149,-0.068] & $-0.030$\,\tiny[-0.092,0.031] &  & Yes & Yes \\
basket vol persistence, month & $0.717$\,\tiny[0.652,0.781] & $0.473$\,\tiny[0.384,0.562] & $\bullet$ & $0.595$\,\tiny[0.516,0.674] & $0.454$\,\tiny[0.358,0.551] &  & No & Yes \\
name $|r|$ autocorrelation, day & $0.243$\,\tiny[0.216,0.269] & $0.205$\,\tiny[0.175,0.234] &  & $0.250$\,\tiny[0.180,0.321] & $0.201$\,\tiny[0.176,0.225] &  & Yes & Yes \\
name $|r|$ autocorrelation, month & $0.157$\,\tiny[0.132,0.181] & $0.059$\,\tiny[0.031,0.088] & $\bullet$ & $0.076$\,\tiny[0.052,0.100] & $0.061$\,\tiny[0.035,0.088] &  & No & Yes \\
name own-vol persistence, month & $0.332$\,\tiny[0.291,0.373] & $0.289$\,\tiny[0.264,0.313] &  & $0.262$\,\tiny[0.220,0.304] & $0.294$\,\tiny[0.260,0.329] &  & Yes & Yes \\
\addlinespace
\multicolumn{7}{l}{\textbf{the rest of the index}} \\
annual drift & $0.076$\,\tiny[0.021,0.131] & $0.132$\,\tiny[0.116,0.147] &  & $0.130$\,\tiny[0.104,0.156] & $0.128$\,\tiny[0.107,0.149] &  & Yes & Yes \\
annual volatility & $0.210$\,\tiny[0.183,0.236] & $0.180$\,\tiny[0.130,0.229] &  & $0.170$\,\tiny[0.151,0.189] & $0.191$\,\tiny[0.163,0.220] &  & Yes & Yes \\
excess kurtosis & $6.8$\,\tiny[4.5,9.2] & $6.7$\,\tiny[4.7,8.8] &  & $14.5$\,\tiny[8.2,20.9] & $21.5$\,\tiny[1.9,41.1] &  & Yes & Yes \\
falls beyond 2\% a year & $12.9$\,\tiny[9.6,16.3] & $8.7$\,\tiny[4.1,13.2] &  & $8.2$\,\tiny[6.4,9.9] & $8.8$\,\tiny[6.9,10.7] &  & Yes & Yes \\
mean depth, 20\% falls & $0.360$\,\tiny[0.314,0.405] & $0.262$\,\tiny[0.215,0.309] & $\bullet$ & $0.257$\,\tiny[0.215,0.300] & $0.277$\,\tiny[0.240,0.313] &  & No & Yes \\
20\% falls per century & $26.7$\,\tiny[20.3,33.1] & $23.4$\,\tiny[8.5,38.2] &  & $21.5$\,\tiny[15.1,27.9] & $28.6$\,\tiny[18.0,39.2] &  & Yes & Yes \\
variance ratio, month & $0.753$\,\tiny[0.709,0.797] & $0.902$\,\tiny[0.817,0.988] & $\bullet$ & $0.768$\,\tiny[0.713,0.823] & $0.853$\,\tiny[0.674,1.032] &  & No & Yes \\
variance ratio, quarter & $0.738$\,\tiny[0.664,0.812] & $0.681$\,\tiny[0.457,0.905] &  & $0.570$\,\tiny[0.510,0.630] & $0.657$\,\tiny[0.539,0.775] &  & Yes & Yes \\
variance ratio, 6 months & $0.852$\,\tiny[0.709,0.995] & $0.548$\,\tiny[0.314,0.782] &  & $0.481$\,\tiny[0.384,0.578] & $0.555$\,\tiny[0.425,0.685] &  & Yes & Yes \\
variance ratio, year & $0.947$\,\tiny[0.744,1.149] & $0.406$\,\tiny[0.185,0.626] & $\bullet$ & $0.457$\,\tiny[0.317,0.597] & $0.425$\,\tiny[0.278,0.573] &  & No & Yes \\
skewness, daily & $-0.155$\,\tiny[-0.251,-0.058] & $-0.099$\,\tiny[-0.465,0.267] &  & $-0.801$\,\tiny[-1.043,-0.559] & $-0.298$\,\tiny[-1.188,0.593] &  & Yes & Yes \\
skewness, weekly & $-0.587$\,\tiny[-0.937,-0.237] & $-0.627$\,\tiny[-1.263,0.010] &  & $-0.431$\,\tiny[-0.813,-0.048] & $-0.918$\,\tiny[-1.603,-0.233] &  & Yes & Yes \\
skewness, monthly & $-0.274$\,\tiny[-0.809,0.261] & $-0.673$\,\tiny[-1.367,0.021] &  & $-1.147$\,\tiny[-2.029,-0.264] & $-0.650$\,\tiny[-1.567,0.267] &  & Yes & Yes \\
skewness, quarterly & $-0.381$\,\tiny[-0.941,0.180] & $-0.668$\,\tiny[-1.191,-0.146] &  & $-1.396$\,\tiny[-1.863,-0.928] & $-0.858$\,\tiny[-1.508,-0.207] &  & Yes & Yes \\
mean pairwise correlation & $0.300$\,\tiny[0.250,0.350] & $0.235$\,\tiny[0.155,0.315] &  & $0.384$\,\tiny[0.346,0.423] & $0.349$\,\tiny[0.295,0.404] &  & Yes & Yes \\
correlation, across pairs & $0.097$\,\tiny[0.090,0.103] & $0.092$\,\tiny[0.076,0.109] &  & $0.119$\,\tiny[0.112,0.126] & $0.125$\,\tiny[0.112,0.137] &  & Yes & Yes \\
correlation, across years & $0.113$\,\tiny[0.090,0.137] & $0.108$\,\tiny[0.067,0.148] &  & $0.124$\,\tiny[0.105,0.143] & $0.147$\,\tiny[0.107,0.186] &  & Yes & Yes \\
correlation against index volatility & $0.797$\,\tiny[0.645,0.949] & $0.966$\,\tiny[0.951,0.981] & $\bullet$ & $0.856$\,\tiny[0.801,0.912] & $0.973$\,\tiny[0.955,0.992] & $\bullet$ & No & Yes \\
within less between industry groups & $0.057$\,\tiny[0.049,0.066] & $0.068$\,\tiny[0.050,0.085] &  & $0.060$\,\tiny[0.053,0.067] & $0.056$\,\tiny[0.039,0.072] &  & Yes & Yes \\
cyclical against sensitive & $0.777$\,\tiny[0.735,0.820] & $0.814$\,\tiny[0.789,0.838] &  & $0.902$\,\tiny[0.878,0.927] & $0.898$\,\tiny[0.884,0.911] &  & Yes & Yes \\
cyclical against defensive & $0.708$\,\tiny[0.665,0.751] & $0.702$\,\tiny[0.677,0.726] &  & $0.816$\,\tiny[0.790,0.843] & $0.849$\,\tiny[0.819,0.878] &  & Yes & Yes \\
sensitive against defensive & $0.629$\,\tiny[0.529,0.728] & $0.693$\,\tiny[0.664,0.722] &  & $0.764$\,\tiny[0.719,0.809] & $0.843$\,\tiny[0.823,0.863] & $\bullet$ & No & Yes \\
market beta, spread & $0.307$\,\tiny[0.277,0.337] & $0.328$\,\tiny[0.311,0.346] &  & $0.281$\,\tiny[0.263,0.299] & $0.345$\,\tiny[0.323,0.367] & $\bullet$ & No & Yes \\
name volatility, mean & $0.371$\,\tiny[0.335,0.407] & $0.376$\,\tiny[0.355,0.397] &  & $0.285$\,\tiny[0.260,0.310] & $0.312$\,\tiny[0.290,0.335] &  & Yes & Yes \\
name volatility, spread & $0.099$\,\tiny[0.084,0.115] & $0.064$\,\tiny[0.053,0.074] & $\bullet$ & $0.067$\,\tiny[0.059,0.074] & $0.063$\,\tiny[0.057,0.069] &  & No & Yes \\
first component, vol changes & $0.290$\,\tiny[0.250,0.330] & $0.359$\,\tiny[0.310,0.408] &  & $0.351$\,\tiny[0.296,0.405] & $0.426$\,\tiny[0.380,0.472] &  & Yes & Yes \\
correlation, vol changes & $0.272$\,\tiny[0.232,0.313] & $0.333$\,\tiny[0.283,0.384] &  & $0.333$\,\tiny[0.276,0.390] & $0.408$\,\tiny[0.360,0.457] &  & Yes & Yes \\
vol variation, median name & $0.813$\,\tiny[0.765,0.861] & $0.802$\,\tiny[0.762,0.841] &  & $0.746$\,\tiny[0.683,0.809] & $0.808$\,\tiny[0.760,0.856] &  & Yes & Yes \\
vol variation, 90th pct & $1.019$\,\tiny[0.938,1.101] & $0.993$\,\tiny[0.949,1.037] &  & $0.876$\,\tiny[0.812,0.939] & $0.898$\,\tiny[0.854,0.941] &  & Yes & Yes \\
spread of log size & $1.072$\,\tiny[1.041,1.102] & $0.933$\,\tiny[0.902,0.964] & $\bullet$ & $0.960$\,\tiny[0.945,0.975] & $0.904$\,\tiny[0.877,0.930] & $\bullet$ & No & No \\
effective names, share & $0.312$\,\tiny[0.295,0.328] & $0.403$\,\tiny[0.386,0.419] & $\bullet$ & $0.308$\,\tiny[0.280,0.336] & $0.353$\,\tiny[0.329,0.376] &  & No & No \\
removals per name-year & $0.058$\,\tiny[0.051,0.064] & $0.050$\,\tiny[0.044,0.057] &  & $0.040$\,\tiny[0.037,0.043] & $0.045$\,\tiny[0.040,0.051] &  & Yes & Yes \\
defaults, bp a name-year & $9.9$\,\tiny[6.3,13.5] & $9.1$\,\tiny[-1.0,19.1] &  & $3.892$\,\tiny[1.675,6.109] & $7.339$\,\tiny[3.314,11.363] &  & Yes & Yes \\
name leverage, week & $-0.064$\,\tiny[-0.074,-0.055] & $-0.071$\,\tiny[-0.083,-0.058] &  & $-0.026$\,\tiny[-0.041,-0.011] & $-0.040$\,\tiny[-0.059,-0.021] &  & Yes & Yes \\
name leverage, month & $-0.058$\,\tiny[-0.091,-0.026] & $-0.041$\,\tiny[-0.061,-0.022] &  & $-0.052$\,\tiny[-0.093,-0.010] & $0.004$\,\tiny[-0.030,0.038] &  & Yes & Yes \\
basket leverage, week & $-0.142$\,\tiny[-0.180,-0.104] & $-0.117$\,\tiny[-0.149,-0.084] &  & $-0.099$\,\tiny[-0.134,-0.064] & $0.031$\,\tiny[-0.022,0.083] & $\bullet$ & No & Yes \\
basket vol persistence, month & $0.746$\,\tiny[0.672,0.820] & $0.566$\,\tiny[0.476,0.655] & $\bullet$ & $0.543$\,\tiny[0.455,0.631] & $0.434$\,\tiny[0.321,0.547] &  & No & Yes \\
name $|r|$ autocorrelation, day & $0.228$\,\tiny[0.203,0.252] & $0.224$\,\tiny[0.206,0.242] &  & $0.230$\,\tiny[0.167,0.293] & $0.273$\,\tiny[0.232,0.314] &  & Yes & Yes \\
name $|r|$ autocorrelation, month & $0.154$\,\tiny[0.123,0.184] & $0.077$\,\tiny[0.055,0.098] & $\bullet$ & $0.105$\,\tiny[0.079,0.131] & $0.071$\,\tiny[0.047,0.094] &  & No & Yes \\
name own-vol persistence, month & $0.383$\,\tiny[0.356,0.409] & $0.398$\,\tiny[0.381,0.414] &  & $0.293$\,\tiny[0.271,0.316] & $0.291$\,\tiny[0.281,0.301] &  & Yes & Yes \\
\addlinespace
\midrule
the ten largest, ranges that meet & \multicolumn{3}{c}{82\%} & \multicolumn{3}{c}{85\%} & 73\% & 91\% \\
the rest of the index, ranges that meet & \multicolumn{3}{c}{78\%} & \multicolumn{3}{c}{88\%} & 70\% & 95\% \\
\bottomrule
\end{tabular}
\caption{The observables of Table~\ref{tab:groupjointfifty} after each group's vector is fine-tuned on one window alone, scored on the same markets no search read.}
\label{tab:grouptuned}
\end{table}

\subsection{I-BBM: the index as a single marketron}
\label{sec:indexone}

We study the market as one tradeable asset with the I-BBM, which is
far cheaper to simulate than the BBM and keeps the states of
Eqs.~\eqref{eq:lmsvx}--\eqref{eq:volmult} exactly rather than as aggregates. Where a
result depends on the model, as the entropy production does, both are reported.

It is fitted to the group objective less what one asset cannot define, fourteen
targets a window: the drift and volatility, the excess kurtosis and the days
falling more than two percent, the depth and number of twenty percent falls, the
variance ratios from a month to a year and the skew from a day to a quarter.
Thirteen parameters are fitted (Table~\ref{tab:indexparams}). The investor's
memory constants are held from the first stage, as the groups hold them
(Table~\ref{tab:modelcompare}), and the size and the decay of self-excitation of a
market revaluation are held at the BBM's capitalization-weighted values. The
vector was fitted with these two at the values of an earlier group calibration,
and at the current values it meets the same targets.

Twelve of the fourteen ranges meet in each window (Table~\ref{tab:indexfit}),
every miss within about one band width. The first window misses the yearly
variance ratio, the market trending over a year where the one marketron reverts,
and the mean depth of a fall. The second misses the skew of a day and of a
quarter, which the one marketron puts near zero. Its mispricing reverts with a
half-life of four months, between the two groups and nearer the rest, which hold three quarters of the capitalization, so the I-BBM is the right object for an
experiment on the market as a whole and not a description of a typical name.

\begin{figure}[!ht]
\centering
\includegraphics[width=0.92\textwidth]{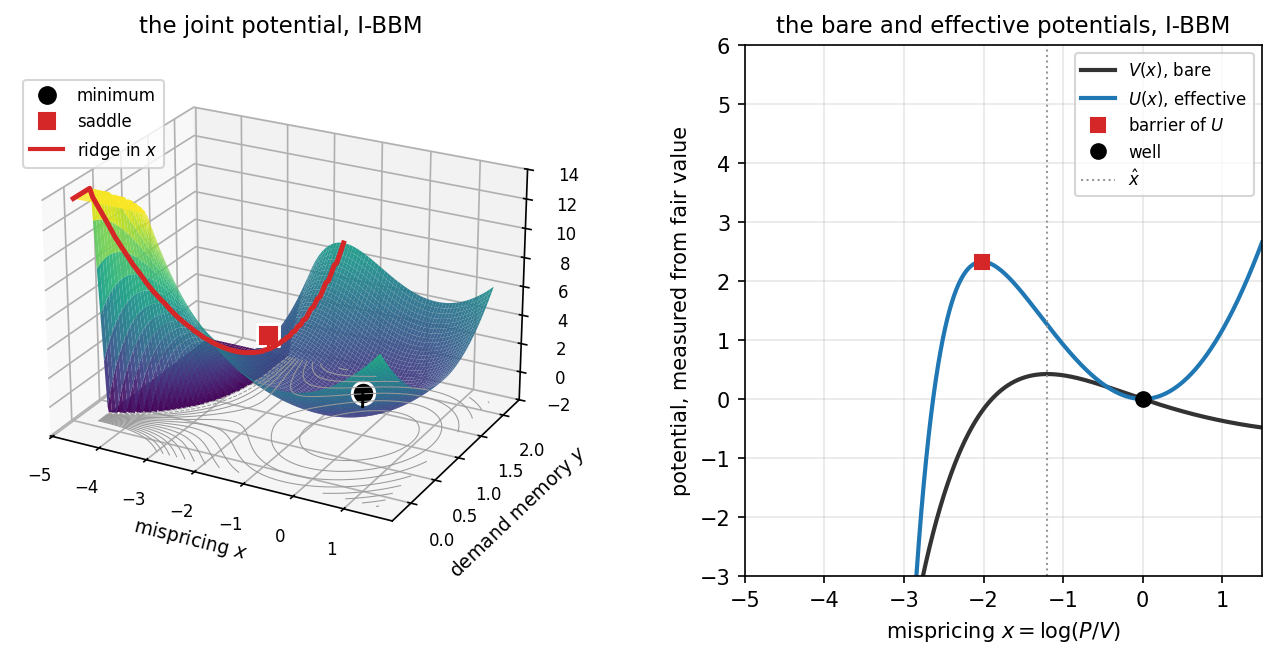}
\caption{The joint potential $\mathcal{V}(x,y)$ of Eq.~\eqref{eq:jointpot} at the
I-BBM's constants (left), and its bare $V(x)$ with the effective $U(x)$ of
Eq.~\eqref{eq:effpot} (right).}
\label{fig:ibbmpot}
\end{figure}

The missing skew is the price of stability. The landscape
(Figure~\ref{fig:ibbmpot}) has one well and one saddle, and the barrier of $U(x)$
lies at a fall of $87$ percent below value, which the index never reaches in
twenty-four runs of two hundred years. Vectors that move the barrier up to a fall
of $55$ to $65$ percent reproduce every target over fifteen years, the skew
included, but their index crosses it in most runs of two hundred years and does
not come back. The objective scores fifteen-year windows and cannot see an event
that arrives once in several decades, so the I-BBM keeps the deep barrier. That
one marketron cannot produce the daily skew without making a total collapse frequent
suggests that the index is hard to describe as one effective asset, since such a
description leaves out the removal and replacement of failing firms on which the
S\&P 500 is built.

\begin{table}[htbp]
\centering
\small
\begin{tabular}{lr}
\toprule
Parameter & Value \\
\midrule
\multicolumn{2}{l}{\emph{fitted}} \\
mispricing mean reversion $\kappa_x$ & $2.1051$ \\
flow impact $c$ & $1.7427$ \\
coupling constant $g$ & $0.8427$ \\
flow reversion rate $\mu$ & $6.1680$ \\
mispricing noise $\sigma_x$ & $0.1842$ \\
demand volatility $\sigma_y$ & $0.0725$ \\
risk penalty $\Lambda$ & $0.1605$ \\
passive share $b$ & $0.3595$ \\
volatility of the log variance $s_v$ & $1.1674$ \\
its reversion rate $\kappa_v$ & $5.9210$ \\
leverage correlation $\rho_v$ & $-0.3455$ \\
market revaluations a year $\lambda_M$ & $0.7516$ \\
share bred by earlier events $\alpha$ & $0.0257$ \\
\addlinespace
\multicolumn{2}{l}{\emph{held, not fitted}} \\
market jump size, the ten largest & $0.0168$ \\
market jump size, the rest & $0.0239$ \\
weight on the first & $0.2467$ \\
excitation decay rate, a year & $39.24$ \\
\bottomrule
\end{tabular}
\caption{The single-index parameters. The size of a market revaluation, a two-point draw over the group fits' sizes, and the decay of the excitation, their capitalization-weighted mean, are held rather than fitted.}
\label{tab:indexparams}
\end{table}

\begin{table}[htbp]
\centering
\footnotesize
\setlength{\tabcolsep}{1.6pt}
\begin{tabular}{l r@{\,}l c r@{\,}l c cc}
\toprule
& \multicolumn{3}{c}{1996--2010} & \multicolumn{3}{c}{2011--2024} & \multicolumn{2}{c}{inside} \\
\cmidrule(lr){2-4}\cmidrule(lr){5-7}\cmidrule(lr){8-9}
Observable & \multicolumn{2}{c}{market \quad model} & & \multicolumn{2}{c}{market \quad model} & & $1\sigma$ & $2\sigma$ \\
\midrule
\multicolumn{7}{l}{\emph{growth and volatility}} \\
index drift & $0.044$ & {\tiny[-0.01,0.10]} $0.094$ {\tiny[0.08,0.11]} &  & $0.115$ & {\tiny[0.09,0.14]} $0.093$ {\tiny[0.08,0.10]} &  & Yes & Yes \\
annualized volatility & $0.207$ & {\tiny[0.18,0.23]} $0.183$ {\tiny[0.16,0.20]} &  & $0.173$ & {\tiny[0.15,0.19]} $0.184$ {\tiny[0.16,0.20]} &  & Yes & Yes \\
\addlinespace
\multicolumn{7}{l}{\emph{tail of daily returns}} \\
daily excess kurtosis & $7.431$ & {\tiny[4.82,10.04]} $8.818$ {\tiny[2.53,15.11]} &  & $15.539$ & {\tiny[8.21,22.86]} $9.006$ {\tiny[2.40,15.62]} &  & Yes & Yes \\
days falling over 2\%, a year & $12.610$ & {\tiny[9.31,15.91]} $8.644$ {\tiny[6.55,10.73]} &  & $8.658$ & {\tiny[6.81,10.51]} $8.686$ {\tiny[6.42,10.95]} &  & Yes & Yes \\
\addlinespace
\multicolumn{7}{l}{\emph{depth and spacing of busts}} \\
mean depth, 20\% falls & $0.380$ & {\tiny[0.33,0.43]} $0.274$ {\tiny[0.23,0.32]} & $\bullet$ & $0.309$ & {\tiny[0.27,0.35]} $0.276$ {\tiny[0.23,0.32]} &  & No & Yes \\
20\% falls per century & $26.688$ & {\tiny[13.03,40.35]} $28.022$ {\tiny[18.20,37.85]} &  & $14.310$ & {\tiny[4.82,23.80]} $28.978$ {\tiny[19.00,38.96]} &  & Yes & Yes \\
\addlinespace
\multicolumn{7}{l}{\emph{mean reversion}} \\
variance ratio, a month & $0.782$ & {\tiny[0.73,0.83]} $0.973$ {\tiny[0.83,1.12]} &  & $0.800$ & {\tiny[0.74,0.86]} $0.982$ {\tiny[0.83,1.14]} &  & Yes & Yes \\
variance ratio, a quarter & $0.809$ & {\tiny[0.71,0.91]} $0.863$ {\tiny[0.60,1.13]} &  & $0.602$ & {\tiny[0.53,0.68]} $0.872$ {\tiny[0.60,1.15]} &  & Yes & Yes \\
variance ratio, half a year & $0.965$ & {\tiny[0.78,1.15]} $0.681$ {\tiny[0.47,0.89]} &  & $0.496$ & {\tiny[0.39,0.60]} $0.691$ {\tiny[0.47,0.91]} &  & Yes & Yes \\
variance ratio, a year & $1.127$ & {\tiny[0.88,1.38]} $0.473$ {\tiny[0.30,0.65]} & $\bullet$ & $0.470$ & {\tiny[0.32,0.62]} $0.480$ {\tiny[0.29,0.67]} &  & No & Yes \\
\addlinespace
\multicolumn{7}{l}{\emph{asymmetry across horizons}} \\
skew, a day & $-0.362$ & {\tiny[-0.50,-0.22]} $0.020$ {\tiny[-0.41,0.45]} &  & $-1.024$ & {\tiny[-1.37,-0.67]} $0.033$ {\tiny[-0.41,0.47]} & $\bullet$ & No & Yes \\
skew, a week & $-0.531$ & {\tiny[-0.90,-0.16]} $-0.122$ {\tiny[-1.13,0.88]} &  & $-0.706$ & {\tiny[-0.96,-0.45]} $-0.102$ {\tiny[-1.14,0.94]} &  & Yes & Yes \\
skew, a month & $-0.825$ & {\tiny[-1.05,-0.60]} $-0.313$ {\tiny[-1.00,0.38]} &  & $-1.146$ & {\tiny[-1.59,-0.70]} $-0.296$ {\tiny[-1.09,0.50]} &  & Yes & Yes \\
skew, a quarter & $-0.482$ & {\tiny[-0.69,-0.27]} $-0.130$ {\tiny[-0.70,0.44]} &  & $-1.423$ & {\tiny[-1.80,-1.05]} $-0.116$ {\tiny[-0.72,0.49]} & $\bullet$ & No & Yes \\
\addlinespace
\midrule
ranges that meet & \multicolumn{3}{c}{86\%} & \multicolumn{3}{c}{86\%} & \multicolumn{2}{c}{at $1\sigma$} \\
\bottomrule
\end{tabular}
\caption{The index as one marketron, against the market it was fitted to. Brackets give one standard deviation either side, the market's block bootstrap error and the model's spread over 20 runs of the same length, and a bullet marks two ranges that do not meet.}
\label{tab:indexfit}
\end{table}

\subsection{Fitted and untargeted quantities}
\label{sec:valleys}

Twenty-seven and thirty-one parameters fitted to thirty-three and forty
observables divide the rest of the paper in two: agreement on a fitted observable
is arithmetic, and everything else the model produces is an \emph{untargeted
in-sample diagnostic}. No parameter was fitted to it, but it is read on the
calibration's history, from fitted baskets and after the choices made in building
the model, so it is not a historical holdout. Black's factor of two, the frequency
and depth of the index's twenty percent drawdowns, the dumb-money effect and the
factor and sector premia are of the second kind and are flagged where they appear.
The fit cannot pin the parameters closely: the loss counts threshold events, so it
is piecewise constant with no usable curvature and cannot tell the constants apart
within a few tenths of their values in log terms.

\subsection{Comparison to other models}
\label{sec:retdist}

Appendix~\ref{sec:comparison} sets the model against models that fit a return
series and claim no cycle, held to the shape of the distribution across horizons,
and models built to generate booms and busts, held to the cycle as well. The
outcome is a division rather than a ranking. Only the leverage GARCH and the BBM produce skew across horizons, the leverage GARCH a little more closely. On how the tails thin the BBM
sits between a L\'evy law and two GARCH variants, and on the variance ratio nobody
succeeds. The parametric family closest to the simulated returns is the normal
inverse Gaussian, the one real returns are usually fitted with. A single name's
downward tail is thinner than the market's for the rest of the index, a Hill slope
of $3.62$ against $2.92$, and meets it for the ten largest.
Table~\ref{tab:checkpoints} collects what each model can do and what it is held to.

\subsection{What drives each result}
\label{sec:sensitivity}

Table~\ref{tab:sensitivity} sets each parameter in turn to three quarters and five
quarters of its calibrated value, in both groups at once, and names for each of
fifteen quantities, eleven targeted and four not, the two parameters that move it
most. One parameter does most of the work. The mispricing noise $\sigma_x$ is the
first driver of thirteen of the fifteen, the names' volatility of volatility
leading the other two, and the reversion $\kappa_x$ comes second on most. Since
$\sigma_x$ is pinned by the volatilities and the tails and also sets Black's
fraction and how far the index's mispricing runs, the quantities that came out
right unasked share their drivers with the fitted ones and are not independent
successes.

A narrow span on a quantity the model misses shows that the miss persists under local
retuning, no single parameter moved by a quarter bringing it into the market's
band. The response to a lift in demand, Black's fraction, the entropy production of
the volatility ranks and the yearly variance ratio persist in this sense. The
mean pairwise correlation and the rate of busts are not, the span over $\sigma_x$
bracketing the market for both. The default rate is measured separately
(Section~\ref{sec:defaultrate}), its count at a sweep point being too small to
read.

\begin{table}[htbp]
\centering
\caption{What drives each result. Each parameter is set to three quarters and five
quarters of its calibrated value (the regularizer to zero and one half) with the
rest held, and the two drivers are those whose swings are largest. Model: the
capitalization-weighted index on the first window's inputs, four markets of two
hundred years. Market: 1996--2024.}
\label{tab:sensitivity}
\vspace{0.4em}
\footnotesize
\setlength{\tabcolsep}{3pt}
\resizebox{\textwidth}{!}{\begin{tabular}{lccc@{\hspace{0.4em}}cc@{\hspace{0.4em}}c}
\toprule
& market & BBM & \multicolumn{2}{c}{driver 1} & \multicolumn{2}{c}{with driver 2} \\
\cmidrule(lr){4-5}\cmidrule(lr){6-7}
& & & par. & range & par. & joint range \\
\midrule
\multicolumn{7}{l}{\emph{Never targeted}} \\
inside Black's factor of two, \% & $\approx 90$ & $99.7$\,\tiny$\pm0.05$ & $\sigma_x$ & $99.3$--$99.9$ & $\kappa_x$ & $98.6$--$100.0$ \\
entropy production, vol.\ ranks & $0.0062$ & $0.0009879$\,\tiny$\pm1.2 \times 10^{-5}$ & $\sigma_x$ & $0.0009149$--$0.001036$ & $s_n$ & $0.0008384$--$0.001076$ \\
largest mispricing, index & n.o. & $0.69$\,\tiny$\pm0.079$ & $\sigma_x$ & $0.51$--$0.82$ & $\mu$ & $0.48$--$0.85$ \\
largest mispricing, name & n.o. & $2.35$\,\tiny$\pm0.37$ & $s_n$ & $2.09$--$2.99$ & $\sigma_x$ & $1.54$--$3.57$ \\
\multicolumn{7}{l}{\emph{Targeted at basket level, untargeted for the index}} \\
index drift & $0.097$ & $0.1080$\,\tiny$\pm0.0027$ & $\sigma_x$ & $0.0971$--$0.1174$ & $\kappa_x$ & $0.0947$--$0.1180$ \\
index volatility & $0.191$ & $0.192$\,\tiny$\pm0.0023$ & $\sigma_x$ & $0.154$--$0.230$ & $g$ & $0.153$--$0.233$ \\
daily skew & $-0.393$ & $-0.242$\,\tiny$\pm0.045$ & $\sigma_x$ & $-0.759$--$0.008$ & $\kappa_x$ & $-0.785$--$0.033$ \\
daily excess kurtosis & $9.8$ & $11.5$\,\tiny$\pm1.7$ & $\sigma_x$ & $11.1$--$13.9$ & $\bar\epsilon$ & $11$--$13.9$ \\
deepest drawdown, log & $-0.792$ & $-0.880$\,\tiny$\pm0.13$ & $\sigma_x$ & $-1.073$--$-0.673$ & $\kappa_x$ & $-1.133$--$-0.647$ \\
years between busts & $4.26$ & $3.65$\,\tiny$\pm0.23$ & $\sigma_x$ & $2.31$--$6.95$ & $\bar\epsilon$ & $2.24$--$6.86$ \\
busts per century & $20.7$ & $26.6$\,\tiny$\pm2.1$ & $\sigma_x$ & $13.9$--$42.6$ & $\bar\epsilon$ & $14$--$44$ \\
variance ratio, 3 months & $0.672$ & $0.724$\,\tiny$\pm0.026$ & $\sigma_x$ & $0.687$--$0.768$ & $\kappa_x$ & $0.659$--$0.786$ \\
variance ratio, 1 year & $0.791$ & $0.441$\,\tiny$\pm0.037$ & $\sigma_x$ & $0.392$--$0.509$ & $\kappa_x$ & $0.358$--$0.541$ \\
mean pairwise correlation & $0.308$ & $0.273$\,\tiny$\pm0.0058$ & $s_n$ & $0.256$--$0.295$ & $\lambda_I$ & $0.256$--$0.299$ \\
single-name volatility & $0.380$ & $0.379$\,\tiny$\pm0.0021$ & $\sigma_x$ & $0.301$--$0.459$ & $s_n$ & $0.299$--$0.465$ \\
\bottomrule
\end{tabular}
}

\end{table}

\begin{figure}[htbp]
\centering
\includegraphics[width=\textwidth]{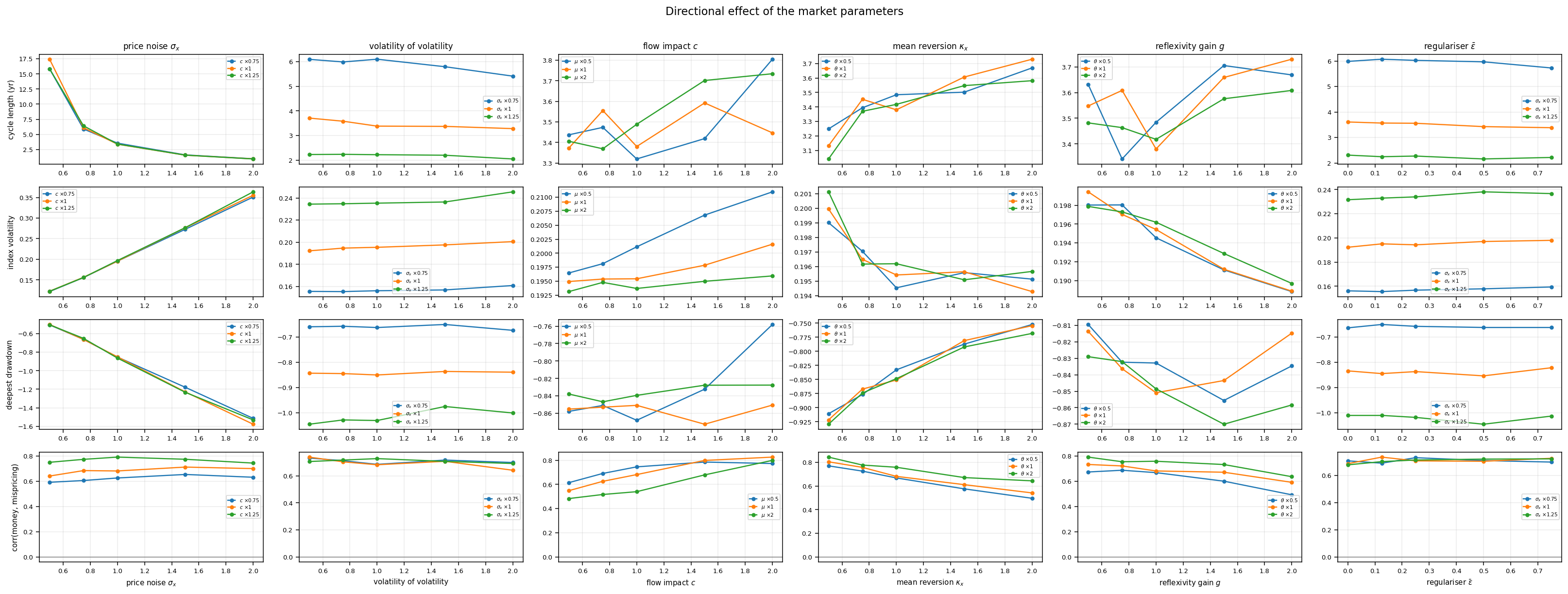}
\caption{Parameter dependence around the calibrated market. Each column sweeps one
parameter from half to twice its fitted value, the regularizer from the bare
potential to $0.75$, at three levels of a second; every point is four markets of a
hundred and forty years.}
\label{fig:sensfig}
\end{figure}

Over the wider range of Figure~\ref{fig:sensfig} the spacing of busts responds
most, from seventeen years between twenty percent falls at half the fitted
$\sigma_x$ to under one at twice it, and the default rate runs from zero to two
percent a name-year, which is why the calibration scores it. Away from the calibrated point a market can
run away, its index moving by more than one half in log in a day, or collapse,
falling by more than $1.5$ in log within fifteen years. Such a market is replaced by
a fresh one drawn from a spare set of seeds, so that these rare realizations do not
dominate the averages. Only two of the ninety points of the sweep needed a spare market, both at doubled volatility
parameters and none at the calibrated vector, which sits a factor of two below that
edge.

\subsection{Stationarity tests}
\label{sec:stationarity}

\begin{figure}[htbp]
\centering
\includegraphics[width=\textwidth,height=0.86\textheight,keepaspectratio]{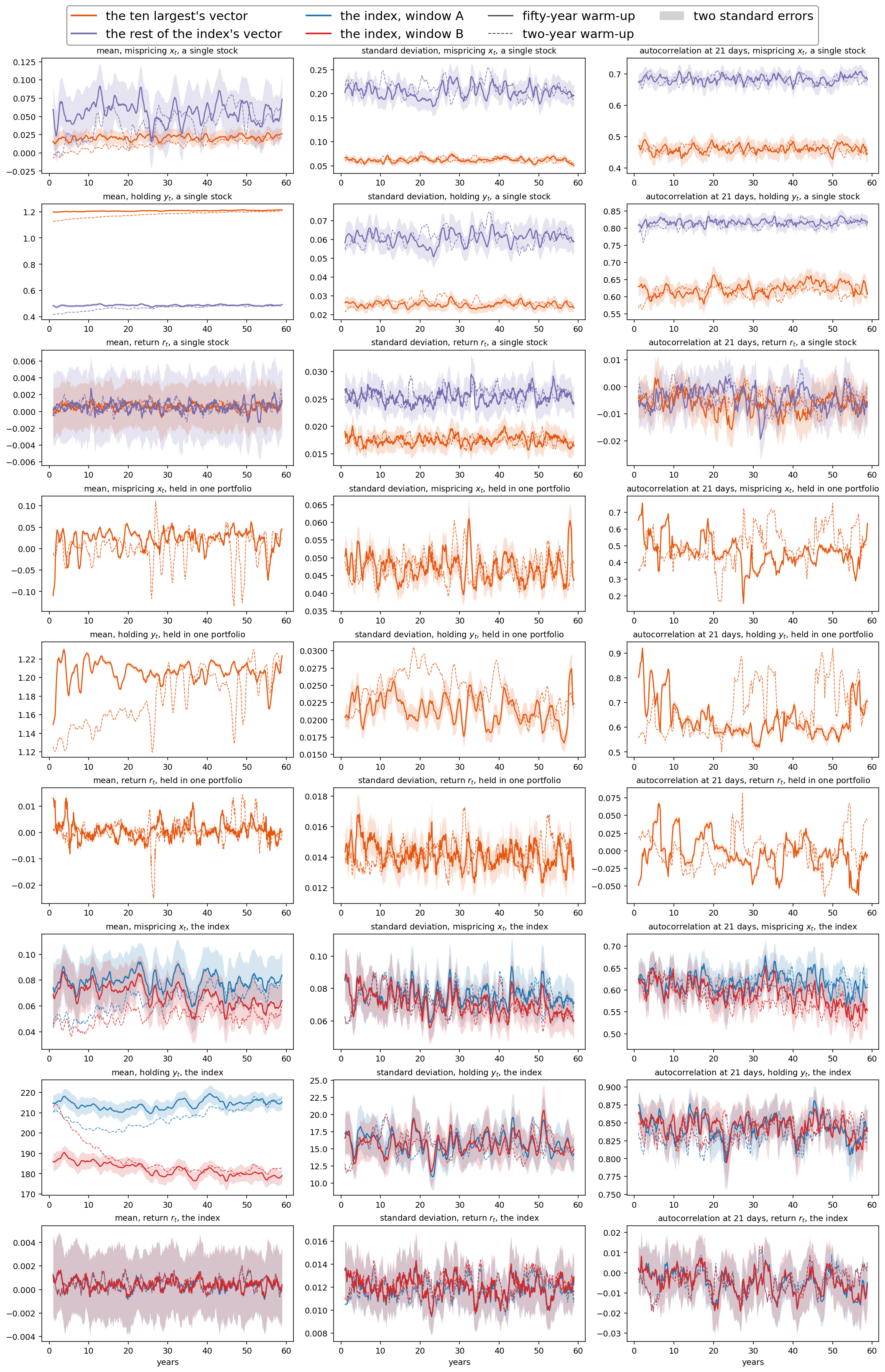}
\caption{The mispricing, the holding and the return across independent paths at
every date. A single stock follows the same paths under both windows' inputs, so
one line is drawn for each group's vector. The middle three rows are the control
that holds every path in one portfolio.}
\label{fig:xystationarity}
\end{figure}

Three questions follow from fitting one model to twenty-nine years: whether the
processes the model is read through are stationary, whether one vector can match
every moment at once, and whether the market needs different parameters in
different periods. The first is about the model alone. A stock is simulated as its
own run with its own investor, every path the same firm, and the index is the
market the rest of the paper measures. Each statistic is taken across independent
paths at every date over sixty years after the warm-up
(Figure~\ref{fig:xystationarity}, Table~\ref{tab:xystationarity}).

After the fifty-year warm-up a single name's return, mispricing and demand memory
are flat to within about a standard error, and so is the index under the first
window's inputs. Under the second the index's mispricing still falls by two to
four and a half percent a decade. After a two-year warm-up the holding is still
relaxing, and held in one portfolio rather than with independent investors the
paths keep relaxing even after fifty years, so it is the shared budget and not the
names that drifts.

The second window's drift comes from how the index holds its two groups together.
Nothing but the shared growth rate holds their relative size, the ten largest being
a fixed set and the rest turning over, so three small differences decide the
shares: the industry growth premia of 2011--2024, which favored the ten largest by
about a point a year, their larger fundamental risk, and the removal rule, which
costs the rest of the index about a point a year. On the first window's inputs
these nearly cancel and the ten largest's share stays between $0.14$ and $0.19$
over a hundred and fifty years, against the market's $0.18$. On the second's they
add and it rises to $0.71$. The construction lacks the rotation of the real index,
where a firm leaving the top ten joins the rest, so the long-horizon results of
this paper are given on the first window's inputs.

The return divides into what the firm earned and what the mispricing moved
(Table~\ref{tab:retvar}). Price against value accounts for most of the variance, nearly
two thirds for a large name, three quarters for a typical one and more than four
fifths for the index.

\begin{table}[htbp]
\centering
\small
\begin{tabular}{lrrrr}
\toprule
 & \multicolumn{2}{c}{a stock} & \multicolumn{2}{c}{the index} \\
\cmidrule(lr){2-3}\cmidrule(lr){4-5}
 & ten largest & rest of the index & A & B \\
\midrule
mean of $x_t$ & +2.6 & -0.6 & -0.3 & -4.4 \\
s.d.\ of $x_t$ & +0.3 & -0.0 & -0.1 & -2.6 \\
autocorrelation of $x_t$ & +0.2 & +0.1 & -0.2 & -1.7 \\
\addlinespace
mean of $y_t$ & +0.2 & +0.1 & +0.2 & -0.9 \\
s.d.\ of $y_t$ & -0.4 & +0.3 & -0.3 & +0.1 \\
autocorrelation of $y_t$ & +0.7 & +0.1 & -0.2 & -0.3 \\
\addlinespace
mean of $r_t$ & -0.1 & +1.7 & -0.5 & +0.4 \\
s.d.\ of $r_t$ & +0.2 & -0.3 & -0.0 & +0.0 \\
autocorrelation of $r_t$ & -6.9 & -7.2 & -32.4 & -29.2 \\
\midrule
\multicolumn{5}{l}{\textit{the same paths held in one portfolio rather than run independently}} \\
s.d.\ of $y_t$ & -1.2 &  &  &  \\
\bottomrule
\end{tabular}
\caption{Does the state of the model drift? The trend of each statistic over sixty years after the warm-up, in percent of its level a decade. The lower panel holds the stock's paths in one portfolio.}
\label{tab:xystationarity}
\end{table}

\begin{table}[htbp]
\centering
\small
\begin{tabular}{lrrrr}
\toprule
 & fundamental & mispricing & covariance & residual \\
\midrule
ten largest & 0.284 & 0.626 & -0.002 & 0.092 \\
rest of the index & 0.049 & 0.750 & -0.000 & 0.201 \\
\addlinespace
the index, A & 0.156 & 0.844 & +0.000 & 0.000 \\
the index, B & 0.166 & 0.831 & +0.001 & 0.003 \\
\bottomrule
\end{tabular}
\caption{Where a return's variance comes from. The share carried by the fundamental, by the mispricing, and by the covariance between them, with the residual left by the fast transitory term. A stock is one name under its group's joint vector, the same in both windows.}
\label{tab:retvar}
\end{table}

The questions about the fit are tested in Appendix~\ref{sec:fittests}, the
calibration being a simulated method of moments in all but name
\cite{mcfadden1989,duffie1993}. The two periods' index moments differ beyond
doubt, at a Wald statistic of $65.3$, most visibly in mean reversion: the two
halves have opposite variance-ratio profiles, so the pooled non-monotone curve is
an average of two opposite ones. Hansen's statistic on the twenty-one unfitted
moments rejects both windows at each window's fine-tuned vector (Section~\ref{sec:groupsummary}) at its nominal
$\chi^2$ level (Appendix~\ref{sec:fittests}), which ignores the fitting pipeline. Whether one vector describes both periods as well
as a vector for each is the structural-change question of Andrews and Fair
\cite{andrewsfair1988,hallsen1999}, which cannot be put in its moment form here,
since no vector was fitted to the index moments.

It is put instead on the observables the fits were made to, on markets neither
search read, and the outcome is asymmetric. The vector tuned to 1996--2010
describes 2011--2024 nearly as well as that window's own fit. The vector tuned to
2011--2024 does clearly worse on 1996--2010, matching twenty-three of thirty-three
and twenty-four of forty observables against twenty-seven and thirty-two for that
window's own fit, and the losses point the same way. The calmer later window, with
one crisis, informs the fit less: a quiet sample nested in a richer one rather
than a change of regime.

The periods therefore differ in their moments without requiring different
parameters, a single vector describing both while its own cycle produces
fifteen-year windows that differ in this way. The one sign of drift that survives a
test sized on the calibrated market, exploratory since it is one of seven, is in the
level of returns (Appendix~\ref{sec:fittests}), where the growth rate of value is
held common by construction. The counts have an error of a few matches and no
formal size, so the model gives no support to a structural break, without
excluding one.

The dominant reading of long-horizon equity data is the opposite. St\v{a}ric\v{a}
and Granger \cite{starica2005} find the dynamics of seventy years of S\&P 500
returns concentrated in shifts of the unconditional variance, and regime-switching
and macroeconomic factor models index the return process by something outside it.
The two readings differ in mechanism more than in what a sample shows. At one fixed
vector this model's volatility rises and falls with the phase of a cycle of decadal
period (Section~\ref{sec:leancycle}), so an observer fitting a locally constant
variance to its path would find breaks and forecast better than a stationary GARCH,
while the process was stationary throughout. The question is decidable: exogenous
shifts should align with the macroeconomic variables supposed to drive them, and
the cycle's with where the market sits in its own boom and bust, which
Section~\ref{sec:leancycle} measures on both sides.

\subsection{Two parameter sets}
\label{sec:twosets}

In what follows, we explore the BBM behavior under two sets of model parameters:
one obtained with the global fit, each group held to both windows at once, and
another one fine-tuned to the second period, 2011--2024. Figures label them by the period each was
fitted to. The tuned set describes the recent window more closely but does not
lead, because a decadal cycle can only be identified on a span holding several of
its excursions and the recent window holds one crisis. Over a hundred and fifty
years on the first window's inputs both are stationary, the index volatility
staying between $0.17$ and $0.21$, and past age seventy the global set's index
falls twenty percent $26.4 \pm 1.9$ times a century and the tuned set's
$23.8 \pm 1.7$. One model history as long as the S\&P's
January 1975 to August 2026 gives
$24.8 \pm 6.4$ under the global set, meeting the S\&P's $15.5 \pm 5.5$ although
its center sits above it. Within the windows the two sets are close
(Table~\ref{tab:sumfit}): the joint pair meets thirty-seven of forty-two index observables and the fine-tunes thirty-eight, and on the groups a window's fine-tune and the joint vector differ by one to three matches.

\begin{table}[htbp]
\centering
\setlength{\tabcolsep}{2.0pt}
\scriptsize
\resizebox{\textwidth}{!}{%
\begin{tabular}{l rrr rrr}
\toprule
& \multicolumn{3}{c}{1996--2010} & \multicolumn{3}{c}{2011--2024} \\
\cmidrule(lr){2-4}\cmidrule(lr){5-7}
Observable & market & joint & tuned & market & joint & tuned \\
\midrule
annual drift & $0.066$\,\tiny[0.02,0.12] & $0.111$\,\tiny[0.09,0.14] & $0.112$\,\tiny[0.09,0.13] & $0.130$\,\tiny[0.10,0.16] & $0.116$\,\tiny[0.09,0.14] & $0.122$\,\tiny[0.10,0.14] \\
annual volatility & $0.207$\,\tiny[0.18,0.23] & $0.204$\,\tiny[0.16,0.25] & $0.202$\,\tiny[0.16,0.24] & $0.172$\,\tiny[0.15,0.19] & $0.208$\,\tiny[0.16,0.25] & $0.200$\,\tiny[0.16,0.24] \\
skewness, daily & $-0.180$\,\tiny[-0.28,-0.07] & $-0.380$\,\tiny[-0.99,0.23] & $-0.164$\,\tiny[-0.58,0.26] & $-0.763$\,\tiny[-1.01,-0.51] & $-0.383$\,\tiny[-1.03,0.26] & $-0.047$\,\tiny[-1.11,1.01] \\
skewness, monthly & $-0.665$\,\tiny[-0.88,-0.45] & $-0.809$\,\tiny[-1.65,0.03] & $-0.859$\,\tiny[-1.58,-0.13] & $-1.061$\,\tiny[-1.47,-0.65] & $-0.911$\,\tiny[-1.89,0.06] & $-1.019$\,\tiny[-2.41,0.38] \\
excess kurtosis & $7.332$\,\tiny[4.79,9.87] & $12.370$\,\tiny[1.37,23.37] & $9.287$\,\tiny[3.12,15.46] & $13.941$\,\tiny[7.41,20.47] & $13.336$\,\tiny[-0.96,27.63] & $24.708$\,\tiny[9.64,39.78] \\
falls beyond 2\% a year & $12.610$\,\tiny[9.36,15.86] & $11.167$\,\tiny[7.66,14.67] & $10.800$\,\tiny[6.86,14.74] & $8.586$\,\tiny[6.80,10.37] & $11.400$\,\tiny[7.62,15.18] & $9.674$\,\tiny[6.90,12.45] \\
deepest drawdown, log & $-0.793$\,\tiny[-1.08,-0.50] & $-0.512$\,\tiny[-0.84,-0.19] & $-0.515$\,\tiny[-0.81,-0.22] & $-0.409$\,\tiny[-0.50,-0.32] & $-0.522$\,\tiny[-0.85,-0.19] & $-0.486$\,\tiny[-0.71,-0.26] \\
20\% falls per century & $26.688$\,\tiny[13.89,39.48] & $30.024$\,\tiny[13.14,46.90] & $30.858$\,\tiny[14.72,47.00] & $14.310$\,\tiny[4.95,23.67] & $29.953$\,\tiny[11.44,48.46] & $27.718$\,\tiny[10.22,45.21] \\
variance ratio, month & $0.750$\,\tiny[0.70,0.80] & $0.891$\,\tiny[0.74,1.04] & $0.899$\,\tiny[0.74,1.06] & $0.761$\,\tiny[0.71,0.81] & $0.877$\,\tiny[0.72,1.03] & $0.872$\,\tiny[0.69,1.06] \\
variance ratio, quarter & $0.737$\,\tiny[0.66,0.81] & $0.693$\,\tiny[0.51,0.88] & $0.705$\,\tiny[0.51,0.90] & $0.557$\,\tiny[0.49,0.62] & $0.672$\,\tiny[0.49,0.85] & $0.660$\,\tiny[0.47,0.85] \\
variance ratio, year & $0.991$\,\tiny[0.79,1.19] & $0.434$\,\tiny[0.25,0.62]$^{\bullet}$ & $0.426$\,\tiny[0.25,0.60]$^{\bullet}$ & $0.439$\,\tiny[0.31,0.57] & $0.389$\,\tiny[0.22,0.56] & $0.417$\,\tiny[0.24,0.60] \\
mean pairwise correlation & $0.272$\,\tiny[0.24,0.31] & $0.292$\,\tiny[0.24,0.34] & $0.244$\,\tiny[0.20,0.29] & $0.335$\,\tiny[0.31,0.36] & $0.292$\,\tiny[0.24,0.35] & $0.308$\,\tiny[0.25,0.37] \\
market beta, spread & $0.440$\,\tiny[0.40,0.48] & $0.374$\,\tiny[0.36,0.39]$^{\bullet}$ & $0.340$\,\tiny[0.33,0.35]$^{\bullet}$ & $0.314$\,\tiny[0.29,0.34] & $0.333$\,\tiny[0.32,0.35] & $0.447$\,\tiny[0.42,0.47]$^{\bullet}$ \\
name volatility, mean & $0.424$\,\tiny[0.39,0.46] & $0.403$\,\tiny[0.37,0.43] & $0.419$\,\tiny[0.39,0.45] & $0.322$\,\tiny[0.30,0.34] & $0.397$\,\tiny[0.37,0.42]$^{\bullet}$ & $0.378$\,\tiny[0.34,0.42] \\
\midrule
ranges that meet & & 12 of 14 & 12 of 14 & & 13 of 14 & 13 of 14 \\
\bottomrule
\end{tabular}}
\caption{The capitalization-weighted index under the joint pair and under each window's fine-tuned pair, fourteen rows of Table~\ref{tab:index}. Both are untargeted, neither fit having seen the index.}
\label{tab:sumfit}
\end{table}

\section{The Market Through the Cycle}
\label{sec:rmmresults}

This section compares the model market with the actual one through the boom-bust
cycle: its stages (Section~\ref{sec:leancycle}), the correlation geometry in calm
and in crises (Section~\ref{sec:leanomd}), and the rotation of factors and sectors
(Section~\ref{sec:factorrotation}), with money flows and firms in
Section~\ref{sec:factorcycle}.

\subsection{The cycle and its stages}
\label{sec:leancycle}
\label{sec:stages}

With the parameters of Table~\ref{tab:bbmparams} the market booms and busts
repeatedly and recovers every time. Over a hundred and fifty years after the
warm-up the capitalization-weighted index falls twenty percent once every $4.0$ years, more often than the S\&P, although one model history of the S\&P's length
still meets the market's range. The fundamentals do not cycle. The index's
mispricing ranges from $-0.43$ to $+0.33$ and the investor's equity share from $0.43$ to $0.77$, so the cycle is a movement of money between the two assets. The
boom raises the extrapolated return, the equity share and the price until the
value term, growing with the mispricing the boom created, overtakes it, and the
same rule then sells until the value term of the opposite sign buys the market
back. Sweeping the coupling against the noise separates a quiet, a cycling and an
attrition regime (Section~\ref{sec:phases}), and the calibrated market sits in the
cycling one.

\label{sec:blackfactor}
Because the model includes an explicit mispricing, Black's bound of
Section~\ref{sec:birdseye}, price within a factor of two of value at least ninety
percent of the time \cite{black1986}, can be tested directly on single stocks, as
$|x_t| < \log 2$. Nothing about it entered the calibration. A large name leaves
the band on one day in several thousand and a typical name on about one in a
thousand, the test being weak for the large names, whose $\log 2$ lies about ten
standard deviations out. A typical name strays about twice as far as a large one
and returns about three times more slowly, at close to the pace of the reversion
$\kappa_x$ alone.

\begin{figure}[htbp]
\centering
\includegraphics[width=\textwidth]{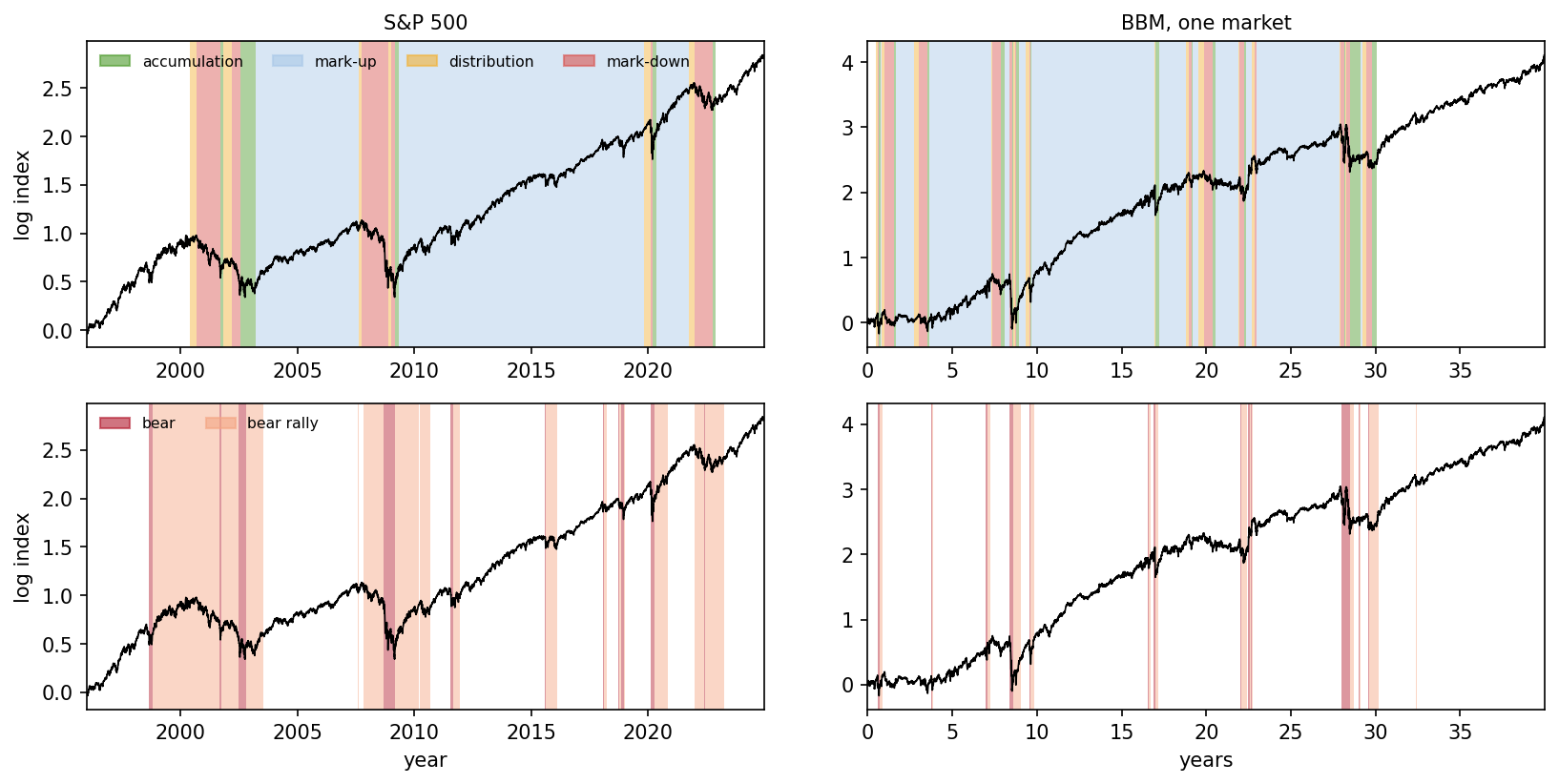}
\caption{Market cycles of the S\&P 500 (left) and one simulated market (right).
Top, the four stages marked on the index price. Bottom, the bear regime of the
four-state model, split into its bear and bear-rally states.}
\label{fig:cycleconv}
\end{figure}

Practitioners divide a cycle into four stages, accumulation at the lows, the
mark-up, distribution at the top and the mark-down, a scheme that goes back to
Wyckoff \cite{wyckoff1931}. The scheme is verbal, so each stage is marked here by
our own rule on the price of the capitalization-weighted index, applied alike to
the model and the S\&P. The mark-down is a bear market by the twenty percent
rule of the literature on bull and bear markets \cite{pagan2003,lunde2004}.
Accumulation runs from the trough until the price closes back above its 200-day
average, a moving-average signal in common use among traders \cite{brock1992}.
Distribution is the last stretch before the next peak in which the price stays
within ten percent of it, and the mark-up is the rest of the bull market. The
second convention is probabilistic, the four-state model of Maheu, McCurdy and Song
\cite{maheu2012}, in which weekly returns switch between a bear regime, made of a
bear state and a bear rally, and a bull regime, made of a bull correction and a
bull state. It is fitted to both sides with their restrictions and priors by Gibbs
sampling \cite{chib1996}. Figure~\ref{fig:cycleconv} shows both conventions.

Table~\ref{tab:cyclesummary} collects the comparison, together with the daily
returns in five phases around each $20\%$ fall, and forty-seven of its fifty-five
scored entries meet. The model reproduces the market-wide signature of every
stage, its length, return, volatility and correlation, the regimes of the
second convention, and the volatility of every phase. The misses are across names: their dispersion
moves too little with the stage, and at the top the S\&P's breadth narrows while
the model's widens, so the narrowing leadership of a late bull market is absent.
Other settings of the two conventions change the count by at most one. Within the
stages the model shows what the market cannot: its mispricing peaks in
distribution, and its investor is a net buyer through the mark-up and distribution
and a net seller through the mark-down, while its latent demand moves the other
way.

An average bust shows how the cycle unfolds (Figure~\ref{fig:leanosc}, bottom),
each bear market aligned at its trough. The cusp at day zero comes from the
alignment, as it would for a random walk, and is not a bifurcation: the parameters
are fixed and a bust is a large excursion about one rest point
(Section~\ref{sec:excitability}). The mispricing and the equity share erode for two
years into the trough and recover within one to two years after it. The investor's
executed trades are near zero until a quarter before the trough, sell into the
last weeks of the fall and buy the rebound.

\begin{table}[htbp]
\centering
\caption{The cycle of the calibrated market against the S\&P 500 over 1996 to 2024. Entries run over the stages, states or phases in order; a bullet marks a model entry whose one-sigma range does not meet the market's.}
\label{tab:cyclesummary}
\vspace{0.4em}
\footnotesize
\setlength{\tabcolsep}{3pt}
\begin{tabular}{llll}
\toprule
 & BBM & S\&P 500 & Meet \\
\midrule
\multicolumn{4}{l}{\emph{Stages: accumulation, mark-up, distribution, mark-down}} \\
Duration, years & 0.15, 2.81, 0.17, 0.41 & 0.21, 5.42, 0.24, 0.59 & 4 of 4 \\
Index return, a year & +1.30, +0.17, +0.79, $-$0.87 & +0.82, +0.14, +0.47, $-$0.71 & 4 of 4 \\
Index volatility & 0.33, 0.18$^{\bullet}$, 0.19, 0.29 & 0.33, 0.14, 0.17, 0.31 & 3 of 4 \\
Mean pairwise correlation & 0.43, 0.28, 0.25, 0.40 & 0.42, 0.31, 0.24, 0.34 & 4 of 4 \\
Names above 200-day average & 0.24$^{\bullet}$, 0.72, 0.80$^{\bullet}$, 0.48 & 0.37, 0.76, 0.66, 0.49 & 2 of 4 \\
Dispersion across names, \% a day & 2.2$^{\bullet}$, 1.9$^{\bullet}$, 1.9, 2.2$^{\bullet}$ & 3.0, 1.5, 2.1, 2.7 & 1 of 4 \\
Top-10 share of capitalization & 0.23, 0.22, 0.21, 0.22 & 0.23, 0.19, 0.23, 0.22 & 4 of 4 \\
\midrule
\multicolumn{4}{l}{\emph{Regimes: bear, bear rally, bull correction, bull}} \\
Weekly mean, \% & $-$1.07, 0.46, $-$0.22, 0.51$^{\bullet}$ & $-$1.22, 0.17, $-$0.08, 0.69 & 3 of 4 \\
Weekly volatility, \% & 6.3, 2.6, 2.4, 1.0 & 5.7, 2.6, 1.9, 1.0 & 4 of 4 \\
Share of time in the bull regime & 0.63 & 0.62 & 1 of 1 \\
Bull and bear regime, years & 2.0, 1.0 & 1.8, 1.1 & 2 of 2 \\
Bear regimes a century & 35 & 35 & 1 of 1 \\
\midrule
\multicolumn{4}{l}{\emph{Phases of a 20\% fall: deep in a boom, late boom, at the top, bust, recovery}} \\
Daily volatility, annualized & 0.21, 0.25, 0.34, 0.29, 0.21 & 0.18, 0.22, 0.33, 0.25, 0.23 & 5 of 5 \\
Skew of daily returns & 0.02, 0.06, 0.19$^{\bullet}$, 0.13, $-$0.04 & $-$0.69, $-$0.20, $-$0.93, $-$0.02, $-$0.06 & 4 of 5 \\
Excess kurtosis of daily returns & 12.5, 7.4, 7.9, 9.0, 12.5 & 13.5, 10.6, 8.9, 2.0, 6.3 & 5 of 5 \\
\midrule
Stage's weeks in the bear regime & 0.81, 0.34, 0.43, 0.64 & 1.00, 0.22, 0.61, 0.91 & -- \\
\midrule
All scored entries & & & 47 of 55 \\
\bottomrule
\end{tabular}
\end{table}

\subsection{Correlation geometry and crisis diagnostics}
\label{sec:leanomd}
\label{sec:empirical}
\label{sec:tests}
\label{sec:regimeaware}

The OMD studies read a market off the geodesic distance matrix
$M(t) = \arccos G(t)$ of its correlation matrix $G(t)$
\cite{halperin2026omdstocks,halperin2026omdportfolio}, whose spectra are the
subject of random matrix theory \cite{bogomolny2003}, and their central finding is
that the market does not learn its distance matrix \cite{halperin2026omdstocks}. A
system that learns concentrates the matrix onto a few stable directions and its
spectrum localizes, as at the grokking transition of a neural network or in
frustrated particles finding an arrangement \cite{halperin2026omd,halperin2026fdm}.
The question here is whether the BBM, whose correlation spectrum is an outcome
rather than a setting, reproduces that finding and the market's crisis geometry.
The delocalized exponent $\beta_M$ is computed from the eigenvalues, the non-Perron
eigenvalues of a distance matrix decaying as a power of their rank,
\begin{equation}
|\lambda_K| \sim K^{-\beta_M}, \qquad 2 \le K \le \sqrt{N},
\qquad d_\beta = \frac{\beta_M}{\beta_M - 1} ,
\label{eq:betam}
\end{equation}
with $\beta_M = d/(d-1)$ for a uniform sample on the hyper-sphere $S^{d-1}$, so a
geometry localized on a manifold has $\beta_M$ above one, and below one the
formula gives no finite dimension, so this spectral estimator cannot detect a
manifold there. The commutator norm $\lVert[\,G(t),G(t-\tau)\,]\rVert_F$
against the matrix a month earlier depends on the eigenvectors, vanishing when the two
share an eigenbasis. We also report the participation ratio and the market-mode share,
\begin{equation}
\mathrm{PR}(t)=\frac{\big(\sum_i \lambda_i(t)\big)^2}{\sum_i \lambda_i(t)^2} , \qquad
S(t) = \frac{\lambda_1(t)}{\sum_{i:\,\lambda_i>0}\lambda_i(t)} ,
\label{eq:modeshare}
\end{equation}
with $\lambda_1(t) \ge \lambda_2(t) \ge \dots$ the eigenvalues of $G(t)$, and the mean absolute correlation, all read from daily simple returns and none in the calibration
objective.

On a fixed panel of names (Table~\ref{tab:crisisgeom}) the model meets the market
on every observable in crises. In calm periods the market-mode share meets, but
the model spreads the rest of its variance more evenly, its participation ratio
higher and its $\beta_M$ several band widths lower, and turns its eigenbasis
faster, a gap that closes through a crisis. Neither relaxes: over a hundred and
fifty years the model's $\beta_M$ and commutator do not drift, the market's own
$\beta_M$ rising only slowly over its twenty-nine years, and both sides sit
well below the regime in which Eq.~\eqref{eq:betam} gives an embedding dimension,
a crisis pushing the geometry only part of the way toward localizing. The model's
lower $\beta_M$ is the clearest miss, but the BBM supports the no-learning finding
in its strong form, the common part of each price innovation being order flow
impacted through the same $V'(x)$ that sets the drift, a state that never
settles.

\begin{table}[htbp]
\centering
\caption{Crisis geometry in the model and in the market: five OMD observables on a six-month window, averaged over calm periods and over crises, falls of twenty percent. Ranges and marks as in Table~\ref{tab:groupjointfifty}.}
\label{tab:crisisgeom}
\vspace{0.4em}
\scriptsize
\setlength{\tabcolsep}{2.5pt}
\resizebox{\textwidth}{!}{%
\begin{tabular}{lcccccc}
\toprule
& \multicolumn{2}{c}{BBM, 1996-2010 inputs} & \multicolumn{2}{c}{BBM, 2011-2024 inputs} & \multicolumn{2}{c}{S\&P 500} \\
\cmidrule(lr){2-3}\cmidrule(lr){4-5}\cmidrule(lr){6-7}
& calm & crisis & calm & crisis & calm & crisis \\
\midrule
participation ratio & $12.88$\,\tiny[11.21,14.56]$^{\bullet}$ & $4.20$\,\tiny[2.74,5.65] & $12.65$\,\tiny[11.18,14.12]$^{\bullet}$ & $3.82$\,\tiny[2.69,4.96] & $9.50$\,\tiny[8.72,10.29] & $3.86$\,\tiny[3.14,4.57] \\
mean absolute correlation & $0.294$\,\tiny[0.275,0.313] & $0.506$\,\tiny[0.412,0.600] & $0.299$\,\tiny[0.283,0.316] & $0.526$\,\tiny[0.446,0.606] & $0.317$\,\tiny[0.300,0.335] & $0.506$\,\tiny[0.460,0.552] \\
market-mode share & $0.320$\,\tiny[0.301,0.339] & $0.531$\,\tiny[0.442,0.619] & $0.324$\,\tiny[0.307,0.341] & $0.549$\,\tiny[0.474,0.625] & $0.337$\,\tiny[0.319,0.355] & $0.522$\,\tiny[0.479,0.565] \\
commutator norm & $316$\,\tiny[292,340] & $540$\,\tiny[444,635] & $321$\,\tiny[300,343]$^{\bullet}$ & $570$\,\tiny[500,639] & $276$\,\tiny[258,294] & $555$\,\tiny[413,698] \\
delocalized exponent $\beta_M$ & $0.307$\,\tiny[0.293,0.321]$^{\bullet}$ & $0.526$\,\tiny[0.378,0.673] & $0.304$\,\tiny[0.292,0.317]$^{\bullet}$ & $0.527$\,\tiny[0.392,0.662] & $0.568$\,\tiny[0.547,0.588] & $0.619$\,\tiny[0.608,0.629] \\
commutator, normalized & $0.122$\,\tiny[0.119,0.125]$^{\bullet}$ & $0.094$\,\tiny[0.075,0.114] & $0.122$\,\tiny[0.119,0.125]$^{\bullet}$ & $0.094$\,\tiny[0.074,0.115] & $0.095$\,\tiny[0.093,0.097] & $0.106$\,\tiny[0.067,0.144] \\
leading eigenvector turn, degrees & $6.4$\,\tiny[6.1,6.8]$^{\bullet}$ & $4.1$\,\tiny[3.2,5.0] & $6.4$\,\tiny[6.1,6.7]$^{\bullet}$ & $4.1$\,\tiny[3.1,5.0] & $5.6$\,\tiny[5.2,6.0] & $5.5$\,\tiny[2.7,8.3] \\
\bottomrule
\end{tabular}}
\end{table}

\begin{figure}[htbp]
\centering
\includegraphics[width=0.85\textwidth]{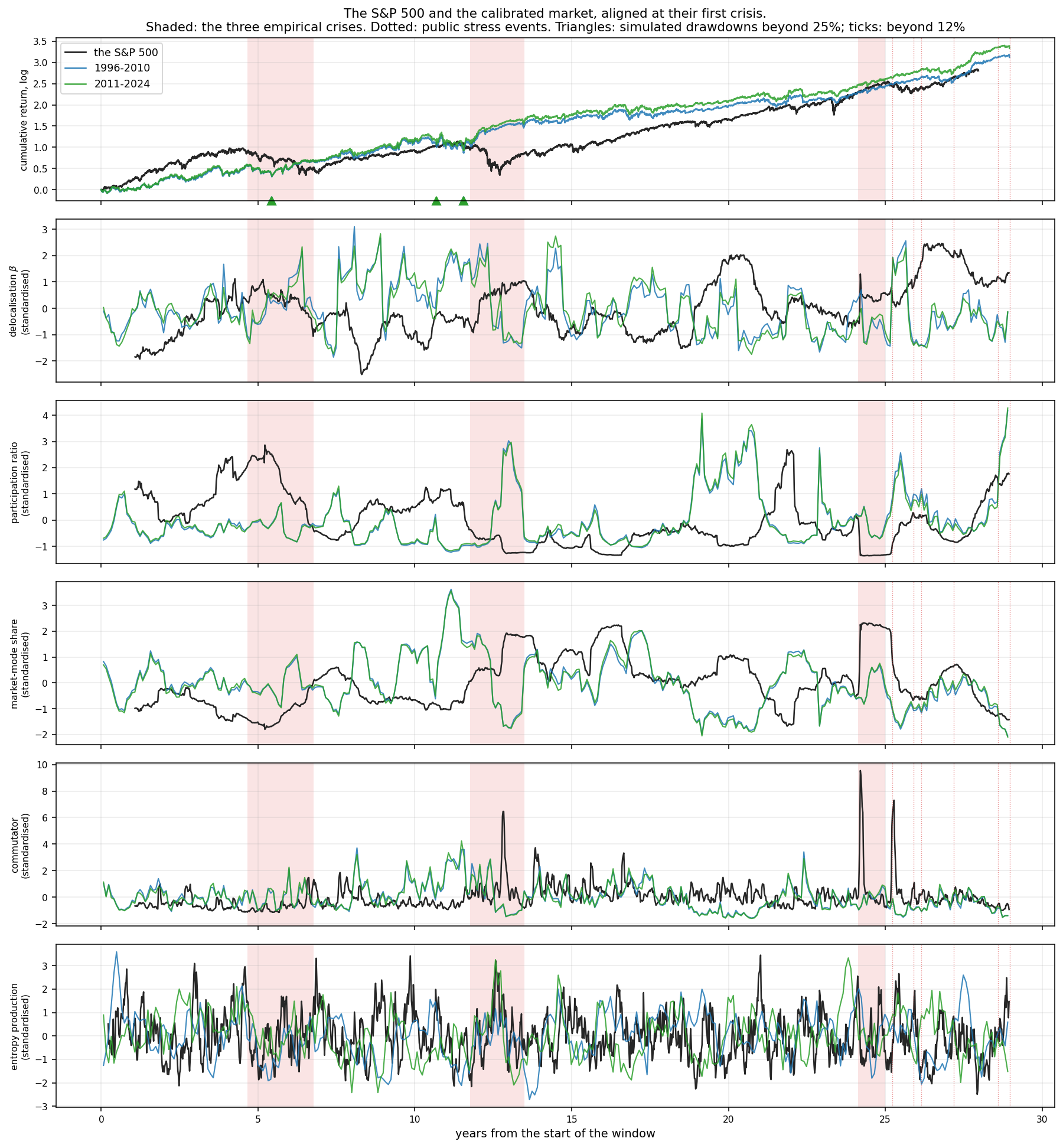}
\caption{The S\&P 500 and the calibrated market over windows of equal length,
aligned at their first crisis: cumulative log return (top) and the five OMD
observables, standardized (below). Shading marks the three empirical crises.}
\label{fig:vshistory}
\end{figure}

On the market alone the same observables, read on a hundred S\&P names over
1996--2024 beside the entropy production of the two ranking chains
\cite{roldan2010}, mark the three crises of 2000--2002, 2007--2009 and 2020
(Table~\ref{tab:contrasts}). The delocalized exponent and the entropy production of
the volatility chain both rise in crises, so the geometry departs from a settled
state and the dynamics grows more irreversible when, in the reflexive account, the
loop tightens. The commutator separates crisis from calm most cleanly, and the
geometry leads the dispersion rather than crowding leading the geometry. Chains
that keep the market mode, as in OMD-Portfolio \cite{halperin2026omdportfolio},
separate crisis from calm more clearly still, their entropy production leading the
geometry by about a quarter. Laid over the model's history at equal length
(Figure~\ref{fig:vshistory}), the market's commutator is a flat baseline broken by
isolated spikes where the model's is noisier.

\begin{table}[htbp]
\centering
\caption{Crisis-versus-calm contrasts of the four OMD observables on the S\&P
500, 1996--2024, with the cross-observable and lead-lag statistics. Crisis marks
the three drawdown periods and calm the rest. The lead-lag rows are partial
Spearman correlations on disjoint windows, controlled for each series' own
persistence. The entropy-production row is standardized.}
\label{tab:contrasts}
\vspace{0.4em}
\small
\begin{tabular}{l c c p{6.4cm}}
\toprule
observable & calm & crisis & reading \\
\midrule
delocalized exponent $\beta$ & $0.518$ & $0.571$ & rises in crises \\
participation ratio (eff.\ factors) & $8.5$ & $9.5$ & inflated by the dispersed
2001 unwind, collapses in the concentrated 2008 and 2020 \\
commutator norm & $68.1$ & $74.5$ & order parameter of the transition \\
entropy production (vol.\ chain) & $-0.16$ & $+0.03$ & the arrow of time
fires in crises \\
\midrule
\multicolumn{4}{l}{\emph{cross-observable and lead-lag statistics}} \\
$\rho(\beta,\ \text{entropy production})$ & \multicolumn{2}{c}{$+0.03$} & two
independent crisis markers rather than one index \\
$\rho(\text{commutator},\ -\text{PR})$ & \multicolumn{2}{c}{$+0.76$} & the
concentration order-parameter link \\
crowding $\to$ PR (partial) & \multicolumn{2}{c}{$-0.11$ to $+0.18$} & over one
to four quarters: crowding does not lead \\
PR $\to$ crowding (partial) & \multicolumn{2}{c}{$+0.19$ to $+0.42$} & over one
to four quarters: the geometry leads the dispersion \\
\bottomrule
\end{tabular}
\end{table}

\begin{figure}[htbp]
\centering
\includegraphics[width=0.88\textwidth]{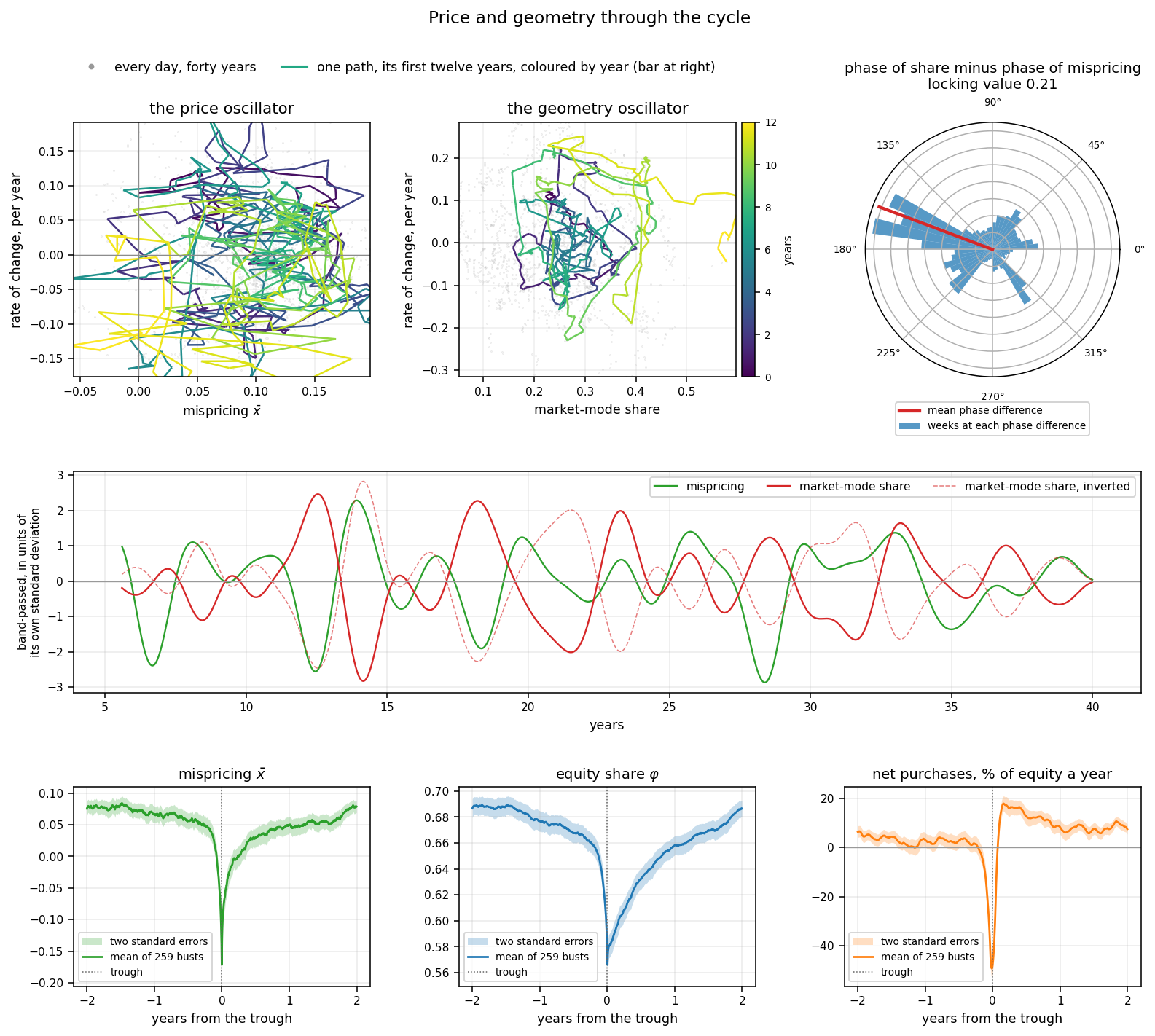}
\caption{Price and geometry through the cycle. Top: phase portraits of the index's
mispricing and of the market-mode share over one market of forty years, its first twelve years colored by year, and the histogram of their phase difference. Middle:
both band-passed to periods of two to ten years. Bottom: mispricing, equity share
and net executed purchases averaged over $259$ bear markets, day zero at the
trough.}
\label{fig:leanosc}
\end{figure}

\label{sec:onecycle}
\label{sec:twooscillators}
If taken literally, Soros's co-determination of perception and price could be read as making the price and
the geometry two coupled oscillators, the geometry here being the market-mode share $S(t)$ of Eq.~\eqref{eq:modeshare}, which
records how far the names move together under the investor's common flow. Both do orbit at three and a half to four
years (Figure~\ref{fig:leanosc}, top). The middle panel shows both series band-passed, keeping only their Fourier
components with periods of two to ten years and leaving the phases unshifted, each
in units of its standard deviation and the share also inverted. Where the inverted
share follows the mispricing, the share peaks as the mispricing bottoms, as in some
cycles and not in others. The polar panel is the histogram, over weekly samples, of
the phase difference $\Delta\Phi_t$ of the two filtered series, read from their
Hilbert transforms. Its mean, $159^\circ$, is near antiphase, but $\Delta\Phi_t$ is
within $45^\circ$ of antiphase in only $43$ percent of the weeks. The phase-locking
value $|\langle e^{i\Delta\Phi_t}\rangle|$, one for a fixed difference and zero
for a uniform spread, is accordingly only $0.21$ ($0.22$ under the tuned set), with no
frequency of its own in either (Section~\ref{sec:relatedmodels}), so the two are coupled
as broadband series rather than as oscillators.

\begin{figure}[htbp]
\centering
\includegraphics[width=\textwidth]{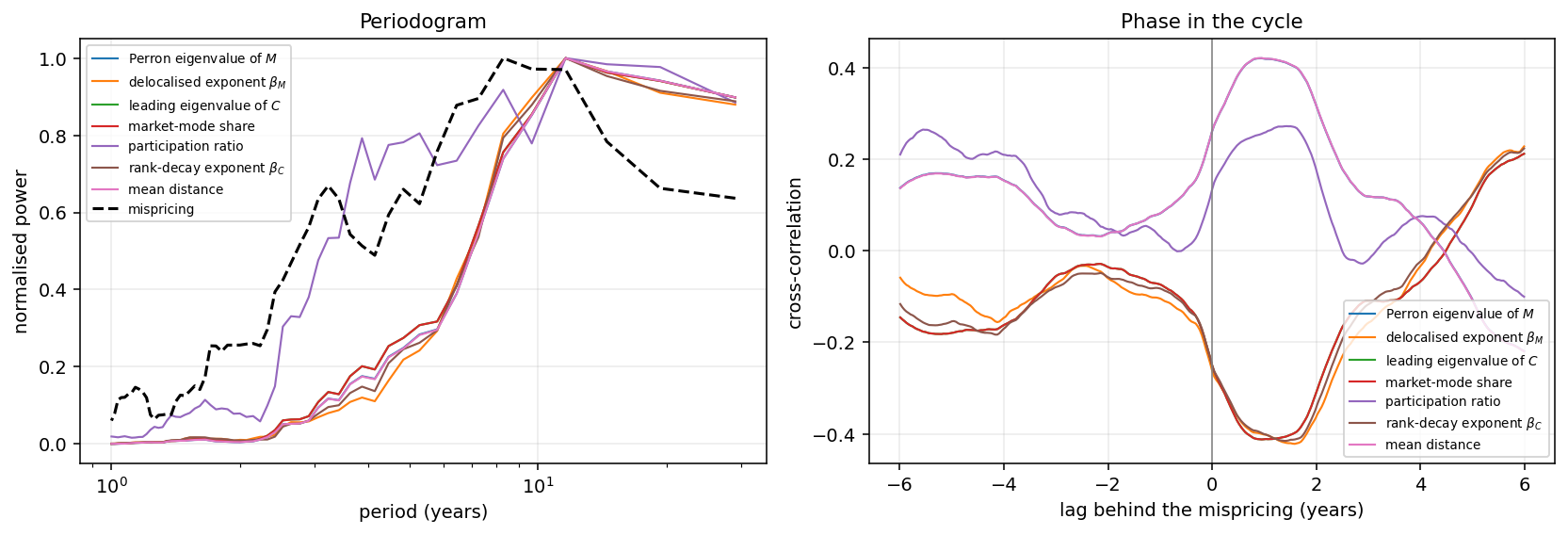}
\caption{Where the spectral measures of the distance matrix sit in the cycle,
on a two-year lookback over one market of sixty years. Left: periodograms
against the mispricing's. Right: cross-correlation with the
mispricing, positive lags placing the measure behind it.}
\label{fig:leanspectra}
\end{figure}

Through the cycle the geometry moves on a broad band of periods near a decade and
follows the mispricing by about a year (Figure~\ref{fig:leanspectra}). The left
panel shows periodograms, the smoothed squared modulus of the Fourier transform of
each detrended series, which spread its variance over frequencies, a sharp peak
marking a dominant period.
As the price falls back toward value the names fold onto their market mode, and
they unfold as it recovers. Nothing was imposed on the correlation,
the same flow moving a depressed market further than a fully valued one.

\subsection{Factors and sectors through the cycle}
\label{sec:factorrotation}

Practitioners read the cycle as a rotation of relative performance across factors
and sectors. Long-short portfolios are formed on the names of the simulated BBM market,
decile-sorted and rebalanced monthly, with the standard definitions applied to the
model's own prices (size, low volatility, reversal, momentum at several formation
windows, and beta), and value run on the exact mispricing and then through noise.
The three industry groups are read as capitalization-weighted indices. No factor
premium or sector return is among the fitted observables.

\begin{figure}[htbp]
\centering
\includegraphics[width=\textwidth]{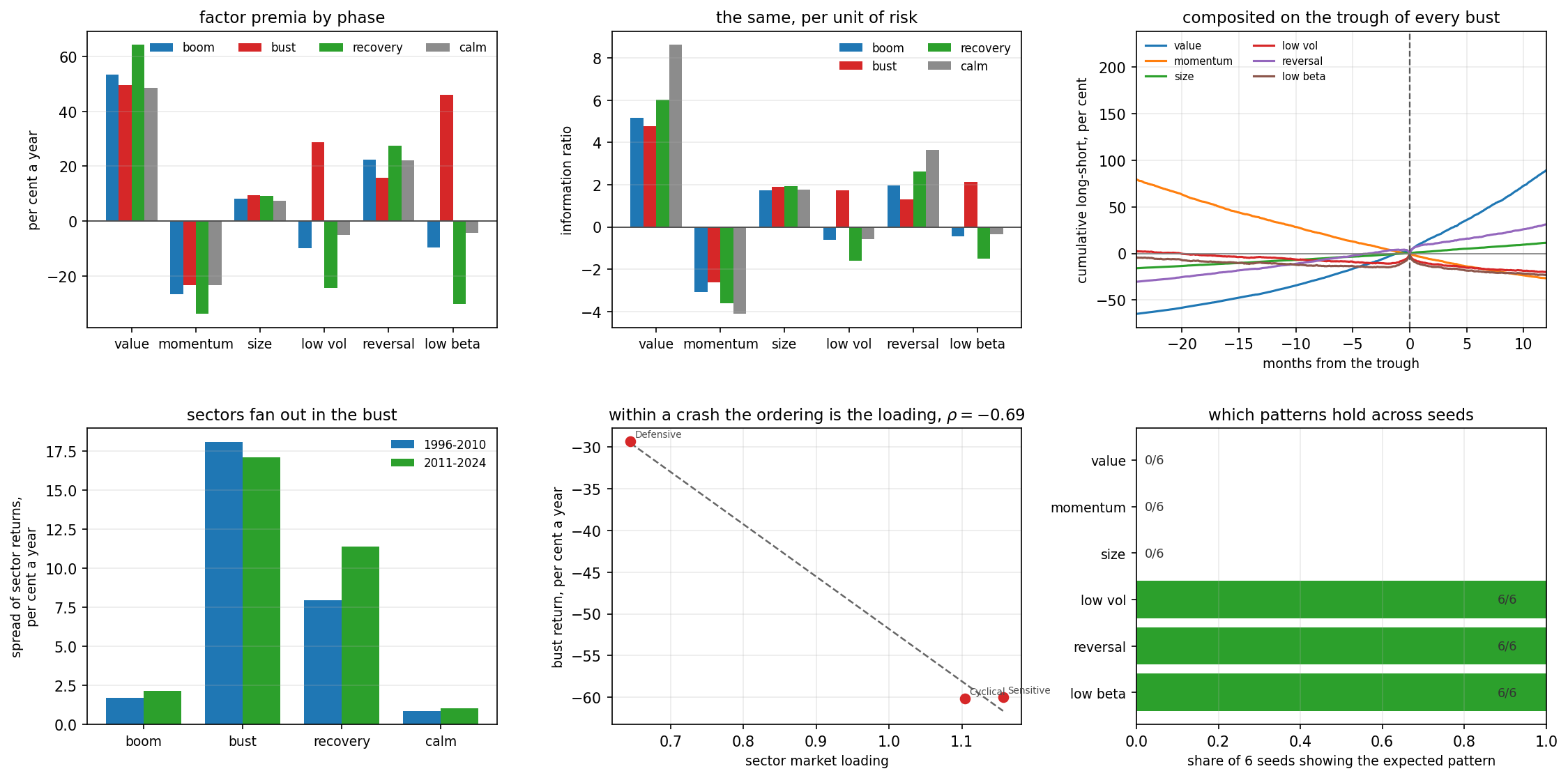}
\caption{Factors and industry groups through the cycle. Above, the six long-short
premia by phase, the same per unit of risk, and the composite around a trough.
Below, the spread of group returns by phase under both parameter sets, each
group's bust return against its market loading, and the share of seeds in which
each factor shows the expected pattern.}
\label{fig:factorsector}
\end{figure}

The factors divide cleanly (Table~\ref{tab:factorrotation},
Figure~\ref{fig:factorsector}). Low volatility and low beta are defensive, earning
in the bust, mostly on the way into the trough, and giving it back in the recovery.
Size and reversal earn throughout. Momentum loses at every formation window and
loses more the longer the window, the sharpest disagreement with the market
anywhere in this paper. The cause is the mean-reverting mispricing, so a past
return predicts a reversal. The monotone dependence on the window identifies it,
a model that merely lacked momentum giving something near zero.
Value on the exact mispricing earns far more than any traded factor, because no
agent here arbitrages across names, and degraded by noise four times the
signal it falls to about what the traded value factor has earned.

\begin{table}[htbp]
\centering
\caption{Long-short factor returns by phase, arithmetic and annualized, in percent and gross of costs. The last column is the information ratio in the bust. Eight markets of 40 years, 98 episodes.}
\label{tab:factorrotation}
\vspace{0.4em}
\small
\begin{tabular}{l rrrr r r}
\toprule
Factor & Boom & Bust & Recovery & Calm & All & IR in the bust \\
\midrule
Value, exact mispricing & $53.4$ & $49.6$ & $64.4$ & $48.7$ & $51.9$ & $4.77$ \\
Size & $8.1$ & $9.4$ & $9.2$ & $7.3$ & $8.0$ & $1.91$ \\
Low volatility & $-10.0$ & $28.7$ & $-24.4$ & $-5.0$ & $-4.8$ & $1.75$ \\
Reversal, one month & $22.3$ & $15.7$ & $27.5$ & $22.2$ & $22.1$ & $1.32$ \\
Low beta, estimated & $-9.7$ & $46.0$ & $-30.2$ & $-4.3$ & $-3.1$ & $2.12$ \\
Low beta, drawn loading & $-6.0$ & $24.1$ & $-17.5$ & $-3.0$ & $-2.4$ & $1.72$ \\
\addlinespace
Momentum, two months & $-13.6$ & $-14.4$ & $-15.7$ & $-12.5$ & $-13.4$ & $-1.61$ \\
Momentum, three months & $-16.9$ & $-16.8$ & $-21.3$ & $-15.6$ & $-16.8$ & $-1.82$ \\
Momentum, six months & $-22.8$ & $-21.0$ & $-28.2$ & $-19.9$ & $-21.8$ & $-2.27$ \\
Momentum, twelve months & $-26.8$ & $-23.5$ & $-33.9$ & $-23.5$ & $-25.6$ & $-2.64$ \\
\bottomrule
\end{tabular}
\end{table}

The industry groups compare directly with the market
(Table~\ref{tab:sectorindex}). The defensive group earns over the market through
the bust, gives it back in the recovery and lags the boom, as in the market, and
the other two follow their market loadings. What the model does not produce is a
rotation of leadership between crises: the group that loses most in one crash
loses most in the next, where the market moves from technology in one crisis to
financials in another. For factors and groups alike, what depends on a name's
loading on the common factor comes out as observed and what depends on a memory of
which led does not.

\begin{table}[htbp]
\centering
\caption{The three industry groups through the cycle, each a capitalization-weighted index, model against market. Each cell is the excess over the market, annualized and arithmetic in percent, and its information ratio.}
\label{tab:sectorindex}
\vspace{0.4em}
\small
\begin{tabular}{ll rr rr rr rr}
\toprule
& & \multicolumn{2}{c}{boom} & \multicolumn{2}{c}{bust} & \multicolumn{2}{c}{recovery} & \multicolumn{2}{c}{calm} \\
\cmidrule(lr){3-4}\cmidrule(lr){5-6}\cmidrule(lr){7-8}\cmidrule(lr){9-10}
Group & & excess & IR & excess & IR & excess & IR & excess & IR \\
\midrule
Cyclical & model & $+1.8$ & $+0.19$ & $-4.1$ & $-0.86$ & $+3.2$ & $+0.33$ & $+0.8$ & $+0.13$ \\
 & market & $-1.9$ & $-0.22$ & $+2.2$ & $+0.48$ & $+0.2$ & $+0.03$ & $-0.9$ & $-0.19$ \\
\addlinespace
Sensitive & model & $+2.6$ & $+0.28$ & $-3.5$ & $-0.72$ & $+11.7$ & $+0.91$ & $+0.8$ & $+0.13$ \\
 & market & $+4.8$ & $+0.54$ & $-5.3$ & $-1.00$ & $+6.1$ & $+0.81$ & $+2.6$ & $+0.50$ \\
\addlinespace
Defensive & model & $-2.9$ & $-0.21$ & $+14.1$ & $+1.76$ & $-14.2$ & $-0.98$ & $-0.8$ & $-0.08$ \\
 & market & $-2.7$ & $-0.20$ & $+14.7$ & $+1.47$ & $-8.9$ & $-0.77$ & $-3.4$ & $-0.43$ \\
\bottomrule
\end{tabular}
\end{table}

\let\savedFloatBarrierFive\FloatBarrier\let\FloatBarrier\relax
\section{Money Flows and Firms Through the Cycle}
\let\FloatBarrier\savedFloatBarrierFive
\label{sec:factorcycle}

This section examines the cycle through money and firms: the price impact of demand
(Section~\ref{sec:inelastic}), how much money moves and where it sits
(Section~\ref{sec:moneyflow}), the return per unit of risk
(Section~\ref{sec:sharpe}), firm sizes (Section~\ref{sec:sizedist}) and failures
(Section~\ref{sec:boomsdefaults}).

\subsection{Market inelasticity and the dumb-money effect}
\label{sec:inelastic}

Classical theory makes the aggregate market close to elastic, a dollar of demand
moving prices by little because other holders stand ready to sell. Gabaix and
Koijen \cite{gabaix2021} argue the opposite, that a dollar invested raises
aggregate value by about five. The model's price equation gives a related but different quantity. Holding the
other inputs fixed, setting the drift of Eq.~\eqref{eq:lmsvx} to zero and
differentiating in the demand memory gives the local susceptibility of the
mispricing to the latent demand,
\begin{equation}
\frac{\partial x^*}{\partial y} = \frac{-c\,V'(x^*)}{\kappa_x + c\,y\,V''(x^*)} ,
\label{eq:multiplier}
\end{equation}
where the denominator does not vanish. Evaluated at the simulated states, off the
nullcline, it is the local drift susceptibility $-\partial_y b_x/\partial_x b_x$, a
log-price response per unit of demand memory, the response to an externally
varied latent state. It is not the value created by an executed dollar, since the
demand memory enters the price directly while executed trades move the holdings
and the cash, and reach the price only through what they do to later demand.

\begin{figure}[htbp]
\centering
\includegraphics[width=0.86\textwidth]{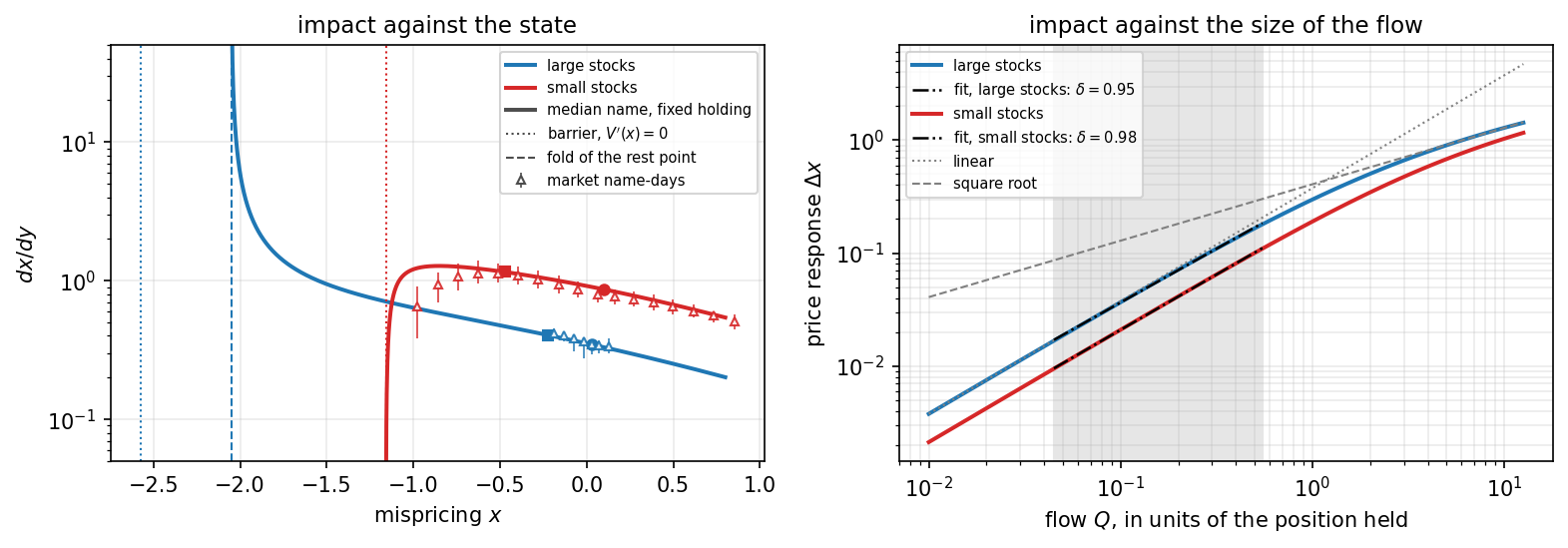}
\caption{Left: the impact of flow on price against the mispricing at each group's
median rest holding, with the simulated names' own values as triangles and the
barriers dotted. Right: the response of the mispricing to a permanent flow against
the size of the flow, with linear and square-root guides, the shaded span being
the range of quarterly flows the market shows.}
\label{fig:impactlaw}
\end{figure}

Under the one-share normalization the susceptibility is in value per unit of
demand memory, evaluated at the simulated states rather than at a solved rest
point. Averaged with the weights of the dollars the investor holds it is $0.64$, a
typical name moving about twice as far as a large one. That is about an eighth of
the Gabaix--Koijen multiplier, but theirs is the price response to an executed
dollar, identified by an instrument, and the model has no derived map from an
executed dollar to its demand memory, the budget and the execution constraints
separating the two. The impact depends on the state (Figure~\ref{fig:impactlaw}).
The restoring force weakens as the price falls, so the market is hardest to move
when it is dear, the reflexive statement in the form of an elasticity. In the size
of the demand the response is close to linear over every level the market reaches,
far from the one half of the square-root law.

Three readings of the model's impact are kept apart. The first is this static
susceptibility, $0.64$. The second is an intervention: paired experiments that lift
the investor's desired holdings move the index by about a seventh of a dollar per
unit of added demand memory within three months. The third is the reduced-form
regression of Gabaix and Koijen, the market's return on the contemporaneous net
flow into equities. Its estimate
depends on which flow is used, $+11$ to $+16$ on the investor's executed purchases and
$-1.5$ to $-1.7$ on the change in the demand memory. Neither flow is exogenous, the
holding moving toward a target that rises with the price and the demand memory
following the preceding quarter's return, and that is the bias their instrument
removes.

The literature offers two answers on whether investor flows chase returns or lean
against them. Studies of individual trades find households contrarian
\cite{kaniel2008,grinblatt2000}, while studies of aggregate fund flows find money
chasing past performance \cite{warther1995} and doing so to its cost, the
dumb-money effect of Frazzini and Lamont \cite{frazzini2008}. The model separates
the two readings. Its belief extrapolates the month just past and its latent demand
follows the preceding quarter's return, but its executed trades do not. The cost of the flows has the data's sign. The third of names the investor bought
most underperforms the third it bought least by $6.0$ and $6.1$ percent a year
over the following year in the two windows. Sorted as Frazzini and Lamont sort,
into value-weighted fifths held for a month, the model's spread is front-loaded.
After a sort on a year's flow it is about $-19$ percent a year in the first month,
against their $-6.0 \pm 2.3$. It falls by about half by the third month and has almost vanished after a
year, while theirs builds over two to three years. Held for three years, a sort on
a quarter's flow loses $8$ to $11$ percent in the model against their $9$. The model
thus gets the sign of the effect and its three-year size, but not its timing. Dumb money here is a property of what
an extrapolating investor holds, not of how fast it trades.

The model reproduces neither square-root regularity of the microstructure
literature, the law found for metaorders \cite{bouchaud2022}, since it has no order
book and no execution, nor the regime that Itkin \cite{itkin2026impact} derives from
depleting latent liquidity, nor the apparent one that Cont, Kukanov and Stoikov
\cite{cont2014} derive from linear impact, since one investor's trades are not
independent events whose signs cancel. The support for the inelastic-markets
picture is therefore partial: the flow sort has the sign of the effect and its
three-year size but not its timing, and the aggregate has the direction but a size
that cannot be set against an executed-dollar multiplier. Neither an elasticity nor
a flow statistic is a calibration target.

\subsection{How much money moves}
\label{sec:moneyflow}

\begin{figure}[htbp]
\centering
\includegraphics[width=0.78\textwidth]{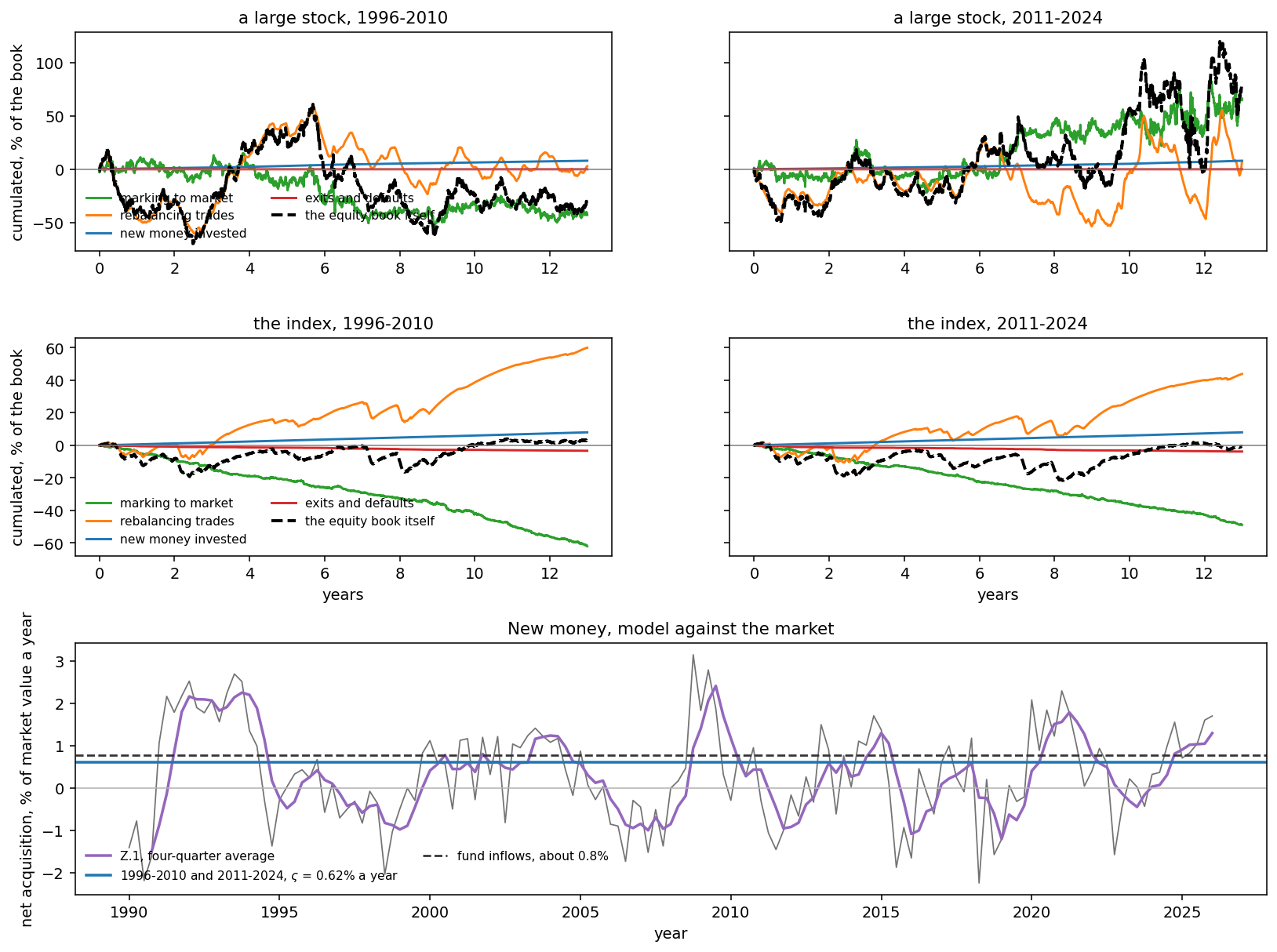}
\caption{Money flows in the calibrated market. Above, the investor's equity book over thirteen years decomposed into new money, rebalancing, marking to market and exits. Below, new money in the model against the net acquisition of corporate equities by all sectors \cite{fedz1}.}
\label{fig:moneyflow}
\end{figure}

The quantity of equity is fixed, so what changes is its price and the money the
investor commits to holding it, every purchase moving money out of the bond and
into the market (Figure~\ref{fig:moneyflow}). Two flows have to be kept apart. The
latent demand memory of Eq.~\eqref{eq:yaccum} is a state and not a trade. On one
denominator, the equity book at the open, its gross movements come to sixteen to twenty-one times the book a year, while the investor's executed trades are fifty to fifty-four percent of it, some thirty to forty times smaller. Against the summed demand memory instead the two are about three times and a twelfth to a tenth of it a year, the same ratio. New money enters at the rate $\varsigma$.

\begin{figure}[htbp]
\centering
\includegraphics[width=0.88\textwidth]{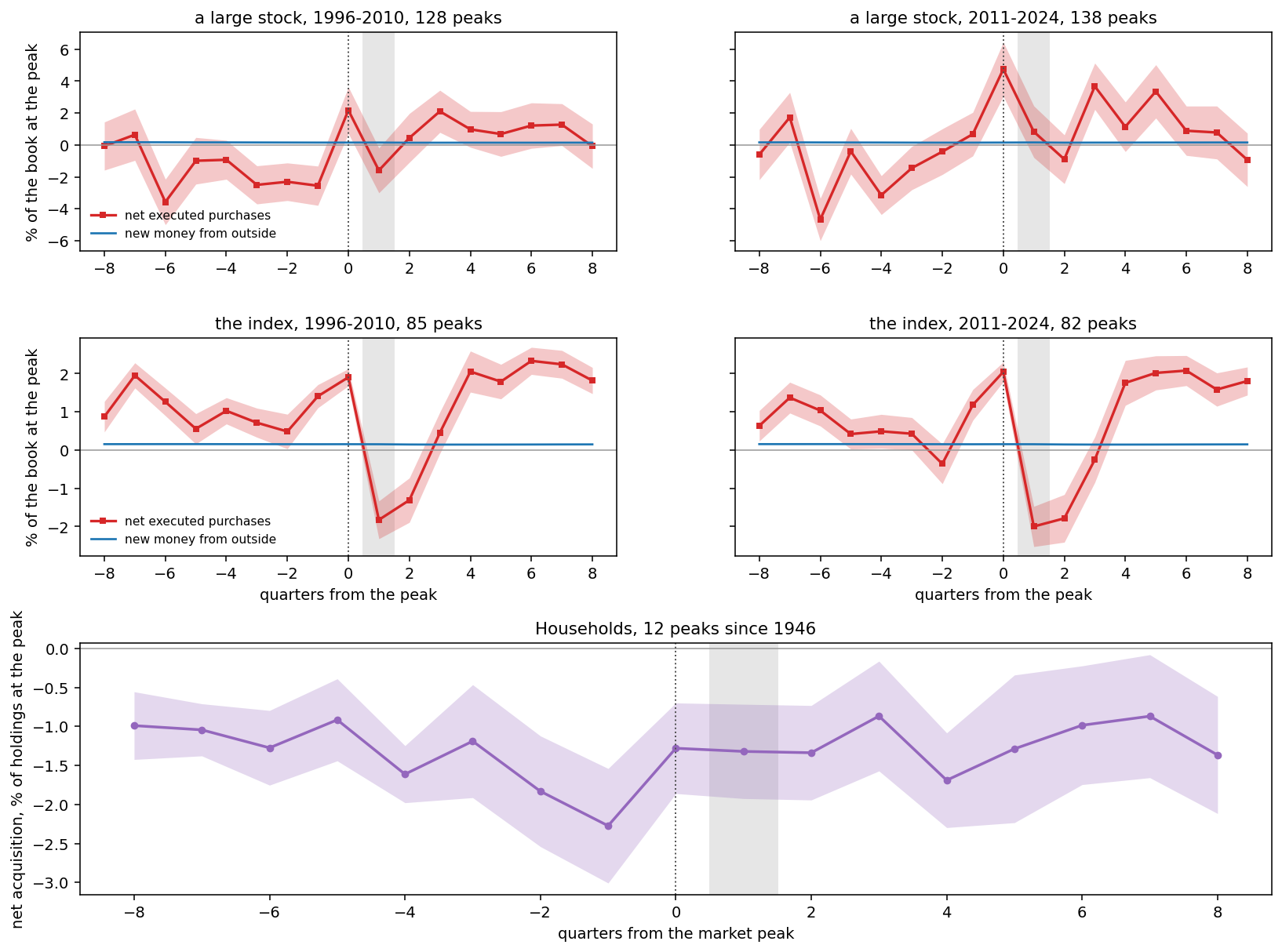}
\caption{Net executed purchases around a market peak, relative to the holding at the peak, in the model (above) and for households around the twelve peaks since 1946 (below) \cite{fedz1}.}
\label{fig:flowpeaks}
\end{figure}

Through the cycle the executed trades follow the price. Around the peaks of
Figure~\ref{fig:flowpeaks} the investor's net purchases fall in the quarter after
a peak by about two and a half percent of its holding, at four to five standard errors, and it buys again from the third or fourth quarter. The
latent demand moves the other way, rising after the top as falling prices open a
gap between the holding and the holding the investor wants, which the partial
adjustment of Eq.~\eqref{eq:partialadj} closes only slowly. The households of the
flow of funds match the sign of the latent demand around the twelve peaks since
1946, though twelve episodes cannot make the difference significant.

\begin{figure}[htbp]
\centering
\includegraphics[width=\textwidth]{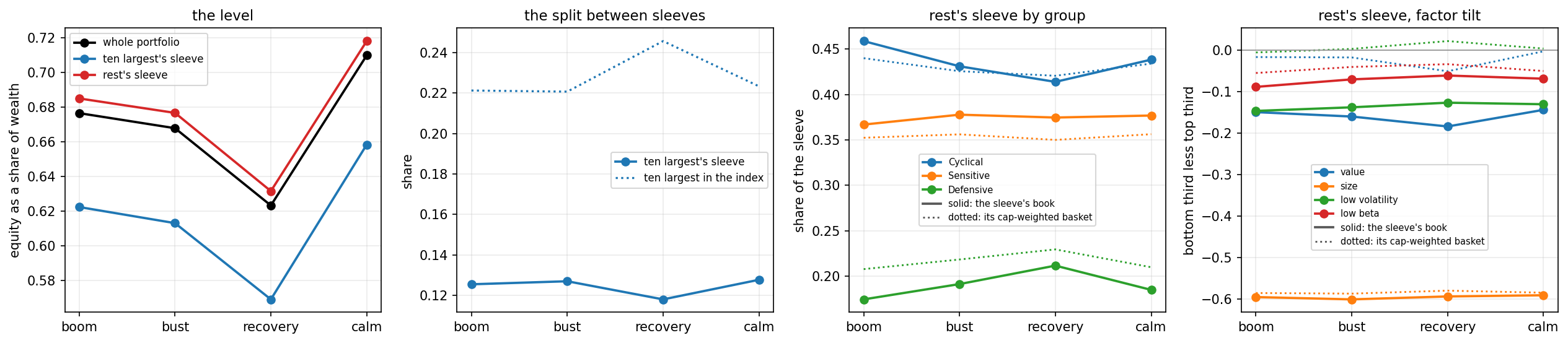}
\caption{The investor's equity book through the cycle, under the first window's
inputs. From left, the equity share of the whole portfolio and of each sleeve,
the ten largest's sleeve against their index weight, the rest's sleeve by industry
group against its basket, and that sleeve's factor tilts against its basket.}
\label{fig:flowcomp}
\end{figure}

Read sleeve by sleeve (Figure~\ref{fig:flowcomp}), the equity share dips through
the bust and the recovery and the two sleeves move together. The one large
departure from the market's weights is the split between the sleeves, each
starting at the index weight and never rebalanced, so as the largest names grow
their sleeve falls behind. Within each sleeve the industry groups and the factor
tilts stay close to the sleeve's basket in every phase. What moves with the cycle
is the level of the book, not its division.

\subsection{What the market and the investor earn per unit of risk}
\label{sec:sharpe}

Table~\ref{tab:sharpe} gives the Sharpe ratio of the market, the model's index and
the investor's book over the two windows, a quantity not among the fitted
observables. The model's index earns its return more smoothly than the market did
over 1996--2010, the dot-com bust and the credit crisis falling in a span the model
reproduces in depth only partly, and the ranges meet over 2011--2024. The investor
keeps less than the index it holds, the shortfall sitting in the sleeve of the rest
of the index, which keeps removed names for up to five years. Both of the
investor's ratios meet the market's.

\begin{table}[htbp]
\centering
\caption{Sharpe ratios over the two windows, in excess of a three percent bond return, with one-standard-deviation intervals. The investor's whole portfolio weights its two sleeves by wealth.}
\label{tab:sharpe}
\vspace{0.4em}
\footnotesize
\setlength{\tabcolsep}{3.5pt}
\begin{tabular}{l rrr rrr c}
\toprule
& \multicolumn{3}{c}{1996--2010} & \multicolumn{3}{c}{2011--2024} & \\
\cmidrule(lr){2-4}\cmidrule(lr){5-7}
Subject & return & vol.\ & Sharpe & return & vol.\ & Sharpe & meet \\
\midrule
S\&P 500 & $0.066$ & $0.207$ & $0.17$\,\tiny[-0.05,0.40] & $0.130$ & $0.172$ & $0.58$\,\tiny[0.35,0.81] & --- \\
BBM index & $0.137$ & $0.200$ & $0.54$\,\tiny[0.46,0.63] & $0.147$ & $0.204$ & $0.58$\,\tiny[0.49,0.67] & No, Yes \\
\addlinespace
Investor, whole portfolio & $0.068$ & $0.138$ & $0.28$\,\tiny[0.20,0.36] & $0.076$ & $0.137$ & $0.34$\,\tiny[0.26,0.42] & Yes, Yes \\
Investor, equity & $0.099$ & $0.211$ & $0.33$\,\tiny[0.26,0.40] & $0.111$ & $0.211$ & $0.39$\,\tiny[0.31,0.46] & Yes, Yes \\
\addlinespace
\quad ten largest, whole sleeve & $0.091$ & $0.119$ & $0.52$\,\tiny[0.39,0.64] & $0.103$ & $0.146$ & $0.51$\,\tiny[0.40,0.63] & Yes, Yes \\
\quad ten largest, equity & $0.141$ & $0.199$ & $0.56$\,\tiny[0.45,0.68] & $0.167$ & $0.245$ & $0.57$\,\tiny[0.47,0.67] & No, Yes \\
\quad rest of the index, whole sleeve & $0.064$ & $0.144$ & $0.24$\,\tiny[0.16,0.31] & $0.070$ & $0.139$ & $0.29$\,\tiny[0.21,0.37] & Yes, Yes \\
\quad rest of the index, equity & $0.092$ & $0.217$ & $0.28$\,\tiny[0.21,0.36] & $0.101$ & $0.211$ & $0.34$\,\tiny[0.27,0.41] & Yes, Yes \\
\bottomrule
\end{tabular}
\end{table}

\subsection{The distribution of firm sizes}
\label{sec:sizedist}

What the size distribution should resemble is disputed. Gibrat's proportionate
growth gives a lognormal \cite{gibrat1931}, Axtell reported a Zipf law across all
US firms \cite{axtell2001}, and Kondo, Lewis and Stella find on Census data a
lognormal, or a lognormal convolved with a non-Zipf Pareto, better far into the
tail \cite{kondo2021}. An index is a selected sample of the largest few hundred, so
the object to reproduce is a lognormal body with a heavy upper tail. The model
inherits its dispersion at the window's opening (Section~\ref{sec:leancal}), and
the dynamics must keep it.

\begin{figure}[htbp]
\centering
\includegraphics[width=\textwidth]{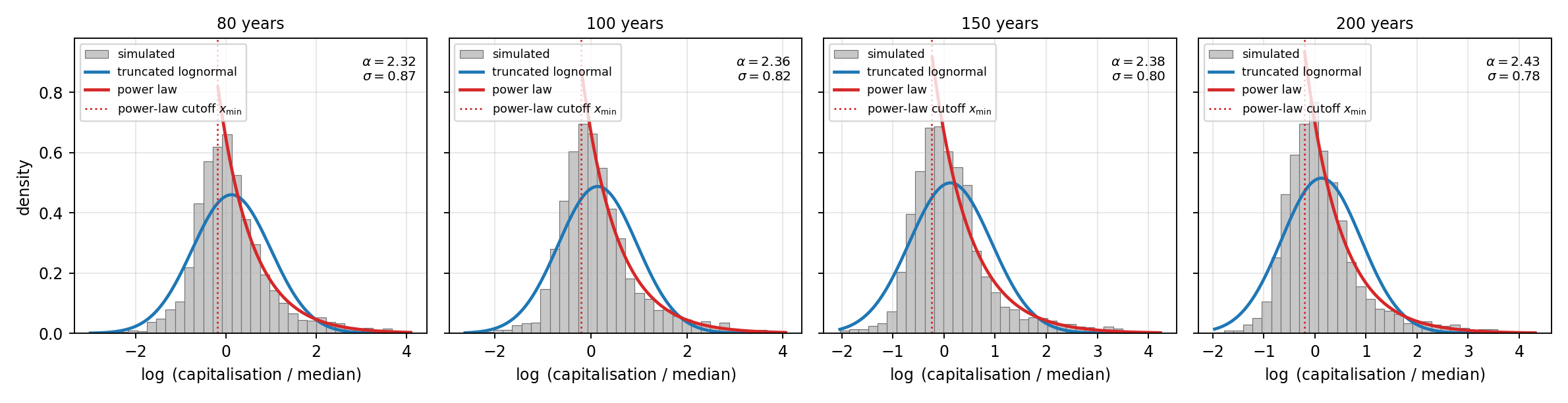}
\caption{The simulated size distribution at four ages, relative to each market's
median and pooled over five markets, with the fitted lognormal and power law.}
\label{fig:sizefits}
\end{figure}

\begin{figure}[htbp]
\centering
\includegraphics[width=\textwidth]{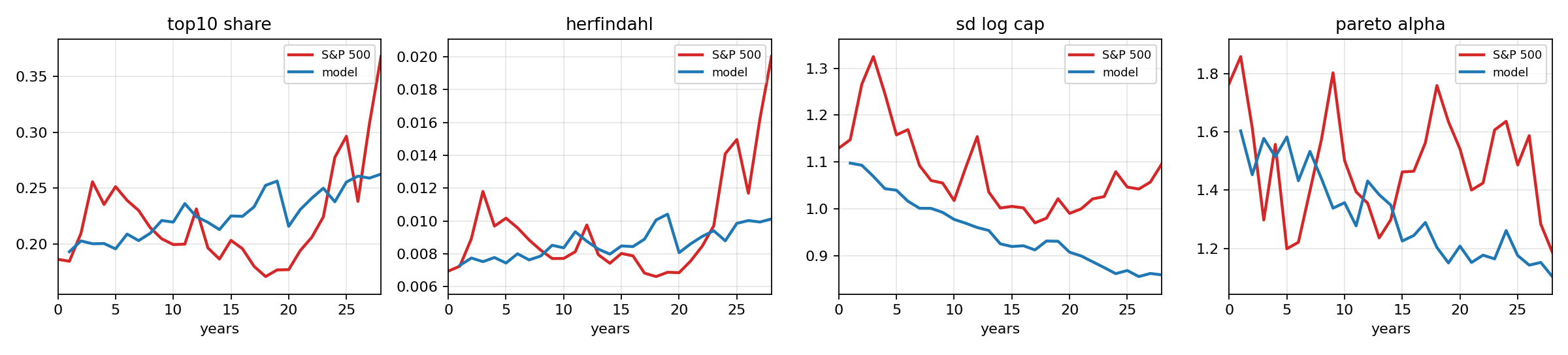}
\caption{Four summaries of the size distribution, model against the S\&P, over
the span the market covers: the share held by the ten largest, the Herfindahl
index, the dispersion of log capitalization and the exponent of the upper tail,
each measured yearly.}
\label{fig:sizedist}
\end{figure}

They do (Figures~\ref{fig:sizefits} and~\ref{fig:sizedist}). At ages of eighty to
two hundred years the family is the S\&P's: the lognormal beats the power law over
all names, a likelihood ratio cannot separate them above the cutoff
of \cite{alstott2014}, and the tail exponent of the density stays near the Zipf
value of two on both sides, the model's tail a little thinner and its sizes
somewhat less dispersed. The body is stationary from an age of about sixty.

Two mechanisms keep it so, against proportionate growth, which alone gives a
lognormal whose spread grows without bound. The first is the mean-field coupling
of Eq.~\eqref{eq:fundamental}. The drift $-\kappa_F\,(l^{(i)}_t - \overline{l}_t)$
draws each name's log fundamental toward the mean over the index's members, the
McKean-Vlasov interaction of Section~\ref{sec:interactions}. A name's log fundamental
relative to that mean is then a mean-reverting jump-diffusion with a stationary
law. Its price departs from the fundamental only by the stationary mispricing and
transitory error, so its log capitalization relative to the market is stationary
too. This coupling makes the distribution of relative size settle and
holds back concentration. Sixty years after the warm-up the largest twenty
names hold $0.37 \pm 0.03$ of the market, against a measured $0.332$, which rises
to $0.47 \pm 0.03$ at $\kappa_F = 0$.

The second mechanism is a barrier near the bottom, which turns Gibrat's law into a
Zipf law \cite{gabaix1999}. The membership rule of Section~\ref{sec:index} is that
barrier, since a retired name's slot goes to an entrant seated at the median
capitalization. It acts on the small names rather than on concentration. Without
it the share of the largest twenty stays at $0.32 \pm 0.02$, while the spread of
log capitalization across the rest of the index widens from $0.73$ to $1.01$.
Rank-based models reach the same stability through the capitalization ordering
\cite{banner2005}, and stochastic portfolio theory is built on a capital
distribution that stays stable while the market compounds \cite{fernholz2002}.

\subsection{Booms and failures}
\label{sec:boomsdefaults}

The barrier of the potential does not time the cycle, but it does govern the
failures. Moving it by any of the parameters that set it moves the default
rate in the direction an escape requires, with an elasticity of the log rate
against the log barrier of about one. That is not a Kramers escape: the barrier
stands at about twenty-five in units of the noise on a typical name's mispricing,
so an escape driven by that noise alone would respond twenty-five times as fast.
The barrier orders the rate, and how often the noise, with its bursts of
volatility, reaches the wall sets its size. Setting the volatility of volatility
of every factor to zero leaves the barrier in place and removes all but one in
forty-five of the failures, so defaults respond to the variance where the cycle
does not. The ten largest face walls far higher, which is why they do not fail.

\begin{figure}[htbp]
\centering
\includegraphics[width=\textwidth]{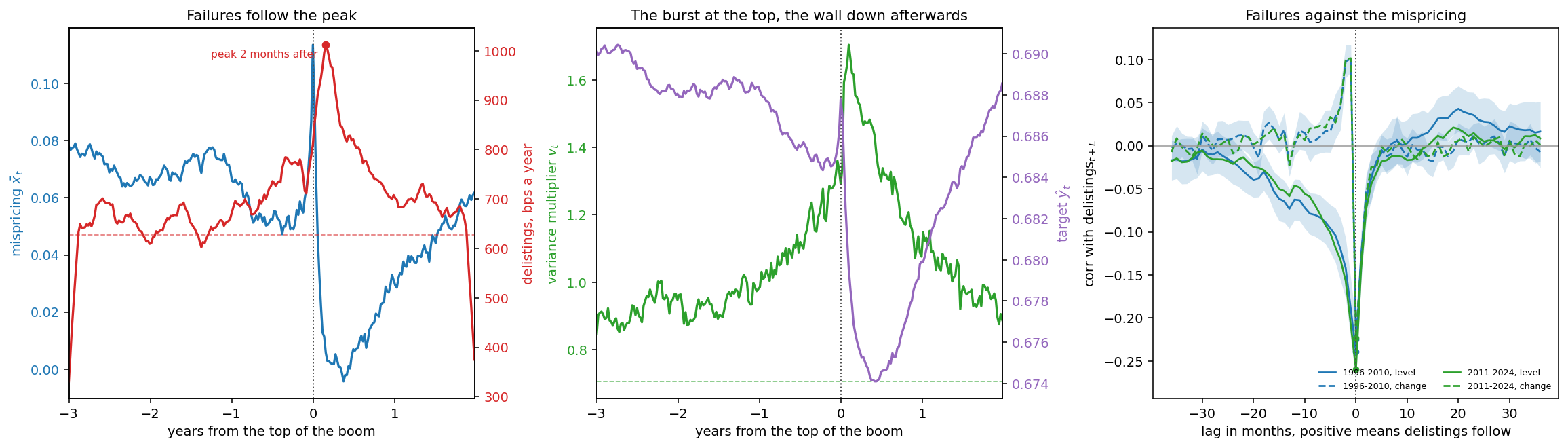}
\caption{Booms, busts and failures, averaged over $250$ bear markets with day zero
at each peak. Left: mispricing and the rate of exits. Center: variance and the
demand memory's target. Right: cross-correlation of monthly exits with the
mispricing and its change.}
\label{fig:boomdefault}
\end{figure}

The two events are ordered in time (Figure~\ref{fig:boomdefault}). The rate at
which names leave the index peaks about two months after the market tops out, and
defaults are four times as frequent in the cheapest third of days as in the
dearest. The market shows the same ordering, with failure waves after the top in
2007--2009 and 2020, and here it falls out of a rule that never refers to the
cycle. The variance peaks in the quarter after the top and the demand memory's
target bottoms about six months after, so the burst of variance comes first and the
withdrawal second.

\section{BBM as an Excitable System}
\label{sec:excitability}

We now return to the second property of markets announced in the introduction,
their excitability, which there was offered only as a qualitative analogy. The
excursion of an excitable system past its threshold has a size of its own,
possessing ``only a weak dependence on the external driving''
\cite{lindner2004}. Izhikevich distinguishes two cases
\cite{izhikevich2000}. In the first the rest state is a global attractor and the
excursion, the spike, returns to it. In the second the rest state's domain of
attraction has a boundary near it, and a large excursion ``does not return to the
equilibrium''.

This section presents the calibrated market as an excitable system at two levels:
the index, read with the rules that retire and replace its members, and a single
firm followed inside it. Section~\ref{sec:relatedmodels} shows that both levels are
driven by noise, with neither an oscillation nor chaos behind them, and
Section~\ref{sec:excitthreshold} that both have a threshold and an excursion of a size
of their own. Section~\ref{sec:bustlaw} shows where they part, on the recovery,
which a firm has and the index does not. Section~\ref{sec:excitmicro} traces the
difference to the geometry: a firm has a critical point, a saddle past which a
deep excursion is absorbed, while the index never comes near one, so the
excitability is structural for a firm and behavioral for the index. The four
stages of a practitioner's cycle map onto that excursion: accumulation is the rest
state, the mark-up the excursion the extrapolative part of the belief sustains,
distribution the turn where the value term overtakes it
(Section~\ref{sec:leancycle}), and the mark-down the return stroke.

The experiments below perturb the market with the instrument that plays the part
of a neuron's applied current: a one-off revaluation of the fundamentals, a firm
marked to a new value on a chosen day while its price is left where it was, so
that the whole revaluation lands on the mispricing. It stands for a revaluation
the market has not yet priced, which the model's own jumps, moving value and price
together, do not produce. Delivered to one firm it measures that firm's threshold,
and delivered to every firm at once it is an exogenous crisis for the whole
index. Every experiment runs after the fifty-year warm-up, and none of the
quantities it measures is a calibration target.

\subsection{Noise-driven dynamics}
\label{sec:relatedmodels}
\label{sec:excitchaos}

The cycle this market produces is not a limit cycle, and the test is cheap. With
every noise source removed the calibrated market produces no bust over a hundred
and fifty years and its index stops moving. Removing the mispricing noise
$\sigma_x$ alone leaves $9$ and $14$ busts in six hundred market-years against
$163$ with it, while removing the demand noise changes nothing. The noise-free
skeleton relaxes to the fixed point of Section~\ref{sec:lmsv} and stays there,
where a skeleton past a Hopf would go on oscillating, so what produces the cycle is
noise pushing the system over a threshold. Nor does the noisy market show a spectral line: band-passed to periods of two to ten years, its mispricing returns
a period of about four years, but so do a phase-randomized surrogate and a matched
Ornstein-Uhlenbeck process, so the number is set by the filter. The market has a
cycle length in the practitioner's sense, since deep drawdowns recur, but no
oscillation.

\begin{figure}[htbp]
\centering
\includegraphics[width=0.92\textwidth]{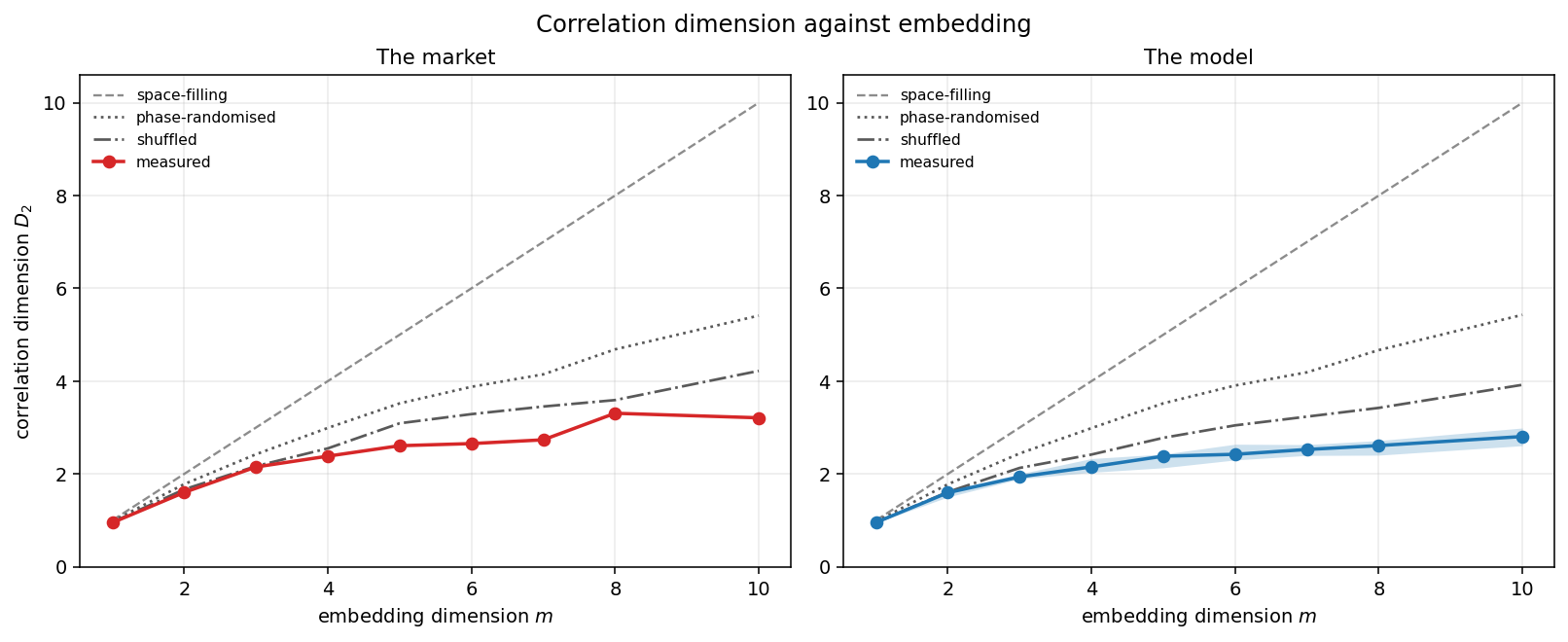}
\caption{Correlation dimension against embedding dimension for the market and the
model, against the space-filling diagonal and two surrogates. The model's band is
ten to ninety percent across twelve runs.}
\label{fig:attractor}
\end{figure}

Irregular firing alone does not separate an excitable system from a chaotic one,
which manufactures its own irregularity, and the noise-free market already shows that the
irregularity here comes from the noise. A second test, which the chaos literature applied to
financial data, checks whether the realized path lies on a low-dimensional attractor. The correlation
dimension of a delay embedding settles near $D$ for a deterministic attractor of
dimension $D$ and never settles for a stochastic process, since noise fills
whatever space it is given. Each series is measured beside surrogates of itself,
shuffled and phase-randomized, because a finite sample bends the curve downward
whatever produced it (Figure~\ref{fig:attractor} and Table~\ref{tab:attractor}). Neither the market's curve nor
the model's settles, as the excitable reading requires, and both run below their
shuffled controls by about the same amount, so there is structure in the order the
returns arrive in and the model has as much of it as the market does.

\begin{table}[htbp]
\centering
\caption{The correlation dimension at each embedding, beside the two surrogates
and the value a space-filling series would return. Delay one day, pairs within a
year excluded, three thousand points per embedding.}
\label{tab:attractor}
\vspace{0.4em}
\small
\begin{tabular}{rccccccc}
\toprule
& \multicolumn{3}{c}{S\&P 500} & \multicolumn{3}{c}{model} & \\
\cmidrule(lr){2-4}\cmidrule(lr){5-7}
$m$ & measured & shuffled & phase & measured & shuffled & phase &
space-filling \\
\midrule
1 & 0.95 & 0.96 & 0.98 & 0.96 & 0.95 & 0.97 & 1 \\
2 & 1.60 & 1.67 & 1.78 & 1.60 & 1.61 & 1.78 & 2 \\
4 & 2.38 & 2.55 & 2.99 & 2.15 & 2.42 & 2.99 & 4 \\
6 & 2.65 & 3.29 & 3.88 & 2.42 & 3.05 & 3.91 & 6 \\
8 & 3.31 & 3.59 & 4.68 & 2.61 & 3.42 & 4.67 & 8 \\
10 & 3.21 & 4.22 & 5.41 & 2.81 & 3.92 & 5.43 & 10 \\
\bottomrule
\end{tabular}

\end{table}

\subsection{Fixed-sized threshold}
\label{sec:excitthreshold}

The market-wide form of the instrument marks every fundamental up or down at once and records
whether a bust follows within nine months, against a paired run identical until
that moment. The window and the depth that counts as firing were
fixed in advance so that the unprovoked rate leaves room for a shock to raise it.

\begin{figure}[htbp]
\centering
\includegraphics[width=0.92\textwidth]{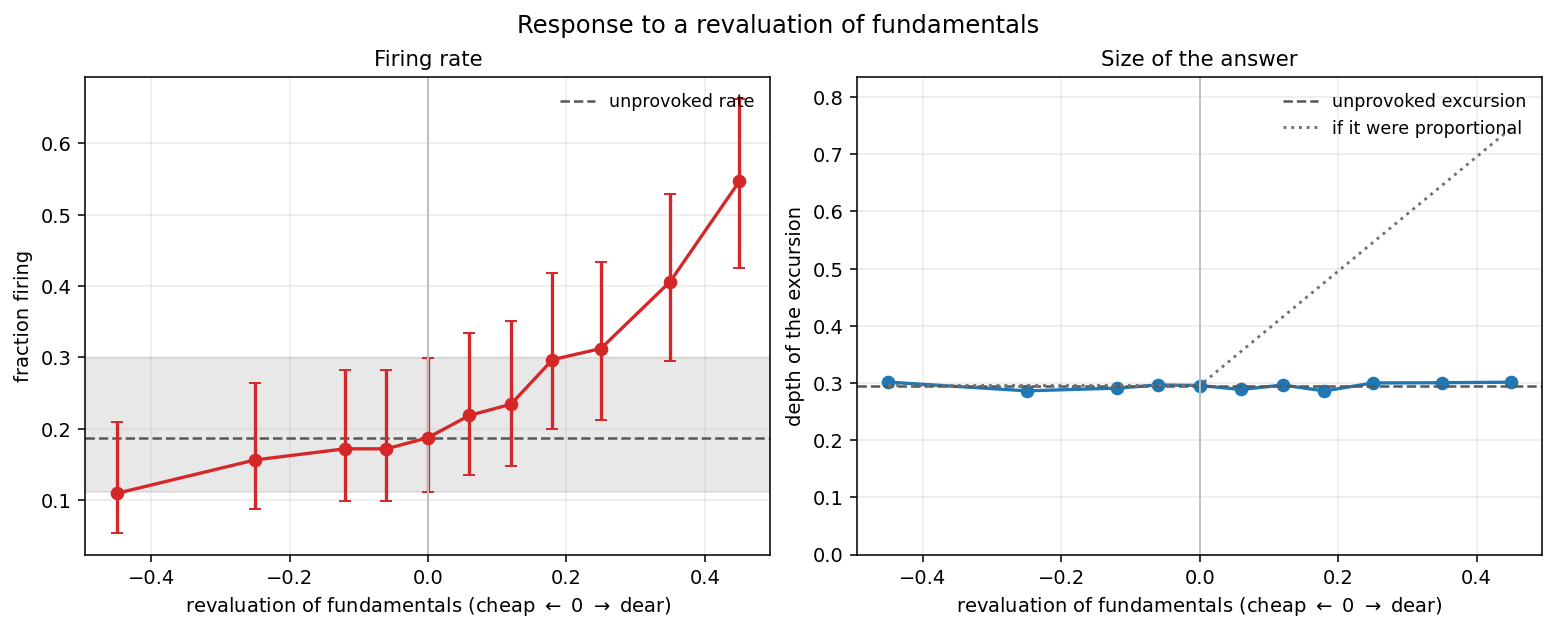}
\caption{Response to a revaluation of every fundamental at once: the fraction of
runs firing a bust, the band being the unprovoked rate (left), and the
depth of the excursion against a proportional response (right).}
\label{fig:threshold}
\end{figure}

Three things come out of the sweep (Table~\ref{tab:threshold} and
Figure~\ref{fig:threshold}). First, a threshold. Small revaluations leave the
firing rate inside its unprovoked interval, and past about a fifth it climbs, to
nearly three times the unprovoked rate at forty-five percent, a market-wide
overvaluation of between forty and fifty-five percent. It acts on one side only:
marking fundamentals down leaves the market dear and provokes it, marking them up
leaves it cheap and does not, where a response in proportion would treat both
alike. The threshold also moves with the state, a market that has risen more over
the two preceding years firing more readily.

Second, the size of the excursion. Once it happens, it has the same depth,
about thirty percent of the price, whether a small piece of news, a large one or
nothing at all set it off, while the drawdown of the runs that do not fire grows
with the shock, as the direct hit should. Third, the timing: a provoked bust
reaches its trough no sooner than one that happens on its own. A larger piece of
news makes a bust more likely without making it bigger or bringing it forward,
which is the all-or-none signature that separates an excitable system from an
amplifier. Conditioning on a crossing truncates the distribution of depths, and a
stochastic null selected the same way has not been run, so the reading stays
behavioral. The rate rises smoothly rather than abruptly because the noise decides
the outcome close to the boundary, while the size of the excursion, which the noise
does not smear, keeps the all-or-none character.

The one-sidedness also bears on the other half of the cycle. Nothing in the
sweep starts a boom. The bust is a threshold crossing followed by an excursion of fixed size,
while the boom is the slow accumulation that brings the market to the threshold
and has no size at which it begins. An oscillator would have two symmetric swings
and no threshold on either.

\begin{table}[htbp]
\centering
\caption{Firing rate and excursion depth against the size and sign of the
revaluation, the depth a fall in price. A star marks a rate whose interval clears
the unprovoked one.}
\label{tab:threshold}
\vspace{0.4em}
\small
\begin{tabular}{rlccc c}
\toprule
revaluation & market made & fired & rate & 95\% interval & depth when it fired \\
\midrule
-0.45 & cheap & 7/64 & 0.109 & [0.05, 0.21] & -0.302 \\
-0.25 & cheap & 10/64 & 0.156 & [0.09, 0.26] & -0.286 \\
-0.12 & cheap & 11/64 & 0.172 & [0.10, 0.28] & -0.291 \\
-0.06 & cheap & 11/64 & 0.172 & [0.10, 0.28] & -0.297 \\
+0.00 & none & 12/64 & 0.188 & [0.11, 0.30] & -0.296 \\
+0.06 & dear & 14/64 & 0.219 & [0.14, 0.33] & -0.289 \\
+0.12 & dear & 15/64 & 0.234 & [0.15, 0.35] & -0.297 \\
+0.18 & dear & 19/64 & 0.297 & [0.20, 0.42] & -0.286 \\
+0.25 & dear & 20/64 & 0.312 & [0.21, 0.43] & -0.300 \\
+0.35 & dear & 26/64 & 0.406 & [0.29, 0.53] & -0.301 \\
+0.45 & dear & 35/64 & 0.547$^{*}$ & [0.43, 0.66] & -0.302 \\
\bottomrule
\end{tabular}

\end{table}

To find what pushes the market across on its own, we remove one candidate
at a time. Market-wide jumps in the fundamentals matter little, the rate of twenty
percent falls barely moving without their clustering or without them at all. The
variance matters for the deep falls. Freezing it at its mean takes falls of thirty
percent from $7.0$ a century to $2.2$, and their clustering goes with them, the
coefficient of variation falling from $1.32$ to $0.78$. Ordinary diffusive shocks therefore push the market across often enough for a shallow bust, and the deep
busts and their clustering come from episodes of high variance, not from a piece
of news large enough to do it alone.

A probe matched to the variance tests that route directly. A process with a random
variance is a process with a constant variance run in a random time, so a one-off
multiple of the market volatility, left to decay at its own rate, speeds up the
market's own clock. The market barely responds (Table~\ref{tab:volthreshold}). Up to
nine times the volatility no level clears the unprovoked interval, while a market
that does fire keeps the fixed depth of its excursion. A burst that fades within weeks
is too short against the months a boom takes to build, and what raises the bust
rate is the variance staying high.

\begin{table}[htbp]
\centering
\caption{Firing rate and excursion depth against a transient multiple applied to
the market volatility, which then decays at the variance process's own rate. A
star would mark a rate whose interval clears the unprovoked one.}
\label{tab:volthreshold}
\vspace{0.4em}
\small
\begin{tabular}{rccc c}
\toprule
volatility & fired & rate & 95\% interval & depth when it fired \\
\midrule
$\times 1.00$ & $12/64$ & $0.188$ & $[0.11, 0.30]$ & $-0.296$ \\
$\times 1.25$ & $11/64$ & $0.172$ & $[0.10, 0.28]$ & $-0.306$ \\
$\times 1.50$ & $12/64$ & $0.188$ & $[0.11, 0.30]$ & $-0.296$ \\
$\times 2.00$ & $12/64$ & $0.188$ & $[0.11, 0.30]$ & $-0.301$ \\
$\times 3.00$ & $13/64$ & $0.203$ & $[0.12, 0.32]$ & $-0.298$ \\
$\times 4.50$ & $14/64$ & $0.219$ & $[0.14, 0.33]$ & $-0.296$ \\
$\times 6.00$ & $15/64$ & $0.234$ & $[0.15, 0.35]$ & $-0.299$ \\
$\times 9.00$ & $15/64$ & $0.234$ & $[0.15, 0.35]$ & $-0.312$ \\
\bottomrule
\end{tabular}

\end{table}

\begin{figure}[htbp]
\centering
\includegraphics[width=\textwidth]{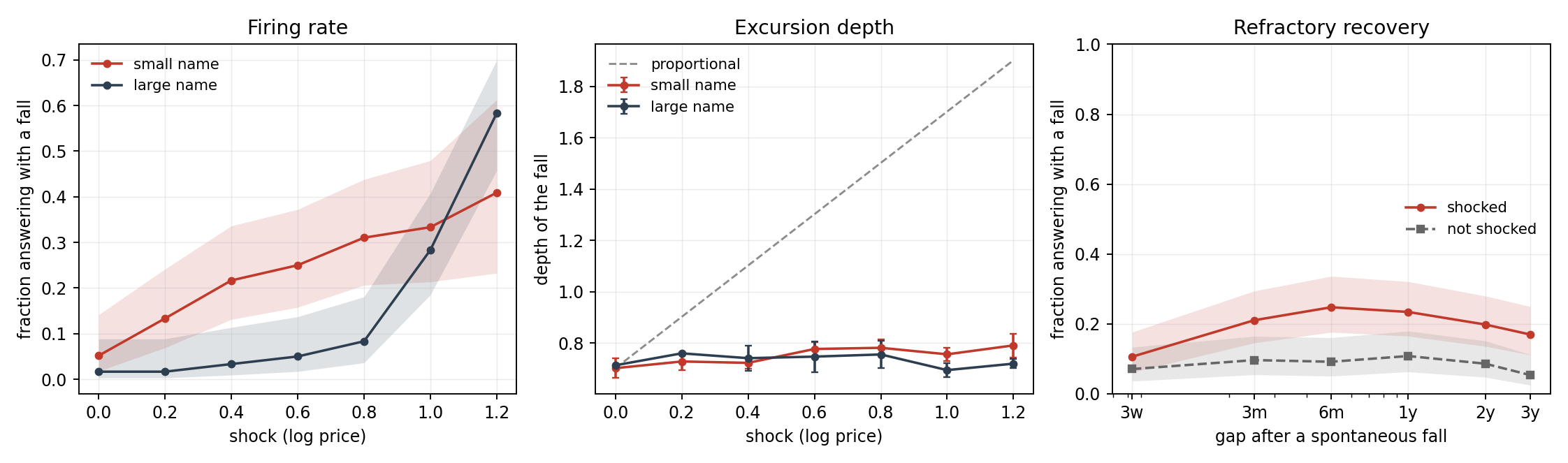}
\caption{Response of a name to a shock delivered to it alone: the fraction of
runs responding with a fall past $0.60$ in the logarithm (left), the depth of that
fall against a proportional response (center), and the response after the name's
own spontaneous fall (right).}
\label{fig:nameresponse}
\end{figure}

A firm has the same two signatures. The instrument is turned on a single stock,
the largest name or the median name of the rest, displaced while the market around
it is left as it was, and the event is that name's own fall past $0.60$ in the
logarithm within nine months (Figure~\ref{fig:nameresponse}). The small name fires
about one time in twenty unprompted, shocks up to forty percent leave it near that
rate, and past them the rate climbs clear, to about eight times the unprompted
rate at the largest shocks. Leaving the market inside the window competes with
falling at the largest shocks, and how a departure is counted changes the size of
the response but not the rise. The large name holds until a shock of a hundred percent, so a bigger name needs a bigger push. Over the whole range the excursion
keeps its size, $0.70$ to $0.79$ in the logarithm, the all-or-none excursion read on
a name's own price.

\subsection{Recovery and spacing of busts}
\label{sec:excitrefractory}
\label{sec:bustlaw}
\label{sec:whatsetsperiod}

The two levels part on the recovery. If the market cannot fire again until it has rebuilt, a shock soon after a bust
should be refused and one later should not, the paired-pulse experiment of the
laboratory. The market's own spontaneous bust is the conditioning event, and the
shock follows at a controlled interval, at two sizes that the sweep places below
the threshold and at its lower edge.

\begin{figure}[htbp]
\centering
\includegraphics[width=0.62\textwidth]{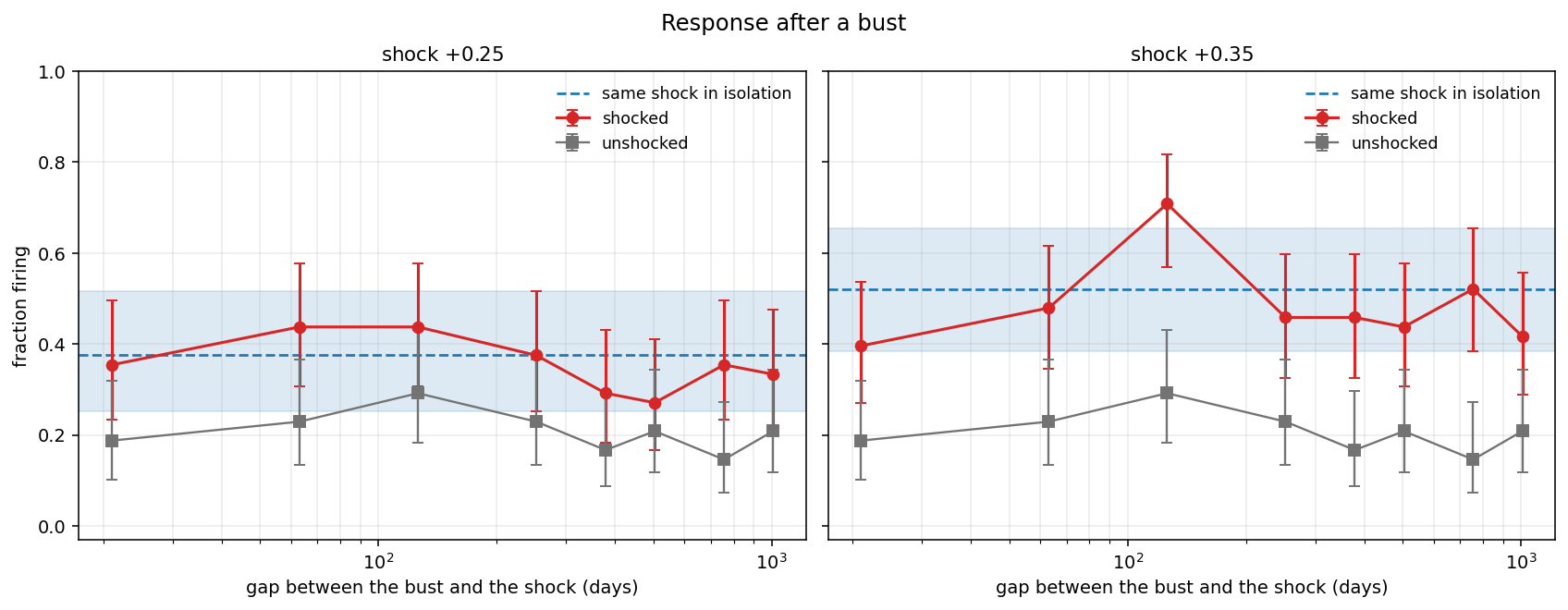}
\caption{Response after a bust: the fraction of runs responding to a shock delivered a
given interval after a spontaneous bust, against the same runs unshocked, at two
shock sizes.}
\label{fig:refractory}
\end{figure}

Neither measurement resolves a refractory period (Table~\ref{tab:refractory} and
Figure~\ref{fig:refractory}). Shocks three and nine weeks after a bust fire about
as often as the same shocks long before it, and left alone, a market that has just
busted busts again more readily, the clustering the spacing below measures.

\begin{table}[htbp]
\centering
\caption{Firing rate against the interval between a spontaneous bust and the
shock, at two shock sizes, with the unshocked rate of the same runs.}
\label{tab:refractory}
\vspace{0.4em}
\small
\begin{tabular}{rccc}
\toprule
gap after the bust (days) & fires, shock $+0.25$ & fires, shock $+0.35$ &
fires unshocked \\
\midrule
21 & 0.354 & 0.396 & 0.188 \\
63 & 0.438 & 0.479 & 0.229 \\
126 & 0.438 & 0.708 & 0.292 \\
252 & 0.375 & 0.458 & 0.229 \\
378 & 0.292 & 0.458 & 0.167 \\
504 & 0.271 & 0.438 & 0.208 \\
756 & 0.354 & 0.521 & 0.146 \\
1008 & 0.333 & 0.417 & 0.208 \\
\midrule
same shock in isolation & 0.375 & 0.521 & -- \\
\bottomrule
\end{tabular}

\end{table}

For a short while after its own fall, a name responds less often to a second shock. Delivered three weeks after the small name's spontaneous
fall, the shock fires $0.11$ of the time, climbing to $0.25$ by six months, where
it meets the rate in isolation, a suppression an exact test separates from chance
at $p = 0.01$. So the threshold and a recovery of a few months are both a name's,
and the simulated BBM market, by retiring and replacing the names that fell, averages the
recovery away.

The busts arrive irregularly, and the law of their spacing discriminates between
mechanisms. A memoryless escape over a barrier gives exponential intervals with a
coefficient of variation of one, and a sharply periodic oscillator gives a
coefficient near zero. The calibrated market gives $1.13 \pm 0.06$ and
$1.20 \pm 0.07$ in the two windows, over about eight hundred and forty intervals
between twenty percent falls, a mean gap of $3.7$ years. That is above one, so
busts cluster, and falls of thirty percent, every twelve to thirteen years, are
within a standard error or two of a memoryless escape. Fitting a shifted
exponential finds no dead time to place.

\begin{figure}[htbp]
\centering
\includegraphics[width=\textwidth]{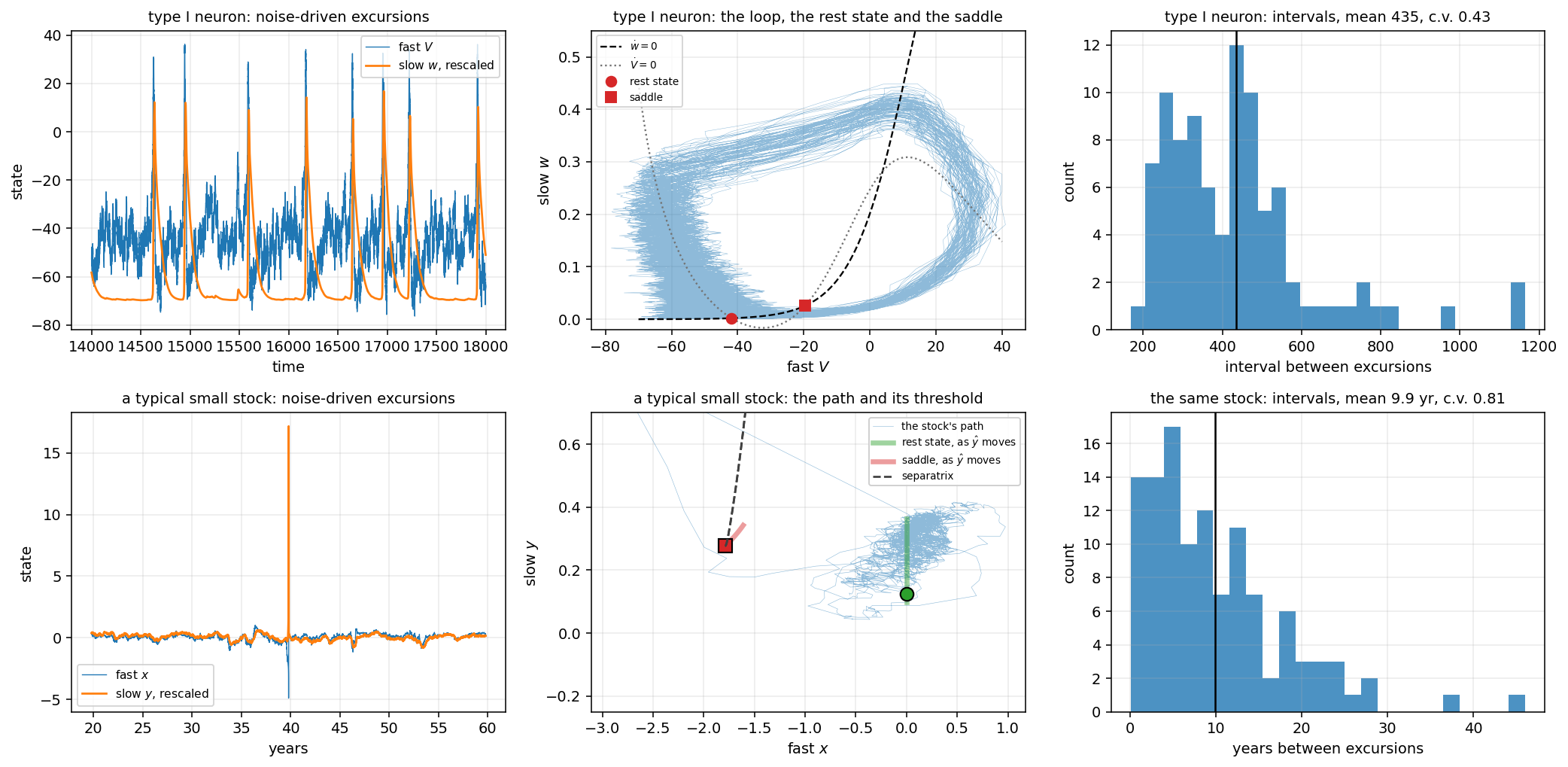}
\caption{The stochastic Morris-Lecar neuron (above) and one typical small stock
inside the calibrated market (below): a stretch of the two states, the plane they
move in with fixed points and separatrix, and the intervals between excursions.}
\label{fig:fhn}
\end{figure}

A firm's falls are more regular. The canonical member of the class
\cite{lindner2004} with a firm's geometry is a stochastic type I model. The stochastic Morris-Lecar model \cite{morris1981} has a stable node beside a saddle exactly as our pair does, while the type II FitzHugh-Nagumo neuron
\cite{fitzhugh1961} sits near a Hopf bifurcation with no clearly defined threshold
\cite{fhnreview2024}. Figure~\ref{fig:fhn} runs the market against the neuron
panel for panel, the market's side being the median single stock of the rest of
the index with the removal rule lifted. A Kramers escape with no recovery gives a coefficient of variation of
one, and a recovery variable pushes it below one. The single stock's large falls
come at $0.81$, as a recovery would make them, and the neuron's firings at $0.43$.

\begin{figure}[htbp]
\centering
\includegraphics[width=0.95\textwidth]{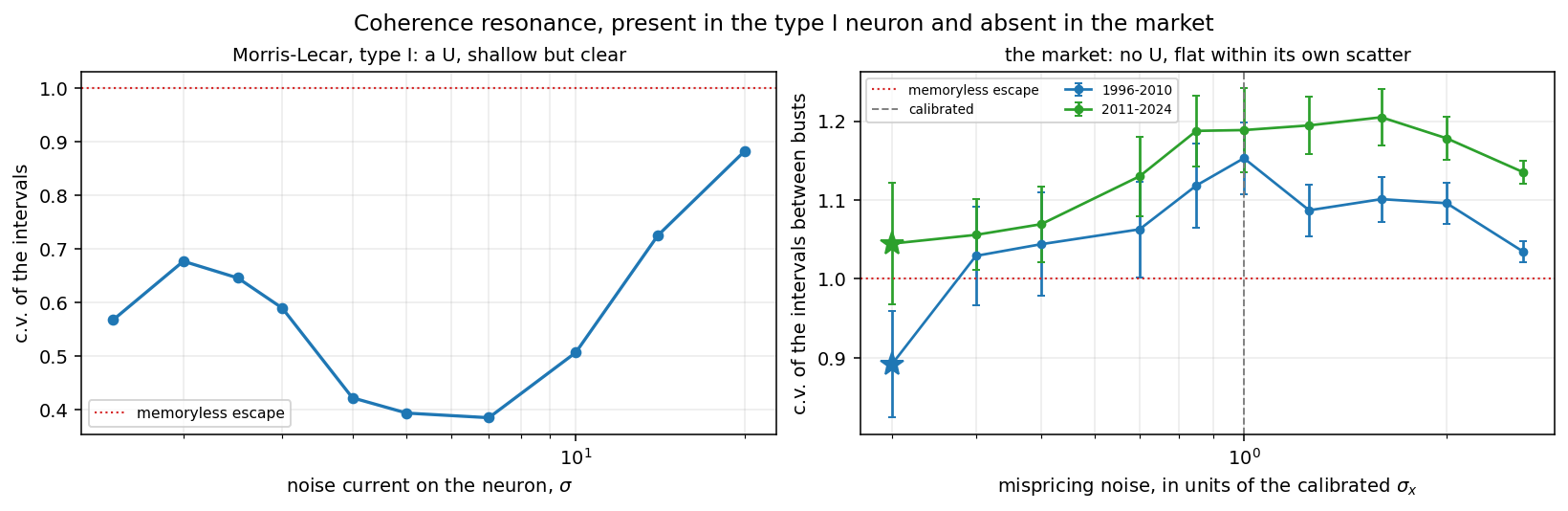}
\caption{Regularity of the intervals against the noise: the stochastic
Morris-Lecar neuron (left) and the market, its noise in units of the calibrated
one (right). The dotted line is a memoryless escape.}
\label{fig:cr}
\end{figure}

Nor does the index show coherence resonance \cite{pikovsky1997}: an excitable system is most
regular at an intermediate noise, since too little never crosses the threshold
and too much fires again before the recovery has run. The Morris-Lecar neuron
shows it, and the index does not (Figure~\ref{fig:cr}). Over a ninefold range of
the price noise its coefficient of variation stays between $1.03$ and $1.20$, with
no interior minimum. Because the interval is set by the noise rather than by a
period, it moves with the noise, the mean gap running from about thirty years at
three tenths of the calibrated noise to under one at two and a half times it. A
market with a Hopf cycle would keep its period and merely blur it.

Nor does a slow pool of resources make them regular. A population of neurons
coupled to a pool that responds to the average activity bursts collectively when
the pool moves slowly \cite{franovic2022}, and this market has that architecture,
since the demand memory's target of Eq.~\eqref{eq:yhat} contains the investor's own holding. Replacing the realized holding there with the fair-value share leaves the
rate, depth and spacing of the busts where they were. Scaling the rate $\mu_i$ of
the demand memory over a factor of sixty-four makes busts rarer only when it is
slowed to an eighth, and the coefficient of variation of the spacing stays between
$1.07$ and $1.11$ throughout.

Palatella and Grigolini \cite{palatella2012} predict noise-induced intermittency
in which the quiet intervals follow an inverse power law with exponent near two.
Neither the I-BBM nor a sleeve of four hundred and ninety names sharing one budget
shows such a tail: the first passage to a twenty-five percent drawdown is
exponential in both, with a median of four and a half years and a coefficient of
variation of one, a Kramers escape. Sharing a budget leaves the escape at about
the pace of the single asset.

To find what times the interval, we move nine constants one at a time over
the narrow range the calibration leaves open (Table~\ref{tab:periods}). The
mispricing noise moves it at an elasticity of $-2.8$, from nearly ten years to
under three across the range, and nothing else reaches a third of that. The
constants that place the barrier move it substantially but move the interval by a
third at most. The barrier is real, nearly every firm that defaults having crossed
its own saddle first (Section~\ref{sec:excitmicro}), but the cycle's interval
follows the noise and the variance rather than the potential, so the cycle and the
defaults respond to different things.

\begin{table}[htbp]
\centering
\caption{What times the cycle: elasticities of the mean interval between twenty
percent bear markets of the index, and of the median barrier of
Eq.~\eqref{eq:effpot}, to each parameter over $0.8$ to $1.25$ times its calibrated
value.}
\label{tab:periods}
\vspace{0.4em}
\small
\begin{tabular}{lrr}
\toprule
Parameter & Interval, 20\% falls & Barrier \\
\midrule
Coupling constant $g$ & $0.09$ & $-1.43$ \\
Flow impact $c$ & $0.10$ & $-0.49$ \\
Regularizer $\bar\epsilon$ & $0.08$ & $0.67$ \\
Flow-memory rate $\mu$ & $0.15$ & $0.01$ \\
Mispricing mean reversion $\kappa_x$ & $0.31$ & $1.47$ \\
Extrapolation weight $\theta$ & $-0.15$ & $0.00$ \\
Mispricing noise $\sigma_x$ & $-2.79$ & $0.00$ \\
Market volatility factor $s_M$ & $0.22$ & $0.00$ \\
Volatility of volatility $s_n$ & $-0.05$ & $0.00$ \\
\bottomrule
\end{tabular}

\end{table}

\begin{figure}[htbp]
\centering
\includegraphics[width=\textwidth]{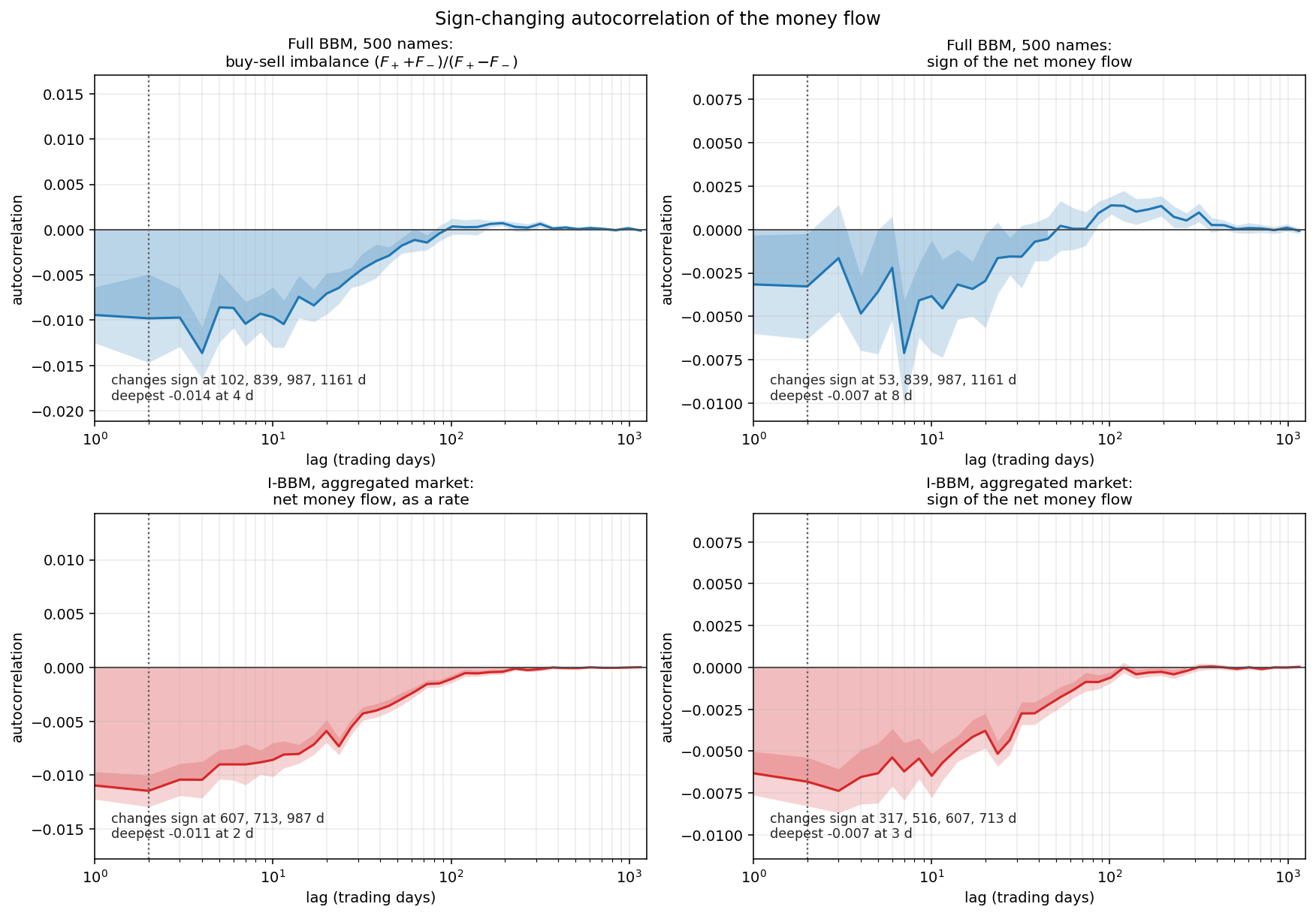}
\caption{Autocorrelation of the money flow, as a rate (left) and its sign
(right), for the simulated BBM market (top) and the I-BBM (bottom), with ninety-five percent intervals across runs.}
\label{fig:excitableflow}
\end{figure}

Read on what the investor does rather than on the price, the autocorrelation
returns something that is not memory. The signed money flow stays positively
correlated for a couple of days, crosses zero and is then anticorrelated with
itself (Figure~\ref{fig:excitableflow}): buying now means selling a few days
later. A single relaxation would decay without crossing and a Poisson process has
no structure to cross with. The crossing fits excitability, though a filter, a
difference, several modes or a constraint can also produce a negative lobe. In the
canonical models of noise-driven excitability the intervals between firings are
independent and the correlation time finite \cite{lindner2004}, which appears as
such a lobe. The market has that time in its flow, at a matter of days, and not in
its busts: nothing in the flow responds to the multi-year recurrence of the deep
busts, so it is the index and not the money moving through it that produces the long cycle.

\subsection{Structural and behavioral excitability}
\label{sec:excitmicro}

\begin{figure}[htbp]
\centering
\includegraphics[width=\textwidth]{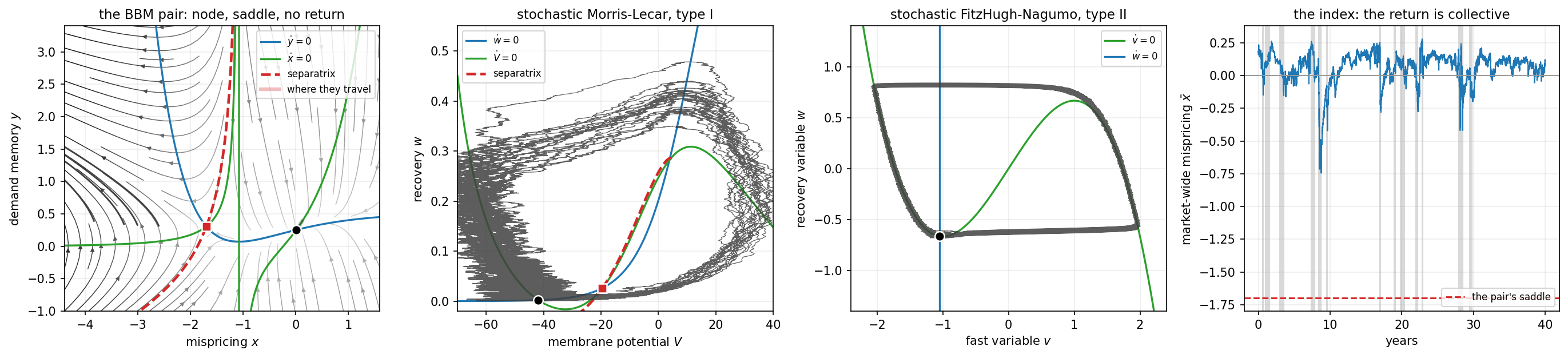}
\caption{Where the threshold is. Left to right: the pair $(x,y)$ of one typical
name with its node, saddle and separatrix; the stochastic Morris-Lecar neuron
(type I); the stochastic FitzHugh-Nagumo neuron (type II); the market-wide
mispricing over forty years against the saddle.}
\label{fig:threshgeom}
\end{figure}

A single firm has the geometry explicitly. Holding the target $\hat y$ fixed, the
drift of the pair $(x_t, y_t)$ is the gradient of the potential $\mathcal{V}$ of
Eq.~\eqref{eq:jointpot}, so the Jacobian of the noise-free flow is minus the
Hessian of a scalar and no mode can rotate. Its trace is far from the zero a Hopf
bifurcation would need. The slice has a stable node at fair value and a saddle
whose stable manifold is a genuine separatrix, the threshold at a fold that
Izhikevich draws \cite{izhikevich2000}. Past it the potential falls away, and a firm pushed far beyond it does not come back but defaults. The saddle sits at
$x = -1.73$ for the median name of the rest of the index. Figure~\ref{fig:threshgeom} puts this geometry beside the two neurons.

The construction departs from the neural models in one structural respect. The
standard excitable neurons are not gradient flows, and their return to rest is driven by a rotating flux \cite{wang2013landscape}, whereas a firm here is a
gradient flow that absorbs past its threshold. The closest neural counterparts to
a default are metabolic: ion pumps whose gradients deplete past a tipping point
\cite{hubel2014,cressman2009} and excitotoxic calcium overload \cite{verma2022}.

What happens past a name's saddle is a question for the dynamics rather than the
accounting, so each small name is followed against a boundary rebuilt from its own
constants at every step, with the removal rule switched off. Most crossings come
back: over forty markets of sixty years, three in four crossings, and three in five
names that crossed, return above the boundary, the median within two days and from
a median depth of a tenth below it. The ones that are lost go deep, and nearly
every failure, $1250$ of the $1267$ names written off at the floor, is an escape
over the barrier, so only a large enough push is irreversible. With the removal
rule live the same markets show $195$ crossings, of which $62$ end with the name
leaving the market, a name that dips toward its boundary having usually been
retired first.
A firm is therefore excitable in Izhikevich's second sense, and the excitability is
its structure: a threshold, an excursion of its own size and a recovery of a few months,
with a far side that absorbs a deep excursion rather than returning it.

\begin{figure}[htbp]
\centering
\includegraphics[width=0.8\textwidth]{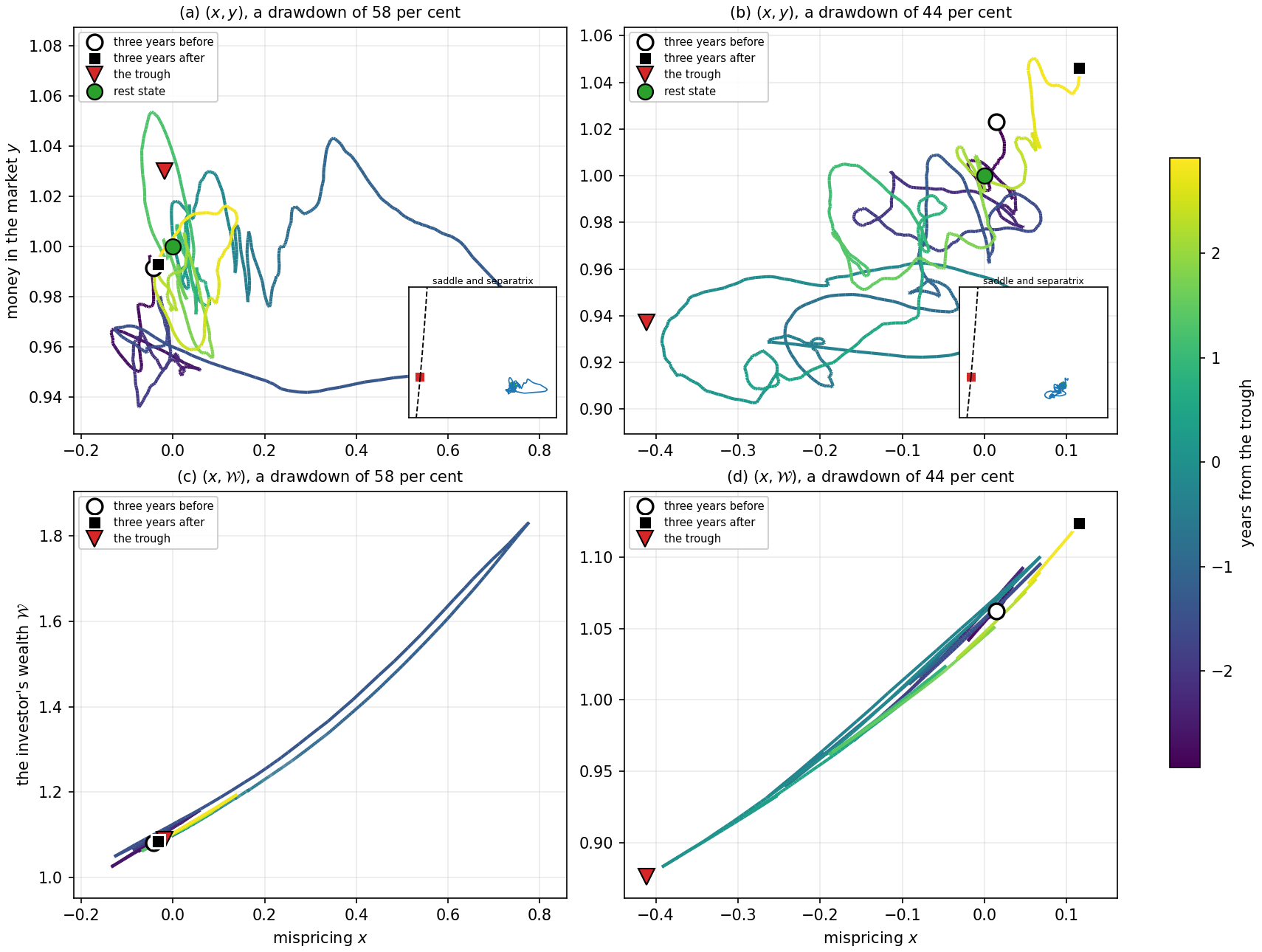}
\caption{Two drawdowns of the I-BBM index over the three years either side of
the trough, colored by time from it, in the plane of the mispricing against the
money held (above, with the rest state, saddle and separatrix inset) and against
the investor's wealth (below).}
\label{fig:phasexy}
\end{figure}

The index never comes near such a point: its mispricing averaged over names
reaches $-0.74$ at its worst over forty years, and the I-BBM's never falls below
$-0.92$ against its saddle at $-2.02$. Its booms and busts are large fluctuations
about its rest point, and what lets the index carry on where a failed firm does
not is the rule that retires a failed name and seats a replacement. Through a bust
the investor's wealth correlates $0.95$ with the mispricing
(Figure~\ref{fig:phasexy}), but the loop is not a path the state retraces, the
individual paths scattering about the mean path by more than its own extent. What
the index has is a circulation the noise sustains rather than a stochastic limit
cycle.

The index inherits the threshold and the fixed size of the excursion as behavior. Its
state reaches under half of the way to its own saddle, so a bust is a large
fluctuation of the circulation in the plane of mispricing against demand
(Section~\ref{sec:currents}), pushed across by ordinary diffusive shocks and deepened
by episodes of high variance. What returns it to rest after a failure is the
membership rule, a reset an integrate-and-fire neuron is given by hand and that this
market's dynamics do not contain, and it turns each firm's absorbing excursion into
a renewal of the index.

\section{The Market as a Non-Equilibrium Steady State}
\label{sec:nesssec}

In this section we explore irreversibility of the BBM dynamics, and measure the
probability current and the entropy production rate of the BBM in its stationary
state, on the mispricing, the demand memory and the log variance. A stationary distribution is an equilibrium only
under detailed balance, where the current vanishes and the dynamics looks the same
run backwards. For the I-BBM the drift and the diffusion of the three states are
known, and the rate is computed from them directly. For the simulated BBM market,
whose states are sums over five hundred names, the dynamics is approximated by a
multivariate Ornstein-Uhlenbeck process fitted to the simulated paths, and the rate
is that of this linear approximation, which we compute for the I-BBM as well. Both
markets break detailed balance. The simulated BBM market produces about fifteen
hundred times the floor of the estimator and the I-BBM $8.5$ nats a year, the
current turns in the direction of the boom-bust cycle, and the irreversibility is
four to five times stronger in busts and recoveries than in the calm.

To see which elements of the construction produce the irreversibility, the linear
approximation is refitted on seven ablations (Table~\ref{tab:ness_sigma}). Four
are of the simulated BBM market: the flow impact $c$ halved, the jumps switched
off, the price and flow noises uncorrelated, and the leverage correlations off.
Three are of the I-BBM: the same without the jumps. Three controls test the
estimates themselves rather than the model: a reversible surrogate that sets the
floor of the linear estimate, the removal of the days on which a constituent is
replaced, and a test on the lagged covariance that assumes neither linearity nor
Gaussianity.

\subsection{Entropy production}
\label{sec:ness}

Every run is recorded after its market's warm-up, so the current belongs to the
stationary state. In the I-BBM the three states are the mispricing $x_t$, the
demand memory $y_t$ and the log variance $z_t$. In the simulated BBM market they
are the capitalization-weighted mispricing, the per-name mean of the demand memory
and the log of the market's volatility factor. A name replaced in the simulated BBM
market restarts its mispricing and demand memory, so that market is reported with
and without the replacement days.

Write the three states of the I-BBM as $u_t = (x_t, y_t, z_t)^{T}$ and their
dynamics as $\mathrm{d}u = b(u)\,\mathrm{d}t + \sqrt{2D(u)}\,\mathrm{d}W$, with $b$
the drift vector, $D$ the diffusion matrix, one half of the instantaneous
covariance of the three noises, and $W$ a vector of independent Brownian motions.
The stationary density $p(u)$ and the probability current
$J = b\,p - \nabla\!\cdot\!(D p)$ satisfy $\nabla\!\cdot\!J = 0$, and detailed
balance is the stronger condition $J = 0$. Here $b$ and $D$ are given directly by
the I-BBM equations (Section~\ref{sec:twomodels}), which removes the need to
estimate the density $p$, the hard part of any such estimate in three dimensions.
With the force $\mathcal{G} = D^{-1}(b - \nabla\!\cdot\!D)$ \cite{seifert2012,chetrite2008},
the entropy production rate $\int J^{T}D^{-1}J\,p^{-1}\,\mathrm{d}u$ splits into
$\int \mathcal{G}^{T}J\,\mathrm{d}u$ and $-\int(\nabla\log p)^{T}J\,\mathrm{d}u$. The second
term integrates by parts to $\int \log p\;\nabla\!\cdot\!J\,\mathrm{d}u$, which is
zero in a steady state. The first is a time average along the path, since the mean
velocity of the process at $u$, read with the Stratonovich (midpoint) product
$\circ$, is $J/p$, so that $\langle \mathcal{G}^{T}\circ\dot u\rangle = \int \mathcal{G}^{T}J\,\mathrm{d}u$.
The rate is therefore the following path average \cite{frishman2020}:
\begin{equation}
\sigma \;=\; \frac{1}{\Delta t}\,\Big\langle \big[\,D^{-1}
\big(b - \nabla\!\cdot\!D\big)\big]^{T}\,\Delta u \Big\rangle ,
\label{eq:ep_direct}
\end{equation}
with the drift and the diffusion taken at the midpoint of each step. For a Markov
diffusion this is the entropy carried into the medium, in a steady state the whole
of the production \cite{seifert2012}. In the I-BBM the drift also depends on the
investor's target $\hat y_t$, which is recorded at every step but is not a function
of $u$, so the triple is not Markov on its own. We therefore read
Eq.~\eqref{eq:ep_direct} as the entropy production conditional on the target rather than as
the path-space entropy production of the triple. Replacing $\hat y_t$ by its
conditional mean given $u$ gives the same functional for a Markovian projection,
about a quarter lower (Table~\ref{tab:eprevision}), and neither is in general a
bound on the path-space rate \cite{dacosta2023}.

The simulated BBM market has no drift that could be written down, and its state is
read as the linear diffusion $\mathrm{d}u = -Au\,\mathrm{d}t + \sqrt{2D}\,\mathrm{d}W$.
As in the estimator of Nartallo-Kaluarachchi and co-authors for multivariate time
series \cite{nartallo2024}, $A$ is the least-squares regression of the daily
increment on the level and $D$ is one half of the covariance of the increments per
unit time. With $C$ the stationary covariance of the level, every stationary
Ornstein-Uhlenbeck process satisfies $AC + CA^{T} = 2D$ \cite{godreche2019}. The
symmetric part of $AC$ is therefore fixed by the noise, and the irreversibility sits
in its antisymmetric part $Q = \tfrac{1}{2}(AC - CA^{T}) = AC - D$. Detailed balance
is the condition $Q = 0$. The current is $J = \Omega' u\,p$ with
$\Omega' = D C^{-1} - A = -Q C^{-1}$, and the rate is
\begin{equation}
\sigma = \mathrm{Tr}\big(\Omega'^{T} D^{-1} \Omega' C\big)
= \mathrm{Tr}\big(Q^{T} D^{-1} Q\, C^{-1}\big),
\label{eq:ep_ou}
\end{equation}
which vanishes exactly when $Q = 0$. For a linear drift and a constant diffusion,
$\mathcal{G} = -D^{-1}Au$ and Eq.~\eqref{eq:ep_direct} reduces to Eq.~\eqref{eq:ep_ou}, so
the two differ only in the drift they are given. For $C$ solving the Lyapunov
equation, Eq.~\eqref{eq:ep_ou} is also the rate $-\mathrm{Tr}(D^{-1}AQ)$ of
Ref.~\cite{nartallo2024} written through the current. We take $C$ as the sample covariance rather than the solution of the
Lyapunov equation and $D$ from the increments rather than the regression residuals,
and on a synthetic three-state process the two versions agree to $0.3$ percent.
The rate is exact for a consistent stationary OU surrogate, approximate for the
empirically fitted matrices, the more so when days are censored, and neither a
controlled estimate nor a bound for the nonlinear market, so we use it as a diagnostic, as broken detailed
balance is measured from recorded trajectories in active matter and living cells
\cite{gladrow2016,gnesotto2018,battle2016}.

The reversible surrogate, an Ornstein-Uhlenbeck process with the same $C$ and $D$
but the drift $DC^{-1}$, returns $0.047$ nats a year, the floor of
Eq.~\eqref{eq:ep_ou}. The simulated BBM market returns $69.0$ on eight markets of
two hundred years, about fifteen hundred times the floor
(Table~\ref{tab:ness_sigma}), and $71.2$ on the six markets of
Section~\ref{sec:reflexverdict}. Removing the increments across a constituent replacement,
about a tenth of the days, raises the latter to $156$, a censored value whose fit
meets $AC + CA^{T} = 2D$ to two percent, against $0.04$ percent on all days (Supplement, Table~S8): those days account for about a third of the demand coordinate's quadratic variation, four times their share of
the days, as reverting jumps, which inflate its diffusion. Every ordering of
Section~\ref{sec:reflexverdict} survives the removal, and switching the jumps off
leaves the figure in place, so what is measured is not the jumps read as noise. On
the I-BBM's thirty-two paths Eq.~\eqref{eq:ep_direct} gives $8.5$ nats a year and
Eq.~\eqref{eq:ep_ou} gives $2.4$ ($2.07$ on the eight markets of
Table~\ref{tab:ness_sigma}). They measure different things, so the gap is not
entropy the linear fit misses, but both have the same sign and circulation. Both markets therefore have a stationary probability current, which is broken detailed
balance. Being separately calibrated models rather than one market and its
projection, their difference is not a measure of dilution by aggregation.

\begin{table}[htbp]
\centering
\caption{Entropy production, in nats a year, with its controls, for the
simulated BBM market of five hundred names and the aggregated I-BBM. Mean and standard
deviation across eight markets of two hundred years, except the row from the model's own drift, over
thirty-two. The ablations are read in the linear approximation and are not a decomposition, each also changing $C$ and $D$.}
\label{tab:ness_sigma}
\vspace{0.4em}
\small
\begin{tabularx}{\textwidth}{p{6.6cm} c X}
\toprule
System & $\sigma$, nats a year & Reading \\
\midrule
Reversible surrogate, same $C$ and $D$ & $0.047 \pm 0.006$ & the estimator's floor \\
\midrule
\multicolumn{3}{l}{\emph{The simulated BBM market, all stocks}} \\
\textbf{Calibrated} & $\bm{69.01 \pm 4.98}$ & $1500$ times the floor \\
Flow impact $c$ halved & $35.89 \pm 2.97$ & removes about half \\
Jumps switched off & $76.17 \pm 12.44$ & the circulation is not the jumps \\
Price and flow noise uncorrelated & $62.61 \pm 6.72$ & removes a tenth \\
Leverage correlations off & $55.44 \pm 6.27$ & removes a fifth \\
\midrule
\multicolumn{3}{l}{\emph{The aggregated market, as one asset}} \\
\textbf{Calibrated, from the model's own drift} & $\bm{8.5 \pm 1.0}$ & Eq.~\eqref{eq:ep_direct}, hidden inputs frozen per step \\
Calibrated, from the fitted process & $2.07 \pm 0.53$ & the linear approximation \\
\quad price and demand noise uncorrelated & $1.21 \pm 0.31$ & removes two fifths \\
\quad leverage correlation off & $1.55 \pm 0.42$ & removes a quarter \\
\quad flow impact $c$ halved & $1.63 \pm 0.39$ & within the error \\
\bottomrule
\end{tabularx}
\end{table}

\begin{table}[htbp]
\centering
\caption{Entropy production of the I-BBM from its own drift, from the drift projected on the noisy state, and from a fitted linear process, with the signed lagged-covariance asymmetries $100\,A_{ij}(\tau)$ of the model and three surrogates, $z$ the log variance. Means and standard errors across runs.}
\label{tab:eprevision}
\vspace{0.4em}
\small
\begin{tabular}{l c c c}
\toprule
\multicolumn{4}{l}{\emph{Entropy production, nats a year, 32 runs of 200 years}} \\
\midrule
Own drift, Eq.~\eqref{eq:ep_direct} & \multicolumn{3}{c}{$8.50 \pm 0.17$} \\
Projected, quadratic $\mathrm{E}[\hat y \mid x,y,z]$ & \multicolumn{3}{c}{$6.12 \pm 0.13$} \\
Projected, binned $16^{3}$ & \multicolumn{3}{c}{$6.24 \pm 0.13$} \\
Ornstein-Uhlenbeck fit & \multicolumn{3}{c}{$2.43 \pm 0.09$} \\
\midrule
\multicolumn{4}{l}{\emph{Signed asymmetry $100\,A_{ij}(\tau)$, 12 runs of 60 years}} \\
\midrule
 & $(x,y)$ & $(x,z)$ & $(y,z)$ \\
\midrule
\multicolumn{4}{l}{One day} \\
\quad I-BBM & $+0.99 \pm 0.05$ & $+0.45 \pm 0.04$ & $+0.88 \pm 0.06$ \\
\quad Reversible gradient flow & $+0.05 \pm 0.04$ & $-0.03 \pm 0.05$ & $+0.05 \pm 0.06$ \\
\quad Symmetrized-spectrum surrogate & $-0.01 \pm 0.03$ & $-0.01 \pm 0.06$ & $+0.03 \pm 0.05$ \\
\quad Phase-randomized surrogate & $+0.99 \pm 0.05$ & $+0.44 \pm 0.05$ & $+0.88 \pm 0.06$ \\
\midrule
\multicolumn{4}{l}{One month} \\
\quad I-BBM & $+14.53 \pm 0.78$ & $+6.82 \pm 0.75$ & $+15.49 \pm 1.07$ \\
\quad Reversible gradient flow & $+0.51 \pm 0.77$ & $-0.37 \pm 1.04$ & $+0.58 \pm 1.02$ \\
\quad Symmetrized-spectrum surrogate & $+0.22 \pm 0.61$ & $-0.07 \pm 1.10$ & $+0.57 \pm 0.72$ \\
\quad Phase-randomized surrogate & $+14.47 \pm 0.79$ & $+6.84 \pm 0.76$ & $+15.42 \pm 1.08$ \\
\bottomrule
\end{tabular}
\end{table}

No single element produces the effect alone (Table~\ref{tab:ness_sigma}). In the
simulated BBM market halving the flow impact removes about half of the rate, the
leverage correlations a fifth and the correlation of the price and flow noises a
tenth. In the I-BBM that correlation removes two fifths and the leverage a quarter,
while halving $c$ stays within the error. The construction explains this. With the
target $\hat y_t$ held fixed the drift of $(x_t, y_t)$ is a gradient field, since
$\partial_y$ of the price drift and $\partial_x$ of the flow drift are both
$-c V'(x)$. Detailed balance would then require that drift to be the gradient with
respect to the metric the noise defines, and it is not: the price and demand noises
differ in size and are correlated at $-0.51$. Three further couplings break it.
Each sleeve's budget tracks its wealth and equity share, so the flow depends on
the past of the price by a route the price has no counterpart for. In the
simulated BBM market the marking of the demand memory puts the price shock into the
flow equation. And the variance multiplies the price noise while its own drift does
not depend on the price, so no stationary density gives the drift the form
$b = D\nabla\log p + \nabla\!\cdot\!D$ that a state-dependent $D$ would allow.

Because Eq.~\eqref{eq:ep_direct} is a per-step quantity, its average over the steps
of one stage of the cycle gives the rate in that stage, which no estimator built
around the rest state can supply. Table~\ref{tab:ness_phase} gives it beside the
change in the system's marginal entropy over the stage, which telescopes to zero
over a run but not within a stage, their sum being a stagewise summary rather
than a balance law. The market is least reversible while falling and while climbing
back, at four and five times the calm rate, and most reversible in the calm, with
the top in between. The cycle thus modulates the arrow of time by a factor of about
five, and it is strongest where the reflexive loop does the most work.

\begin{table}[htbp]
\centering
\caption{Entropy production from the model's own drift by stage of the cycle on the
aggregated market, in nats a year: the stage average of Eq.~\eqref{eq:ep_direct},
the change in the system's marginal entropy, and their sum. Thirty-two runs of two
hundred years, standard deviation across runs.}
\label{tab:ness_phase}
\vspace{0.4em}
\small
\begin{tabularx}{\textwidth}{p{4.0cm} c c c X}
\toprule
Stage & Eq.~\eqref{eq:ep_direct} & $+$ system & $=$ sum & The market factor \\
\midrule
Recovery & $22.8$ & $-1.71$ & $21.1 \pm 5.7$ & low and rising \\
Bust & $19.7$ & $-2.53$ & $17.2 \pm 2.6$ & low and still falling \\
Boom & $9.4$ & $-2.10$ & $7.3 \pm 1.9$ & high and still rising \\
The break & $7.6$ & $-1.01$ & $6.6 \pm 2.7$ & high and no longer rising \\
Calm & $2.2$ & $+2.02$ & $4.2 \pm 0.6$ & in the middle band \\
\midrule
All steps & $8.5$ & $0.00$ & $8.5 \pm 1.0$ & \\
\bottomrule
\end{tabularx}
\end{table}

The last control needs no estimator. A reversible stationary process has a lagged
covariance obeying $C(\tau) = C(\tau)^{T}$ at every lag, so any asymmetry in that
matrix proves irreversibility outright, with neither linearity nor Gaussianity
assumed (Table~\ref{tab:ness_reversal}). Against a reversible nonlinear null on
runs of the same length, the I-BBM separates by seven to eight and a half standard
errors at every short lag and the simulated BBM market by six to seventeen. The
signed antisymmetric component of each pair (Table~\ref{tab:eprevision}) is
positive at every lag, the mispricing leading the demand and the volatility and the
demand leading the volatility, five to eighteen standard errors from two reversible
nulls. The coordinates are primitive states, even under time reversal, so the test
applies to them as it stands.

\begin{table}[htbp]
\centering
\caption{The test that assumes nothing: the relative antisymmetric part of the
lagged covariance, $\|C(\tau) - C(\tau)^{T}\| / 2\|C(\tau)\|$, which reversibility
forces to vanish, against a reversible nonlinear null. Twelve runs of sixty
years.}
\label{tab:ness_reversal}
\vspace{0.4em}
\small
\begin{tabularx}{\textwidth}{X c c c}
\toprule
Lag & Market & Reversible null & $z$ \\
\midrule
\multicolumn{4}{l}{\emph{The aggregated market, as one asset}} \\
One day & $0.00032 \pm 0.00008$ & $0.00008 \pm 0.00005$ & $8.4$ \\
One week & $0.00165 \pm 0.00046$ & $0.00042 \pm 0.00028$ & $7.9$ \\
One month & $0.00799 \pm 0.00224$ & $0.00235 \pm 0.00151$ & $7.2$ \\
\midrule
\multicolumn{4}{l}{\emph{The simulated BBM market, all stocks}} \\
One day & $0.00121 \pm 0.00023$ & $0.00008 \pm 0.00005$ & $16.7$ \\
One week & $0.00213 \pm 0.00071$ & $0.00043 \pm 0.00022$ & $7.9$ \\
One month & $0.00929 \pm 0.00378$ & $0.00230 \pm 0.00157$ & $5.9$ \\
\bottomrule
\end{tabularx}
\end{table}

\subsection{Probability currents}
\label{sec:currents}

The probability current shows where the irreversibility runs and how fast. The
antisymmetric part $Q$ of $AC$ is only a few percent of its symmetric part
$D$, but that ratio is dominated by the fast, large variance direction. Normalized by the standard deviations of each pair it
becomes a circulation rate, which can be set beside the system's own relaxation
rates. In the I-BBM the fastest loops turn at about three quarters of the rate the
slowest direction relaxes, and in the simulated BBM market all three loops turn
faster than it, a driven steady state rather than a marginal one.

\begin{figure}[htbp]
\centering
\includegraphics[width=\textwidth]{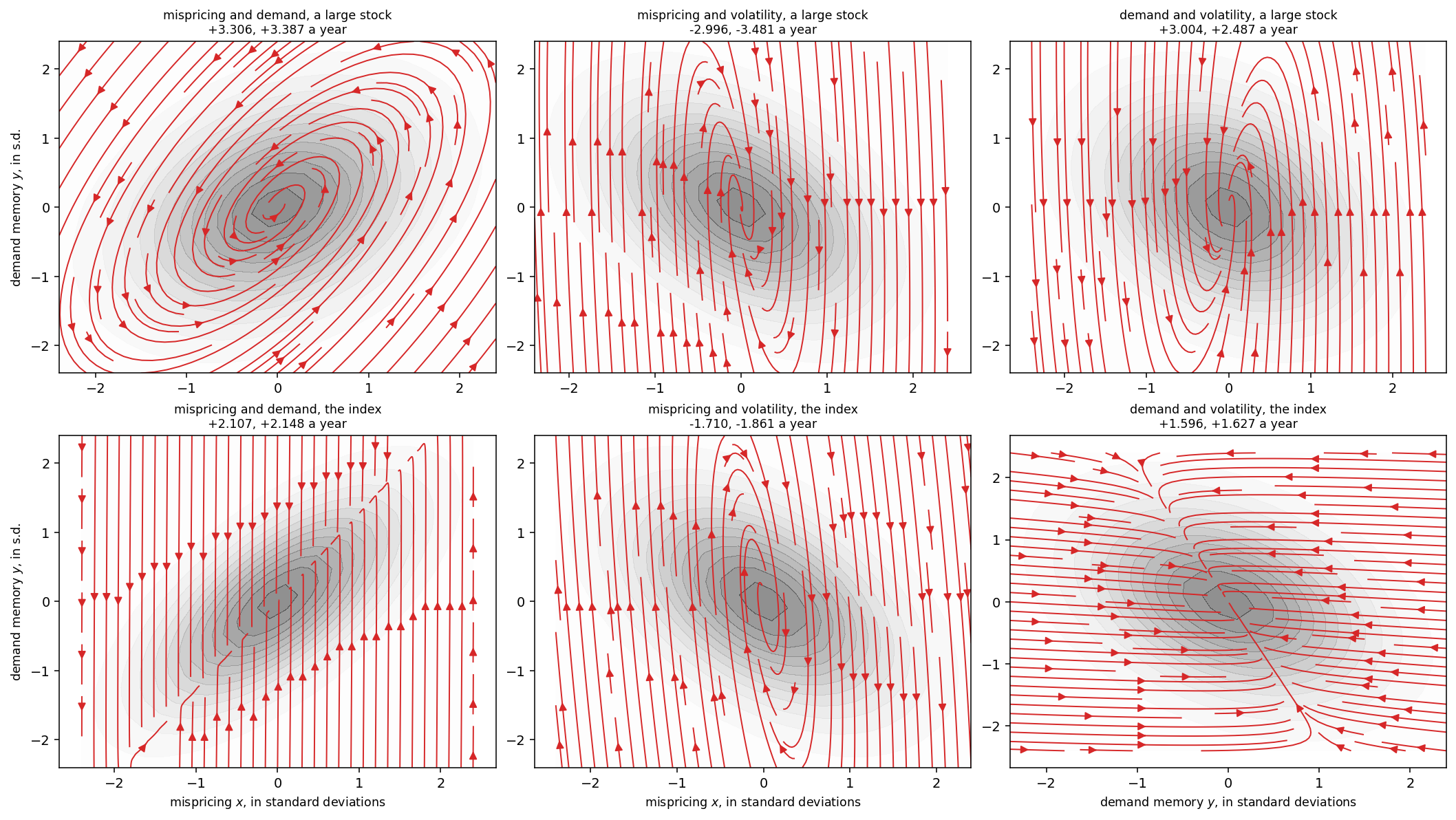}
\caption{The stationary probability current in the three planes of the market
factor over the stationary density, for a large stock and for the index, with the
circulation rates under the two windows' inputs above each panel.}
\label{fig:current}
\end{figure}

\begin{figure}[htbp]
\centering
\includegraphics[width=\textwidth]{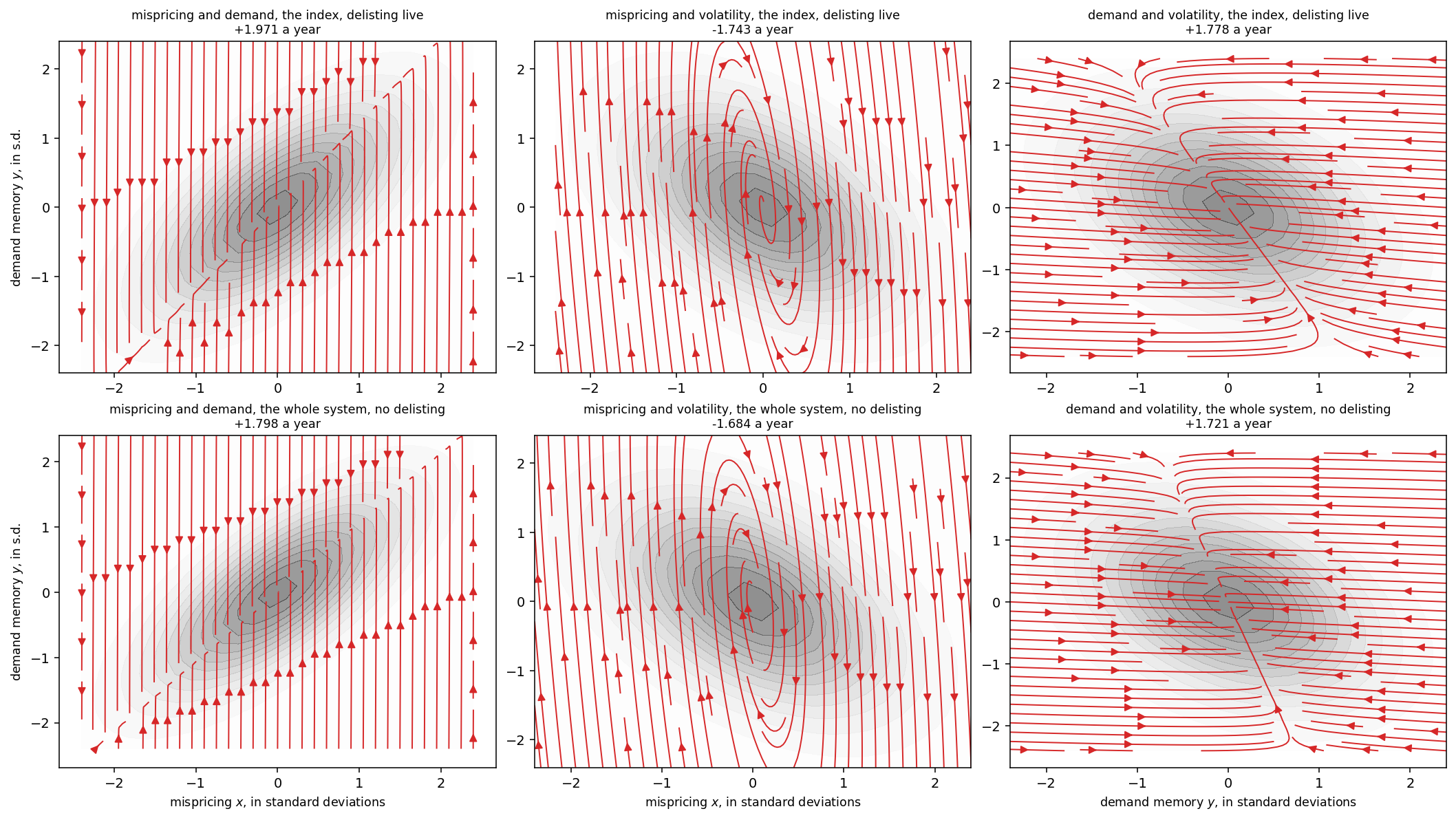}
\caption{The stationary probability current of the market factor with the index's
delisting rule live (above) and switched off (below): stationary density in gray,
current as streamlines, the circulation rate of each plane in its title.}
\label{fig:currentfree}
\end{figure}

Figure~\ref{fig:current} draws it. In the plane of mispricing and demand the current
turns steadily around the stationary density, at about two a year for the index
and three for a large stock, and the direction is the cycle: demand accumulates
while the mispricing is high, the mispricing falls, demand is withdrawn, and the
mispricing recovers with demand low. The state is stationary and is \textbf{not} an
equilibrium: fifty years after the construction the density has stopped moving,
and the current through it does not vanish. A gradient system cannot produce such a
closed loop, and the rest of the paper follows this loop in the
time domain. Lifting the removal rule, so that only a default retires a name,
leaves the plane turning the same way at the same rate
(Figure~\ref{fig:currentfree}), so the sum of firms shows the cycle too.

\begin{figure}[!htbp]
\centering
\includegraphics[width=0.9\textwidth]{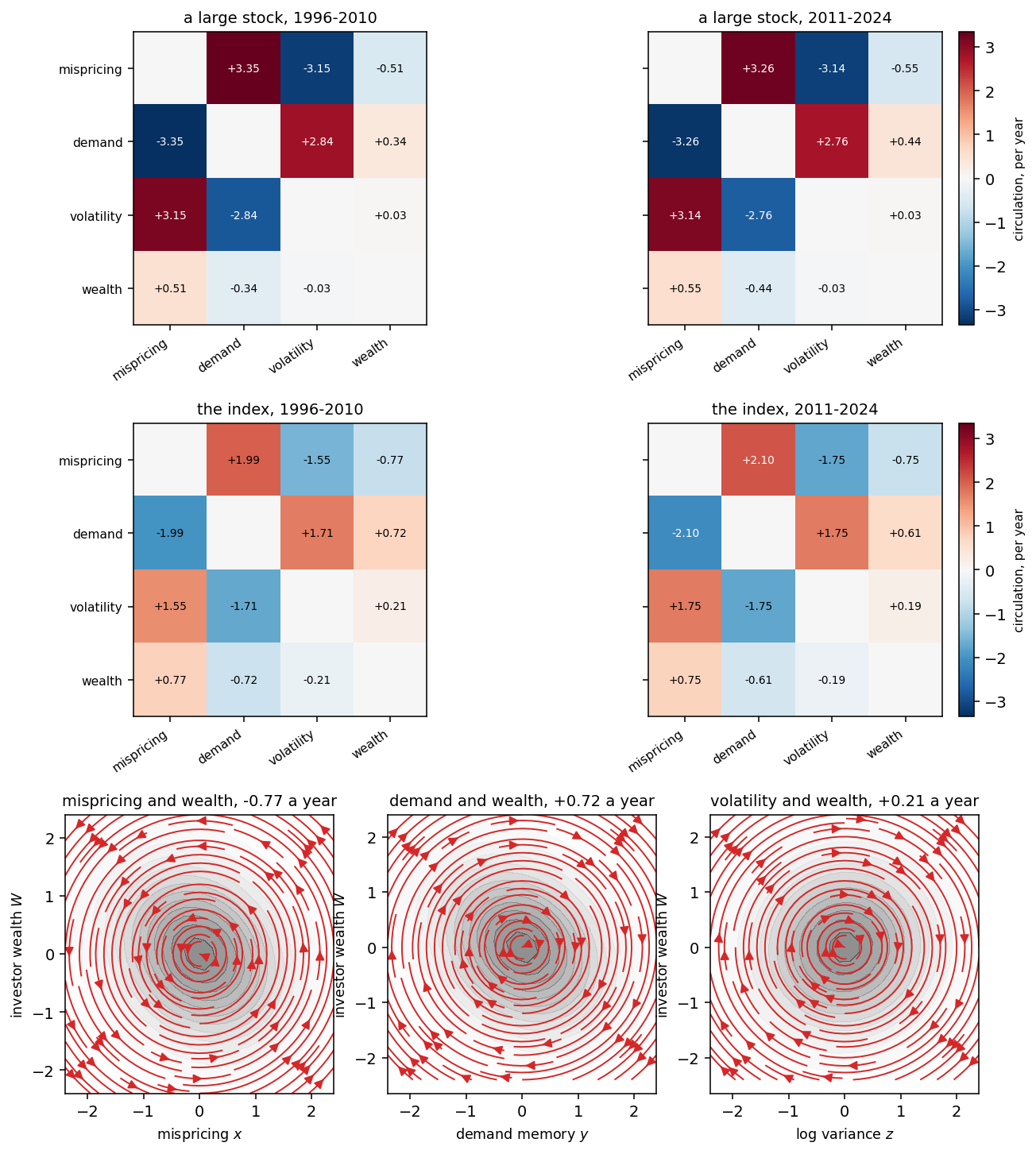}
\caption{The circulating current once the investor is counted: circulation rates
between every pair of the four states, in inverse years (upper rows), and the three
planes that exist only with the investor's wealth in the state (below).}
\label{fig:investorcurrent}
\end{figure}

We next include the investor itself in the analysis of irreversibility and
probability currents. In both models the investor's state includes, besides the demand
memory, the wealth $\mathcal{W}_t$, the momentum signal $s_t$ and
the two trailing variances $\hat v^{s}_t$ and $\hat v^{l}_t$
(Section~\ref{sec:lmsv}). The trailing variances are exponential averages of the
squared return with no noise of their own. Refining the clock of the I-BBM from
$252$ to $2016$ steps a year, their increment variance falls with the step while
that of the other states stays flat (Table~\ref{tab:ness_states}). With them
included the diffusion matrix is degenerate and their relaxation lies outside its
range, so the rate is infinite \cite{dacosta2023}. The momentum signal filters $dx$ in the
simulated BBM market (Eq.~\eqref{eq:memdef}) and $dx$ plus the fundamental's jumps
in the I-BBM, so its continuous part is degenerate with the mispricing, and a
five-state figure is regularized by jumps rather than a rate. The linear surrogate is therefore
well posed only on the three market states.

Adjoining the wealth and the momentum signal to the three states changes
Eq.~\eqref{eq:ep_ou} by an eighth to a third in the I-BBM and by about a
seventeenth in the simulated BBM market. The circulation needs no inverse of $D$
(Figure~\ref{fig:investorcurrent}). The
investor's wealth turns with the mispricing and the demand at a ninth to two fifths
of their rate, a correction to the cycle rather than a larger loop hidden behind it.

\begin{table}[!htbp]
\centering
\caption{Degrees of freedom and readouts on the I-BBM, the clock refined from
$252$ to $2016$ steps a year: exponents of each state's increment variance over the
step and of the entropy production in the linear approximation.}
\label{tab:ness_states}
\vspace{0.4em}
\small
\begin{tabularx}{\textwidth}{X c c c}
\toprule
State & $D_{kk}$ exponent & Driven by & $\sigma$ exponent \\
\midrule
Mispricing $x$, log variance $z$, momentum signal $s$ & $0.00$, $+0.01$, $0.00$
& noise & \\
Demand memory $y$, investor wealth $\mathcal{W}$ & $0.00$, $+0.16$ & noise, and jumps & \\
\midrule
The five states together & & & $-0.05$ \\
\midrule
Short trailing variance $\hat v^{s}$ & $-0.54$ & an ODE & $+0.50$ \\
Long trailing variance $\hat v^{l}$ & $-0.68$ & an ODE & $+0.67$ \\
All seven together & & & $+0.70$ \\
\bottomrule
\end{tabularx}
\end{table}

Broken detailed balance has been measured on financial time series before, in the
OMD studies of stocks \cite{halperin2026omdstocks} and portfolios
\cite{halperin2026omdportfolio}, which bin returns and volatilities into ranks
and measure the irreversibility of the resulting chain, and by
Nartallo-Kaluarachchi and co-authors, who fit the linear process of
Eq.~\eqref{eq:ep_ou} to the prices of $119$ NYSE stocks \cite{nartallo2024}. Their results are not
directly comparable with ours, but the qualitative conclusion agrees: the market
breaks detailed balance. Here
the state is continuous, it includes the investor's demand, and the current is
located between named variables, as in active matter \cite{battle2016}. We are not
aware of a calibrated generative market model shown to sit in a non-equilibrium
steady state with its current located this way.

\section{Discussion}
\label{sec:discussion}

\subsection{Crisis geometry of the distance matrix}
\label{sec:boombust}

The geometry here is that of the OMD studies, the spectrum and the eigenvectors of
the geodesic distance matrix $M(t) = \arccos G(t)$ of the correlation matrix
(Section~\ref{sec:leanomd}). In the S\&P 500's three crises since 1996 the
delocalized exponent $\beta_M$ rises, the entropy production of the volatility
ranks turns positive, and the commutator norm separates crisis from calm most
clearly. The simulated BBM market reproduces this geometry, every crisis value of
Table~\ref{tab:crisisgeom} falling within the market's range. It misses only in
calm periods, where it spreads its variance more evenly across names than the
market does. The data show three further things. The concentration of the
correlation structure leads the dispersion of the ranking chains, rather than
crowding leading the geometry. Not every drawdown concentrates the market onto one
mode. And the delocalized exponent and the entropy production are separate crisis
markers rather than two readings of one.

\subsection{Stationarity and the non-equilibrium steady state}
\label{sec:stationarydisc}

The primitives of the model are stationary: the
mispricing, the holding and the log variance do not drift over sixty years after
the warm-up, four times any horizon at which the model is calibrated or read
(Section~\ref{sec:stationarity}). Two derived observables are not. The price inherits the fundamental's growth, and the index is built through a membership that
retires names as they fall behind while its size distribution concentrates. A
stationary mechanism can thus produce a non-stationary aggregate, where
St\v{a}ric\v{a} and Granger \cite{starica2005} and the regime-switching literature
take non-stationary data to require a non-stationary generating process.
Twenty-nine years of data cannot separate the two readings, but longer data would.

The stationary state, however, is not an
equilibrium but a non-equilibrium steady state, with a current that circulates
between mispricing, demand and volatility as fast as the state relaxes, confirmed
by the estimator-free test on the lagged covariance and unchanged without the
delisting rule (Sections~\ref{sec:ness} and~\ref{sec:currents}). Stationarity and equilibrium, often treated in finance as
nearly the same thing, are thus separate here. To our knowledge this has not been shown for
an equity market with a calibrated generative model behind it.

\subsection{Reflexivity in the generative market}
\label{sec:reflexverdict}
\label{sec:threeway}

Soros's claims become testable once they are mechanisms with numbers attached,
and three of them can be read off the calibrated market: fallibility, the
co-determination of perception and price, and the boom-bust sequence. What the reflexive loop itself contributes is then tested by making ablation experiments on the
model. Such experiments will be presented next.

Fallibility holds: the investor's belief is wrong in a systematic way, and yet
prices still return to value. The calibrated investor extrapolates the trend and
holds the market's equity as a mean-variance investor, buying less when risk rises
rather than when risk is high. Its belief follows the market's past, not its
future. It correlates with the month just past and slightly negatively with the
month ahead, at $-0.14$ for the largest name and $-0.10$ for the median one, and as
a forecast of the next month's return it does worse than the unconditional mean
(Appendix~\ref{sec:beliefapp}). Acting on this belief, the investor buys the wrong
names, as retail fund flows do \cite{warther1995,frazzini2008}: the third of names
it buys most underperforms the third it buys least by six percent a year
(Section~\ref{sec:inelastic}), although its executed trades do not chase quarterly
returns (Section~\ref{sec:moneyflow}). This is the fallibility Soros took from
Popper \cite{popper1957,soros1987}, here a measured property of the calibrated
investor rather than an assumption. It does not stop the price from correcting.
The mispricing reverts mainly through its own dynamics, at the rate $\kappa_x$ of
Eq.~\eqref{eq:lmsvx}, and a name returns to value at about that pace whether or not
the investor misjudges it (Section~\ref{sec:blackfactor}).

Co-determination is tested only through the coupling of the price with the correlation geometry, which is weak, the two moving as two broadband series rather than as two oscillators (Section~\ref{sec:twooscillators}, Figure~\ref{fig:leanosc}).
Lastly, the boom-bust sequence comes out in part. Its asymmetry is reproduced,
a mark-up lasting $2.8$ years on average against $0.4$ for the mark-down, as in the
S\&P's $5.4$ and $0.6$ (Table~\ref{tab:cyclesummary}), and the market-wide
mispricing has a single peak a little above value with a long tail toward value
that the busts supply. Its timing is not: the turn is set by the noise in the price
(Section~\ref{sec:bustlaw}) rather than by a misconception being tested and
eventually failing.

The reflexive loop adds little to the boom-bust statistics but changes the
stationary state of the market substantially. We show this with ablations that
replace one part of the loop, keep everything else and refit nothing
(Table~\ref{tab:interventions}). Setting $c = 0$ also removes the drift constant
$a_i$, which is set from $c$, while with $\theta = 0$ the investor still reads the
price through its value term. None of the ablations moves the cycle (Table~\ref{tab:rulecycle}). The rate,
spacing and depth of the falls stay within about one and a half standard errors of
the calibrated market. An index fall is mostly a fall in the mispricing of the rest
of the index, where the potential is close to linear and its nonlinearity is
reached only by small firms near the saddle, so $c$ acts on removals and defaults
rather than on the index.

\begin{table}[htbp]
\centering
\caption{The interventions of this section. None refits a parameter; each keeps
every noise, correlation and constant not listed, at the first window's calibrated
values.}
\label{tab:interventions}
\vspace{0.4em}
\footnotesize
\begin{tabularx}{\textwidth}{p{2.7cm} X p{4.2cm}}
\toprule
Intervention & What is replaced & What else changes \\
\midrule
$c = 0$ & impact term and drift constant $a_i$ in the price; value term in the
belief and the demand target & the potential no longer acts on the price \\
$\theta = 0$ & the belief's trend term & none; the value term still depends on the price \\
contrarian, $2\theta$ & $\theta$ by $-\theta$ or $2\theta$ & none \\
random investor & each day's momentum increment given a random sign & none \\
no investor flow & the demand's drift, leaving $dy = \sigma_y\,dW$ & the mispricing
is not stationary \\
matched exogenous demand & the $y$ in the price, by a stationary process matched in
mean, spread, persistence and cross-basket correlation & its noise is uncorrelated
with the price's \\
open-loop replay & the $y$ in the price, by another market's recorded path on each
name's rest level & as above \\
\bottomrule
\end{tabularx}
\end{table}

\begin{table}[htbp]
\centering
\caption{The index's cycle with the investor's rule replaced, over 8 markets of 150 years each on the first window's inputs, mean and standard error across markets. Busts are falls of the capitalization-weighted index from its running peak.}
\label{tab:rulecycle}
\vspace{0.4em}
\footnotesize
\setlength{\tabcolsep}{3.5pt}
\begin{tabular}{l rr rr rr}
\toprule
& \multicolumn{2}{c}{falls a century} & \multicolumn{2}{c}{interval between 20\% falls} & \multicolumn{2}{c}{drawdown} \\
\cmidrule(lr){2-3}\cmidrule(lr){4-5}\cmidrule(lr){6-7}
investor & 20\% & 30\% & mean, years & c.v. & mean depth & deepest \\
\midrule
calibrated investor & $25.8 \pm 1.8$ & $7.0 \pm 0.7$ & $3.7$ & $1.32$ & $0.280$ & $-0.562$ \\
no trend term, $\theta = 0$ & $25.8 \pm 1.6$ & $7.5 \pm 0.7$ & $3.7$ & $1.36$ & $0.280$ & $-0.559$ \\
contrarian, $\theta \to -\theta$ & $25.5 \pm 1.5$ & $6.9 \pm 0.7$ & $3.8$ & $1.33$ & $0.277$ & $-0.551$ \\
trend doubled, $\theta \to 2\theta$ & $25.8 \pm 1.7$ & $7.4 \pm 0.6$ & $3.7$ & $1.34$ & $0.280$ & $-0.567$ \\
random investor & $25.2 \pm 1.6$ & $7.4 \pm 0.7$ & $3.8$ & $1.34$ & $0.280$ & $-0.555$ \\
no feedback onto prices, $c = 0$ & $26.7 \pm 2.0$ & $8.1 \pm 0.7$ & $3.6$ & $1.36$ & $0.282$ & $-0.569$ \\
\bottomrule
\end{tabular}
\end{table}

\begin{table}[htbp]
\centering
\caption{The entropy production in the linear approximation of the simulated BBM market's three states, the capitalization-weighted mispricing, the per-name mean of the demand memory and the log of the market's volatility factor, with the investor's rule replaced, over 6 markets of 200 years each on the first window's inputs, by the estimator of Section~\ref{sec:ness}, mean and standard error across markets. The first column reads every daily increment, the second removes the increments across a constituent replacement. The estimator's floor, from a reversible process with the same covariance and diffusion, is $0.05$.}
\label{tab:ruleep}
\vspace{0.4em}
\footnotesize
\begin{tabular}{l r r}
\toprule
investor & all days & replacements removed \\
\midrule
trend doubled, $\theta \to 2\theta$ & $92.7 \pm 5.0$ & $197.1 \pm 6.2$ \\
calibrated investor & $71.2 \pm 3.0$ & $156.2 \pm 5.7$ \\
no trend term, $\theta = 0$ & $40.8 \pm 3.5$ & $107.5 \pm 5.1$ \\
random investor & $42.8 \pm 1.8$ & $109.5 \pm 5.5$ \\
contrarian, $\theta \to -\theta$ & $27.3 \pm 1.0$ & $78.0 \pm 3.2$ \\
no feedback onto prices, $c = 0$ & $12.1 \pm 0.5$ & $36.9 \pm 1.3$ \\
no feedback and no leverage & $6.6 \pm 0.5$ & $16.7 \pm 1.3$ \\
\bottomrule
\end{tabular}
\end{table}

Instead of affecting the boom-bust cycle, the reflexive loop changes how far the
market sits from equilibrium (Table~\ref{tab:ruleep}). With $c = 0$ about four fifths of the entropy production
in the linear approximation of Section~\ref{sec:ness} disappear, and $\theta = 0$
removes nearly a half, a third with the replacement days removed. Across the other rules the rate grows with the strength of
the extrapolation. Dropping the replacement days raises every value two- to
threefold but keeps the ordering. With neither feedback nor leverage the rate is
still far above the estimator's floor, since the investor reads the price one way
through its trend term and the downward jumps are not reversible either. The market
is thus out of equilibrium without the feedback, and the feedback multiplies its
entropy production about sixfold, or fourfold with the replacement days removed.

Three further ablations replace the investor's demand in the price. Removing the
flow altogether, the demand memory keeping its noise and losing its drift
(Table~\ref{tab:flowdrives}), leaves the cycle and the return statistics
unchanged and lowers the entropy production about ninefold, but the mispricing then
drifts away from value, so this case mixes the loss of the feedback with the loss
of stationarity. The other two keep $c$ and the stationarity and replace the demand
memory by one that does not respond to prices, a matched stationary process or another
market's recorded demand. They also drop the correlation of the demand's noise with
the price's (Eq.~\eqref{eq:corr}), so they remove somewhat more than the feedback.
The entropy production falls from $68$ nats a year to $1.5$ and $1.7$, still some
thirty times the floor, while the twenty percent falls stay at $26$ to $28$ a
century. Most of the irreversibility thus comes with the closing of the loop rather
than with the impact alone, and the booms and busts need neither. Firm failures
need the investor in the price, while the dumb-money spread survives every case.

\begin{table}[htbp]
\centering
\caption{What the investor's money flows drive, window A inputs: with no investor flow, the demand memory a pure diffusion, and with no price impact. The drifts are from the first to the last third of a century, per firm held throughout (4 markets of 100 years). The other rows are means and standard errors over 6 markets of 100 years, the last two over 8 of 40 years. n.m.\\ marks a quantity not measured.}
\label{tab:flowdrives}
\vspace{0.4em}
\scriptsize
\setlength{\tabcolsep}{2.5pt}
\begin{tabular}{l ccc}
\toprule
& calibrated & no investor flow & no price impact \\
\midrule
drift of $|x|$, large names, a century & $-0.000 \pm 0.001$ & $0.043 \pm 0.016$ & n.m. \\
drift of $|x|$, small names, a century & $-0.006 \pm 0.002$ & $0.098 \pm 0.024$ & n.m. \\
drift of time inside a factor of two, small names, \% & $0.2 \pm 0.1$ & $-4.3 \pm 2.7$ & n.m. \\
entropy production, linear approximation, nats a year & $71.1 \pm 2.8$ & $7.8 \pm 1.1$ & $14.0 \pm 0.9$ \\
firm failures, bp a year & $8.9 \pm 1.0$ & $11.3 \pm 1.3$ & $0.0 \pm 0.0$ \\
dumb-money spread, \% a year & $-6.0 \pm 0.2$ & $-2.8 \pm 0.2$ & $-6.8 \pm 0.2$ \\
flow on the preceding quarter's return & $0.035 \pm 0.004$ & $0.006 \pm 0.006$ & $0.023 \pm 0.004$ \\
\bottomrule
\end{tabular}
\end{table}

The same cuts leave the paper's comparisons with the market in place. Of the $315$
ranges on three ledgers, the $174$ calibration targets, the $42$ index rows of
Table~\ref{tab:index} and the $99$ banded untargeted comparisons of
Section~\ref{sec:inelasticdisc}, $253$ meet as calibrated, $245$ with $c = 0$ and
$249$ with $\theta = 0$.\footnote{Four of the nine point comparisons hold in all three
cases.} No observable moves by a full band, and the median move is a few
hundredths of one. What $c = 0$ loses is the defaults, which fall to zero, and a
handful of variance-ratio, volatility, correlation, leverage and factor rows. The
sweep of Table~\ref{tab:sensitivity} agrees, ranking $c$ first only on the response
to a lift in demand and $\theta$ no higher than sixth. The fit and the boom-bust
statistics are therefore made by the noise, the variance and the potential, while
the impact $c$ produces the failures, the response to demand and most of the entropy
production, and the trend term nearly a half of it. These are comparisons within
the simulator at a fixed calibration, and they do not identify the mechanism of the
real market.

The same division places the model against its closest relative, the extended
Chiarella model of Majewski, Ciliberti and Bouchaud \cite{majewski2020}. There
noise traders do the work our demand and price noise do, and its oscillations, like
the cycle here, are driven by that noise rather than by a limit cycle. What the BBM
adds is a loop through one investor's budgeted stock of money, and the ablations
above measure what that loop contributes, which has not been done for the extended
Chiarella model. Table~\ref{tab:newelements} sets out the shared and the
different elements of the two models and what each new element of the BBM contributes.

\begin{table}[htbp]
\centering
\caption{The extended Chiarella model \cite{majewski2020}, the other models with a
fundamental of Appendix~\ref{sec:comparison}, and the BBM, element by element,
with what each new element of the BBM contributes, measured by removing it from the
calibrated market. The Chiarella figures are from their equity estimates, with the
trend's rate $\alpha = 1/7$ a month that they set, and their $\kappa$,
$\kappa_3$, $\beta$ and $\gamma$ are in their notation. Ours are the calibrated
values of Table~\ref{tab:bbmparams} for the ten largest names and the rest of the
index.}
\label{tab:newelements}
\vspace{0.4em}
\scriptsize
\setlength{\tabcolsep}{4pt}
\renewcommand{\arraystretch}{1.12}
\begin{tabular}{>{\raggedright\arraybackslash}p{2.2cm} >{\raggedright\arraybackslash}p{3.9cm} >{\raggedright\arraybackslash}p{4.3cm} >{\raggedright\arraybackslash}p{4.3cm}}
\toprule
Element & Extended Chiarella and other models & BBM & What it contributes \\
\midrule
\multicolumn{4}{l}{\emph{Shared}} \\
Log price & A fundamental with drift $g$ plus a mispricing
& \multicolumn{2}{>{\raggedright\arraybackslash}p{8.88cm}}{A fundamental with drift $\gamma$ plus a mispricing $x$ and a transitory error $\eta$} \\
Price formation & Impact linear in the traders' demand (Kyle's $\lambda$), noise
traders $\sigma_N$
& \multicolumn{2}{>{\raggedright\arraybackslash}p{8.88cm}}{Impact of the demand memory, $c\,y\,V'(x)$, with demand noise $\sigma_y$ and price noise $\sigma_x$} \\
Trend term & Exponential average of log returns over six months, demand
$\beta\tanh(\gamma M)$
& \multicolumn{2}{>{\raggedright\arraybackslash}p{8.88cm}}{Exponential average of the mispricing's moves over sixteen years, $\theta s$ in the belief, divided by the perceived risk} \\
Value term & $\kappa x + \kappa_3 x^3$, nearly flat near value and steep far
from it, two wells for $\kappa < 0$
& \multicolumn{2}{>{\raggedright\arraybackslash}p{8.88cm}}{Reversion $\kappa_x$ and $-c\,V(x)$, close to linear near value and turning at the saddle, a second minimum for a larger $c$} \\
Price against value noise & $0.043$ against $0.018$ a month, US index
& \multicolumn{2}{>{\raggedright\arraybackslash}p{8.88cm}}{$0.159$ against $0.061$ and $0.266$ against $0.034$ a year} \\
Noise-free skeleton & Stable, $\alpha + \kappa - \alpha\gamma\beta = 0.079$
(linear) and $0.038$ (cubic) for the US
& \multicolumn{2}{>{\raggedright\arraybackslash}p{8.88cm}}{A fixed point, noise pushing the mispricing over a threshold} \\
Black's factor of two & Root-mean-square mispricing about $0.5$
& \multicolumn{2}{>{\raggedright\arraybackslash}p{8.88cm}}{$99.7$ percent inside it} \\
\midrule
\multicolumn{4}{l}{\emph{Different}} \\
Reflexive loop through a budgeted stock of money
& Three trader groups with fixed weights; demand enters by a linear impact
\cite{majewski2020} or a clearing condition \cite{brock1998}, with no money
& One investor, trend, value and crowding in one belief, moving the price through
$c\,y\,V'(x)$; wealth, bonds, an inflow of $0.62$ percent a year, an equity cap,
no shorting
& Most of the entropy production, $68$ nats a year against $12$ at $c = 0$; every
firm failure; the price response to demand \\
\addlinespace
Nonlinear potential with a barrier
& A symmetric restoring force, a cubic value demand \cite{majewski2020}, or a
deterministic route to chaos \cite{brock1998}
& An excitable rest point on a one-sided potential, with a barrier beyond which a
firm fails
& The cycle, $25$ to $27$ twenty percent falls a century under every investor
rule, and $9$ to $14$ against $163$ in six hundred market-years without the price
noise; failures as escapes over the barrier \\
\addlinespace
Stochastic variance with leverage at three levels
& Gaussian monthly shocks \cite{majewski2020}, a symmetric law
\cite{alfarano2008}, or volatility from the shares of trader types
\cite{franke2012}
& A daily step, the exponential of an Ornstein-Uhlenbeck log variance with leverage
for the name, the common factor and the market and a slow market mode, and jumps
& The deep busts and their clustering, thirty percent falls dropping from $7.0$ to
$2.2$ a century with the variance frozen; a quarter of the I-BBM's entropy
production \\
\addlinespace
Many names with renewal
& One asset at a time, $27$ series in four classes from 1800
& $500$ names in two size groups and three industry groups, firm and market jumps,
removals and entrants, 1996--2024
& The $146$ group targets, and the correlation, sector, geometry, size, membership
and default quantities a single asset cannot define \\
\addlinespace
Fundamental & Hidden, recovered by Kalman and unscented filters
& Known to the investor, with jumps and a reversion across names & \\
Estimation & Likelihood, six parameters a series
& Moments in two windows, $27$ and $31$ parameters & \\
\bottomrule
\end{tabular}
\end{table}

\subsection{Default rate and the price floor}
\label{sec:defaultrate}

Equity in this framework can default, which is unusual. The standard models of
equity dynamics, from geometric Brownian motion through local and stochastic
volatility, keep the price positive and the firm alive, leaving default to
structural credit models rarely asked to reproduce equity returns. The marketron
potential of Eq.~\eqref{eq:leanpot} does both: below the threshold $\hat x$ the
drift reverses, the flow that supported the price drives it down, and a name that
falls far enough does not come back.

We define a default as a barrier crossing in price, the first close below ten
cents, and count it by the same rule in the model and in the market. A default is
distinct from a removal from the index, which the committee decides while the firm
goes on trading and to which the model's removal threshold is fitted, and from a
delisting, which the model does not represent. Each firm is followed from
its entry into the index for as long as it trades, since a failing firm leaves the
index long before it reaches the floor, giving seventeen thousand name-years of the
market from 1996 to 2023. The ten largest never default in the model, so every
model rate below is that of the rest of the index, annual and per name-year.

Counted this way the calibrated market fails $5.0$ and $4.0$ basis points of its
firms a year in the two windows, against $8.2$ for every firm that was ever a
member. The calibration target of Table~\ref{tab:groupjointfifty} counts a
different population, each window's members with leavers followed for five years,
and meets the market in both windows. A close under ten cents is a distress proxy,
not
an actual credit event as a rating agency defines it, so the comparison with
investment-grade issuers, which default at of order ten basis points a year
\cite{spratings2020}, is indicative only.

\begin{figure}[htbp]
\centering
\includegraphics[width=0.62\textwidth]{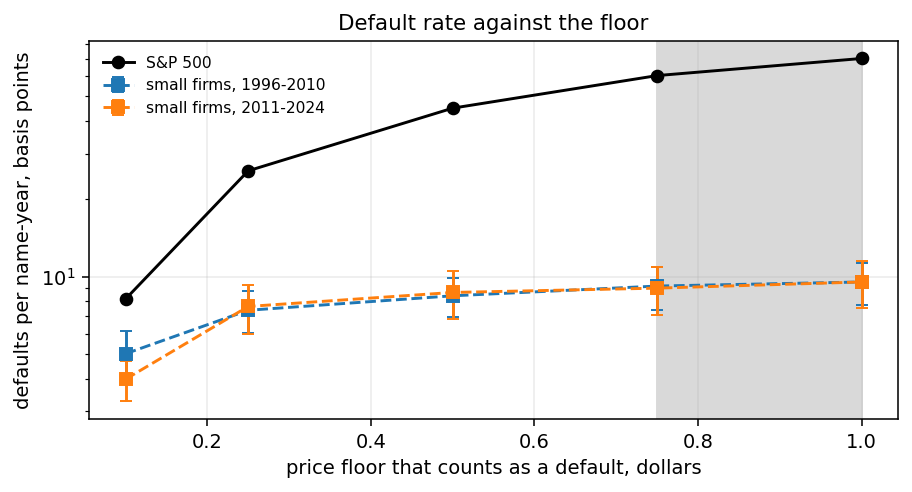}
\caption{The default rate against the price floor that defines it, for the
market and for each parameter set. The shaded span is the delisting range,
seventy-five cents to a dollar. Rates are annual, per name-year.}
\label{fig:floor}
\end{figure}

The ten-cent level is a convention which can be varied to explore the sensitivity
of the model's predicted default rate to this price level. Raising the floor toward a
dollar, the range where an exchange delists, counts
distress that a firm survives as well as failure, and no term of the objective
covers these shallower floors (Figure~\ref{fig:floor}). The market's count
climbs ninefold from ten cents to a dollar, while the model's barely doubles. A name that fails in the model falls all the way, so the
floor at which it is counted hardly matters, and what the model under-produces are
the shallower collapses a firm survives.

\subsection{How well the BBM matches the market}
\label{sec:inelasticdisc}
\label{sec:scorecard}

Every comparison in this paper checks whether a range of one standard deviation on
the model overlaps one on the market, the market's range a block bootstrap of its
single history and the model's the spread across simulated markets of the same
length. Where the market supplies only a point or a sign, the model misses when the
two differ by more than a factor of two or in sign. The calibrated observables meet
on $119$ of $146$ group ranges, the I-BBM on twelve of its fourteen in each window,
and the unfitted index on $18$ and $19$ of its twenty-one
(Section~\ref{sec:leancal}).

Table~\ref{tab:scorecard} collects the other untargeted
comparisons of the paper, each with
the section that computes it. Of its $99$ banded comparisons $73$ meet at one
standard deviation and $89$ at two, and four of its nine point comparisons hold.
The banded misses, in the calm geometry, the volatility bursts of single names and
the shallow price falls, are grouped by cause in
Section~\ref{sec:notquiteright}. Among the point comparisons the model misses
the aggregate price multiplier (on a latent-demand scale, not like for like),
momentum, the persistence of group leadership
across busts, the trend of $\beta_M$ and the strength of the arrow of time, which
has the right sign on the ranks of volatility but a sixth of the market's
strength. Black's factor of two holds $99.7$ percent of the time against his
ninety, the flow sort has the sign of the dumb-money effect and its three-year size, and the log
volatility of the index and the ten largest is about as rough as the market's with
no fractional driver anywhere in
the construction, as it is in GARCH and lognormal stochastic-volatility controls.

{\scriptsize
\setlength{\tabcolsep}{2.0pt}
\begin{longtable}{>{\raggedright\arraybackslash}p{4.3cm} >{\raggedright\arraybackslash}p{3.4cm} >{\raggedright\arraybackslash}p{3.9cm} c@{\hspace{8pt}}c@{\hspace{6pt}}c}
\caption{The untargeted comparisons of the paper against the S\&P 500, the index moments of
Table~\ref{tab:index} aside, with single-name volatility counted over the rows of
Table~\ref{tab:volpredicted}. Brackets give one standard deviation either side,
the market's block bootstrap and the model's spread across markets of the same
length, for 1996--2010 and 2011--2024 where two entries are given. Percentile
bands are converted to one standard deviation, and those of the industry groups
are the tracking error over the square root of the market's years in each phase.
The $1\sigma$ and $2\sigma$ columns say whether the ranges meet, window by window.
An asterisk marks a point comparison, which misses at a factor of two or in sign.}
\label{tab:scorecard}\\
\toprule
Quantity & Market & Model & $1\sigma$ & $2\sigma$ & Section \\
\midrule
\endfirsthead
\toprule
Quantity & Market & Model & $1\sigma$ & $2\sigma$ & Section \\
\midrule
\endhead
\multicolumn{6}{l}{\emph{Single-name volatility}} \\
The 16 untargeted rows of Table~\ref{tab:volpredicted} & \multicolumn{2}{l}{ranges that meet, of 16} & 7, 16 & 15, 16 & \ref{sec:leancal} \\
\addlinespace
\multicolumn{6}{l}{\emph{Roughness and memory of the volatility}} \\
Hurst exponent of log vol, the index & $0.143$\,\tiny[0.112,0.174] & $0.130$\,\tiny[0.102,0.158] & Yes & Yes & \ref{sec:roughness} \\
Hurst exponent of log vol, the ten largest names & $0.115$\,\tiny[0.093,0.138] & $0.127$\,\tiny[0.099,0.154] & Yes & Yes & \ref{sec:roughness} \\
Hurst exponent of log vol, names from the middle & $0.103$\,\tiny[0.080,0.125] & $0.154$\,\tiny[0.127,0.180] & No & Yes & \ref{sec:roughness} \\
Detrended fluctuation $\alpha$ of the index's $|r|$, I-BBM & $0.944$\,\tiny[0.893,0.978] & $0.913$\,\tiny[0.881,0.943] & Yes & Yes & \ref{sec:comparison} \\
Fractional integration $d$ of the index's $|r|$, I-BBM & $0.422$\,\tiny[0.335,0.498] & $0.305$\,\tiny[0.242,0.410] & Yes & Yes & \ref{sec:comparison} \\
\addlinespace
\multicolumn{6}{l}{\emph{Correlation geometry of the market}} \\
Participation ratio, calm & $9.50$\,\tiny[8.72,10.29] & $12.88$\,\tiny[11.21,14.56], $12.65$\,\tiny[11.18,14.12] & No, No & Yes, Yes & \ref{sec:leanomd} \\
Participation ratio, crisis & $3.86$\,\tiny[3.14,4.57] & $4.20$\,\tiny[2.74,5.65], $3.82$\,\tiny[2.69,4.96] & Yes, Yes & Yes, Yes & \ref{sec:leanomd} \\
Mean absolute correlation, calm & $0.317$\,\tiny[0.300,0.335] & $0.294$\,\tiny[0.275,0.313], $0.299$\,\tiny[0.283,0.316] & Yes, Yes & Yes, Yes & \ref{sec:leanomd} \\
Mean absolute correlation, crisis & $0.506$\,\tiny[0.460,0.552] & $0.506$\,\tiny[0.412,0.600], $0.526$\,\tiny[0.446,0.606] & Yes, Yes & Yes, Yes & \ref{sec:leanomd} \\
Market-mode share, calm & $0.337$\,\tiny[0.319,0.355] & $0.320$\,\tiny[0.301,0.339], $0.324$\,\tiny[0.307,0.341] & Yes, Yes & Yes, Yes & \ref{sec:leanomd} \\
Market-mode share, crisis & $0.522$\,\tiny[0.479,0.565] & $0.531$\,\tiny[0.442,0.619], $0.549$\,\tiny[0.474,0.625] & Yes, Yes & Yes, Yes & \ref{sec:leanomd} \\
Commutator norm, calm & $276$\,\tiny[258,294] & $316$\,\tiny[292,340], $321$\,\tiny[300,343] & Yes, No & Yes, Yes & \ref{sec:leanomd} \\
Commutator norm, crisis & $555$\,\tiny[413,698] & $540$\,\tiny[444,635], $570$\,\tiny[500,639] & Yes, Yes & Yes, Yes & \ref{sec:leanomd} \\
Delocalized exponent $\beta_M$, calm & $0.568$\,\tiny[0.547,0.588] & $0.307$\,\tiny[0.293,0.321], $0.304$\,\tiny[0.292,0.317] & No, No & No, No & \ref{sec:leanomd} \\
Delocalized exponent $\beta_M$, crisis & $0.619$\,\tiny[0.608,0.629] & $0.526$\,\tiny[0.378,0.673], $0.527$\,\tiny[0.392,0.662] & Yes, Yes & Yes, Yes & \ref{sec:leanomd} \\
Commutator, normalized, calm & $0.095$\,\tiny[0.093,0.097] & $0.122$\,\tiny[0.119,0.125], $0.122$\,\tiny[0.119,0.125] & No, No & No, No & \ref{sec:leanomd} \\
Commutator, normalized, crisis & $0.106$\,\tiny[0.067,0.144] & $0.094$\,\tiny[0.075,0.114], $0.094$\,\tiny[0.074,0.115] & Yes, Yes & Yes, Yes & \ref{sec:leanomd} \\
Leading eigenvector turn, degrees, calm & $5.6$\,\tiny[5.2,6.0] & $6.4$\,\tiny[6.1,6.8], $6.4$\,\tiny[6.1,6.7] & No, No & Yes, Yes & \ref{sec:leanomd} \\
Leading eigenvector turn, degrees, crisis & $5.5$\,\tiny[2.7,8.3] & $4.1$\,\tiny[3.2,5.0], $4.1$\,\tiny[3.1,5.0] & Yes, Yes & Yes, Yes & \ref{sec:leanomd} \\
Trend of $\beta_M$, a decade & $+0.080$ & $-0.0006$ & \multicolumn{2}{c}{No$^{*}$} & \ref{sec:leanomd} \\
\addlinespace
\multicolumn{6}{l}{\emph{Shape of the return tail}} \\
Hill slope of a name's downward tail, one of the ten largest & $2.92$\,\tiny[2.90,2.94] & $3.19$\,\tiny[2.85,3.53] & Yes & Yes & \ref{sec:retdist} \\
Hill slope of a name's downward tail, the rest of the index & $2.92$\,\tiny[2.90,2.94] & $3.62$\,\tiny[3.39,3.85] & No & No & \ref{sec:retdist} \\
\addlinespace
\multicolumn{6}{l}{\emph{Mispricing and failure}} \\
Names whose price falls under 25 cents, bp a year & $25.7$\,\tiny[21.8,29.6] & $7.4$\,\tiny[3.8,11.0], $7.7$\,\tiny[3.3,12.0] & No, No & No, No & \ref{sec:defaultrate} \\
Names whose price falls under a dollar, bp a year & $70.1$\,\tiny[63.7,76.5] & $9.5$\,\tiny[4.9,14.2], $9.5$\,\tiny[4.3,14.7] & No, No & No, No & \ref{sec:defaultrate} \\
Time inside Black's factor of two, percent & $\ge 90$ \cite{black1986} & $99.7$ & \multicolumn{2}{c}{Yes$^{*}$} & \ref{sec:blackfactor} \\
\addlinespace
\multicolumn{6}{l}{\emph{Flows}} \\
Price response to permanent latent demand & $\approx 5$ \cite{gabaix2021} & $0.64$ & \multicolumn{2}{c}{No$^{*}$} & \ref{sec:inelastic} \\
High-flow less low-flow names, \%/yr & negative \cite{frazzini2008} & $-5.9$, $-6.0$ & \multicolumn{2}{c}{Yes, Yes$^{*}$} & \ref{sec:inelastic} \\
\addlinespace
\multicolumn{6}{l}{\emph{Factors and industry groups}} \\
Cyclical over the market, boom, \%/yr & $-1.9$\,\tiny[-5.5,1.7] & $1.8$\,\tiny[-2.0,5.6], $-1.0$\,\tiny[-5.0,2.9] & Yes, Yes & Yes, Yes & \ref{sec:factorrotation} \\
Cyclical over the market, bust, \%/yr & $2.2$\,\tiny[-0.9,5.2] & $-4.1$\,\tiny[-7.4,-0.8], $1.5$\,\tiny[-1.6,4.7] & Yes, Yes & Yes, Yes & \ref{sec:factorrotation} \\
Cyclical over the market, recovery, \%/yr & $0.2$\,\tiny[-2.8,3.2] & $3.2$\,\tiny[-1.0,7.4], $-6.3$\,\tiny[-11.7,-1.0] & Yes, Yes & Yes, Yes & \ref{sec:factorrotation} \\
Cyclical over the market, calm, \%/yr & $-0.9$\,\tiny[-2.1,0.3] & $0.8$\,\tiny[-0.7,2.4], $0.1$\,\tiny[-1.6,1.7] & Yes, Yes & Yes, Yes & \ref{sec:factorrotation} \\
Sensitive over the market, boom, \%/yr & $4.8$\,\tiny[1.2,8.3] & $2.6$\,\tiny[-1.2,6.5], $4.7$\,\tiny[0.3,9.2] & Yes, Yes & Yes, Yes & \ref{sec:factorrotation} \\
Sensitive over the market, bust, \%/yr & $-5.3$\,\tiny[-8.9,-1.7] & $-3.5$\,\tiny[-6.9,-0.2], $-4.4$\,\tiny[-7.8,-1.1] & Yes, Yes & Yes, Yes & \ref{sec:factorrotation} \\
Sensitive over the market, recovery, \%/yr & $6.1$\,\tiny[2.9,9.4] & $11.7$\,\tiny[6.2,17.2], $19.5$\,\tiny[12.8,26.2] & Yes, No & Yes, Yes & \ref{sec:factorrotation} \\
Sensitive over the market, calm, \%/yr & $2.6$\,\tiny[1.3,4.0] & $0.8$\,\tiny[-0.7,2.2], $1.6$\,\tiny[0.0,3.1] & Yes, Yes & Yes, Yes & \ref{sec:factorrotation} \\
Defensive over the market, boom, \%/yr & $-2.7$\,\tiny[-8.1,2.7] & $-2.9$\,\tiny[-8.4,2.6], $-4.2$\,\tiny[-11.0,2.7] & Yes, Yes & Yes, Yes & \ref{sec:factorrotation} \\
Defensive over the market, bust, \%/yr & $14.7$\,\tiny[7.9,21.6] & $14.1$\,\tiny[8.6,19.5], $13.1$\,\tiny[7.0,19.3] & Yes, Yes & Yes, Yes & \ref{sec:factorrotation} \\
Defensive over the market, recovery, \%/yr & $-8.9$\,\tiny[-13.9,-4.0] & $-14.2$\,\tiny[-20.4,-8.0], $-21.4$\,\tiny[-29.8,-13.0] & Yes, Yes & Yes, Yes & \ref{sec:factorrotation} \\
Defensive over the market, calm, \%/yr & $-3.4$\,\tiny[-5.4,-1.4] & $-0.8$\,\tiny[-3.2,1.6], $-1.8$\,\tiny[-4.3,0.7] & Yes, Yes & Yes, Yes & \ref{sec:factorrotation} \\
Momentum premium, two to twelve months & positive & $-13.4$ to $-25.6$ & \multicolumn{2}{c}{No$^{*}$} & \ref{sec:factorrotation} \\
Short-term reversal premium & positive & $+22.1$ & \multicolumn{2}{c}{Yes$^{*}$} & \ref{sec:factorrotation} \\
Same group hit hardest in successive busts & no & $+0.71$, $+0.74$ & \multicolumn{2}{c}{No$^{*}$} & \ref{sec:factorrotation} \\
\addlinespace
\multicolumn{6}{l}{\emph{Return per unit of risk}} \\
Sharpe ratio of the index & $0.17$\,\tiny[-0.05,0.40], $0.58$\,\tiny[0.35,0.81] & $0.54$\,\tiny[0.46,0.63], $0.58$\,\tiny[0.49,0.67] & No, Yes & Yes, Yes & \ref{sec:sharpe} \\
Sharpe ratio of the investor's portfolio & $0.17$\,\tiny[-0.05,0.40], $0.58$\,\tiny[0.35,0.81] & $0.28$\,\tiny[0.20,0.36], $0.34$\,\tiny[0.26,0.42] & Yes, Yes & Yes, Yes & \ref{sec:sharpe} \\
\addlinespace
\multicolumn{6}{l}{\emph{Irreversibility}} \\
Entropy production of the volatility ranks & $0.0062$ & $0.0010$ & \multicolumn{2}{c}{No$^{*}$} & \ref{sec:sensitivity} \\
\midrule
Ranges that meet at $1\sigma$ & \multicolumn{5}{l}{73 of 99} \\
Ranges that meet at $2\sigma$ & \multicolumn{5}{l}{89 of 99} \\
Point comparisons that hold$^{*}$ & \multicolumn{5}{l}{4 of 9} \\
\bottomrule
\end{longtable}
}

\subsection{BBM as a stochastic neuron ensemble}
\label{sec:neurons}

The resemblance between the marketron and a spiking neuron was noticed for a single
asset in the original Marketron paper \cite{marketron2025}, which set the
mispricing beside the neuron's voltage and the accumulated flow beside its recovery
variable, and observed that expanding the exponentials to quartic order leaves
something close to the stochastic FitzHugh-Nagumo equations. The BBM extends this
view to many assets. Each name is an excitable unit, the units are coupled through
the mean field of the log fundamentals $\overline{l}_t$
(Section~\ref{sec:interactions}), the shared market shocks and one investor's
budget, and the index is the ensemble's readout, renewed by the membership rule.
Section~\ref{sec:excitability} tested this picture at both levels, and sorted by
geometry \cite{izhikevich2007} the analogy is falsifiable and has passed its tests.

The analogies hold at both levels. Both are driven by noise, with neither a limit
cycle nor chaos behind them (Section~\ref{sec:relatedmodels}). Both have a
threshold and an excursion of a size of their own, so that a larger shock makes a
bust more likely without making it bigger or bringing it forward
(Section~\ref{sec:excitthreshold}). A firm also has the geometry of a type I
neuron such as Morris-Lecar \cite{morris1981}, a stable node beside a saddle, a
recovery of a few months after its own fall, and falls more regular than a
memoryless escape (Sections~\ref{sec:bustlaw} and~\ref{sec:excitmicro}).

The differences are as informative. A firm is a gradient flow that absorbs past
its threshold, so its far side is a default rather than a return to rest, closer to
the metabolic failures of a neuron than to its spike. The index is excitable in
behavior only, since its state never comes near its own saddle. It has no
refractory period, and its busts cluster rather than arrive regularly, the
variance supplying the clustering as a second slow variable supplies the bursts of
the Hindmarsh-Rose neuron \cite{hindmarsh1984}, with counts of busts over ten to
forty years over-dispersed at a Fano factor near one and a half. The depth of a
bust varies with the variance state rather than being fixed by a nullcline. Nor
does the index show coherence resonance, collective bursting through its slow pool
of money, or the power-law quiet intervals of noise-induced intermittency. What
returns the index to rest after a failure is the membership rule, the hand-made
reset of the integrate-and-fire neuron \cite{burkitt2006}, which turns each firm's
absorbing excursion into a renewal of the ensemble. What the market adds to the
family is the loop between the investor and the price. It does not drive the
firing, which the noise, the variance and the jumps supply as an external current
drives a neuron, but it holds each name's mispricing to a stationary law, keeps
that state out of equilibrium and sets the rate at which firms fail.

\subsection{Phases of the market}
\label{sec:phases}

\begin{figure}[htbp]
\centering
\includegraphics[width=0.72\textwidth]{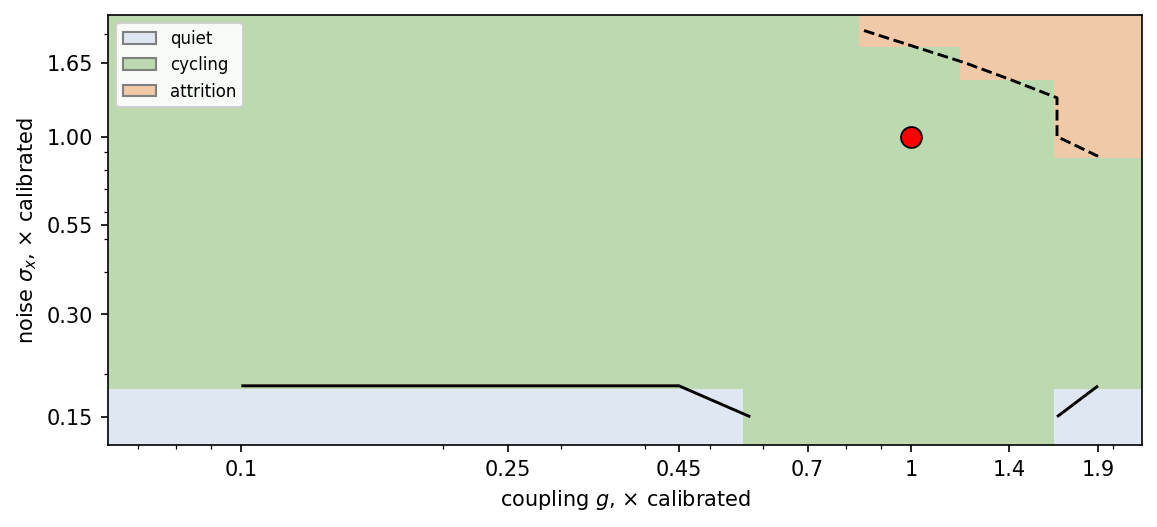}
\caption{The regime the market settles into, over the coupling and the mispricing
noise as multiples of their calibrated values, with the calibrated market in red.
The solid line is where busts begin to arrive and the dashed one where the default
rate passes a hundred basis points a year.}
\label{fig:leanphase}
\end{figure}

A construction with this many coupled states should have a phase structure far
richer than one study can chart. We take a two-dimensional projection of it, on
two key parameters of the model, the coupling $g$ of the potential and the
mispricing noise $\sigma_x$. Two
order parameters separate the regimes on it: the rate at which busts arrive, and
the default rate, which distinguishes a market that recovers from its drawdowns
from one that survives by replacing the names that fail.
Figure~\ref{fig:leanphase} maps them over the two parameters, each a multiple of
its calibrated value, with everything else held.

Three phases appear. Below about a fifth of the calibrated mispricing noise no bust
ever arrives and the market drifts without punctuating. Above it the busts resolve
into countable episodes and the default rate stays low, the cycling phase the rest
of this paper is about. Further out the default rate runs into hundreds of basis
points a year, a phase we call attrition, since the index persists there by
exchanging its constituents rather than recovering them. Within the cycling phase
the market's memory of volatility comes from a spread of relaxation rates
rather than from a fractional driver (Appendix~\ref{sec:returnshape}).

The lower boundary is nearly flat in the coupling, so the reflexive coupling
barely moves the point at which a market begins to cycle. It governs what a bust
does rather than whether one happens, the opposite of what the subject would lead
one to expect. The upper boundary runs across the plane, so a louder mispricing or
a stronger coupling pushes the market into attrition, the default rate climbing
steadily with the coupling. The calibrated market sits about five times its noise
above the quiet edge and about two thirds of its noise below attrition, and the
window fits move it within the phase. No market of the map had a runaway day or
needed a spare, so a phase of terminal collapse, which the sensitivity sweep shows
only where the names' volatility of volatility is doubled, lies outside this slice
and is left open.

\subsection{What the BBM does not get quite right}
\label{sec:notquiteright}

The misses of Table~\ref{tab:scorecard} fall into four groups. The first comes
from the reverting mispricing. Momentum has the wrong sign, the index's trend effect
is a reversal (Appendix~\ref{sec:comparison}), and the leadership of the industry
groups does not rotate between crises, all for one reason: the dynamics keep no
record of which name or group led, so a past return predicts a reversal
(Section~\ref{sec:factorrotation}). The second is failure and impact. Prices fall
under twenty-five cents or a dollar far less often than in the market, because a
firm that fails in the model falls all the way (Section~\ref{sec:defaultrate}), and the price's
susceptibility to latent demand, $0.64$, cannot yet be compared with the
inelastic-markets multiplier (Section~\ref{sec:inelastic}). The third is the index over long horizons,
whose skew beyond a day is too shallow and not concentrated at the top of the
cycle. The I-BBM shows the trade-off: the vectors that reproduce the skew place the
barrier where the index collapses within decades (Section~\ref{sec:indexone}). The
fourth is the names of the simulated BBM market, whose calm geometry is too spread out while every
crisis row meets, whose names' volatility bursts in the rest of the index are too
small and short-lived under the first window's inputs, and whose downward tails are
thinner than the market's. The index's
multifractal spectrum is too wide, $0.53$ against $0.32$, its bands apart over
$-2 \le q \le 2$, and its volatility of volatility too high
(Sections~\ref{sec:roughness} and~\ref{sec:excitmf}).

One miss persists across the sweep of Section~\ref{sec:sensitivity}: the variance
ratio at a year reaches at most $0.54$ against the market's $0.79$, whose band
starts at $0.62$, so no single parameter moved by a quarter brings it into range. The market's variance-ratio profile is non-monotone over the whole
sample and monotone within each half, in opposite directions, and the I-BBM
reproduces the second half and not the first. Against the standard alternatives estimated on the same index, the
BBM is first of five on the kurtosis, second on the variance ratio across horizons
and third on the skew (Appendix~\ref{sec:classone}), while also producing the
names, the cycle, the default rate and the geometry none of them attempts.

\subsection{Comparison of BBM with other models}
\label{sec:whatitamounts}

Here we summarize the qualitative and quantitative characteristics of the market
dynamics in the BBM, against those matched by other models in the literature
(Table~\ref{tab:checkpoints}), with the BBM read after the fifty-year warm-up.
Qualitatively, every model in the table is dynamic and most produce a cycle or
switching regimes, but only the BBM resolves the market into stocks and includes a budgeted flow of money, an index membership, a correlation and sector structure and
factor dynamics through the cycle. It is also the only one to reproduce the
asymmetry of returns at every horizon and the depth and spacing of busts together.
Its volatility clustering is imposed through the stochastic variance, where Franke
and Westerhoff and Alfarano and co-authors make it endogenous, and only for the BBM
has the arrow of time been tested.

Quantitatively, the published models report few of the table's quantities, so for
them the table records mostly what each is built to produce. Where they do report a
daily moment, the skew and excess kurtosis of Franke and Westerhoff, $-0.005$ and
$1.89$, and of Alfarano and co-authors, $0$ and $1.50$, fall far short of the
market's $-0.39$ and $9.8$, which the BBM meets within one band. Of the BBM's
targeted rows all but the variance ratio at a year lie within about one band. Of the
rows never calibrated, the Hurst exponents of the index and the ten largest, Black's
factor of two and the inflow of
new money meet the market, the price multiplier is not yet comparable
(Section~\ref{sec:notquiteright}), and the entropy production of the volatility
ranks is about a sixth of the market's. On the return statistics across horizons the
BBM is level with the best of the statistical models (Appendix~\ref{sec:classone}).
Everything beyond the fitted observables was read off afterwards, from four
literatures that do not usually meet: the stylized facts of returns, corporate
credit, the microstructure of price impact and the flow of household saving. None
of the generative models in the table targets more than one of them, and a
single-asset model cannot produce the many-name quantities at all.

The BBM fits $146$ quantitative characteristics of the market and is then matched
against $141$ more that it was never fitted to. The fitted ones are the ranges of
the two groups' calibration targets, $119$ of which meet. The matched ones are the
$42$ index rows of Table~\ref{tab:index} and the $99$ banded comparisons of
Table~\ref{tab:scorecard}. This is achieved
without any exotic stochastic processes, using a Markovian system of stochastic
differential equations that grows only linearly with the number of names. Its memory is held in states rather than kernels, so it simulates at the speed of an
ordinary diffusion. Features often argued to require exotic processes come out of it
as well. Rough volatility, long memory and multifractality are usually built in with
fractional Brownian motion or fractional integration, a generalized Langevin
equation whose kernel has separate short-time and tail exponents, rough volatility models, multifractal cascades or
L\'evy laws \cite{granger1980,calvet2004,gatheral2018,itkin2026gle}. At daily resolution and over
the ranges studied, the BBM reproduces several persistence and roughness statistics
of returns and volatility proxies without such drivers, its roughness being the
realized estimate of a diffusive volatility and its memory a spread of relaxation
rates. These finite-sample matches do not establish asymptotic long memory or
roughness of latent volatility, the roughness and memory statistics are reproduced as
well by GARCH and lognormal stochastic-volatility controls
\cite{corsi2009,contdas2024}, and the estimated multifractal spectrum is
broader than the market's, a quantitative miss (Appendices~\ref{sec:returnshape}
to~\ref{sec:excitmf}). They come from the stochastic variance: unchanged with
the reflexive loop cut, they are strongly reduced with the variance frozen. The BBM fits sixty-six constants, twenty-seven and thirty-one for the two groups and eight in a first
stage, and the fifteen constants set by hand are not counted. That is more than two
fitted quantities per constant, and more than four counting the matched ones,
although many of these quantities share their drivers
(Section~\ref{sec:sensitivity}).

{\scriptsize
\setlength{\tabcolsep}{3pt}
\begin{longtable}{R{4.0cm}R{2.3cm}R{2.3cm}R{1.42cm}R{1.42cm}R{1.42cm}R{1.42cm}}
\caption{What each model is capable of and what it is held to. The first column
is the S\&P 500 from 1996 to 2024, banded by a stationary block bootstrap of
that sample in blocks of about a year, over days and over names. The bracket
beside each BBM figure is its signed distance from the market, in units of that
band and the model's own spread combined, not an error bar. Model figures are means over eight markets of five hundred names and a hundred and fifty years, read on the capitalization-weighted index
of the model of Sections~\ref{sec:excitability}
and~\ref{sec:nesssec}, so they describe the settled market and not the window
the calibration uses. A bust is a drawdown beyond thirty percent.}
\label{tab:checkpoints}\\
\toprule
& \textbf{The Market} & \textbf{BBM}
& Majew.\ \cite{majewski2020} & Fr.--W.\ \cite{franke2012}
& Alfar.\ \cite{alfarano2008} & Br.--H.\ \cite{brock1998} \\
\midrule
\endfirsthead
\toprule
& \textbf{The Market} & \textbf{BBM}
& Majew.\ \cite{majewski2020} & Fr.--W.\ \cite{franke2012}
& Alfar.\ \cite{alfarano2008} & Br.--H.\ \cite{brock1998} \\
\midrule
\endhead

\multicolumn{7}{l}{\emph{Provenance}} \\
Fitted parameters  & ---  & 27 and 31, 8 more in a first stage, 15 set  & 6, 2 set  & 6--8  & 3--4  & $\sim$5 \\
Observables fitted to  & ---  & 33 and 40, in each of two windows  & likelihood  & 9  & 3  & none \\
Fitted over more than one period  & ---  & yes  & no  & no  & no  & no \\
\midrule
\multicolumn{7}{l}{\emph{Scope and capability}} \\
Stocks resolved into size groups  & yes  & yes  & no  & no  & no  & no \\
Budgeted flow of money, small inflow  & ---  & yes  & no  & no  & no  & no \\
Reproduces the stylized facts  & yes  & yes  & partly  & yes  & partly  & partly \\
Asymmetry at every horizon  & yes  & yes  & no  & no  & no  & no \\
Fat tails that thin with horizon  & yes  & yes  & no  & yes  & no  & no \\
Bust depth and spacing together  & yes  & yes  & no  & no  & no  & no \\
Index membership and turnover  & yes  & yes  & no  & no  & no  & no \\
Correlation and sector structure  & yes  & yes  & no  & no  & no  & no \\
Factor dynamics through the cycle  & yes  & yes  & no  & no  & no  & no \\
\midrule
\multicolumn{7}{l}{\emph{Stylized facts}} \\
Dynamic market  & yes  & yes  & yes  & yes  & yes  & yes \\
Multi-asset market  & yes  & yes ($N{=}500$)  & no  & no  & no  & no \\
Equilibrium model  & no  & no  & no  & no  & no  & clears each period \\
Business cycles  & yes  & yes  & yes  & regimes only  & no  & yes \\
Volatility clustering  & yes  & yes, imposed  & no  & yes, endog.  & yes, endog.  & qualit. \\
Time-reversal symmetry  & broken in crises  & broken  & not tested  & not tested  & not tested  & not tested \\
\midrule
\multicolumn{7}{l}{\emph{Quantitative metrics, targeted at basket level}} \\
Index drift  & $0.097 \pm 0.034$  &  $0.114$ ($+0.5\sigma$)  & n.r.  & n.r.  & n.r.  & n.r. \\
Index volatility  & $0.191 \pm 0.018$  &  $0.195$ ($+0.2\sigma$)  & n.r.  & n.r.  & n.r.  & n.r. \\
Daily skew  & $-0.393 \pm 0.151$  &  $-0.232$ ($+0.7\sigma$)  & undef.  & \emph{$-0.005$}  & \emph{$0$ exactly}  & n.r. \\
Daily excess kurtosis  & $9.8 \pm 2.98$  &  $13.3$ ($+1.0\sigma$)  & undef.  & \emph{$1.89$}  & \emph{$1.50$}  & n.r. \\
Deepest drawdown  & $-0.792 \pm 0.247$  &  $-0.844$ ($-0.2\sigma$)  & n.r.  & n.r.  & n.r.  & n.r. \\
Years between busts  & $9.71 \pm 4.25$  &  $9.4$ ($-0.1\sigma$)  & period n.r.  & n.r.  & n.r.  & period n.r. \\
Busts per century  & $10.3 \pm 4.62$  &  $10.8$ ($+0.1\sigma$)  & n.r.  & n.r.  & n.r.  & n.r. \\
Variance ratio, 3 months  & $0.672 \pm 0.064$  &  $0.701$ ($+0.4\sigma$)  & n.r.  & \emph{$0.977$}  & n.r.  & n.r. \\
Variance ratio, 1 year  & $0.791 \pm 0.174$  &  $0.435$ ($-2.0\sigma$)  & n.r.  & n.r.  & n.r.  & n.r. \\
Single-name volatility  & $0.386 \pm 0.025$  &  $0.378$ ($-0.3\sigma$)  & undef.  & undef.  & undef.  & undef. \\
Mean pairwise correlation  & $0.308 \pm 0.025$  &  $0.276$ ($-1.1\sigma$)  & undef.  & undef.  & undef.  & undef. \\
\midrule
\multicolumn{7}{l}{\emph{Quantitative metrics, never calibrated}} \\
Hurst exponent, price  & $0.470 \pm 0.034$  &  $0.439$ ($-0.9\sigma$)  & n.r.  & n.r.  & n.r.  & n.r. \\
Hurst exponent, $|r|$  & $0.944 \pm 0.055$  &  $0.869$ ($-1.2\sigma$)  & n.r.  & n.r.  & n.r.  & n.r. \\
Entropy production, vol.\ ranks  & $0.0062$  &  $0.0010$  & undef.  & undef.  & undef.  & undef. \\
\quad on the continuous state  & not observable  &  $69$ nats a year  & undef.  & undef.  & undef.  & undef. \\
Largest mispricing, name  & not observable  &  $2.4$ in log  & n.r.  & n.r.  & n.r.  & n.r. \\
Inside Black's factor of two  & yes \cite{bouchaud2017}  &  $99.7\%$  & r.m.s.\ $0.5$  & n.r.  & n.r.  & n.r. \\
Price response to latent demand  & $\approx 5$ \cite{gabaix2021}  & $0.64$, not comparable  & n.r.  & n.r.  & n.r.  & n.r. \\
Default intensity, bps a year  & $4$--$10$ \cite{spratings2020}  &  $5.0$ and $4.0$, fitted  & none  & none  & none  & none \\
New money, \% a year  & $0.35$ \cite{fedz1}, $\approx 0.8$ funds  & $0.62$  & n.r.  & n.r.  & n.r.  & n.r. \\
\bottomrule
\end{longtable}
}

A composite model assembled from standard parts needs more parameters for the
fitted quantities alone. A GJR-GARCH with Student $t$ shocks for each size group
takes about twelve parameters, a normal inverse Gaussian law at five horizons
twenty, a two-regime switching model for the timing of busts about six, a
square-root default intensity for each group six, a lognormal size law with Gibrat
growth three, a factor model with sectors for the correlations about twenty, a
hazard of entry and exit for the index four, and an impact law and a flow equation
about five. That is some seventy-five parameters, more than the sixty-six of the
BBM, and the composite would still have nothing to say about the $141$ matched
quantities. None of its parts generates the others, so the cycle, the defaults and
the geometry stay unexplained, and nothing makes the parts consistent with one
another. A
network trained to generate market paths \cite{ni2021} replaces the count with one
several orders of magnitude larger and gives up the question of what any of it
means. Ours can be steered rather than only fitted, since
Table~\ref{tab:sensitivity} shows which two constants move each quantity. Every
observable used so far is realized, and option surfaces, credit spreads and the
term structure of variance swaps are tests it has not yet been put to.

\section{Summary and Outlook}
\label{sec:conclusion}

Our objective in this paper was to develop a reasonably simple model of both the
longitudinal and the cross-sectional dynamics of the US equity market, its booms,
busts and cycle included, using only an autonomous system of stochastic
differential equations with no external factors beyond its noises. Our BBM model is
bipartite, with a single rational investor and the capital market forming a
reflexive loop. The market is heterogeneous, and made of $500$ names in two size
groups and three industry groups. Its sixty-six fitted constants are calibrated to $146$ ranges
of the S\&P 500 over 1996--2024, $119$ of which meet, and the calibrated model meets
the market on most of $141$ further comparisons it was never fitted to
(Section~\ref{sec:inelasticdisc}).

Once reflexivity is written as a dynamics with parameters, the calibration determines
how much of it the data support. Fallibility holds: the investor's belief extrapolates
the past and anti-predicts the next month, yet prices return to value through the
reversion in the price equation. The coupling of the price with the correlation
geometry is weak (Section~\ref{sec:twooscillators}), and the boom-bust sequence comes out in part, its asymmetry reproduced
and its timing set by the noise in the price rather than by a misconception tested
and failing. The reflexive loop adds little to the boom-bust statistics, which come
from the noise, the variance and the fundamental's jumps and survive a contrarian
or a random investor and the removal of its flow or of its effect on prices. What
the loop changes is how far the market sits from equilibrium, and the failure of
firms (Section~\ref{sec:reflexverdict}).

The market dynamics are noisy self-excitable, with the cycle driven largely by
noise, and without any oscillatory or chaotic dynamics that would underlie it. The
market responds to a shock only past a threshold and only on one side, with an
excursion of its own size. A firm is excitable in its structure: its mispricing and
demand memory move down the gradient of one potential, a stable node sits beside a
saddle as in a type I neuron, a name recovers for a few months after its own fall,
and a default is a noise-driven escape past the saddle. The index never comes near
its saddle and is excitable in behavior only. Its busts cluster through the
variance, and the membership rule rather than the dynamics returns it to rest after
a failure (Sections~\ref{sec:excitability} and~\ref{sec:neurons}).

The stationary state is not an equilibrium but a non-equilibrium steady state. A
probability current circulates between mispricing, demand and volatility in the
direction of the cycle, the entropy production from the model's own drift is four to five
times larger in busts and recoveries than in the calm, and the feedback multiplies the entropy
production of the simulated BBM market in the linear approximation about sixfold (Section~\ref{sec:nesssec}). In the geometry of the distance matrix the
simulated market reproduces the crisis values of the S\&P 500 and, like it, never
relaxes (Section~\ref{sec:boombust}). These observables give Soros's claim of a
market away from equilibrium a measurable form, although a statistical
non-equilibrium does not exclude an economic equilibrium, and a simulator cannot
identify the real market's causal mechanism.

Among the $141$ market observables against which the BBM is checked without being
calibrated to them, a few appear especially interesting from the point of
comparison with the literature. The market stays inside Black's factor of two of
value $99.7$ percent of the time, against the ninety percent or more Black
proposed. The names the investor buys most underperform those it buys least by
about six percent a year, the dumb-money effect of Frazzini and Lamont with its
sign and its three-year size, though not its timing. In every crisis the
correlation geometry of the market (participation ratio, market-mode share,
commutator norm and delocalized exponent) falls within the S\&P 500's range. The
BBM is also broadly consistent with the roughness of the market's log volatility, a
Hurst exponent of $0.13$ against $0.14$ on the index, and with the memory of its
magnitudes, at least in finite-sample measurements, without resorting to any exotic
stochastic process. Simple stochastic-volatility controls are consistent with them
too (Appendix~\ref{sec:comparison}).

The most interesting practical implication of the BBM is an unlimited supply of
simulated market data over long horizons. Long-horizon allocation is otherwise
judged against one twentieth century and a handful of crises, while a model that
keeps the market's moments, correlation structure and cadence of busts can supply
as many independent histories as are wanted, for tasks such as optimizing and
stress-testing portfolios. The claim
is limited to what was tested. On the first window's inputs the statistics hold
steady for a century and a half once the market has settled, while on the second
window's the mispricing and the share of the ten largest names still drift slowly.
The model's factor returns, a momentum premium of the wrong sign and a value return
earned only on the exact mispricing, limit what it can say about allocation across
factors. The model can be extended in several directions, for example by adding a
second investor with a different rule, or explicit predictive signals that the
investor trades on. The following three directions along the lines suggested in this
paper could also be interesting. First, beyond the cross-window checks made here, a
specification frozen on one period could be scored on later data and other markets
by its forecasts of volatility, drawdowns and portfolio outcomes, with the
calibration's parameter uncertainty carried into the predictions. Second, the latent demand that transmits the feedback could be tied to observed fund flows, so that the
feedback can be identified in the data. Third, the excess volatility of volatility
and multifractal width could be corrected together with the weak leverage, checking
that one change improves several of them without spoiling the fit.

\appendix

\section{The Belief Function Against the Actual Market}
\label{sec:beliefapp}

\begin{figure}[htbp]
\centering
\includegraphics[width=0.78\textwidth]{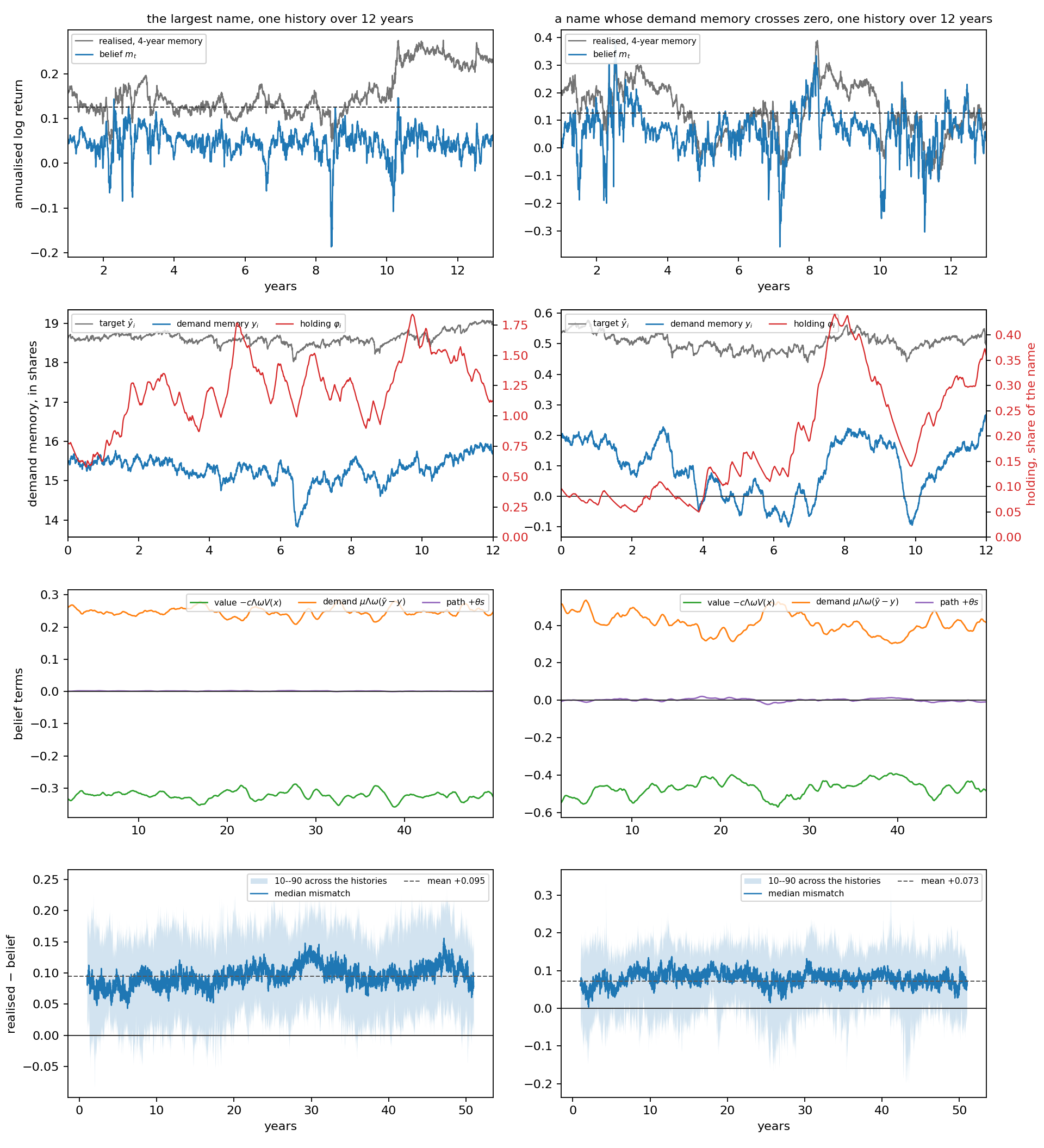}
\caption{What the investor expects against what the market delivered, for the
largest name (left) and a name whose demand memory crosses zero (right): the belief
against the smoothed realized return over one history, the demand memory and
holding, the three terms of the belief, and the mismatch with its band across
histories.}
\label{fig:perceived}
\end{figure}

The reading of Section~\ref{sec:reflexverdict} rests on paths of the belief
function set against the smoothed returns the same model produces, drawn for single
names, since averaging across names removes the idiosyncratic variance the belief
was formed about (Figure~\ref{fig:perceived}), and correlated history by history in
Table~\ref{tab:belief}. Against the month just past the belief correlates at
$+0.53$ for the largest name and $+0.71$ for the median one, and against the month
that follows at $-0.14$ and $-0.10$. Its squared correlation stays between $0.9$ and
$2.8$ percent at every horizon, and as a forecast it does worse than the
unconditional mean, mostly because it is a required return of order $0.10$ rather
than an expectation of the realized one. Realized returns smoothed over one to three
months match it best, and at the belief's own long memory the correlation falls to
$0.15$ and $0.32$. What is left over is a level rather than a failure of tracking,
the belief being a required return and the realized side what the name delivered.

\begin{table}[htbp]
\centering
\small
\begin{tabular}{l rr c rr}
\toprule
& \multicolumn{2}{c}{the largest name} & & \multicolumn{2}{c}{a median name} \\
\cmidrule(lr){2-3}\cmidrule(lr){5-6}
& ahead & behind & & ahead & behind \\
\midrule
\multicolumn{6}{l}{\emph{non-overlapping windows}} \\
\quad 1 month & $-0.137$\,\tiny$\pm$0.004 & $+0.527$\,\tiny$\pm$0.007 & & $-0.098$\,\tiny$\pm$0.003 & $+0.706$\,\tiny$\pm$0.003 \\
\quad 3 months & $-0.052$\,\tiny$\pm$0.007 & $+0.482$\,\tiny$\pm$0.009 & & $-0.119$\,\tiny$\pm$0.007 & $+0.713$\,\tiny$\pm$0.005 \\
\quad 6 months & $-0.009$\,\tiny$\pm$0.009 & $+0.326$\,\tiny$\pm$0.011 & & $-0.082$\,\tiny$\pm$0.010 & $+0.577$\,\tiny$\pm$0.008 \\
\quad 1 year & $+0.005$\,\tiny$\pm$0.013 & $+0.239$\,\tiny$\pm$0.014 & & $-0.064$\,\tiny$\pm$0.014 & $+0.413$\,\tiny$\pm$0.014 \\
\addlinespace
\multicolumn{6}{l}{\emph{exponential memory, same day}} \\
\quad previous day & \multicolumn{2}{c}{$+0.153$\,\tiny$\pm$0.004} & & \multicolumn{2}{c}{$+0.199$\,\tiny$\pm$0.003} \\
\quad 1 month & \multicolumn{2}{c}{$+0.541$\,\tiny$\pm$0.014} & & \multicolumn{2}{c}{$+0.767$\,\tiny$\pm$0.008} \\
\quad 3 months & \multicolumn{2}{c}{$+0.531$\,\tiny$\pm$0.017} & & \multicolumn{2}{c}{$+0.796$\,\tiny$\pm$0.007} \\
\quad 6 months & \multicolumn{2}{c}{$+0.455$\,\tiny$\pm$0.019} & & \multicolumn{2}{c}{$+0.712$\,\tiny$\pm$0.009} \\
\quad 1 year & \multicolumn{2}{c}{$+0.369$\,\tiny$\pm$0.021} & & \multicolumn{2}{c}{$+0.604$\,\tiny$\pm$0.010} \\
\quad 2 years & \multicolumn{2}{c}{$+0.289$\,\tiny$\pm$0.019} & & \multicolumn{2}{c}{$+0.502$\,\tiny$\pm$0.011} \\
\quad 4 years & \multicolumn{2}{c}{$+0.235$\,\tiny$\pm$0.018} & & \multicolumn{2}{c}{$+0.414$\,\tiny$\pm$0.015} \\
\quad 8 years & \multicolumn{2}{c}{$+0.183$\,\tiny$\pm$0.024} & & \multicolumn{2}{c}{$+0.363$\,\tiny$\pm$0.015} \\
\quad $1/\kappa$ = 16.5 years & \multicolumn{2}{c}{$+0.146$\,\tiny$\pm$0.022} & & \multicolumn{2}{c}{$+0.315$\,\tiny$\pm$0.019} \\
\bottomrule
\end{tabular}
\caption{Correlation of the belief with the realized return, for the largest name and a median one. Above, on non-overlapping windows of the horizon named, \emph{ahead} taking the period that follows the belief and \emph{behind} the period before it, over 120 histories. Below, against the return smoothed by an exponential memory of the horizon named and ending at the same date, over 30 histories. Each entry is the mean across histories and its standard error.}
\label{tab:belief}
\end{table}

\let\savedFloatBarrier\FloatBarrier\let\FloatBarrier\relax
\section{Comparison to Other Models}
\label{sec:comparison}

The BBM's cycle is driven by the noise in the price and the reversion of the
mispricing, with no regimes, time-varying parameters or fitted kernels
(Section~\ref{sec:relatedmodels}). We compare it with two classes of models in the
literature. The first are statistical models that fit a return series and claim no
cycle. Against a GARCH with Student $t$ shocks, its GJR leverage extension with and
without an autoregressive mean, and a normal inverse Gaussian law, the BBM is the
closest on the kurtosis across horizons, second on the variance ratio and third on
the skew (Section~\ref{sec:classone}). The second are agent-based models that
generate booms and busts from inside. They share with the BBM a trend and a value
term and fluctuations about value driven by noise, but none resolves the market into stocks, includes a budget of money or produces skewed returns, and on the trend
effect the BBM shows a reversal where the extended Chiarella model finds a bend
(Section~\ref{sec:classtwo}).

\subsection{Models of the return distribution}
\label{sec:classone}

Table~\ref{tab:horizonbench} gives the skew, excess kurtosis and variance ratio at
five horizons. The BBM's range contains the market's skew and kurtosis at every
horizon except the skew at a year, and misses the variance ratio, above the
market's at a week and below it at a year (Section~\ref{sec:sensitivity}). Among
the alternatives only the leverage GARCH produces skew across horizons, a little more
closely than the BBM. On the kurtosis the BBM sits between the normal inverse
Gaussian law, whose tails thin too fast with the horizon, and the GARCH variants,
whose tails thin too slowly. On the variance ratio no model succeeds, the AR-GJR
coming closest. The BBM's baskets have variance ratios in their objective while
the alternatives were fitted to the daily series alone, so the comparison is not
quite even, and a reader who wants only a return distribution should use a GARCH.
GARCH and L\'evy models can also produce drawdowns of the size the cycle is scored
on, but neither has a state that is over-extended and then corrected. Among the
stylized facts, the leverage effect and the coarse-fine asymmetry imply a direction of time, which the entropy production of Section~\ref{sec:nesssec} measures
directly.

\begin{table}[htbp]
\centering
\caption{Moments across horizons against models estimated on the same index and
simulated forward, each the median over windows of twenty-nine years. GJR adds a
leverage term to the GARCH and AR-GJR an autoregressive mean as well.}
\label{tab:horizonbench}
\vspace{0.4em}
\small
\setlength{\tabcolsep}{3.6pt}
\begin{tabular}{lrrrrrr}
\toprule
Horizon & S\&P 500 & BBM & GARCH-$t$ & GJR & AR-GJR & NIG \\
\midrule
\multicolumn{7}{l}{\emph{Skew}} \\
A day & $-0.39$ & $-0.25$ & $0.00$ & $0.14$ & $0.14$ & $-0.49$ \\
A week & $-0.91$ & $-0.53$ & $0.00$ & $-0.38$ & $-0.39$ & $-0.21$ \\
A month & $-1.36$ & $-0.56$ & $0.06$ & $-0.90$ & $-0.82$ & $-0.08$ \\
A quarter & $-0.68$ & $-0.44$ & $-0.01$ & $-1.11$ & $-1.04$ & $-0.05$ \\
A year & $-1.18$ & $-0.18$ & $0.03$ & $-0.68$ & $-0.65$ & $-0.01$ \\
\midrule
\multicolumn{7}{l}{\emph{Excess kurtosis}} \\
A day & $9.77$ & $9.40$ & $15.31$ & $11.56$ & $12.06$ & $6.53$ \\
A week & $6.64$ & $6.18$ & $10.92$ & $8.24$ & $7.40$ & $1.26$ \\
A month & $6.18$ & $3.80$ & $7.65$ & $5.37$ & $5.38$ & $0.21$ \\
A quarter & $0.16$ & $1.81$ & $4.31$ & $3.66$ & $3.29$ & $-0.01$ \\
A year & $0.76$ & $0.19$ & $0.91$ & $0.55$ & $0.48$ & $-0.31$ \\
\midrule
\multicolumn{7}{l}{\emph{Variance ratio}} \\
A week & $0.845$ & $0.962$ & $0.997$ & $0.993$ & $0.949$ & $1.001$ \\
A month & $0.755$ & $0.846$ & $0.973$ & $0.965$ & $0.917$ & $0.989$ \\
A quarter & $0.672$ & $0.677$ & $0.935$ & $0.898$ & $0.873$ & $0.986$ \\
A year & $0.792$ & $0.416$ & $0.874$ & $0.813$ & $0.790$ & $0.953$ \\
\midrule
\multicolumn{7}{l}{\emph{Summed absolute error against the market}} \\
Skew & --- & $2.56$ & $4.59$ & $\mathbf{2.46}$ & $2.49$ & $3.88$ \\
Excess kurtosis & --- & $\mathbf{5.4}$ & $15.6$ & $7.9$ & $7.3$ & $15.8$ \\
Variance ratio & --- & $0.588$ & $0.715$ & $0.605$ & $\mathbf{0.468}$ & $0.864$ \\
\bottomrule
\end{tabular}
\end{table}

\subsection{Models that generate a cycle}
\label{sec:classtwo}

The closest relative in this class is the extended Chiarella model of Majewski,
Ciliberti and Bouchaud \cite{majewski2020}, which takes the trend-following and
value-investing demands of Chiarella \cite{chiarella1992}, adds noise traders, a
drifting fundamental and a cubic fundamentalist demand, and is estimated by Kalman
filtering on monthly prices across four asset classes back to 1800.
Table~\ref{tab:newelements} of Section~\ref{sec:reflexverdict} sets the two models
side by side. Like the BBM's, its noise-free skeleton is stable for every asset
they fit, so its fluctuations come from the noise. The trend terms are close
relatives with very different memories. Ours averages the moves of the mispricing
over sixteen years, and integrating Eq.~\eqref{eq:memdef} by parts gives
$s_t = \kappa\,(x_t - \bar x_t)$, with $\bar x_t$ the exponential average of the
mispricing at the same rate. On the timescales of the cycle the trend term
therefore responds to the level of the mispricing and pushes the investor's demand away
from value, the part the negative linear coefficient plays in the extended model.
The two models differ in the law of the mispricing. Majewski and co-authors find a
second mode for the US and Canadian indices, though not on average across their
assets, and Schmitt and Westerhoff \cite{schmitt2017} report a bimodal law for the
S\&P 500, while the BBM's market-wide mispricing has a single peak in every market.

\begin{figure}[htbp]
\centering
\includegraphics[width=0.62\textwidth]{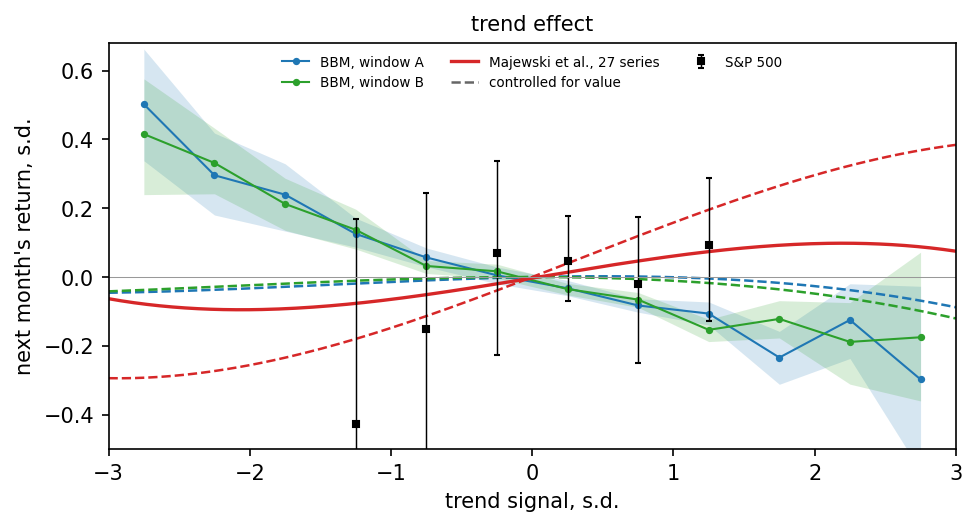}
\caption{Next month's index return against the trend signal of
Majewski and co-authors \cite{majewski2020}, both in standard deviations, raw
(solid) and controlled for the mispricing (dashed). Bands are 95 percent across
sixteen markets, and the S\&P bars a block bootstrap.}
\label{fig:trendeffect}
\end{figure}

Their central target is the response of the next month's return to the trend
signal, positive for moderate signals and turning back for large ones. We measure
it on the BBM's capitalization-weighted index with their signal and regressions
over sixteen markets of a hundred and fifty years, and their own calibrated US
model, put through the same estimator, returns their shape. On the BBM the response
is a reversal instead (Figure~\ref{fig:trendeffect}), with slopes of $-0.12$ and
$-0.11$ in the two windows against their $+0.068$. Controlled for the index's
mispricing the BBM's slope vanishes, while the value coefficient, $0.27$ and
$0.30$, sits beside their $0.25$. The BBM's trend effect is therefore its value
effect read through the trend signal. The S\&P's twenty-nine years are too short to distinguish
the two pictures, its slope of $+0.03$ with a 95 percent band from
$-0.13$ to $+0.16$.

Of the other models of this class, the structural stochastic volatility of Franke
and Westerhoff \cite{franke2012} produces switching between calm and turbulent
regimes from the shifting shares of fundamentalists and chartists, the herding
models of Alfarano, Lux and Wagner \cite{alfarano2008} and Lux and Marchesi
\cite{luxmarchesi1999} reproduce the stylized facts of returns \cite{cont2001}, and the adaptive
belief systems of Brock and Hommes \cite{brock1998} obtain cycles from a
deterministic map that period-doubles into chaos. None was built for a skewed
return distribution. Franke and Westerhoff exclude skewness from the nine moments
they match, the law of Alfarano and co-authors is symmetric, and both shocks of the
extended Chiarella model are Gaussian, while the BBM's asymmetry at every horizon
falls out of the same construction as its cycle. On value itself, Bouchaud and
co-authors do without a fundamental \cite{bouchaud2017}, building a proxy from a
trend fitted over the previous twenty years. Their anchor is estimated from the
price it judges, while ours fixes $\gamma$ in advance and never confronts its law
for value with a measurement of value. The two bracket the result, and Black's
factor of two survives both.

\subsection{The shape of returns, tails and memory}
\label{sec:returnshape}

\begin{figure}[htbp]
\centering
\includegraphics[width=0.88\textwidth]{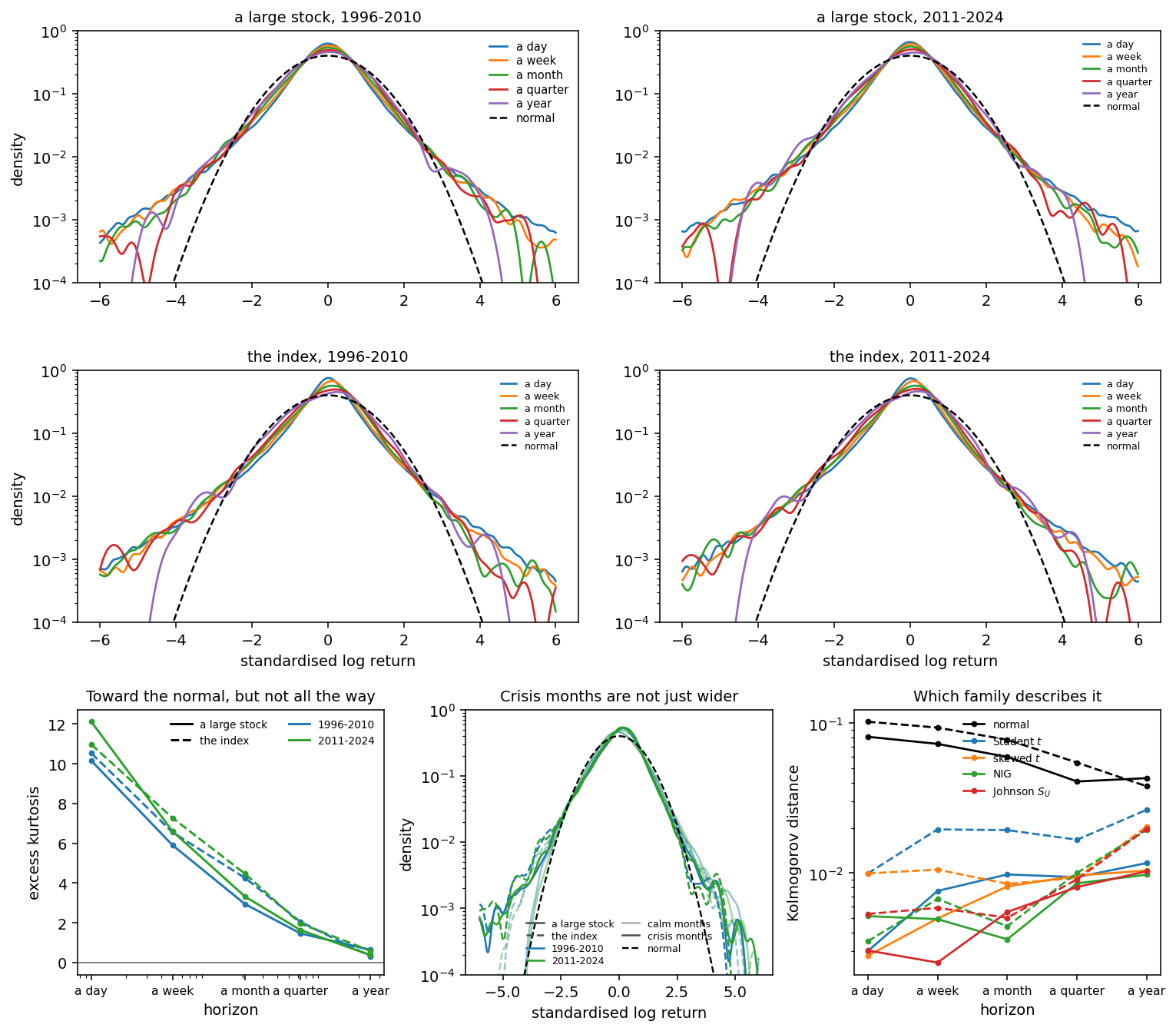}
\caption{The distribution of log returns for a large stock and for the index under
each parameter set: standardized densities at five horizons against the normal
(above), and the excess kurtosis by horizon and the fit of each family (below).}
\label{fig:retdist}
\end{figure}

Of the families the literature fits to returns (Figure~\ref{fig:retdist}), the
normal inverse Gaussian and Johnson's $S_U$ describe the model's index best, each
narrowly ahead at some horizons, and the normal is rejected everywhere. Around each
$20\%$ fall (the phase rows of Table~\ref{tab:cyclesummary}) the index's volatility
follows the market's through the boom to its peak at the top, but stays higher
through the bust, where the market's has already eased. The market's skew is
negative in every phase and deepest at the top. The model's sits near zero and
meets the market's in every phase but the top, so the model does not concentrate
the asymmetry at the top. Through the bust its kurtosis is higher, so the market
delivers a crash as an extended turbulent stretch and the model as a few large days
on a quieter background. The return distributions are unimodal in every phase.

\begin{figure}[htbp]
\centering
\includegraphics[width=\textwidth]{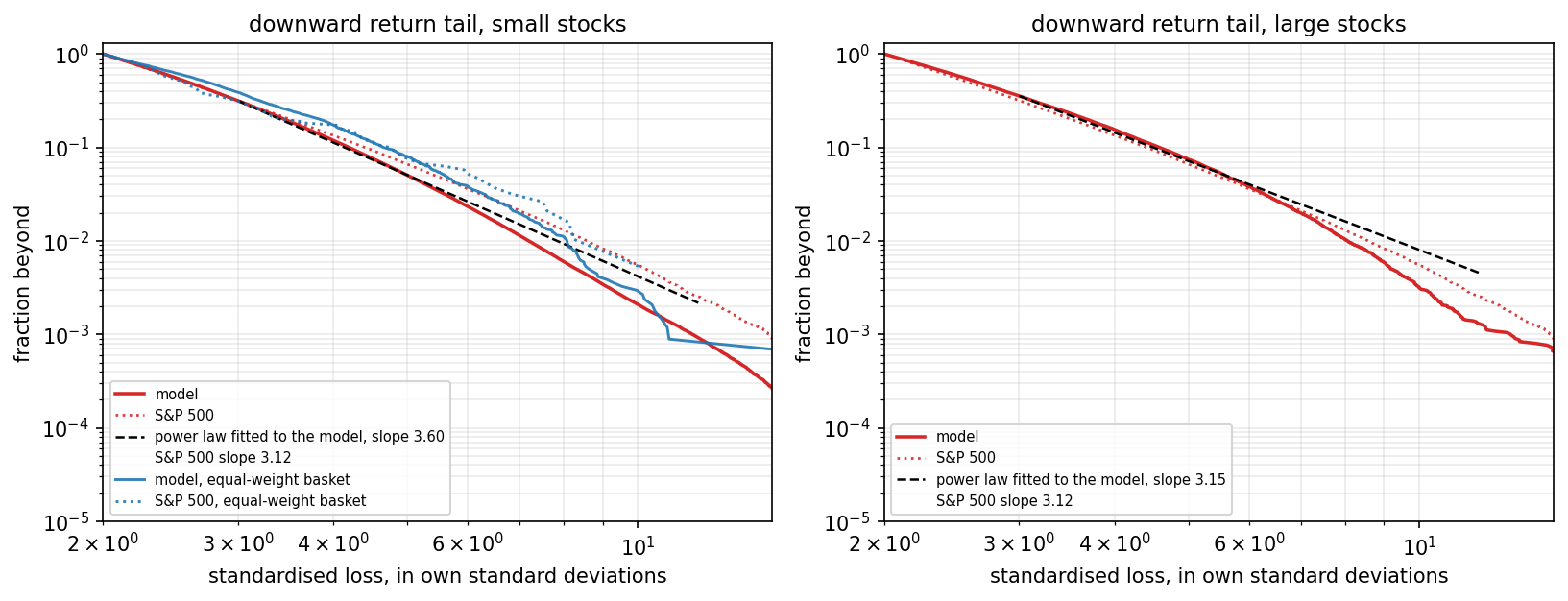}
\caption{The downward tail of standardized daily returns, model against market,
for the rest of the index and the ten largest, with an equal-weighted basket in the
first panel and power laws fitted beyond three standard deviations dashed.}
\label{fig:rettail}
\end{figure}

No tail statistic is among the targets. Standardized by its own volatility and fitted beyond three standard
deviations \cite{alstott2014}, the market's tail follows the inverse-cubic law
\cite{gopikrishnan1998}, and the model's sits at or above its exponent, the ten
largest close to it and the rest of the index thinner (Figure~\ref{fig:rettail}). Pooled over names, the model's frequency is within ten
percent of the market's out to four standard deviations and under half of it at ten, so
the deepest single days are too rare. On an equal-weighted basket of names the comparison reverses, the model's
tail matching the market's within its sampling error out to six standard
deviations. On both sides a lognormal is preferred to a power law, so the tail is
fat, and over the tested range not scale-free, a result of the comparison rather than
of the construction alone.

Models that fit the persistence of $|r|$ build it in, by fractional integration
\cite{granger1980}, the heterogeneous autoregressive cascade \cite{corsi2009}, the
Markov-switching multifractal \cite{calvet2004}, rough volatility
\cite{gatheral2018} or a generalized Langevin equation that separates scaling from memory \cite{itkin2026gle}. The BBM has no fractional driver and its state
is Markovian. Its volatility factors are exponentials of Ornstein-Uhlenbeck
processes, and its persistence comes from a spread of relaxation rates set by other
targets, from the six trading days of the transitory term to the ten years of the
growth reversion. In kernel language this is a sum of exponentials with no period
among them, which mimics a power law across the band its rates cover, a choice the latent-liquidity model of \cite{itkin2026impact} also makes. Beyond the slowest of
them the covariance decays exponentially, a plateau in the spectrum at the lowest
frequencies. Conversely, a sum of exponentials is a finite-dimensional Markovian realization
of a generalized Langevin equation (GLE). A GLE whose kernel is a finite sum of exponentials
has an exact Markovian state built from auxiliary variables \cite{itkin2026glelsv},
and for the original Marketron \cite{marketron2025} Itkin \cite{itkin2026lean}
shows that the log-price obeys a GLE with a closed-form, state-modulated memory
kernel, of which the demand memory $y_t$ is the exact Markovian lift. Holding memory
in auxiliary states therefore does not by itself set the BBM apart from GLE
formulations. The differences lie in the economic variables, the feedback and the
noise. In the BBM the covariance of the market factor's
volatility follows a power law of exponent $0.39$ to within five percent from ten
to about a hundred and forty days, and steepens beyond (Supplement, Section~S3). The I-BBM meets the market on the scaling estimators
below, while the autocorrelation of its $|r|$ is $0.055$ at a quarter and $0.009$
at half a year, against the market's $0.109$ and $0.067$, both above the model's
band. The flattening is not a defect to be repaired by adding memory but the
signature of a system with a characteristic time, which
Section~\ref{sec:excitability} tests directly.

In the market returns have no memory, their signs little, and their magnitudes a
great deal. Table~\ref{tab:longmemory} measures all
three with three estimators that fail differently, the detrended fluctuation
exponent $\alpha$ \cite{peng1994}, which does not read a drift as memory, the
log-periodogram order $d$ of Geweke and Porter-Hudak \cite{gph1983}, and the decay
exponent of the autocorrelation, every band the tenth to ninetieth percentile at the
data's length. Both BBM constructions reproduce the contrast, meeting the market on
the detrended fluctuation and the fractional order of the magnitudes, the market's
own order sitting in the range reported since Ding, Granger and Engle
\cite{dinggrangerengle1993}. Their one miss is a mild anti-persistence of the
aggregated market's returns. What the table cannot do is separate the
constructions: a GARCH$(1,1)$-$t$ meets it on all three estimators while its
autocorrelation decays geometrically by construction, so twenty-nine years cannot
tell genuine long memory from a short-memory process fitted across shifts in the
level of its variance.

\begin{table}[htbp]
\centering
\small
\setlength{\tabcolsep}{4pt}
\begin{tabular}{l ccc}
\toprule
& DFA $\alpha$ & GPH $d$ & ACF $\gamma$ \\
\midrule
\multicolumn{4}{l}{\emph{Returns $r_t$}} \\
\quad S\&P 500, 1996--2024 & $0.470$\,\tiny[0.435,0.504] & $-0.016$\,\tiny[-0.125,0.063] & --- \\
\quad BBM, aggregated market & $0.453$\,\tiny[0.422,0.494] & $-0.263$\,\tiny[-0.364,-0.168]\,$\bullet$ & --- \\
\quad BBM, simulated market & $0.436$\,\tiny[0.407,0.466] & $-0.166$\,\tiny[-0.288,-0.081] & --- \\
\quad GARCH$(1,1)$-$t$, fitted to the market & $0.486$\,\tiny[0.433,0.561] & $-0.009$\,\tiny[-0.116,0.158] & --- \\
\addlinespace
\multicolumn{4}{l}{\emph{Signs $\mathrm{sgn}\,r_t$}} \\
\quad S\&P 500, 1996--2024 & $0.519$\,\tiny[0.487,0.556] & $0.144$\,\tiny[0.042,0.220] & --- \\
\quad BBM, aggregated market & $0.507$\,\tiny[0.482,0.530] & $-0.017$\,\tiny[-0.118,0.075] & --- \\
\quad BBM, simulated market & $0.510$\,\tiny[0.482,0.547] & $0.041$\,\tiny[-0.069,0.183] & --- \\
\quad GARCH$(1,1)$-$t$, fitted to the market & $0.515$\,\tiny[0.485,0.543] & $0.028$\,\tiny[-0.062,0.102] & --- \\
\addlinespace
\multicolumn{4}{l}{\emph{Magnitudes $|r_t|$}} \\
\quad S\&P 500, 1996--2024 & $0.944$\,\tiny[0.879,0.988] & $0.422$\,\tiny[0.311,0.520] & $0.872$\,\tiny[0.667,1.258] \\
\quad BBM, aggregated market & $0.913$\,\tiny[0.872,0.952] & $0.305$\,\tiny[0.224,0.440] & $1.358$\,\tiny[0.874,1.884] \\
\quad BBM, simulated market & $0.869$\,\tiny[0.827,0.916] & $0.409$\,\tiny[0.285,0.536] & $0.560$\,\tiny[0.350,1.094] \\
\quad GARCH$(1,1)$-$t$, fitted to the market & $0.972$\,\tiny[0.899,1.058] & $0.512$\,\tiny[0.390,0.723] & $1.115$\,\tiny[0.656,1.593] \\
\addlinespace
\midrule
\quad \emph{no memory} & $0.5$ & $0$ & --- \\
\quad \emph{long memory} & $>0.5$ & $>0$ & $<1$ \\
\bottomrule
\end{tabular}
\caption{Long memory in the three series, by three estimators that fail in different ways. DFA removes a local linear trend at every scale, so a drift does not read as persistence; GPH reads the log-periodogram near zero frequency; the last column fits the decay of the autocorrelation itself over 20 to 250 days, and is reported only for the magnitudes. They agree through $\alpha = d + 1/2$ and $\gamma = 1 - 2d$ when the memory is genuine; otherwise they are finite-band summaries rather than one law. A bullet marks a band that does not meet the market's. Brackets are the tenth to ninetieth percentile, of a block bootstrap for the market and of 60 independent runs the length of the sample for the models, a separate protocol from the sixteen-run scaling validation of the Supplement.}
\label{tab:longmemory}
\end{table}

The leverage effect is weaker than the market's. Read as the correlator of
Bouchaud, Matacz and Potters \cite{bouchaud2001lev},
$L(\tau) = \langle r_t r_{t+\tau}^{2}\rangle / \langle r^{2}\rangle^{2}$, the
simulated BBM market's index has the right sign over the first month at about a third of
the market's size and none over the second, and the I-BBM at about a fifth. The
simulated BBM market's autocorrelation of $|r|$ meets the market's at three and six
months. The asymmetry of its returns, negative at every horizon, is thus reached
with a short-lag leverage well below the market's.

\subsection{How rough is the volatility?}
\label{sec:roughness}

Two exponents go by the name Hurst, and the rough volatility literature reports
the one Table~\ref{tab:longmemory} does not: the scaling of the increments of the
log volatility over short horizons, $\mathbb{E}|\log\sigma_{t+\Delta} -
\log\sigma_t|^{q} \sim \Delta^{qH}$, with $H$ near $0.1$ the published value
\cite{gatheral2018}. Measured on this market, it reflects mostly the
estimator rather than what generated the series. An Ornstein-Uhlenbeck
control at the model's own reversion rate is a diffusion by construction and
still returns $0.31$ rather than $0.5$, and the model's latent log volatility
gives $0.32$. A five-day realized estimate of the same series falls to $0.13$,
and the market's index sits at $0.14$, the two bands, from the tenth to the
ninetieth percentile over eight-year sub-windows, running from $0.10$ to $0.17$
and from $0.10$ to $0.18$; banded instead over independent runs and a block
bootstrap of the market's series, the estimator gives $0.13$ [$0.11$, $0.15$] and
$0.15$ [$0.14$, $0.18$]. On single names the same estimate gives $0.127$ for the
ten largest against the market's $0.115$, and $0.154$ for the median names of the
rest against $0.103$. The roughness an observer reads is therefore made by
the realized estimate of a diffusive volatility, with no fractional driver anywhere
in the construction.

The same reading through one daily pipeline, on twenty-nine years of every subject
and over lag ranges fixed in advance, shows how little the exponent identifies. The
local slope of the second moment of a five-day estimate over five to twenty days is
$0.13$ for the S\&P, with a bootstrap band from $0.12$ to $0.16$, $0.10$ for the
BBM, and $0.11$ for a fitted GARCH(1,1)-$t$ and for a lognormal stochastic-volatility
control. Over one to five days every subject gives close to a half, and a twenty-day
estimate gives about $0.6$ on all of them. The low exponent is common to every
stochastic-volatility model tried, it is unchanged with the reflexive loop cut,
$c = 0$, and with the BBM's variance frozen the local slope falls to about zero
and the published estimator to $0.07$. The models differ
instead in the level of the increments. The second moment of the BBM index's log-volatility
increments is $40$ to $48$ percent larger than the market's at every lag, $0.091$
against $0.065$ at a day and $0.59$ against $0.40$ at twenty days, outside the
market's bands, the root-mean-square increments $18$ to $21$ percent larger, while the I-BBM and
both controls are generally closer, with misses at single horizons (Supplement,
Table~S4), so the volatility of
the index's volatility is too high (Supplement, Table~S3). Cont and Das \cite{contdas2024} show that realized estimates
of a diffusive spot volatility look rough, Fukasawa, Takabatake and Westphal
\cite{fukasawa2022} still find roughness after correcting for estimation error, and
Markovian multifactor models are used to approximate rough ones
\cite{abijaber2019}. Itkin \cite{itkin2026gle} argues that such estimators measure the
scaling of the proxy and not the memory of the volatility, which a generalized
Langevin kernel sets separately. The daily exercise here does not adjudicate between them.

\subsection{Multifractality test}
\label{sec:excitmf}

A series is multifractal when its small and large fluctuations scale differently
with the time scale. Multifractal detrended fluctuation analysis
\cite{kantelhardt2002} measures this through the $q$-th order fluctuation function
$F_q(s)$ of the integrated returns in windows of $s$ days, which scales as
$s^{h(q)}$. A monofractal has one generalized Hurst exponent, $h(q) = H$ at every
order, while for a multifractal $h(q)$ varies with $q$, the large fluctuations
($q > 0$) scaling differently from the small ones ($q < 0$). With
$\tau(q) = q\,h(q) - 1$, the Legendre transform $\alpha = \tau'(q)$,
$f(\alpha) = q\,\alpha - \tau(q)$ gives the singularity spectrum, and its width
$\Delta\alpha$ over $-5 \le q \le 5$ is the standard measure of multifractality,
zero for a monofractal.

Multifractality is a stylized fact of returns that no calibration here touched,
and the test checks whether the model reproduces it, and for the same reason. A broad spectrum has two sources the width cannot tell
apart, heavy tails and the order in which returns arrive, so the same spectrum is
also measured on surrogate series built to keep one source and remove the other.
A shuffled surrogate permutes the daily returns at random in time. It keeps every
value, and so the distribution and its tails, but destroys any dependence between
days. A phase-randomized surrogate takes the Fourier transform of the returns,
replaces the phases by random ones and transforms back. It keeps the power
spectrum, and so the linear autocorrelation, but brings the values close to Gaussian and
removes most nonlinear dependence, such as the clustering of large moves. An IAAFT
surrogate keeps the distribution exactly and the Fourier amplitudes to within $0.3$
percent.

\begin{figure}[htbp]
\centering
\includegraphics[width=0.92\textwidth]{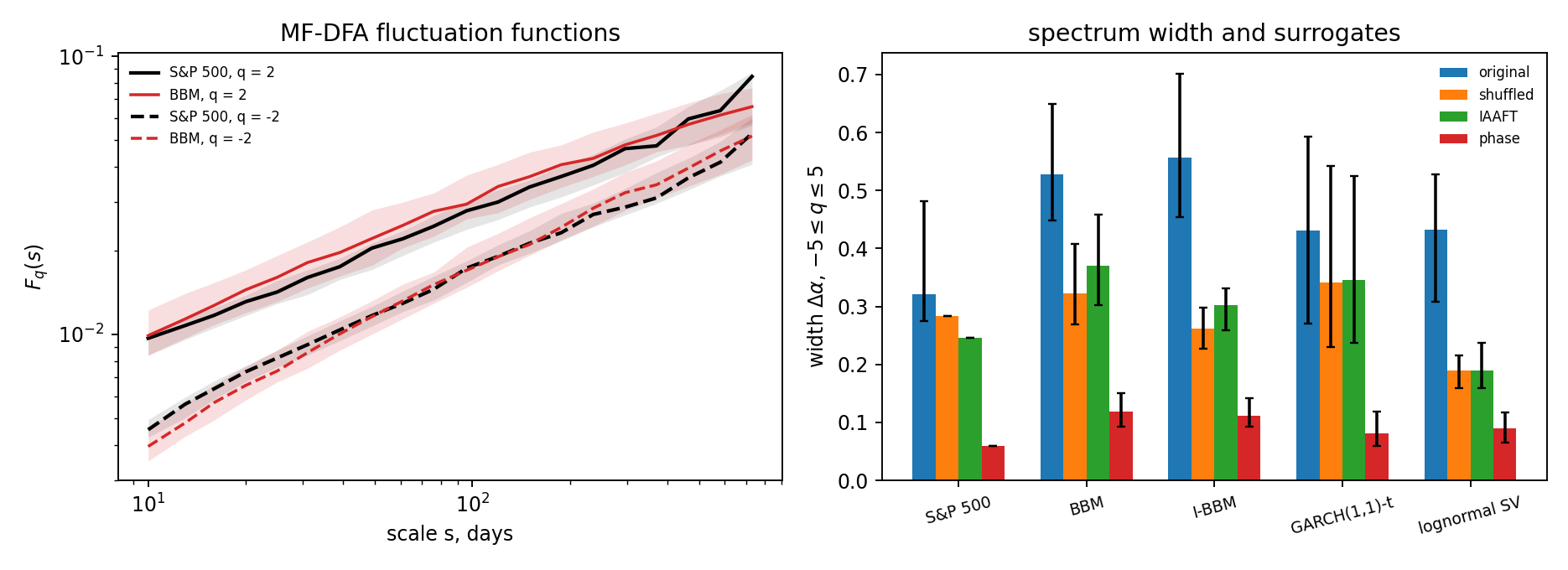}
\caption{Multifractality of the market and the models: MF-DFA fluctuation functions
at $q = \pm 2$ (left), and the spectrum width over $-5 \le q \le 5$ with its
shuffled, IAAFT and phase-randomized surrogates (right), bands from runs and from a
block bootstrap of the market.}
\label{fig:multifractal}
\end{figure}

Both the market and the BBM show apparent multifractal scaling over the tested
range, and the BBM too much of it (Figure~\ref{fig:multifractal}), on twenty-nine
years of each. Over $-5 \le q \le 5$ the market's spectrum is $0.32$ wide, with a
bootstrap band from $0.28$ to $0.48$, and the BBM's $0.53$, from $0.45$ to $0.65$.
Over $-2 \le q \le 2$, where rare extremes weigh less, the widths are $0.11$ and
$0.25$ and the bands are apart. The scales run from ten days to a tenth of the
sample with first-order detrending, and the excess survives second-order detrending
and either end of the range moved by a factor of two. A GARCH(1,1)-$t$ and a lognormal
stochastic-volatility control give $0.43$ each, between the two. Shuffling leaves
the market almost where it was, at $0.28$, a difference of $0.04$ whose paired
bootstrap band runs from $0.00$ to $0.21$, and an IAAFT surrogate at $0.25$, so most
of the market's estimated width survives the removal of temporal dependence. The
BBM shuffles to $0.32$, agreeing with the market on what the distribution
contributes, and loses its excess. That sensitivity to the order of the returns is
not specific to the BBM, since the lognormal control loses as much, and in the BBM it
is unchanged with $c = 0$ and is removed with the variance frozen, the width falling
to $0.18$ against $0.13$ for that run's own shuffled surrogate. It belongs to the stochastic variance rather than to the reflexive loop
or the excitable dynamics. The difference of widths is a descriptive contrast rather
than an additive decomposition. The protocol, the controls and the paired contrasts
are in the Supplement (Tables~S1, S2, S6 and~S7).

\section{Testing the Fit for Parameter Drift}
\label{sec:fittests}

This appendix gives the formal tests behind Section~\ref{sec:stationarity}, with four findings. The market's index moments differ between the two windows,
at a Wald statistic of $65$ on twenty-one moments, $p < 10^{-5}$. Hansen's statistic rejects the model's index as an exact description of either window at every shrinkage of the covariance. And followed through rolling five-year windows with the
parameters fixed, the model's error shows no drift in any moment of the shape of
returns once the test is sized on the model's own stationary runs, the one
surviving sign of drift being a rise in the market's mean return against the
model's.

The two tests on all twenty-one moments keep the covariance between them, since
several are near duplicates, four variance ratios being four readings of one curve
\cite{mcfadden1989,duffie1993}. Hansen's statistic \cite{hansen1982} is formed with
the bootstrap covariance shrunk toward its diagonal, at each window's fine-tuned
vector (Table~\ref{tab:stationarity}). No parameter was fitted to these moments
directly, but neither the vector's uncertainty nor the shrinkage enters the
reference distribution, so the chi-squared comparison is nominal and measures how
closely the index matches the market rather than stationarity. It rejects both
windows at every shrinkage. The structural-change test of Andrews and Fair
\cite{andrewsfair1988,hallsen1999} is not available, since no vector was fitted to
the index moments, and Section~\ref{sec:stationarity} addresses that question with the
fitted observables instead. The Wald test of the difference between the windows uses the market alone and gives $49$ to $65$ across shrinkages, $p < 10^{-3}$.

The variance ratios of the whole sample dip and rise again, from $0.755$ at a month
to $0.672$ at a quarter and $0.792$ at a year, a shape no single relaxation rate
gives. Cut at the midpoint, the ratio at a year sits above the ratio at a quarter in
the first half and below it in the second, the change at half a year being $-0.398$
with a 95 percent interval of $[-0.647, -0.014]$. The model's index, fitted to no
variance ratio, falls monotonically (Table~\ref{tab:index}), the shape of the second
period.

\begin{figure}[htbp]
\centering
\includegraphics[width=\textwidth]{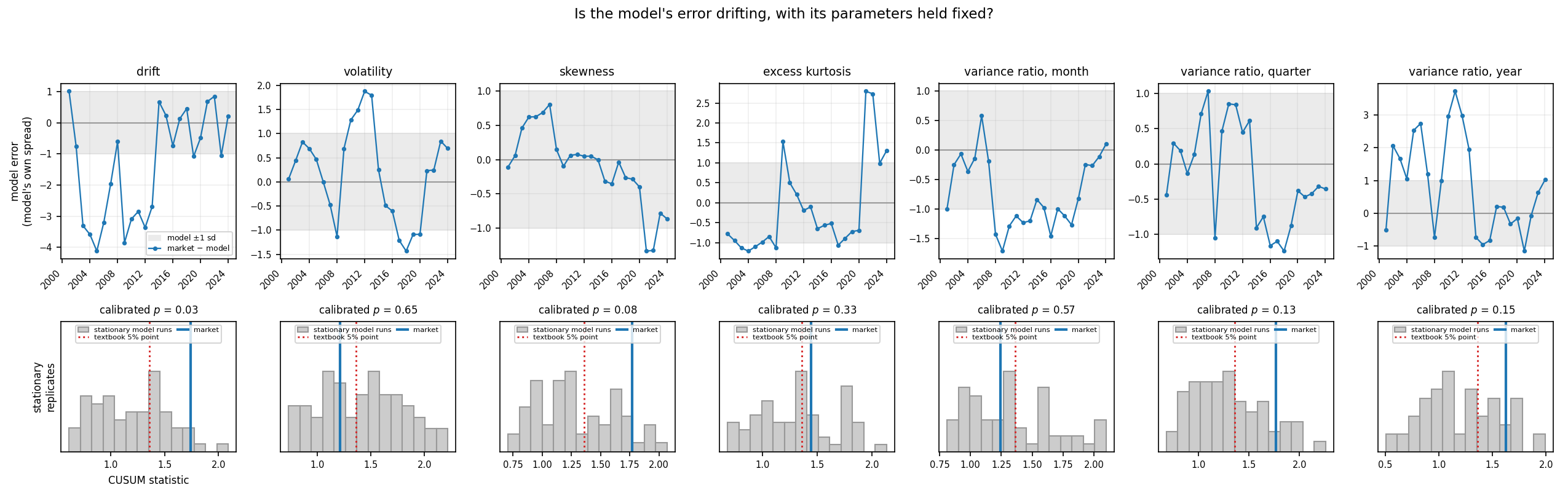}
\caption{Is the model's error drifting at fixed parameters? The error in each
moment across rolling five-year windows, in units of the model's spread (above),
and its cumulative-sum statistic against the distribution it takes on stationary
runs (below).}
\label{fig:roll}
\end{figure}

The rolling test follows the error in each of seven moments across rolling five-year
windows at one fixed vector, testing it for a trend and by a cumulative-sum
statistic \cite{brown1975} (Figure~\ref{fig:roll}). The textbook distribution does
not apply, the windows overlapping by four years and a five-year skewness or
kurtosis moving in steps as single extreme days enter and leave: on sixty
stationary runs of the calibrated market its five percent point rejects a third to
a half of the time. Against that simulated null only the drift's statistic falls
below $0.05$, at $p = 0.033$, which is one of seven tests and would not survive a
correction for multiplicity, so we treat it as exploratory; and the drift is also the one moment with a trend, the market's
five-year return rising against the model's. With so little power the test's
silence on the shape of returns is weak evidence, and its one finding is in the
level of returns.

\begin{table}[htbp]
\centering
\small
\begin{tabular}{lrrrr}
\toprule
\multicolumn{5}{l}{\textit{Can one vector match all the moments at once? Hansen's $J$, by shrinkage of the covariance}} \\
shrinkage & \multicolumn{2}{c}{1996--2010} & \multicolumn{2}{c}{2011--2024} \\
 & $J$ & $p$ & $J$ & $p$ \\
\midrule
0.0 & 140 & $<10^{-3}$ & 163 & $<10^{-3}$ \\
0.2 & 54 & $<10^{-3}$ & 76 & $<10^{-3}$ \\
0.4 & 44 & 0.002 & 57 & $<10^{-3}$ \\
0.6 & 40 & 0.007 & 47 & 0.001 \\
0.8 & 41 & 0.006 & 41 & 0.006 \\
\addlinespace
\midrule
\multicolumn{5}{l}{\textit{Where does the error drift? Rolling five-year windows, parameters fixed}} \\
moment & CUSUM & nominal $p$ & calibrated $p$ & \\
\midrule
annual drift & 1.74 & 0.005 & 0.033 & \\
annual volatility & 1.21 & 0.105 & 0.650 & \\
skewness, daily & 1.77 & 0.004 & 0.083 & \\
excess kurtosis & 1.44 & 0.031 & 0.333 & \\
variance ratio, month & 1.24 & 0.092 & 0.567 & \\
variance ratio, quarter & 1.76 & 0.004 & 0.133 & \\
variance ratio, year & 1.63 & 0.010 & 0.150 & \\
\bottomrule
\end{tabular}
\caption{Two tests of the fit to the index: Hansen's $J$ on the twenty-one index
moments, compared with $\chi^2$ only nominally (above), and the rolling CUSUM test of
each moment at the textbook and at the simulated stationary distribution (below).}
\label{tab:stationarity}
\end{table}

\end{document}